\documentstyle[uabib,amsmath,supertabular,rotating,epsfig,portland,float,feyn]{uathes}

\input{psfig}

\newcommand{\simlt}{\lower.5ex\hbox{$\; \buildrel < \over \sim \;$}}
\newcommand{\simgt}{\lower.5ex\hbox{$\; \buildrel > \over \sim \;$}}

\newcommand{\be}{\begin{equation}}
\newcommand{\ee}{\end{equation}}

\def\AuthorName{Giorgio Torrieri}
\def\GradYear{2004}
\author{\AuthorName\ }
\title{}
\def\ThesisType{dissertation}
\thesistitle{Statistical hadronization phenomenology in heavy ion collisions at SPS and RHIC energies}
\degreetitle{DOCTOR OF PHILOSOPHY} 
\department{DEPARTMENT OF PHYSICS} 
\copyrightyear{2004}

\begin{document}

\maketitle


\begin{figure}[h]
\centering
  \psfig{figure=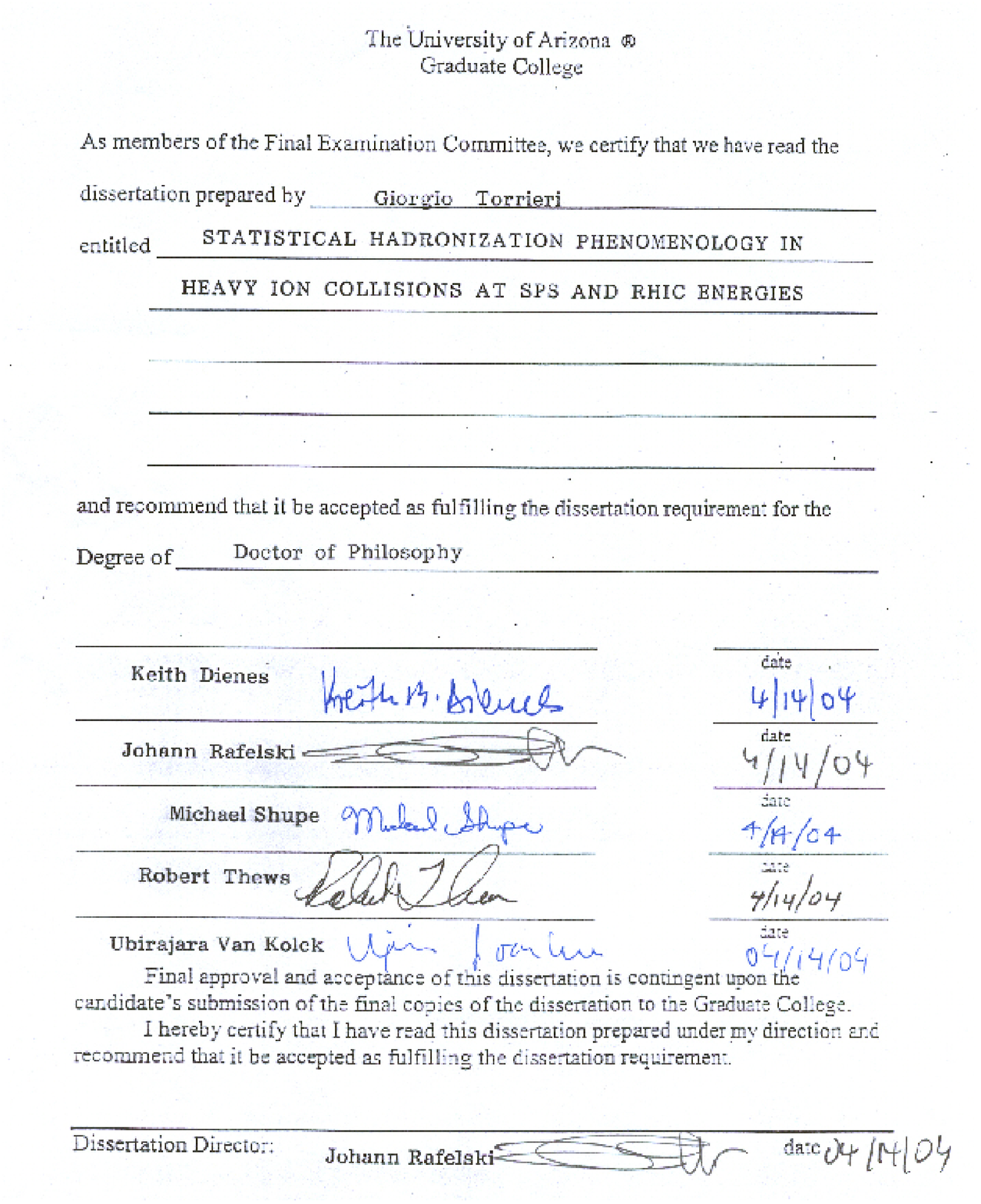,width=15cm}
\end{figure}

\newpage

\statementbyauthor


\acknowledgements
\noindent I would like to thank my advisor, committee, a lot more 
people 
in and outside the department and, above all, my family.

\noindent See Appendix E for details.

The research presented in this dissertation has been supported in part by grants from the U.S. Department of Energy, DE-FG03-95ER40937 and \\ DE-FG02-04ER41318, as well as NATO Science Program, PST.CLG.979634.

\begin{dedication}
To mum.  

\end{dedication}

\headsep 0.5in

\textheight 7.8in  
\tableofcontents

\textheight 8.3in  
\listoffigures

\listoftables

\headsep 0.3in

\begin{abstract}

This dissertation examines the phenomenology of statistical hadronization at ultrarelativistic
energies.
We start with an overview of current experimental and theoretical
issues in Relativistic heavy ion physics.
We then introduce statistical hadronization, and show how it gives a
description of particle abundances and spectra through
relativistic covariance and entropy maximization.

We argue that several statistical hadronization models are possible;
In particular, a distinction can be made between equilibrated staged freeze-out in which
post-formation hadron interactions play an important role in determining
final-state observables, and non-equilibrium sudden freeze-out where spectra and abundances
get determined at the same time and further
interactions are negligible.

We attempt to falsify sudden freeze-out by examining 
whether particle abundances and spectra can be
described using the same formation temperature.   This is done both
in the chemical equilibrium framework, and using a chemical non-equilibrium
ansatz.    Our fits to experimental data suggest that the sudden
freeze-out model explains both the particle abundances and spectra.

We then try to extract the particle formation temperature,
and quantify post-freeze-out hadronic interactions using experimentally
observable resonances.   We discuss observed resonances
and suggest further measurements that have the potential to distinguish between the possible
freeze-out scenarios experimentally.

Finally, we infer from experimental data how particle
formation proceeds in spacetime, in particular whether freeze-out
dynamics agrees with the sudden freeze-out expectation.
We examine particle spectra, and show that they are not sensitive enough to pick the right freeze-out dynamics.   
We suggest  resonances
and azimuthal anisotropy  as experimental probes for this task. 
\end{abstract}


\setcounter{figure}{0}
\setcounter{equation}{0}
\setcounter{table}{0}
\chapter{The strong interaction and quark-gluon plasma}
\label{cha:intro}
\section{Elementary and collective phenomena}
Physics research within the last fifty years can be divided into two classes of problems:  figuring out the fundamental rules which govern
physical processes at the microscopic level, and examining how these 
 fundamental rules give rise to the phenomena we observe macroscopically.
Particle physics is concerned with finding
the fundamental rules.  Condensed matter physics studies how
these  rules generate complex structure.

This division is somewhat simplistic.
Quantum mechanics, together with relativity, imply that every interacting system will always have an
infinite number of degrees of freedom.   Even the vacuum state will always contain
particle-antiparticle pairs popping in and out of it at all times.
This means that the vacuum, just like any material studied in condensed
matter physics, may exhibit collective phenomena.   If the interaction
is strong enough, the collective
phenomena will dominate ``elementary'' physics, and the observed degrees of
freedom might be, in their nature and behavior, different from the fundamental ones.
In other words, to understand the physics of one particle, we need to understand the collective medium through which this particle will propagate.

As an arena to understand general collective systems, condensed matter physics has some limitations.
While the structures and properties of condensed matter systems are enormous
in their variety, the ``elementary rules'' they are based on are limited and understood:
electromagnetism and quantum mechanics.
It is not at all certain that by introducing a different set of fundamental rules
 we will not arrive at a qualitatively novel macroscopic system.   The fact that an analogue of strong interaction
confinement has never been found in a condensed matter system reinforces the need to investigate
how the ``macroscopic phenomena'' change when a qualitatively different
set of ``microscopic rules'' is introduced.

The purpose of heavy ion physics is to do just that:  to create and study a ``condensed matter system'' using a different fundamental interaction than electromagnetism.   
\section{The strong interaction}
\subsection{Quarks}
According to the current theory of the strong interaction, all hadrons are composed of fermionic constituents called quarks.
The non-interacting quark Lagrangian therefore comprises several massive fermion fields, denoted by a quantum number, called
flavor ($f$), conserved in strong interactions
\begin{equation}
\label{lnoint}
L= \sum_{f} \overline{\psi_f}(i \gamma^{\mu} \partial_{\mu}-m_f)\psi_{f}.
\end{equation}
Quarks are arranged in three flavor doublets, of which the
top member has electromagnetic charge $+2/3$ and the bottom member has charge $-1/3$ (in units where the proton has charge +1).\\
The quarks in the first doublet, called up and down ($u$ and $d$ from now on), have a mass that is negligible with respect to the mass of the observed hadrons, and are therefore often called ``light quarks''.   These quarks are the only ones
stable against flavor-changing weak decays.

The quarks forming the second doublet, called strange and charm  ($s$ and $c$) are more massive.   The strange quark is of particular interest to the subject of this dissertation, since, as we will see, its mass (50 to 150 ${\rm MeV}$) is comparable to both the mass of the lightest hadron and the energy scale of the QCD phase transition.  The charm quark is much more massive ($\sim 1$ ${\rm GeV}$), though it may be produced in the initial processes of a heavy ion collision.  The third doublet, top and bottom, is too massive to be produced in present heavy ion collision experiments (though  the $\sim 4$ ${\rm GeV}$ $b$ quarks could
be studied in future heavy ion collisions).
\subsection{Color and gluons}
The strong interaction, responsible for holding hadrons and nuclei together, is fundamentally very similar to QED in that
interactions maintain the Lagrangian invariant under a local phase
transformation of the quark wave function  
\begin{equation}
\label{gauge1}
\psi \rightarrow U \psi 
\end{equation}
\[\  UU^{+}=1.  \]
However, we postulate that quarks transform as vectors
in a 3-dimensional space called color space, while anti-quarks are co-vectors
\begin{equation}
\label{q_color}
\psi = \left( \begin{array}{c} \psi_{r} \\ \psi_{b} \\ \psi_{g}  \end{array}  \right) \; , \; \overline{\psi}=( \overline{\psi_{r}},\overline{\psi_{b}}, \overline{\psi_{g}})
.\end{equation}
Hence, the most general form of Eq.~(\ref{gauge1}) allows $U$ to be a  special unitary matrix.  ($U U^{+}=1$ , $|U |=1$).   
Such matrices form a group, called $SU(3)$ \footnote[1]{The results described in this section can be generalized to $SU(N)$ invariance, corresponding to physically
$N$ colors and $N^2-1$ generators. Historically, the first example of this theory  to be examined was the Yang-Mills Lagrangian, corresponding to $SU(2)$. }.
It can be shown that for any $U$ eight real numbers $\alpha_j$ can be chosen
 such that
\begin{equation}
\label{gellman}
U = e^{i \sum_{j=1}^{8} \alpha_j t_j }
\end{equation}
where the generators of the $SU(3)$ group $t_j$ can be represented by the Gell-Mann matrices
\begin{equation}
\label{t123}
\begin{array}{ccc}
t_1=\frac{1}{2} \left( \begin{array}{ccc} 0 & 1 & 0 \\ 1 & 0 & 0\\ 0 & 0 & 0 \end{array} \right) &
t_2=\frac{1}{2} \left( \begin{array}{ccc} 0 & -i & 0 \\ i & 0 & 0\\ 0 & 0 & 0 \end{array} \right)  & 
t_3=\frac{1}{2} \left( \begin{array}{ccc} 1 & 0 & 0 \\ 0 & -1 & 0\\ 0 & 0 & 0 \end{array} \right) \\
t_4=\frac{1}{2} \left( \begin{array}{ccc} 0 & 0 & 1 \\ 0 & 0 & 0\\ 1 & 0 & 0 \end{array} \right) & 
t_5=\frac{1}{2} \left( \begin{array}{ccc} 0 & 0 & -i \\ 0 & 0 & 0\\ i & 0 & 0 \end{array} \right)  & 
t_6=\frac{1}{2} \left( \begin{array}{ccc} 0 & 0 & 0 \\ 0 & 0 & 1\\ 0 & 1 & 0 \end{array} \right) \\
t_7=\frac{1}{2} \left( \begin{array}{ccc} 0 & 0 & 0 \\ 0 & 0 & -i\\ 0 & i & 0 \end{array} \right) &  
t_8=\frac{1}{2 \sqrt{3}} \left( \begin{array}{ccc} 1 & 0 & 0 \\ 0 & 1 & 0\\ 0 & 0 & -2 \end{array} \right). & \\
\end{array}  
\end{equation}
Just as in the case of electromagnetism, imposing the phase symmetry locally requires 
the introduction of a covariant derivative to subtract the effect of a varying  $U$ on neighboring points.
This covariant derivative, however, now contains eight terms corresponding to each Gell-Mann matrix
\begin{equation}
\label{cov_dev}
\partial_{\mu} \rightarrow D_{\mu}= \partial_{\mu}-i g \sum_{j=1}^8 A_{\mu j} t_{j}.
\end{equation}
Physically, substituting $D_{\mu}$ into Eq~(\ref{lnoint}) means that quarks can interact through eight bosons called gluons.    However, unlike QED, each gluon
also carries a color charge, which, in the Gell-Mann representation Eq.~(\ref{t123}) corresponds to
\begin{equation}
\label{gluons}
  \frac{r \overline{b}+b\overline{r}}{\sqrt{2}},  \frac{r \overline{b}-b\overline{r}}{i \sqrt{2}},\frac{r \overline{r}-b\overline{b}}{\sqrt{2}},\frac{r \overline{g}+g\overline{r}}{\sqrt{2}},\frac{r \overline{g}- g\overline{r}}{i \sqrt{2}},\frac{b \overline{g}+g\overline{b}}{\sqrt{2}}, \frac{b \overline{g}-g\overline{b}}{i \sqrt{2}},\frac{r \overline{r}+b\overline{b}-2 g\overline{g}}{2 \sqrt{3}}
\end{equation}
for $t_{1..8}$ respectively.

Gauge invariance also allows for a dimension four gluon field contribution to the Lagrangian, equivalent to a curvature tensor in gauge space
\begin{equation}
\label{FQED}
L \rightarrow L+\sum_{j=1}^8 {\rm Tr}\left( [D_{\mu},D_{\nu}] [D^{\mu},D^{\nu}] \right)=L+\frac{1}{4} \sum_{j=1}^{8} F_{\mu \nu j} F^{\mu \nu}_{j}.
\end{equation}
A qualitative difference with respect to QED emerges due to the fact that the matrices $t_{1-8}$ do not commute.
Their commutators, represented by the $SU(3)$ structure constants $f_{j,j_1,j_2}$
\begin{equation}
[t_{j_1},t_{j_2}]=i \sum_{j=1}^{8} f_{j,j_1,j_2} t_j
\end{equation}
will therefore appear in the QCD field strength since
\begin{equation}
\label{selfinteraction}
[D_{\mu} D_{\nu} ] = ...+\sum_{j_1,j_2=1}^{8} [t_{j_1} t_{j_2}] A^{\mu}_{j_1} A^{\nu}_{j_2} = ...+i \sum_{j,j_1,j_2=1}^{8} f_{j j_1 j_2}  t_j A^{\mu}_{j_1} A^{\nu}_{j_2}
\end{equation} 
Hence,  gluons can also interact, and their interaction manifests itself in a physical
gauge-invariant way through a
term in the gluon field strength.

In summary the Lagrangian of interacting QCD is
\begin{equation}
\label{LQCD}
L=\frac{1}{4} \sum_{j=1}^{8} F_{\mu \nu}^{j} F^{\mu \nu j} + \sum_{f} \sum_{c,\overline{c}=1}^3 \overline{\psi_{\overline{c},f}}\left[ i \gamma_{\mu} \left(\partial^{\mu}-i g\sum_{j=1}^8 A^{\mu}_{j} t_{j c \overline{c}}\right)-m_f \right]\psi_{c,f} 
\end{equation}
with
\begin{equation}
\label{FQCD}
F^{\mu \nu}_{j} = \partial^{\mu} A^{\nu}_j - \partial^{\nu} A^{\mu}_j - g \sum_{j_1,j_2=1}^{8} f_{j j_1 j_2} A^{\mu}_{j_1} A^{\nu}_{j_2}.
\end{equation}
Here, the index $c$ runs over the color
quantum number, while $j$ runs over the gluon type.   The gluon matrices $t_{1-8}$ contract the anti-quark co-vector with the quark vector. 

While the symmetry principles used in deriving the QCD Lagrangian follow the QED case closely, the self-interaction term has
profound consequences not seen in QED.   This becomes evident when the theory is quantized.
\subsection{Quantization and the running coupling constant}
Expanding Eq.~(\ref{LQCD}) in terms of $A_{\mu j}$ leads to the following Feynman rules\footnote[3]{We will ignore the additional ``ghost'' particle needed in some gauges.  See \cite{peskin} for more details.}
\begin{eqnarray}
\Diagram{\vertexlabel^a \\fd \\
& g\vertexlabel_{\mu,j_1} \\
\vertexlabel_b fu\\
} = i g \gamma_{\mu} t_{j_1}\\
\Diagram{\vertexlabel^{p^{\alpha},j_1} \\gd \\
& g\vertexlabel_{q^{\beta},j_2} \\
\vertexlabel_{r^{\gamma},j_3} gu\\
} = g f_{j_1 j_2 j_3} (g_{\beta\gamma} (q-r)_{\alpha}+g_{\gamma \alpha} (r-p)_{\beta}+g_{\alpha\beta} (r-p)_{\gamma}) \\
\Diagram{
 \vertexlabel^{\mu,j_1}  gd 
& gu \vertexlabel^{\nu,j_2} \\
\vertexlabel_{\rho,j_3} gu 
& gd \vertexlabel_{\sigma,j_4} 
} = -i g^2 \left[  \begin{array}{c}
\sum_{j'}\left\{ f_{j_1 j_2 j'} f_{j_3 j_4 j'} (g^{\mu \rho} g^{\nu \sigma} -g^{\mu \sigma} g^{\nu \rho}) \right. \\
+f_{j_1 j_3 j'} f_{j_2 j_4 j'} (g^{\mu \nu } g^{\rho \sigma} -g^{\mu \sigma} g^{\nu \rho})\\
\left. +f_{j_1 j_4 j'} f_{j_2 j_3 j'} (g^{\mu \nu} g^{\rho \sigma} -g^{\mu \rho} g^{\nu \sigma}) \right\}
\end{array} \right]
\end{eqnarray}
Where the indices correspond to color for quarks ($a,b=1..3$), type for gluons ($j_{1..4}=1..8$) and Greek letters for gluon Lorentz indices.

While the first term is similar to the QED case, the next two, which come out of the $f_{a b c} A^{\mu}_{b} A^{\nu}_{c}$ contribution to $F^{\mu \nu}$ are new.
Their profound effect can be seen when one attempts to calculate quantum
corrections to the coupling constant $g$ ($\mu$ is the momentum transfer at which the coupling
constant is measured)
\begin{equation}
 g^2 (q^2) = \frac{g^2 (\mu^2)}{1- [\Pi(q^2)-\Pi (\mu^2)]}\\
\end{equation}
\begin{equation}
 \Pi (q^2)=\left[ \feyn{ g fl g}+\feyn{ g g gl g g} + \Diagram{& \smallbosonloop \\  gd 
& gu  \\
 gu 
& gd  } +... \right]
\end{equation}
The explicit calculation of this $\Pi (q^2)$ is a somewhat involved endeavor beyond the scope of this thesis \cite{peskin}.
However, one can see that since boson loops have the opposite sign from fermion loops, the contribution of the last two diagrams will in general have an effect on
the running of the coupling constant opposite to that of the first term.   
Intuitively, while the first term screens the charge, the gluon loops
increase the field strength while maintaining the attractive potential,
resulting in a larger effective charge than the bare one (anti-screening).
Combining all the diagrams, one can see that in a general Yang-Mills theory with $N$ colors and $n$ light flavors, 
the coupling constant will be, to a leading order correction \cite{peskin}\footnote[1]{higher order corrections are generally not thought to change the direction of the running.} 
\begin{equation}
g^2 (k^2) = \frac{g^2 (\mu^2)}{1+\frac{g^2 (\mu^2)}{(4 \pi)^2} (\frac{11}{3} N- \frac{2}{3} n)\ln(k^2/\mu^2)}.
\end{equation}
Provided a scale $\mu^2$ can be found where $g^2<<4 \pi$ this is a self-consistent result.

In the QED case ($N=0$,$n=1$ for the light electron) $\alpha(k^2)$ increases at increased momentum transfer, and it is found
that in low energy experiments $\frac{ g^2 (k^2 \rightarrow 0)}{4 \pi}= \alpha =\frac{1}{137}\ll 1$.
Hence, 
\begin{equation}
g^2_{QED} (k^2) = \frac{ 4 \pi \alpha}{1- \frac{2}{3} 4 \pi \alpha \ln(k^2/m_{e}^2)}
\end{equation}
with the theory remaining perturbative until a very high energy scale ($\sim e^{700} m_e$) at which QED is not expected
to hold anyways.

In the QCD case, however ($N=3$,$n=3$ for three light quarks) the situation
is reversed.     $g^2$ decreases with increasing momentum transfer, and eventually becomes perturbative
at a large scale $\Lambda$  (  the experimentally measured $\frac{g^2(100 {\rm GeV})}{4 \pi} = 0.12$ )
\begin{equation} 
\label{gqcd}
g^2_{QCD} (k^2) = \frac{g^2 (\Lambda^2)}{1+ 9 \frac{g^2 (\Lambda^2)}{(4 \pi)^2} \ln(k^2/\Lambda^2)}.
\end{equation}
The highly energetic quarks produced at this energy scale
become bundles of approximately collinear
fast particles known as jets, with each jet having the momentum of the original quark.

However, at smaller $k^2$ $g^2$ increases to the point where perturbation
theory ceases to be a useful tool.
Thus, while high-energy scattering cross-sections can be computed through
perturbative QCD, low-energy effects such as vacuum structure and
infrared corrections to particle propagators are governed by non-perturbative
physics.
\section{The structure of the QCD vacuum at zero temperature}
The non-perturbative nature of the strong interaction in the low-energy limit
should not come as a surprise:   the fundamental degrees of freedom of QCD (quarks, gluons) are very different from the fundamental degrees of freedom observed
in the strong interactions at low energies (the 100+ hadrons in the particle data book).
For instance, every hadron observed so far is not charged under color, and does
not interact manifestly through gluon exchange. It must be the case, therefore,
that somehow only color singlets survive as physical states.
 Moreover, QCD predicts
gluons to be massless, and high energy scattering processes have determined
light quarks to be extremely light ($\sim$ a few ${\rm MeV}$).
Yet the observed hadrons are much heavier than that ($\sim$ ${\rm GeV}$).
Both of these phenomena are not as yet rigorously understood.  However, 
there are phenomenological ways to see how they arise.
\subsection{Confinement}
A simple picture of what happens when two color charges (for instance, a quark-antiquark pair $q \overline{q}$) become separated is provided by relativistic quantum mechanics \cite{gribov}:  As the separation of the  $q \overline{q}$ pair  increases, so does the effective charge.
At a certain point it becomes energetically convenient
to create a new  $q \overline{q}$ pair so that a separated $q \overline{q}$ becomes two tightly bound $q \overline{q}$.

An analogous situation exists in atomic physics: Dirac energy levels of an electron orbiting a $Z>137$ point charge become complex \cite{rafvac}
as it becomes energetically possible to create a real  $e^+ e^-$ pair out
of the vacuum (  $A_{Z} \rightarrow A_{Z}+e^{+}+e^{-}$ ).  The positron
escapes to infinity, and the energy contained in the $(A_{Z}+e^{-})_{bound}+e^{+}$ state is
less than that of the field of the isolated $A_Z$ ion.
The anti-screening of QCD means that the field strength necessary for this vacuum instability will be
reached when any two color charges become sufficiently separated \cite{gribov} .

This picture is reinforced by evidence from lattice gauge theory \cite{wilson}, a calculational technique valid when asymptotic freedom 
applies.   In this approach, the continuous spacetime is replaced by a discrete lattice of points, as it is assumed that 
smaller-scale degrees of freedom become
perturbative.
It is then possible to represent the pure-gauge action using links from one lattice point to the next
\begin{equation}
\label{latticelink}
U(x_1,x_2)_a \sim e^{i g \int_{x_1}^{x_2} dx_{\mu} A^{\mu}_a t_{a}}
\end{equation}
and calculate physical observables in euclidean spacetime ($t \rightarrow it$) by evaluating a trace over the lattice points (which approaches the functional integral
in the continuum limit) using Monte-Carlo techniques.
For instance,  the potential energy between two static color
sources 
\[\ <V(r)> \sim \sum_{a=1}^{8} g A^0_{a} (r) = \sum_{a=1}^{8} <j_{\mu a} A^{\mu}_a>\]
corresponds to the expectation value of  
a loop in Euclidean spacetime spanning the particle's separation $r$ 
\cite{wilson,confin_review}.
\begin{equation}
\label{wilson_pot}
<V(r)>=\lim_{t \rightarrow \infty} \frac{1}{t}\log \left( <W(r)> \right) 
\end{equation}
\begin{equation}
\label{wilson_loop}
<W(r,t)>=\frac{\int d A \exp \left[ \oint_{0,0}^{r,t}\left( j_{\mu} A^{\mu} -  L \right) d^4 x \right]}{ \int dA \exp \left[ -  \oint_{0,0}^{r,t} L d^4 x\right] }
\end{equation}
\begin{figure}[h]
\centering
  \psfig{figure=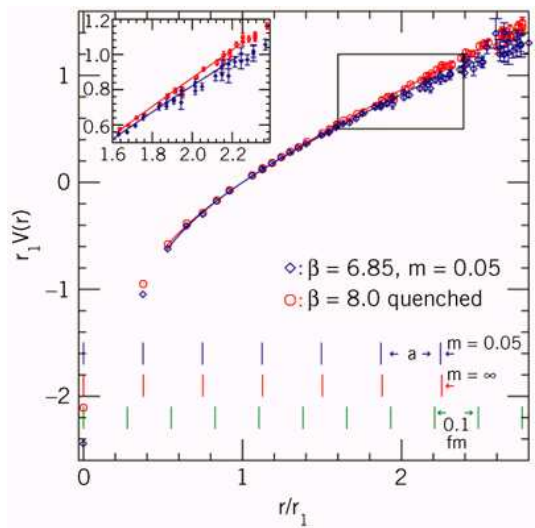,width=7cm}
  \caption{Potential calculated with full three flavor QCD \label{wilsonloop}, rescaled by a constant $r_1$ at which confinement effects become significant.}
\end{figure}
In the strong coupling limit and neglecting quark loops \footnote{This is known as the quenched approximation.  Without it, the potential
will saturate at the value necessary to create a quark-antiquark pair, as per the Gribov mechanism.} it can be proven \cite{creutz} that 
\begin{equation}
V(r) \propto r.
\end{equation} 
A numerical calculation with a more realistic picture, with three 
quark flavors, does not change this conclusion;  as Fig.~\ref{wilsonloop} \cite{touissaint} shows, the effective QCD potential between heavy
quark charges can be fitted by a function increasing linearly with $r$.
Heavy quark spectoscopy($c \overline{c}$ $J/\Psi$ states and $b \overline{b}$ $\Upsilon$ states) has experimentally confirmed that the QCD potential can indeed be effectively modeled by a coulomb and a linear term \cite{charmpot1,charmpot2}.

The Gribov argument and lattice evidence point to a coherent qualitative picture of what
happens as the strong interaction coupling constant increases beyond the perturbative limit.
Instead of decreasing at large distances, as the electromagnetic potential
does, the strong interaction potential continues to increase monotonically until the vacuum decays into more quarks and gluons in such a way as
to make every state ultimately observed a color singlet.

This phenomenon, called confinement, 
has not been fully understood.    Nothing similar has, as yet, been observed within many-body systems
interacting via electromagnetism.
Thus, the study of how confinement arises is a good example of the problems
outlined in section 1.1.

\subsection{Chiral symmetry breaking}
A considerable number of hadron masses ($\rho$,nucleons,$\Delta$,$\Lambda$,$\Sigma$,$\Sigma^*$,$\Xi$,$\Xi^*$,$\Omega$) can be fitted reasonably well by assuming hadrons to be made up of ``constituent'' quarks, whose number is given by the hadron's flavor
content.   If the constituent quark mass 
is set to $\sim$~300~${\rm MeV}$ for $u,d$ and~$\sim$~500~${\rm MeV}$ for $s$, observed masses for these hadrons can be described in terms of the constituent quark masses plus spin couplings and electromagnetic
effects.

To see how such constituent quark states might arise, consider decomposing 
the light quark part of the Lagrangian into left-handed and right-handed components 
\begin{eqnarray}
\psi=\left(\begin{array}{c} u \\ d \\ s \end{array}  \right)\\
\psi_L=\frac{1}{2} (1-\gamma_5) \psi\\
\psi_R=\frac{1}{2} (1+\gamma_5) \psi
\end{eqnarray}
The quark part of the Lagrangian then becomes
\begin{equation}
\label{chiral1}
L_q = \overline{\psi_L} i \gamma^{\mu} D_{\mu} \psi_L+\overline{\psi_R} i \gamma^{\mu} D_{\mu} \psi_R -\sum_{i=u,d,s} m_{\psi_i} (\overline{\psi_L} \psi_R +  \overline{\psi_R} \psi_L).
\end{equation}
We see that the only mixing term is provided by the quark mass.
Therefore, in  ``bare'' QCD, the left and right light quark currents
should be separated with only a small correction ($\sim 10 {\rm MeV}$ for $u,d$, $\sim 100 {\rm MeV}$ for s).

However, if the QCD vacuum exhibits quark condensation
\[\ <vac|\overline{\psi} \psi|vac>= <vac| \overline{\psi_L} \psi_R +  \overline{\psi_R} \psi_L |vac>  \ne 0  \] 
the low energy effective Lagrangian will exhibit an additional quark mass, since, as implemented in effective models such as  
Nambu-Jona-Lasinio \cite{NJL1,NJL2}, 
interactions between quarks will generate an effective mass term.   

We can see that this is the case in QCD
since the Lagrangian in Eq.~(\ref{chiral1}) is, in the $m_{u,d,s} \rightarrow 0$ limit, invariant under two separate symmetries
\begin{equation}
\label{transforms}
\psi_L \rightarrow e^{i (\sum_{i=1,2,3} \alpha_{Li} \sigma_i+\beta_L I) }  \psi_L \;  , \; \psi_R \rightarrow e^{i (\sum_{i=1,2,3} \alpha_{Ri} \sigma_i+\beta_R I)} \psi_R
\end{equation}
where $I$ is the identity matrix, $\sigma_i$ are the Pauli matrices and $\alpha_{Li,Ri}$,$\beta_{L,R}$ are real numbers.
Using the transformations of Eq.~(\ref{transforms}), corresponding
to the
\[\
U(3)_L \times U(3)_R= SU(3)_L \times SU(3)_R\times U(1)_L\times U(1)_R
\]
\[\ =SU(3)_V \times SU(3)_A\times U(1)_V\times U(1)_A\]
symmetry group, it should be possible
to transform any hadron into a hadron of inverse parity (e.g., a negative-parity vector meson, with a current given by
$\overline{\psi} \gamma_{\mu} \psi$, into a positive parity pseudo-vector $\overline{\psi} \gamma_{\mu} \gamma_5 \psi$).
Hence, the mass difference between such states (e.g. $\rho$ and $f_1 (980)$, or the $\Lambda$ and the $\Lambda(1405)$) should
be of the order of the quark mass.   In fact, these differences
are $\sim 300 {\rm MeV}$, the same order of magnitude as the constituent quark mass.

Hence, it appears that $U(3)_L\times U(3)_R$ is spontaneously broken to a lower subgroup.   Identifying the lower subgroup as $SU(3)_V\times U(1)_V$ (Isospin$\times$Baryon number)\footnote[1]{$U(1)_A$, is broken through quantum corrections \cite{thooft}, which is why no light isospin=0 pseudo-scalar meson exists, the $\eta'$ having a mass of $\sim 1 {\rm GeV}$.} we are left with eight Goldstone modes, the $\pi,K,\eta$ mesons (which are all considerably lighter than the constituent quark prediction, although the large
strange quark bare mass makes the $K$ and $\eta$ much heavier than the $\pi$).

The nature of the chiral symmetry breaking, and its relationship with confinement, are under intense study.   However, numerical results described in the next section suggest the two phenomena are related.
\section{The QCD phase transition}
The previous two sections make it clear that in empty space
light colored particles become confined in massive (at least 140 ${\rm MeV}$)
color-neutral composite particles.
The question which immediately follows is:  What happens if temperature
or quark-density increases to such an extent that inter-hadron separation
becomes comparable to the separation of quarks within a hadron?
Intuitively, hadrons should then break down, and at the very high temperature
or density limit the system should approach the asymptotically free limit
of a relativistic ideal gas.
However, does this occur as a phase transition (ie, a discontinuity in
some order parameter) or is it rather a continuous cross-over, much like
the formation of the electromagnetic plasma?   At what temperature
do the quark and gluon degrees of freedom start manifesting themselves?
What does a strongly interacting QGP look like away from the infinite temperature
limit?   

Studying these questions, at both theoretical and experimental levels, is important for a variety of reasons, some of which are discussed in section 1.1.
The Universe underwent this phase transition shortly ($t \sim 10^{-6}$ seconds) after  the big bang, and
evolved as a quark-gluon plasma before that time.
Hence, the QCD phase transition might have had consequences for the evolution and structure of the Universe.    Moreover, there are astronomical objects today,
such as neutron stars, where a QGP might be expected to exist naturally.
Finally, to reiterate the point made at the start of this chapter, the QCD phase transition (if it is indeed a phase transition) is the
only vacuum phase transition which is at energies experimentally
accessible for the foreseeable future  (It is very unlikely, for instance, that we will be able to create  any-time soon a  thermalized system  hot enough  for electroweak symmetry restoration).
The study of the QCD phase transition, therefore, is invaluable for our understanding of quantum field theory in general, especially in the regime where
standard perturbation theory methods fail.
\subsection{A rough calculation (which will prove useful later)}
We shall proceed to calculate the QCD phase transition parameters from the MIT bag model \cite{mitbag1}, an ansatz which can fit reasonably well the spectra of the
observed non-Goldstone hadrons.
The zero-temperature QCD vacuum is assumed to have a ``bag constant'' term due to deconfinement/chiral symmetry breaking, i.e. a vacuum energy with positive
energy density ($\epsilon$) but negative pressure ($P$).
The energy-momentum tensor is then
\begin{equation}
T^{\mu \nu}=T^{\mu \nu}_{matter}+B g^{\mu \nu}  
\end{equation}
with
\begin{eqnarray}
 \epsilon&=&\epsilon_{matter}+B  \\ P&=&P_{matter}-B
\end{eqnarray}
Hadrons arise as ``bags'' filled with free or weakly interacting quarks, trapped in volumes
where the pressure exerted by the quarks in the bag balances the negative
pressure exerted by the bag.

Through solving the Dirac equation, one can find the quark wave-functions inside the bag and the hadron masses in terms of the bag constant. Fits to the $\rho$ meson, nucleon and $\Delta$ yield a bag constant of $B^{1/4}=150 {\rm MeV}$ \cite{mitbag2}.
It can be seen that in this model a first order phase transition may occur 
when a gas of hadrons has the same pressure
as a quark-gluon plasma \cite{rafbag}.
We calculate the pressures of the particles using the grand canonical ensemble for a non-interacting ideal gas (see the next chapter for details)
\begin{equation}
P=\frac{\partial F}{\partial V} = T \sum_{particles} \int_0^{\infty} 
\frac{g_i 4 \pi p^2 dp}{(2 \pi)^3} (\pm 1) \ln \left( 1+(\pm1) e^{\frac{\mu-\sqrt{m^2+p^2}}{T}} \right)
\end{equation}
where $\pm$ applies, respectively, to fermions and bosons and $g_i$ is the degeneracy.
For the QGP phase we used 2 massless and one massive quark (fermions) with a degeneracy of 6 (spin$\times$color) and eight gluons (bosons) with a degeneracy of 2 
(polarizations).
For the hadron gas, in addition to the vacuum pressure given by the bag constant, we used all particles in the particle data book with a mass lighter than 2 ${\rm GeV}$ \cite{pdg}.
Light flavor chemical equilibrium was assumed ($\mu_{q}=-\mu_{\overline{q}}$) and the strangeness chemical potential was put to 0 (no strange quarks).

\begin{figure}[h]
\centering
  \psfig{figure=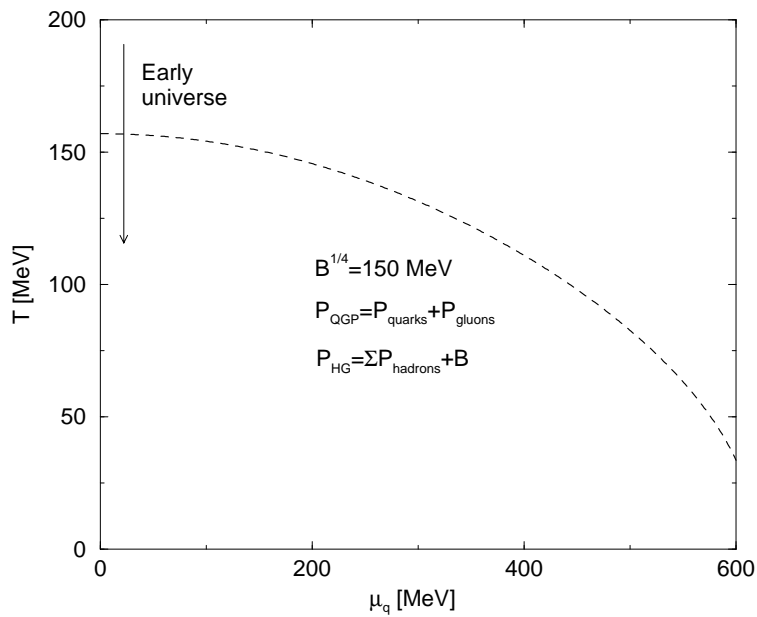,width=10cm}
  \caption{The bag model phase diagram \label{phasediag_bag}.}
\end{figure}

Fig. ~\ref{phasediag_bag} shows the phase diagram calculated in this
ansatz.
As can be seen, our  estimate predicts a critical temperature roughly equivalent to the mass of the lightest hadron.
While many assumptions used in this estimate can be considered naive at
first glance,
calculations on the lattice give very similar results.
\subsection{Results from lattice QCD}
A phase diagram calculation can also be performed from first principles QCD through lattice techniques, introduced in section 1.3.1.
The fact that lattice field theory can be used to describe finite temperature systems is evident from the fact that the QFT partition function
\begin{equation}
\label{ZQFT}
Z_{QFT}=\int D \psi D A e^{i \int L(\psi,A) d^3 x dt}
\end{equation}
is related to the thermodynamic partition function
\begin{equation}
\label{ZTH}
Z_{TH}=\int D \psi D A e^{- \int L(\psi,A) d^3 x/T}
\end{equation}
by the substitution $t \rightarrow i \frac{1}{T}$.
This substitution, together with the fact that $e^{A}=e^{A+2 \pi i}$,
means that a quantum field theory with a periodic boundary conditions in the time direction (anti-periodic for fermions) \cite{matsubara,kapusta_field}
\begin{eqnarray}
\label{periodic}
\psi \left(x,t+\frac{2 \pi}{T} \right)=-\psi(x,t)\\
A \left(x,t+\frac{2 \pi}{T} \right)=A(x,t)
\end{eqnarray}
will effectively model a finite-temperature quantum field in equilibrium
with a heat bath.

\begin{figure}[h]
\centering
  \psfig{figure=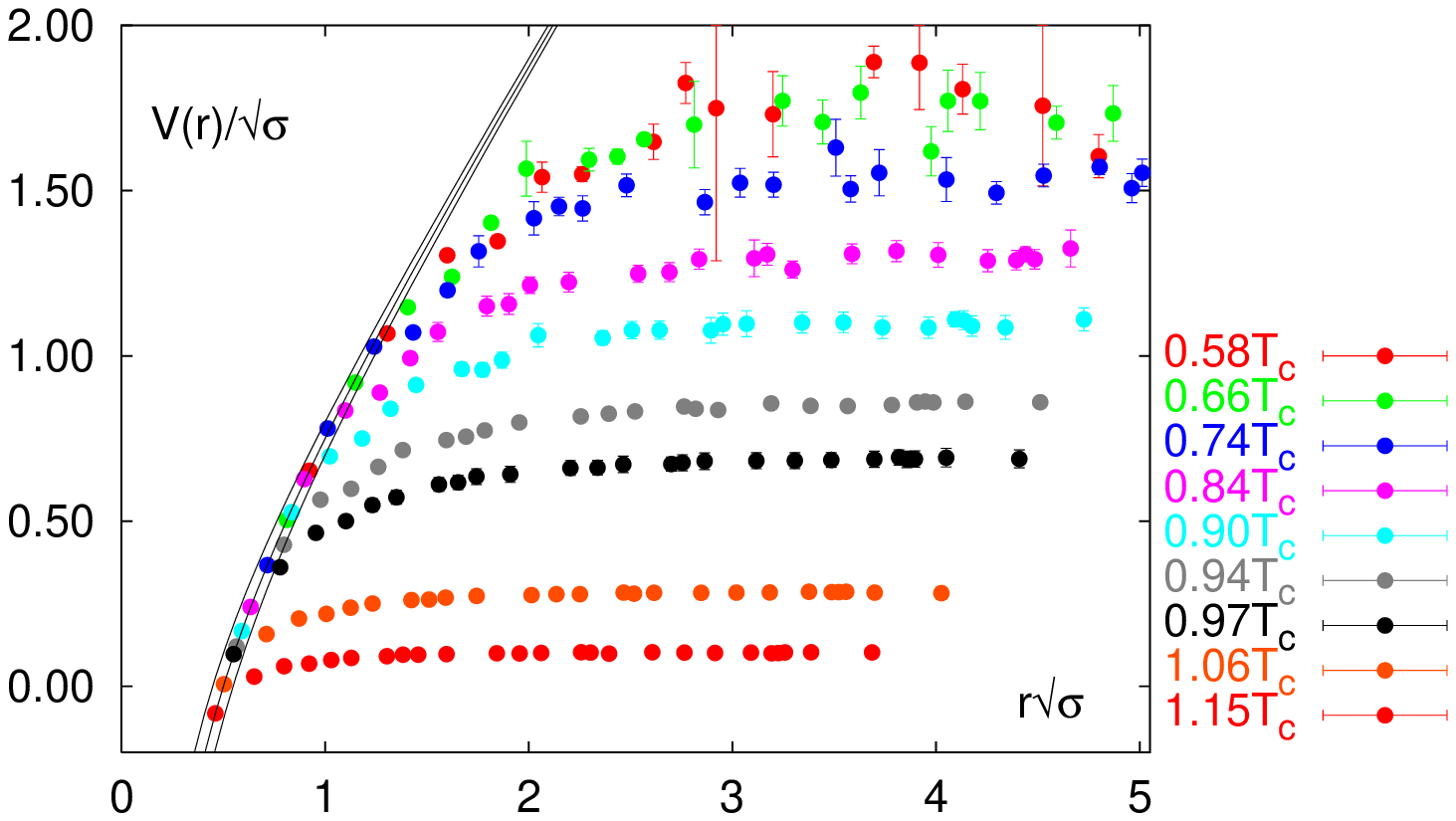,width=7cm}
  \psfig{figure=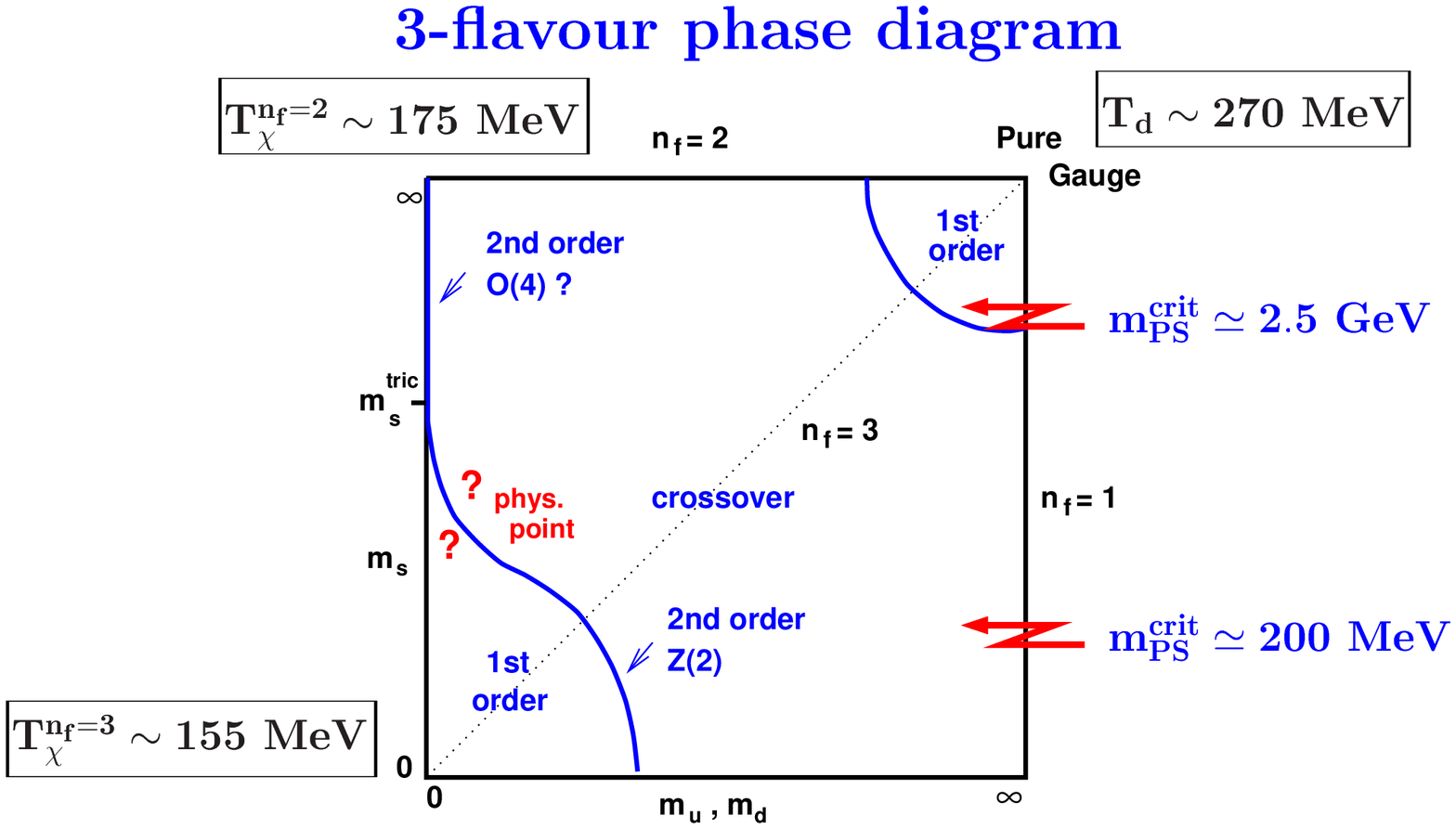,width=7cm}
  \caption{Left: The Polyakov loop expectation value as a function of distance at several temperatures.  $\sigma$ refers to the best fit zero temperature flux tube tension.
Right: The order of the deconfinement phase transition as a function of the number of flavors and quark masses (physical QCD is somewhere in between the extrema of the diagram). \label{lattice_order} }
\end{figure}

In such a formulation the Wilson loop definition given in Eq.~(\ref{wilson_loop}) has to be modified, since
no $t \rightarrow \infty$ limit is possible.
Instead, the relevant variable becomes the expectation value of the Polyakov Loop \cite{polyakov1,polyakov2}, where the
quark propagator closes on itself in the periodic time:
\footnote{Further evidence that the Polyakov loop is 
the order parameter for deconfinement is the fact that it is not gauge-invariant  for color non-singlets 
\cite{polyakov2}.  Hence, unless its expectation value becomes independent of quark separation
(which it does above a certain temperature as shown in Fig.~\ref{lattice_order} ), only
color singlets survive as physical states above the lattice scale.}
\begin{equation}
<W_{T\ne0}(T,r)>=\frac{\int d A \exp \left[ \oint_{0,0}^{r,\frac{2 \pi}{T}} \left( j_{\mu} A^{\mu} - L \right) dt d^3 x \right] }{\int dA \exp \left[ - \oint_{0,0}^{r,\frac{2 \pi}{T}} L  dt d^3 x \right] }.
\end{equation}

Its far from obvious, however, if this calculation will be relevant to the physical world.   Eq.~(\ref{ZTH}) refers to a system in perfect equilibrium, something
which did not apply either to the expanding early Universe or to the conditions
in which quark-gluon plasma is created.    Nevertheless, a lattice calculation can serve as an indication of what we can expect the quark-gluon plasma phase transition to be like.

Fig.~\ref{lattice_order} \cite{karsch2} shows that the lattice does indeed
exhibit something which looks like a phase transition.   As the temperature
increases, the distance dependence of the Polyakov loop decreases, to vanish
at a critical value as expected in a deconfining phase.
We still, however, do not know what order characterizes this phase
transition, or even if a phase transition is present at all.
As the right panel of Fig.~\ref{lattice_order} shows, the order of the phase
transition is strongly dependent on theory input, such as the number of flavors
and the values of the masses of the light and strange quark.
Since performing calculations using realistic inputs for the light and strange quark masses (the realistic values for these  
quantities are somewhere in the middle of those in the panel) is beyond the range
of today's computing power, the order of the phase
transition is still unknown.
\begin{figure}[h]
\centering
  \psfig{figure=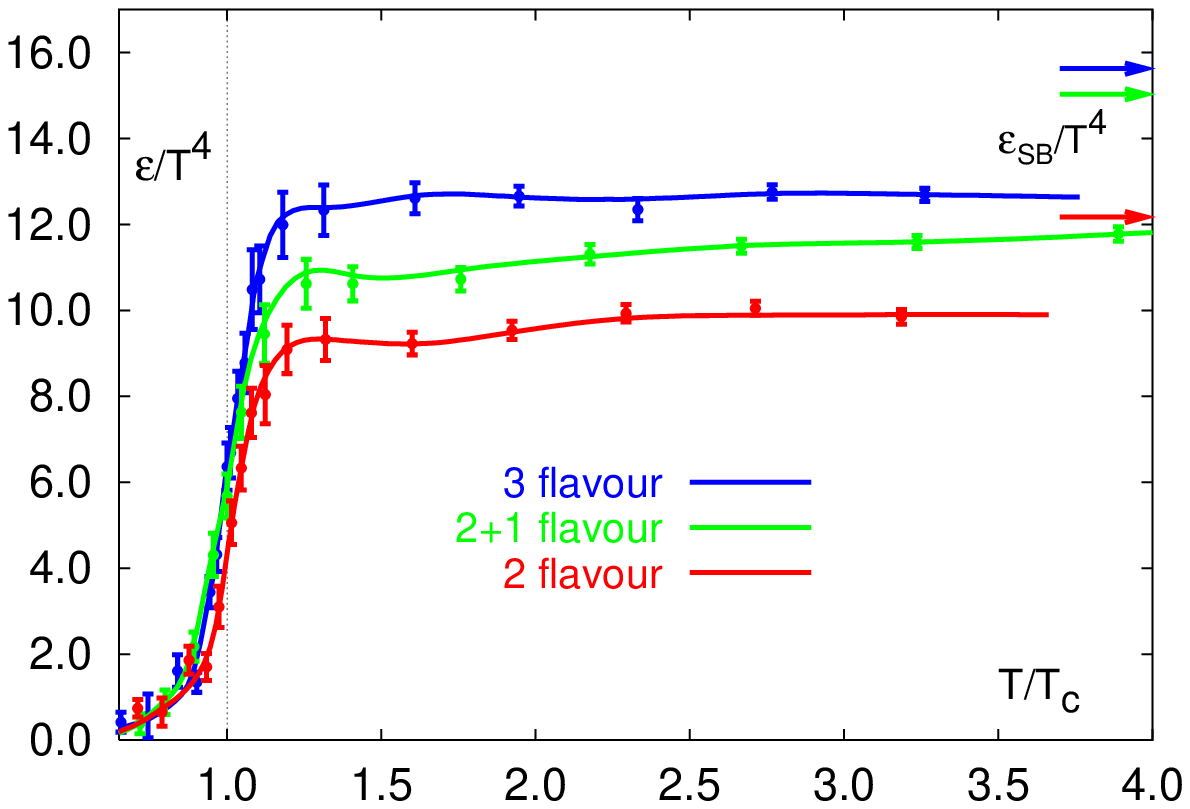,width=7cm}
  \psfig{figure=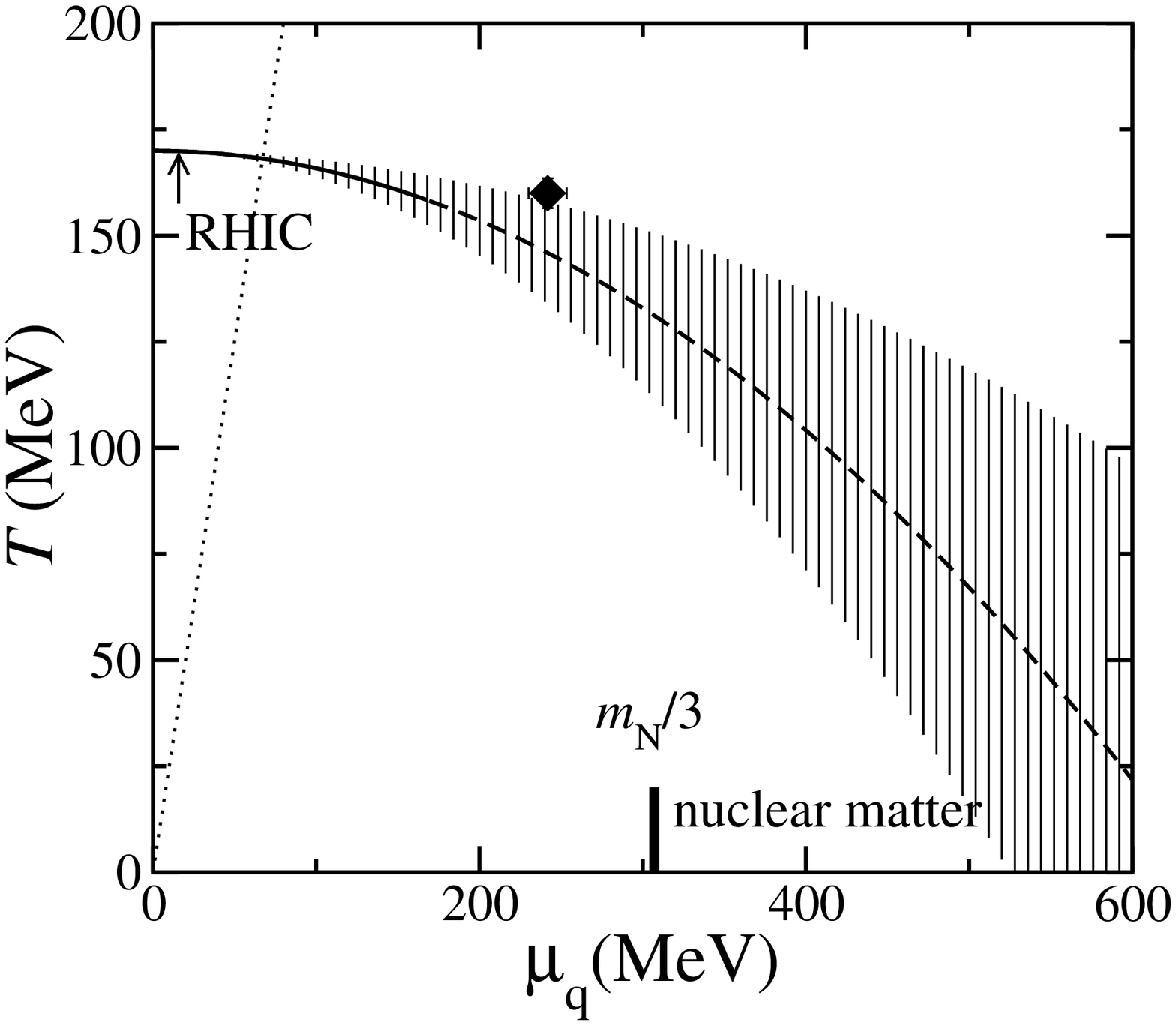,width=6cm}
  \caption{Left: Energy density over $T^4$ (should be flat for a relativistic ideal gas), calculated on the lattice as a function of
temperature, scaled with the critical temperature (where the sharp step occurs).  Right:  The phase diagram on the lattice. 
The diagonal dotted line shows the limit beyond which the curvature of $p/T^4$ can not be fitted.  The diamond refers
to the estimated critical point.
\label{lattice_phasediag}}
\end{figure}
Fig.~\ref{lattice_phasediag} (right) \cite{karsch3}  shows that the phase diagram mirrors very closely that we obtained in the  bag model picture
of Fig.~\ref{phasediag_bag}.
To compound this, Fig. ~\ref{lattice_phasediag} (left) shows that, above the rather sharp
critical temperature jump, the thermodynamics of the strongly interacting
system seems to be close to that of a relativistic ideal gas with perturbative corrections \cite{lattice_rafelski}.

In fact, as calculated in \cite{lattice_reso}, the hadron gas phase seems to be modeled closely by the resonance gas
ansatz used in this thesis, while perturbation theory to a few orders can describe the QGP
just above critical temperature reasonably well \cite{lattice_rafelski}.
\begin{figure}[h]
\centering
  \psfig{figure=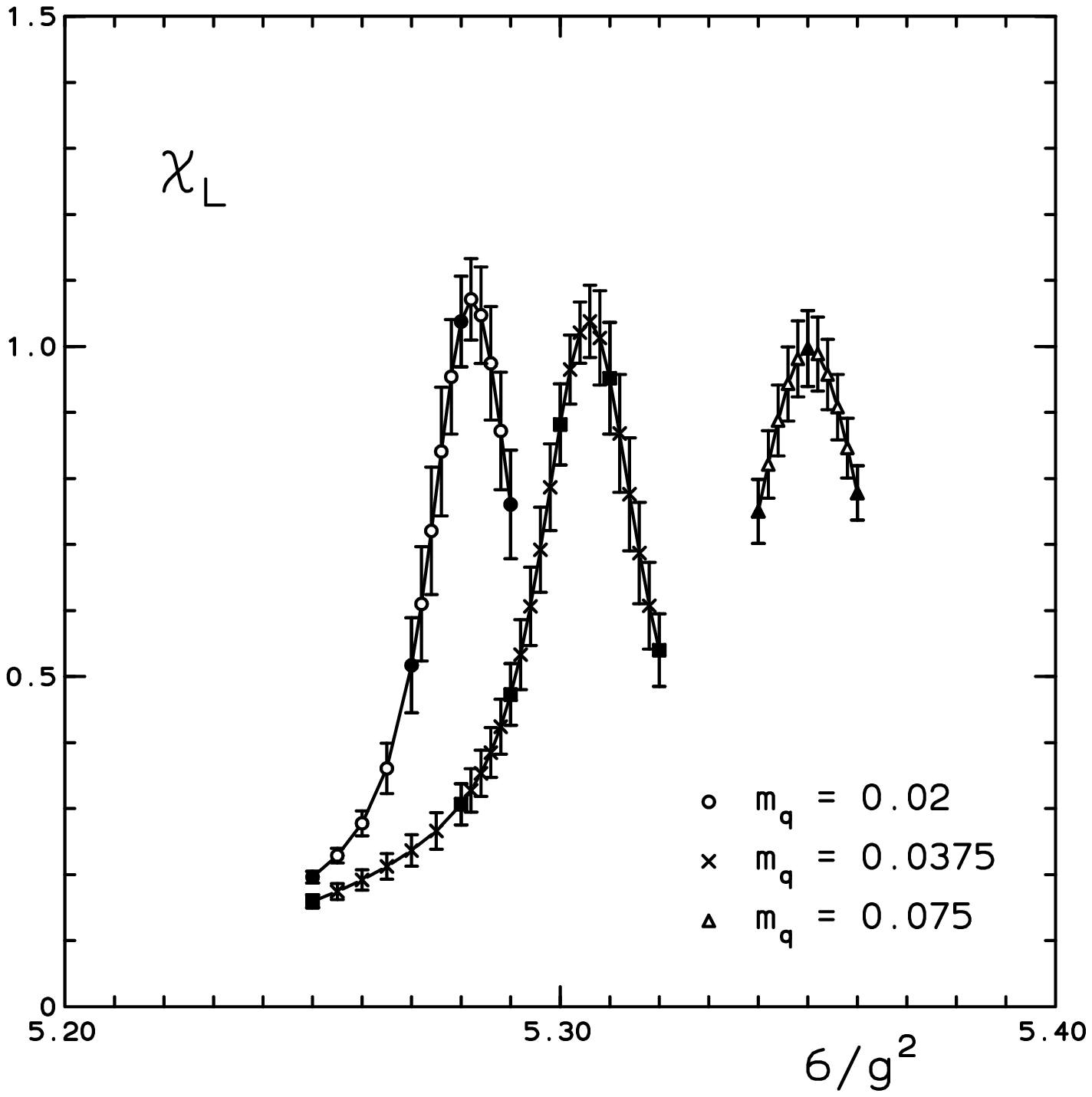,width=7cm}
  \psfig{figure=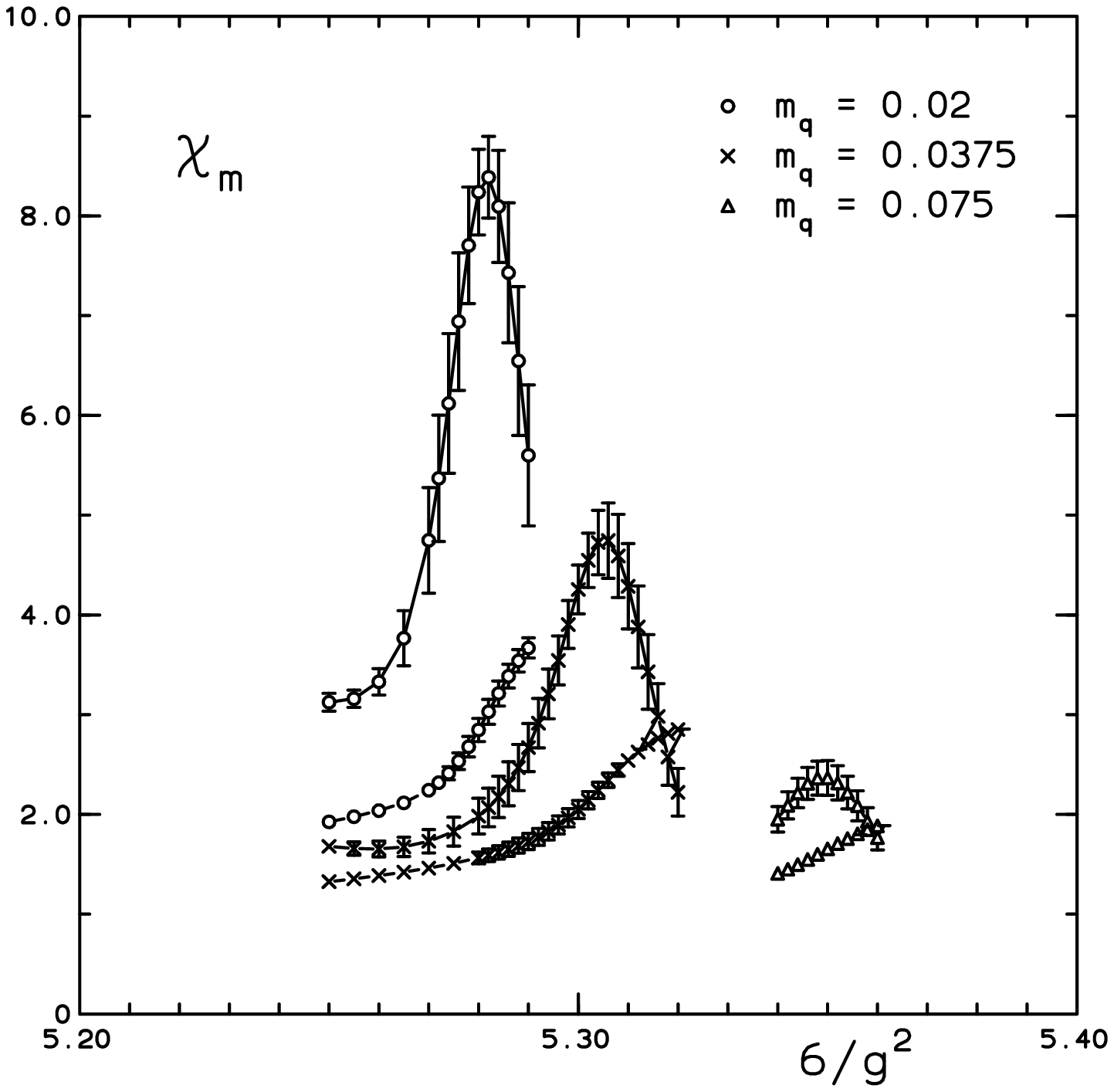,width=7cm}
  \caption{Polyakov loop susceptibility (left) and $\overline{\psi} \psi$ susceptibility vs $1/g^2$ ($\sim T$) for various quark masses (expressed
in terms of the inverse of the lattice size).  
\label{deconf_chiral}}
\end{figure}
Fig.~\ref{deconf_chiral} \cite{karsch1} can perhaps provide an indication of why this is the case.
As can be seen, the discontinuity in the Polyakov loop susceptibility $\chi_L$ and the $\overline{\psi} \psi$ susceptibility $\chi_m$
seem occur at the same temperature for a range of quark masses.   Hence, the onset of deconfinement also means that the
quarks lose their dynamically acquired ``constituent'' mass and become nearly massless.   This,
together with the color screening in a deconfined medium, means that
the QGP can be described by the ideal ultra-relativistic gas ansatz at temperatures
much lower than originally expected. 
\section{How to study the quark-gluon plasma experimentally}
We now proceed to give an overview of the main experimental issues raised
in the study of QGP.
The only way we can think of to make QGP in a laboratory is to collide two
energetic heavy ions (ideally large enough to make an equilibrated system described by the grand canonical ensemble).
Hopefully, the collisions will transform some of the energy of the heavy ion
into heat, resulting in a hot hadronic system in the center of mass frame.
If enough energy is produced, the system will then undergo a phase transition, evolve as quark-gluon plasma, change back into hadrons, and we will be able to tell  that a phase transition has taken place and to study the properties of deconfined matter through a careful analysis of
the decay products.

It is easy to see that in practice telling whether quark-gluon plasma has
been produced is not a simple matter.   Indeed, the two claims of quark-gluon plasma production ( \cite{cern_evidence} and \cite{rhic_evidence} ) are both based
on an assessment of different kinds of signatures. (And, it should be said, the kinds of signatures claimed by   \cite{cern_evidence} differ significantly from
those used in \cite{rhic_evidence}).
\begin{table}
\caption{Heavy ion Experimental program. \label{accellerators}}
\begin{tabular}{|l|l|l|l|l|}
\hline
Accelerator &Type & System & Energy (${\rm GeV}$/A) & Place \\
&&under study&(Center of Mass) &
\\ \hline
SIS & Fixed target & Various & 2-4  & GSI, Darmstadt
\\ \hline
AGS & Fixed target & Au-Au, S-Au & 2-11  & BNL, NY
\\ \hline
SPS  &Fixed target & C-C, S-S, Pb-Pb & 40, 80, 158 & CERN, Geneva 
\\ 
& & & (8.73, 12.3, 17.3 ) &
\\ \hline
RHIC &Collider  &Au-Au & 19.5, 130, 200  & BNL,NY
\\ \hline
LHC (2007) &Collider  & From p-p to Pb-Pb & 7000 & BNL,NY
\\ \hline
GSI (200...)& Collider & High $\mu_B$/density & In development & GSI, Darmstadt 
\\ \hline
\end{tabular}
\end{table} 
Table~\ref{accellerators} summarizes the experimental heavy ion collision programs
energies and nuclei.
Some accelerators are geared toward fixed-target experiments and
others are colliders.    The  advantage of the collider is that much
more energy is available to make a thermalized system.
The disadvantage, aside from the cost, is that it is a lot more difficult
to have an acceptance covering a large solid angle (rather than some region,
for example mid-rapidity).
Table~\ref{expts} provides a list of some heavy ion experiments.
In the next subsections, we will proceed to give a summary of the issues involved in each experiment.
\begin{table}
\caption{Selected heavy ion Experiments.   ``Telescope'' refers to a detector narrow in both angle and rapidity.  ``mid-rapidity'' to a 
detector covering rapidity region where the center of mass momentum vanishes.   ``Wide acceptance'' means an area outside
the mid-rapidity is covered. \label{expts}}
\begin{tabular}{|l|l|l|l|}
\hline
\underline{Name} & \underline{Location} & \underline{Predominant phase space} & \underline{What it measures}   
\\ \hline
KaoS, FOPI & SIS & Wide acceptance & Hadrons, Strangeness
\\ \hline
E864,E878,E886 & AGS & Wide acceptance & Hadrons, Strangeness  
\\ \hline
NA57  &  SPS & Telescope & Strangeness 
\\  \hline
NA49  &  SPS & Wide acceptance & All hadronic signatures 
\\  \hline
NA50, NA60 &  SPS & Telescope & leptons, $J/\Psi$
\\ \hline
WA98 & SPS & Telescope & $\gamma$ s, hadrons 
\\ \hline
CERES/NA45 & SPS & Mid-rapidity &$\gamma$ s, leptons 
\\ \hline
STAR & RHIC & Mid-rapidity & Hadrons, jets, strangeness
\\ \hline
PHENIX & RHIC & Mid-rapidity and telescope & Hadrons, jets, leptons, $\gamma$ s 
\\ \hline
PHOBOS & RHIC & Wide, mid-rapidity parts & Hadrons, jets 
\\ \hline
BRAHMS & RHIC & Telescope, wide rapidity & Hadrons, jets, strangeness
\\ \hline
ALICE & LHC & Mid-rapidity & Everything 
\\  \hline
CMS  & LHC & Mid-rapidity & $J/\Psi$, $\Upsilon$, $\gamma$ s, jets, leptons   \\ \hline
ATLAS  & LHC & Mid-rapidity & $\Upsilon$, jets   \\ \hline
\end{tabular}
\end{table} 
\subsection{Comparison benchmarks \label{comparison}}
The first issue to think about is a comparison benchmark.
To test for qualitatively new physics, the system under consideration has 
to be compared to a system in which the physics can be understood in terms of 
phenomena encountered previously.   In our case, this means a system with
low energy, or very few hadrons, in which we know tat quark-gluon plasma can not
be produced, and the dynamics can be understood either by
perturbative QCD or as a superposition of hadron-hadron collisions.
The latter can be analyzed in terms of elementary collisions measured
in control experiments, either through bulk analytical calculations
(e.g. the wounded nucleon model \cite{wnm})
or microscopic kinetic models  (e.g. Quantum molecular dynamics, or uRQMD \cite{uRQMD}
or hadronic string dynamics, or HSD \cite{hsd}).

The most obvious choice for a benchmark experiment is proton-proton collisions.
Any effects peculiar to Nucleus-Nucleus ($A-A$ or $A-B$) collisions can be parametrized through the nuclear modification
factor (normally referred to as $R_{AB}$, sometimes as ``enhancement'' or ``suppression'', depending on what the measured
or expected physics looks like):  
the ratio of a quantity observed in the nuclear collision to the same quantity observed in a $p-p$ collision normalized
by the product of the two nuclei mass numbers $A \times B$, or by the number of participants.   

Proton-proton collisions might not always be a good benchmark, since there can be large
volume or many-body effects which are not associated with the phase
transition but still have to be accounted for.   Hence, proton-nucleus, deuteron-nucleus or collisions between light nuclei 
are also studied.   In these, there is a large volume of excited matter and many collisions, but (presumably) no phase transition.
\begin{figure}[h]
\centering
  \psfig{figure=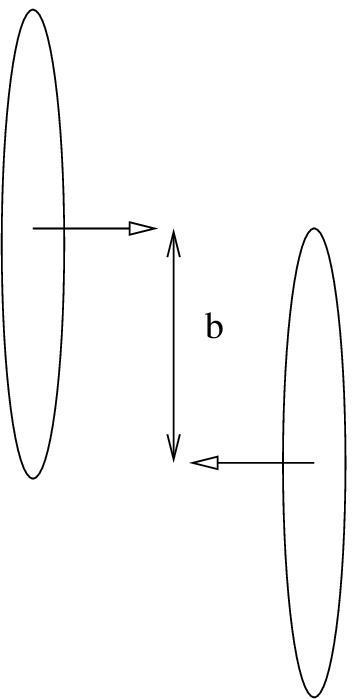,width=3cm}
  \psfig{figure=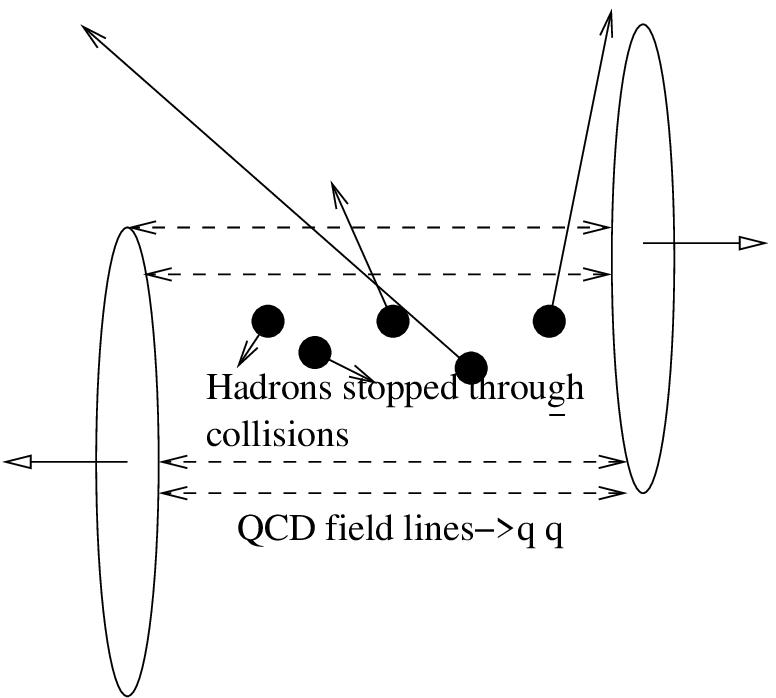,width=6cm}
  \caption{Left: The definition of impact parameter in a colliding system.
Right: How energy is distributed after the collision. ``Stopping'' refers to the dissipation of the system's initial longitudinal momentum.\label{collision} }
\end{figure}
More generally, it is possible to analyze observables in terms of
collision centrality (number of participants) or collision energy.
This way, one can investigate if there is a critical energy (roughly correlated
with temperature) or number of participants (roughly correlated
with reaction volume and evolution time) at which qualitatively new behavior (e.g., a phase transition) occurs.
A measure of the collision centrality is provided by the number of participating nuclei, related to the impact parameter, $b$, Fig. (\ref{collision} Left).
$b$ can be found either directly, by having a detector placed downstream from
the collision point to measure how many nucleons escaped the primary collision,
or indirectly, by simply measuring the total multiplicity (which is related to the volume of the thermal system).

It should be noted, however, that increasing the collision energy leads to some
non-trivial consequences which should be taken into account.
First of all, higher collision energy increases the role of perturbative
QCD within the system.
In the non-perturbative limit, the particles will undergo repeated soft interactions after the primary collision.  These interactions will stop the particles
in the collision region and transform their initial colliding energy into
thermal energy.
Hence, the size of the colliding system, and ultimately the total number of particles produced, will scale as the number of participants
\begin{equation}
\label{glauber}
N_{prod} \propto N_{participants} 
\end{equation}
As the collision energy increases, perturbative interactions
will play a bigger role.
This means that collisions, rather than participants, will play a larger role in stopping the system, and the scaling
becomes (parametrized by $\alpha$,$\beta$)
\begin{eqnarray}
\label{hard_glauber}
N_{prod} = \alpha N_{participants}+ \beta N_{collisions}\\
N_{collisions} \sim N_{participants}^{4/3}.
\end{eqnarray}

Due to the large center of mass momentum and asymptotic freedom, stopping in primary collisions will decrease with increasing energy.
Two highly energetic heavy ions will pass each-other nearly transparently.
Their energy will be released into the medium due to confinement, as the color
fields generated by the initial collision interactions become stronger and ultimately
melt into incoherent $q \overline{q}$ pairs  (see Fig.~\ref{collision} (right). The two highly Lorentz-contracted nuclei can be thought of as capacitors of the color field).
For this reason, heavy ion systems colliding at high energy also tend to
have a lower baryon density and chemical potential.

These considerations also control the likely initial conditions
for hydrodynamic evolution.   A system with high baryon stopping will evolve
from the pancake-shaped collision region, localized in rapidity
space around the center of mass frame.   This system has been originally analyzed as a starting point for
hydrodynamic evolution by Landau \cite{Lan53}.  In the infinite time
limit, it should give rise to an elliptical fireball with comparable
transverse and longitudinal flow profiles, which at mid-rapidity can be well approximated by a spherically symmetric freeze-out.

By contrast, the thermalized system formed when the two very energetic nuclei pass through each other
will have a high degree of boost-invariance.  As argued in \cite{bjorken_boost}, two highly Lorentz-contracted
``pancakes'' passing nearly transparently through each-other will look the same in a boosted reference frame.
Hence, the dynamics of this collision should exhibit a rapidity plateau.
As shown in \cite{bjorken_boost} this initial condition is hydro-dynamically stable.

\begin{figure}[h]
\begin{center}
\epsfig{width=7cm,clip=1,figure=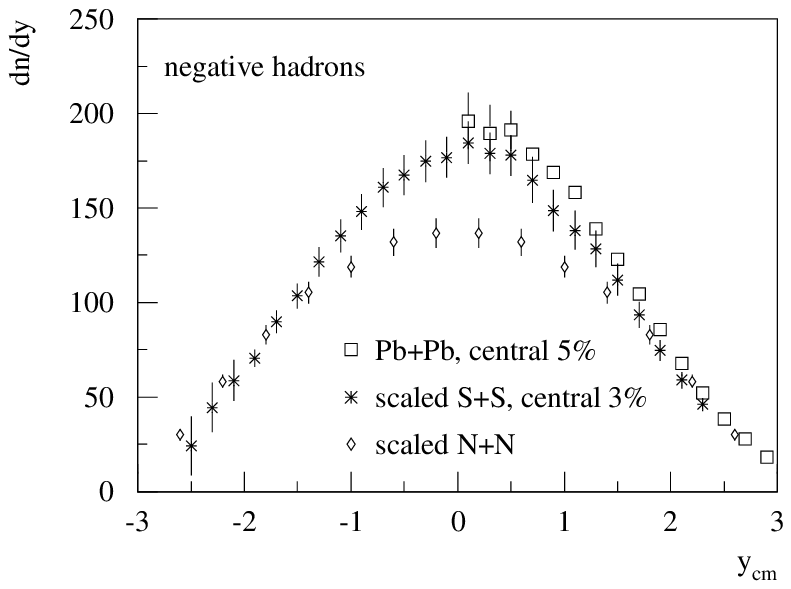}
\epsfig{width=7cm,clip=1,figure=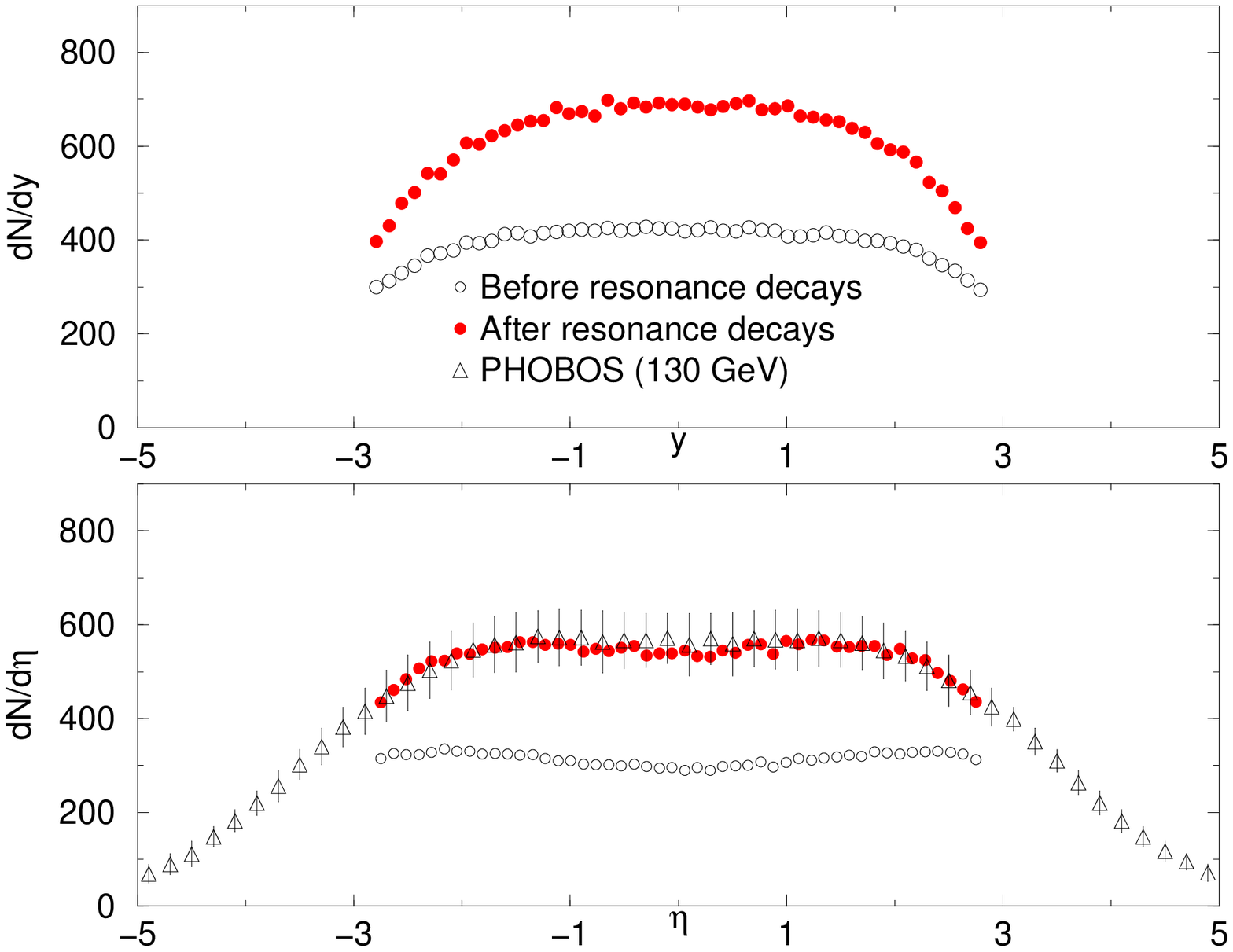}
\vspace*{-0.21cm}
\caption{Left: SPS 19.4 ${\rm GeV}$/A charged particles pseudorapidity distribution measured by the NA49 collaboration  Right:  RHIC 130 ${\rm GeV}$/A charged particle pseudorapidity distribution measured by PHOBOS, and compared to a boost-invariant statistical model (see chapter 4).  \label{landaubjorken}}
\vskip -0.5cm
\end{center}
\end{figure}

Fig.~\ref{landaubjorken} shows that the two limits can be used as an ansatz for, respectively, SPS and RHIC energies.   At the SPS  \cite{NA49stop}, the pseudorapidity distribution is approximately Gaussian, corresponding to a sharp longitudinal structure.
Hence, we have used a spherical flow profile to model SPS mid-rapidity data
in chapter 3.
As the right panel of Fig.~\ref{landaubjorken} shows, the situation is perhaps
different at RHIC.   We have used the Monte-Carlo described in detail in chapter 4 to verify that the flat pseudorapidity distribution observed by PHOBOS  \cite{phobos} can be reproduced by a boost-invariant source.   Such a source
is described in detail, and used to model RHIC data, in chapter 4\footnote{Recent data from BRAHMS \cite{brahmsgauss} may have thrown the validity of this picture in doubt, as the rapidity distribution of identified $\pi$ is approximately a Gaussian, similar to \cite{NA49stop} and predicted by the Landau model.}.
\subsection{Jet quenching}
Jets, or energetic streams of hadrons resulting from a perturbative QCD interaction,
are a promising test of whether the dense system formed in a heavy ion collision exhibits hadronic or partonic degrees of freedom.
A parton propagating through a quark-gluon plasma can lose a lot of energy
quickly due to repeated interactions with soft gluons.
This energy loss is especially pronounced due to the quantum
interference of interactions between many coherent gluons (this is known as the Landau-Pomeranchuk-Migdal effect) \cite{gyulassy1}.
The quark-gluon plasma is therefore expected to be much more opaque to jets than ordinary
nuclear matter.
One way to quantify such opacity is to measure jet azimuthal correlation, and compare
it to elementary (proton-proton or proton-nucleus) collisions. 
\begin{figure}[h]
\centering
  \psfig{figure=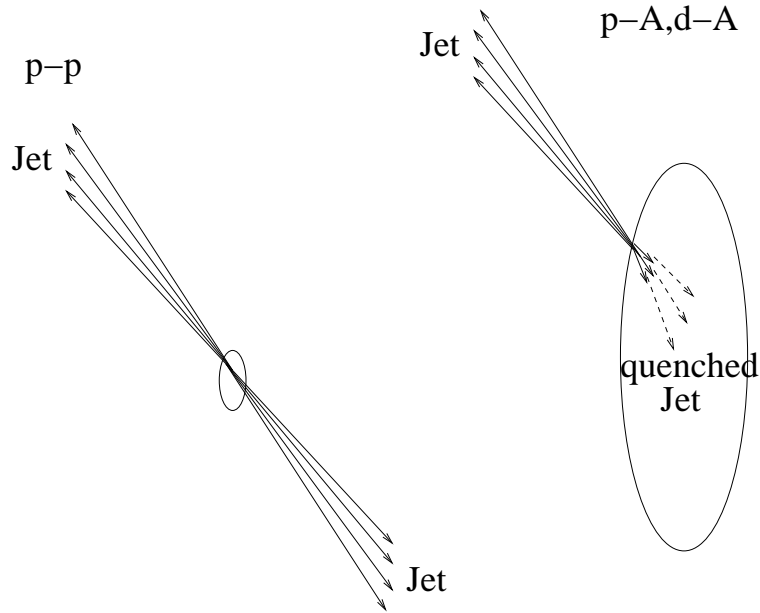,width=10cm}
  \caption{How an opaque system leads to jet correlations \label{jet_correl}. }
\end{figure}
As Fig.~\ref{jet_correl} shows, in an opaque system one of a pair of jets should become quenched, and hence unobservable.
For this reason, jets should lose the expected azimuthal correlation.
In case of non-central collisions, this effect will also acquire an angular
dependence \cite{azijets}.  
\begin{figure}[h]
\centering
  \psfig{figure=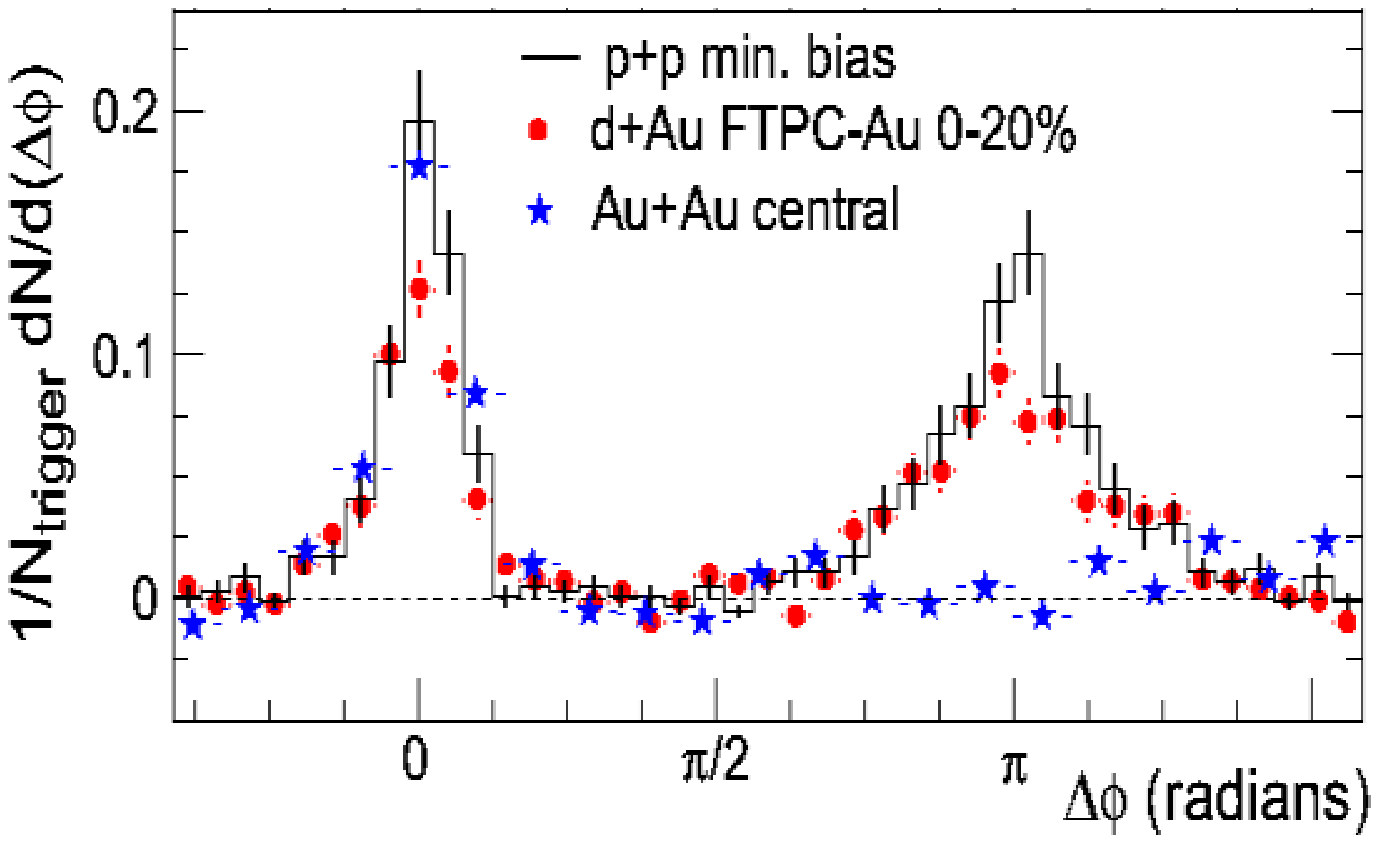,width=7cm}
  \psfig{figure=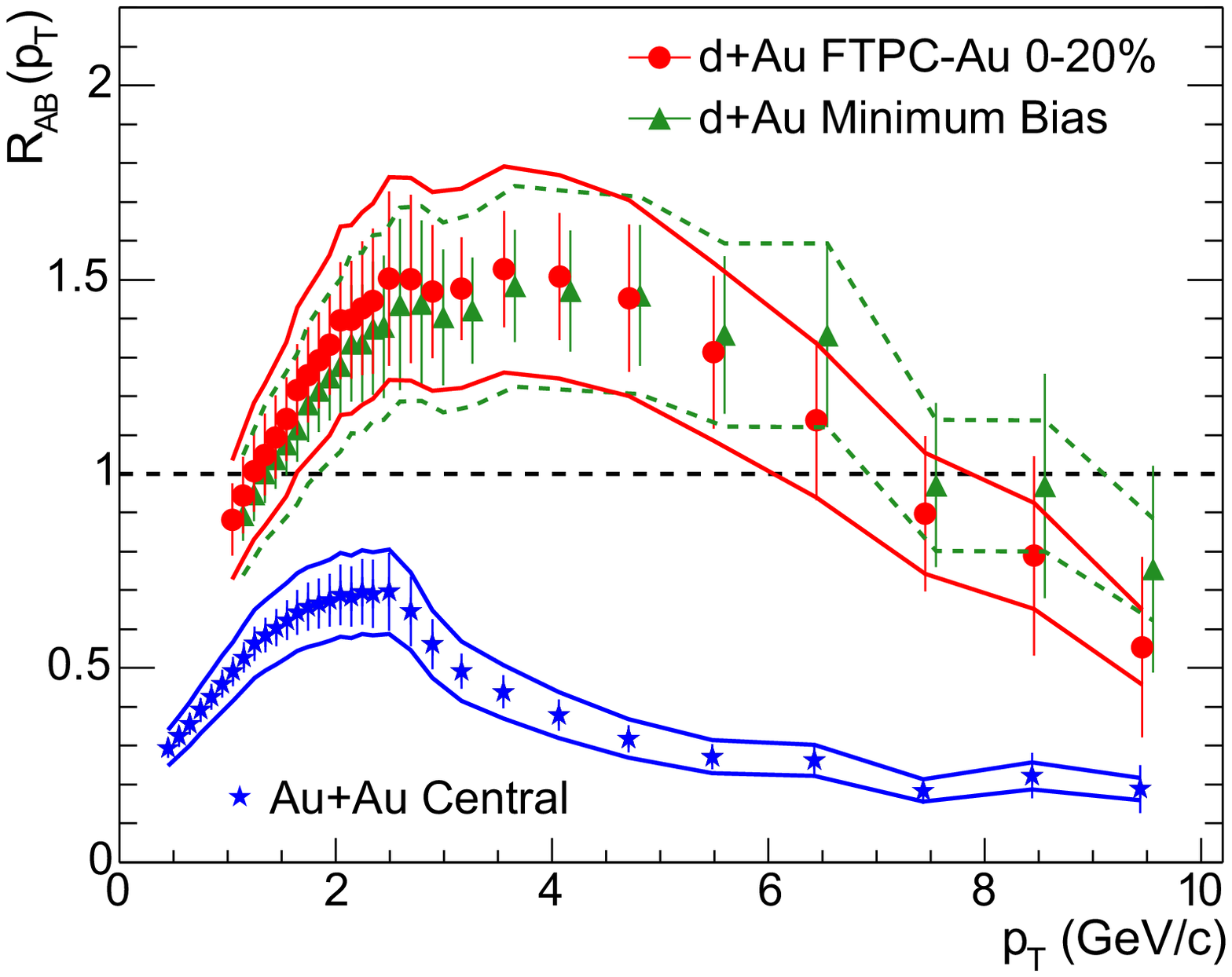,width=7cm}
  \caption{Left: jet azimuthal correlation (number of jet pairs emitted at an angular separation $\Delta \phi$) normalized by the number of triggered jets, as measured by the STAR collaboration in p-p, d-Au and Au-Au collisions and 200 ${\rm GeV}$. Right:  The nuclear modification factor (as defined in section \ref{comparison}) for d-Au and Au-Au collisions, measured by STAR.  The curves refer to calculations done within a model where
partonic thermalized degrees of freedom are assumed.   Minimum bias means that no centrality selection was performed.\label{quenchfigs} }
\end{figure}

As Fig.~\ref{quenchfigs} (left) shows, this effect has received a remarkable
experimental confirmation \cite{jet1}.
The observed loss of azimuthal correlation is qualitatively
unique to Au-Au collisions, the jets in both p-p and d-Au collisions
being perfectly correlated.
As  Fig.~\ref{quenchfigs} (right) shows, the loss of azimuthal correlation
is accompanied by the loss of hard (jet-energy) particles, with respect to
an extrapolation of hard particles produced in p-p collisions.  This is also unique to Au-Au, hard particles in d-Au being enhanced.

While some features of the data have recently been modeled within a hadronic scenario
with strong in-medium modifications \cite{capella}, the full high $p_T$ dataset
from p-p, Au-Au and d-Au collisions has only been convincingly described
in a model which assumes that the matter
at the center of Au-Au collisions exhibits colored degrees of freedom and is 100 times the density of normal nuclear
matter. Furthermore,  the non-decrease of the jet suppression with $p_T$ could
only be modeled by taking the Landau-Pomeranchuck-Migdal
effect into account \cite{gyulassy2}.

While the evidence described above makes jet quenching widely regarded as the definitive proof
of QGP formation  \cite{rhic_evidence}, jet quenching as a QGP diagnostic has limitations.    The information
it provides about QGP equilibration, kinetics and equation of state
is rather limited.   Jets will not evolve with the bulk of the matter,
and their hadronization, due to asymptotic freedom, will not impact the observed energy-momentum distribution significantly.
While jets can sample some extensive quantities, such as the density of matter they traverse, soft physics
is needed to understand if the system having those extensive
quantities is equilibrated in a particular phase, or if it is in a different phase from the usual hadronic matter
\cite{gyulassy1}.   Soft physics is also needed as a complement to
perturbative QCD to account for the missing jet energy.
Modeling quantitatively how the jet energy is distributed among the soft degrees of freedom, and what the effect of the deposited energy is, is an interesting
theoretical problem potentially rich in insights into fundamental physics.     However, an understanding
of the soft degrees of freedom is essential to solve it.

Hence, while jet quenching might convince us that QGP is there, it is
not a very good tool for measuring its properties and how it changes
into normal matter.     For this, probes more dependent on soft physics
are required.
\subsection{Direct photon and dilepton production}
Historically, the standard way to measure the temperature of a hot medium has been to measure the thermal spectrum of its emitted light.
This can be expected to be true for a QGP as well, given the abundance
of scattering reactions with quarks and gluons \cite{kapusta1}.
Of course, the processes under consideration here are energetic enough to allow
the emitted photons to produce lepton-antilepton pairs \cite{kapusta2}.
Photons ($\gamma$) and leptons antilepton pairs ($l \overline{l}$) will be emitted in reactions such as
\begin{equation}
\Diagram{\vertexlabel^{g} gd &  f & hu \vertexlabel^{\gamma} \\\vertexlabel^{q}  fu &  &  fd \vertexlabel^{q}
}  \; \; \; \;
\Diagram{\vertexlabel^{q} fd &  h  & fu \vertexlabel^{l} \\ \vertexlabel^{\overline{q}} fu &\vertexlabel^{\gamma} &  fd \vertexlabel^{\overline{l}} } 
\; \;  \; \;
\Diagram{\vertexlabel^{g} gd & \vertexlabel_{q}  f & fu \vertexlabel^{q} \\ fu &  & hd  \vertexlabel_{\gamma} & h &  fu \vertexlabel^{l} \\
 & & & & fd \vertexlabel^{\overline{l}} } \label{dilep_reactions}
\end{equation}
The particularly attractive feature of this signature is that photons and
dileptons always emerge from the system without undergoing further not-well understood
interactions (energy loss, hadronization etc.)
However, the enormous background due to the many short-lived mesons that decay into
 photons ($\rho, \omega, \eta,...$) makes extricating the signal from the background experimentally very difficult.
\begin{figure}[h]
\centering
 \psfig{figure=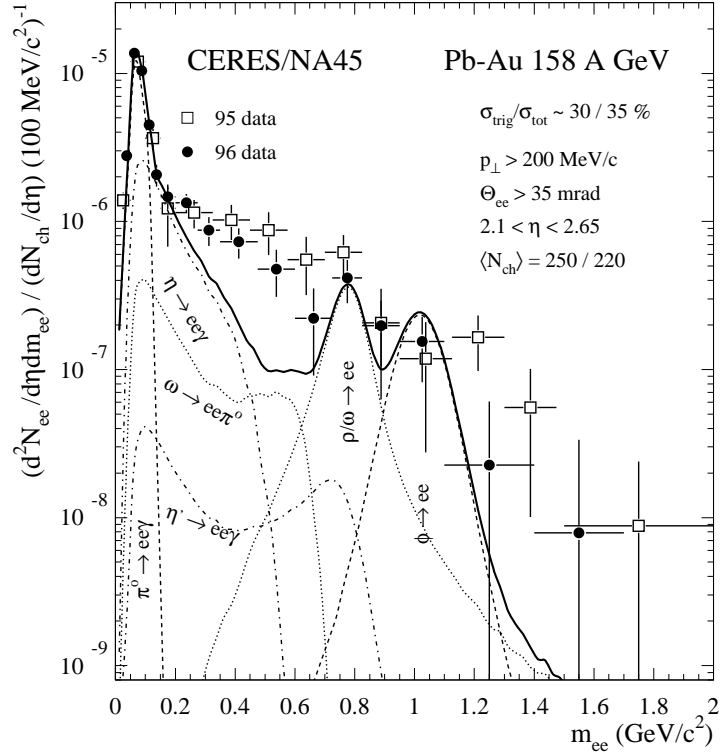,width=10cm}
  \caption{The di-electron invariant mass ($m_{ee}$) distribution observed at CERES, per unit of pseudorapidity ($\eta$) and scaled with the total
pseudorapidity density of observed charged particles $d N_ch/d \eta$. The dotted, dashed and dot-dashed lines represent the expected distribution
from electromagnetic decays of known resonances.    An excess is observed at $m_{ee} \sim 0.7$. \label{ceres} }
\end{figure}
The most interesting result which has come out of this approach so far is CERES's
report of observing a dilepton excess \cite{ceres} (see Fig.~\ref{ceres} ).
This result was presented by CERN as part of the evidence for its announcement of having found deconfined matter \cite{cern_evidence}.
However, it is unclear whether it signals a dilepton excess due to QGP or, rather, a shift in the $\rho$ peak due to partial chiral symmetry restoration.
Measurement of photons/dileptons is part of the planned ALICE program, where
it can be a very useful probe if correlated with jet-quenching \cite{gale}. The energetic
reactions in Eq.~(\ref{dilep_reactions}) could produce a detectable photon/dilepton
pair together with either a detectable, partially quenched, or totally quenched jet.  The correlation
will then be a powerful tomography tool.
\subsection{Charmonium suppression}
\begin{figure}
\centering
  \psfig{figure=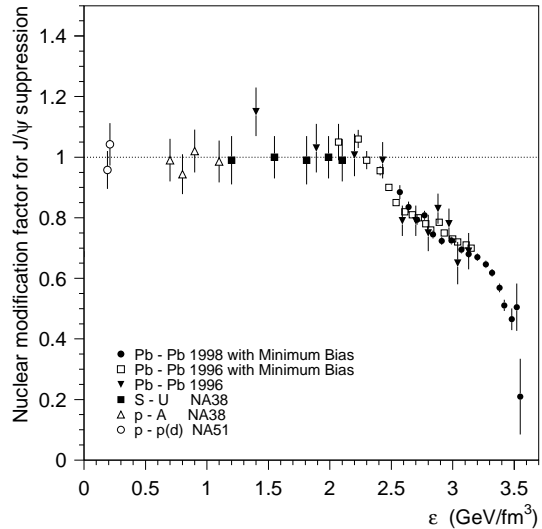,width=7cm}
  \caption{Nuclear modification factor (See section\ref{comparison}) of charmonium particles (detected through the $J/\Psi \rightarrow \mu^+ \mu^-$ decay)
  It is plotted against the energy density calculated within the Bjorken model ($\propto R_0^2 dN/dy$, where $R_0$, the overlapping nuclear area, depends
on centrality).
Minimum bias means that no centrality selection was performed. \label{charmexp}}
\end{figure}
One of the original ways to look for deconfinement is through Charmonium
suppression \cite{charm0}. 
Charmonium (bound $c \overline{c}$ states, such as $J/\Psi,\Psi',...$) can be produced in the initial reactions of an intermediate energy heavy ion collision.
Normally, these particles are very narrow (tightly bound) states. 
However, in a deconfined medium, charmonium pairs will melt due to the color screening provided by the free quarks
and gluons.     Moreover, different bound states will
melt at different temperatures/densities, thus providing an effective
QGP thermometer.

The NA50 experiment has indeed detected such a suppression \cite{charm1} (see Fig.~\ref{charmexp})
which is absent in proton-proton collisions and can not be explained through standard nuclear absorption models.
This absorption was one of the pieces of evidence CERN put forth in its
announcement of quark-gluon plasma discovery \cite{cern_evidence}.

However, the NA50 experiment cannot detect total charm, but only $J/\Psi \rightarrow \mu \mu$ reactions.
Hence, it is possible to adjust nuclear absorption models through in-medium
mass modification rather than deconfinement.
$J/\Psi$ production is particularly sensitive to such mass shifts since a small modification
of its mass will make the  $J/\Psi \rightarrow D \overline{D}$ decay possible.
In fact, a nuclear absorption model with $J/\Psi$ mass modification
does manage to explain the observed $J/\Psi$ suppression \cite{charm2}.
The story is not completely over.   Other than the fact that the $J/\Psi$ in-medium modification, though reasonable, has never been observed, the nuclear
absorption model still can not account for the suppression of the $\Psi'$.

The NA60 experiment, which will also measure open charm, might shed
some light on what exactly causes $J/\Psi$ absorption at SPS energies. 
Charm data is forthcoming from RHIC and, when it turns on, the LHC (both of which also measure open charm).

As energy increases, we will reach a point where several charm pairs
will be produced in each collision.   This might change things considerably,
since a QGP phase will dissolve the existing $c \overline{c}$ pairs, but will
also allow $J/\Psi$ s to form from initially uncorrelated  $c$ and $\overline{c}$ at hadronization \cite{charm3}.
This means that the number of $J/\Psi$ will become
\begin{equation}
\label{charm_enhancement}
N_{J/\Psi}=A N_{c \overline{c}} + B N_{c \overline{c}}^2
\end{equation}
($B=0$ if only one  $c \overline{c}$ pair is expected per collision).
If enough  $c \overline{c}$ s are produced, we might observe an enhancement
of charmonium produced in heavy ion collisions instead of a suppression \cite{charm3}.
This is especially true if charm and bottom quark states (such as the $B_c$) are also considered, since production of such a state
requires initially uncorrelated quark pairs to fuse \cite{thews2}.
\subsection{Strangeness enhancement}
Strangeness enhancement has long been considered one of the most promising
signatures of QGP formation \cite{strange_rafelski1}, as well as a useful tool to study soft matter
produced in a heavy ion collision \cite{strange_rafelski2}.
The basic idea is that $s \overline{s}$ pairs should form much more readily in a QGP than in nuclear collisions through reactions
such as
\begin{equation}
\Diagram{\vertexlabel^{q} fd &  g  & fu \vertexlabel^{s} \\ \vertexlabel^{\overline{q}} fu & \vertexlabel^{g} &  fd \vertexlabel^{\overline{s}} } 
\; \;  \; \; \; \;  \;
\Diagram{\vertexlabel^{g} gd &  g  & fu \vertexlabel^{s} \\ \vertexlabel^{g} gu &\vertexlabel^{g} &  fd \vertexlabel^{\overline{s}} } 
\; \;  \; \; \; \;  \;
\Diagram{\vertexlabel^{g} gd  & fu \vertexlabel^{s} \\& fv   \\\vertexlabel_{g} gu  &  fd \vertexlabel^{\overline{s}} } 
\; \;  \; \; \; \;  \;
\end{equation}
Since gluons are massless and the mass of the light quarks is much less than the QGP temperature, the threshold for forming strange quarks is $2 m_s \sim 100-300 {\rm MeV}$,
boosted by chiral symmetry restoration and the abundance of $q \overline{q}$ pairs.    
In a hadron gas, on the other hand, the threshold energy for the leading strange producing processes
\begin{equation}
\Diagram{\vertexlabel^{p} fd & fu \vertexlabel^{\Lambda} \\ \vertexlabel^{\pi} hu &  hd \vertexlabel^{K} } 
\; \;  \; \; \; \;  \;
\Diagram{\vertexlabel^{\pi} hd & hu \vertexlabel^{K} \\ \vertexlabel^{\pi} hu &  hd \vertexlabel^{K} } 
\end{equation}
is, respectively, 600 and 800 ${\rm MeV}$ (note that the first process is also suppressed in a baryon-poor environment).

\begin{figure}[h]
\centering
  \psfig{figure= 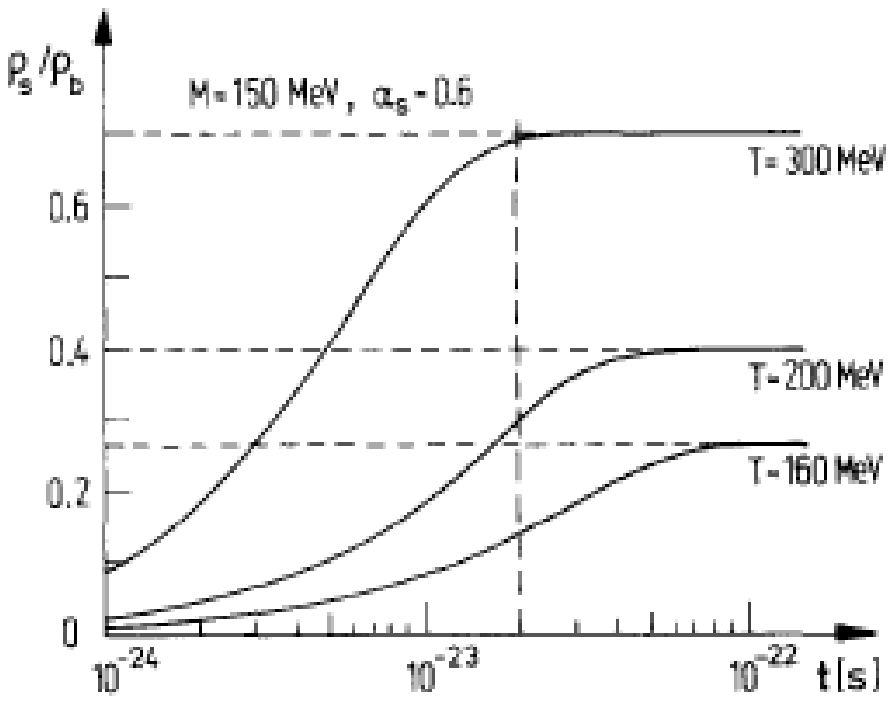,width=5cm}
  \psfig{figure=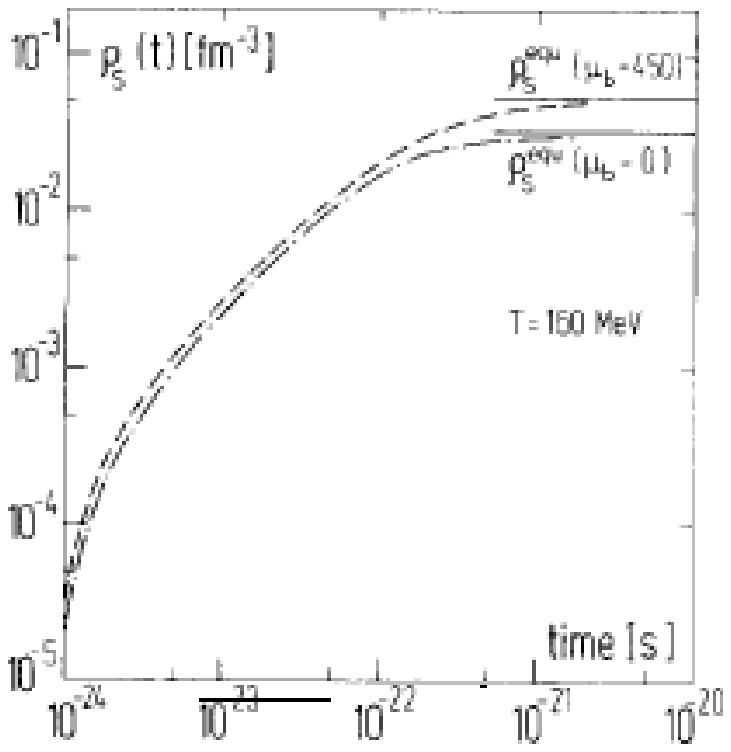,width=5cm}
  \psfig{figure=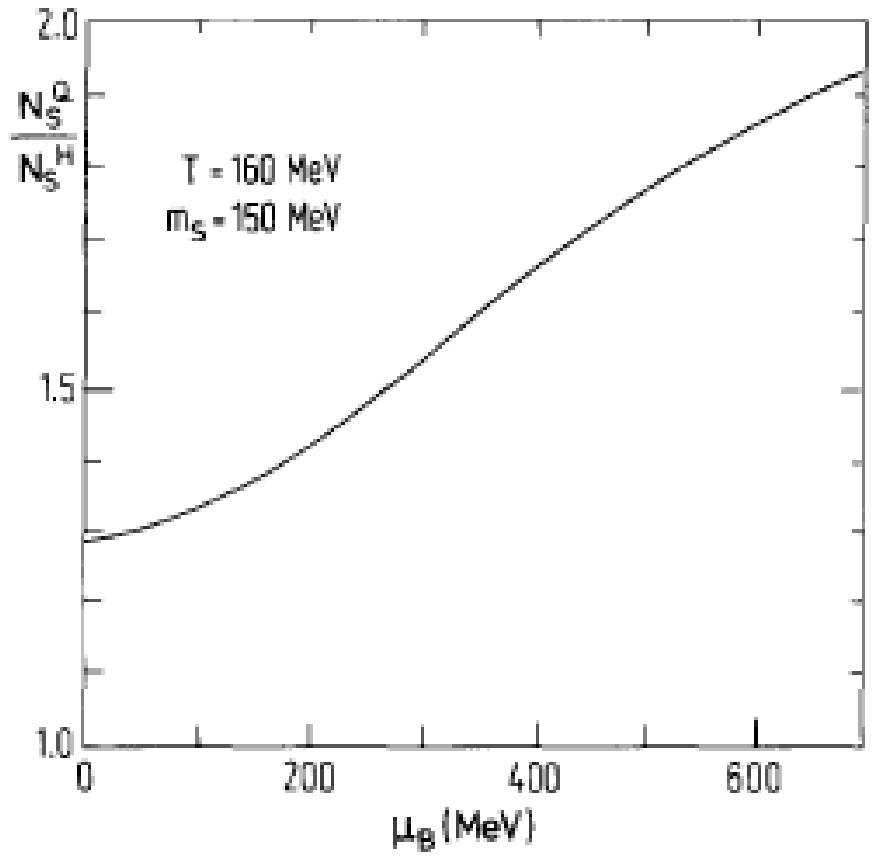,width=4cm}
  \caption{Density of strange quarks with respect to time calculated in a QGP (left, normalized to baryon density with $m_s$=150MeV) and a hadron gas (middle) kinetic
production models.   As can be seen, equilibration time is significantly shorter for the QGP phase.  In addition, as the panel on the right shows, equilibrium strangeness quark content will be stronger in the QGP phase at all chemical potentials.\label{strangeness_formation} }
\end{figure}

\begin{figure}[h]
\centering
  \psfig{figure=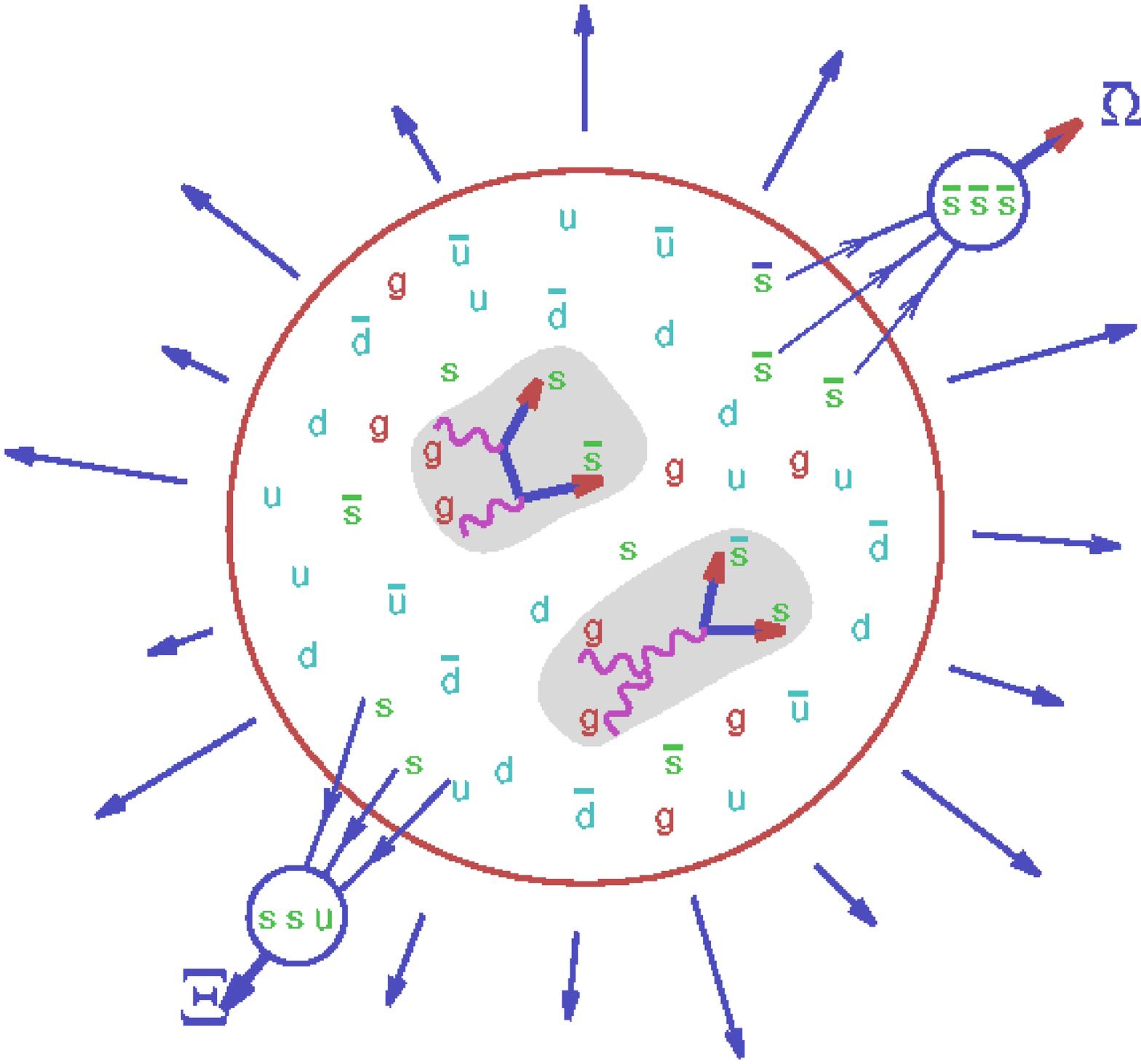,width=6cm}
  \psfig{figure=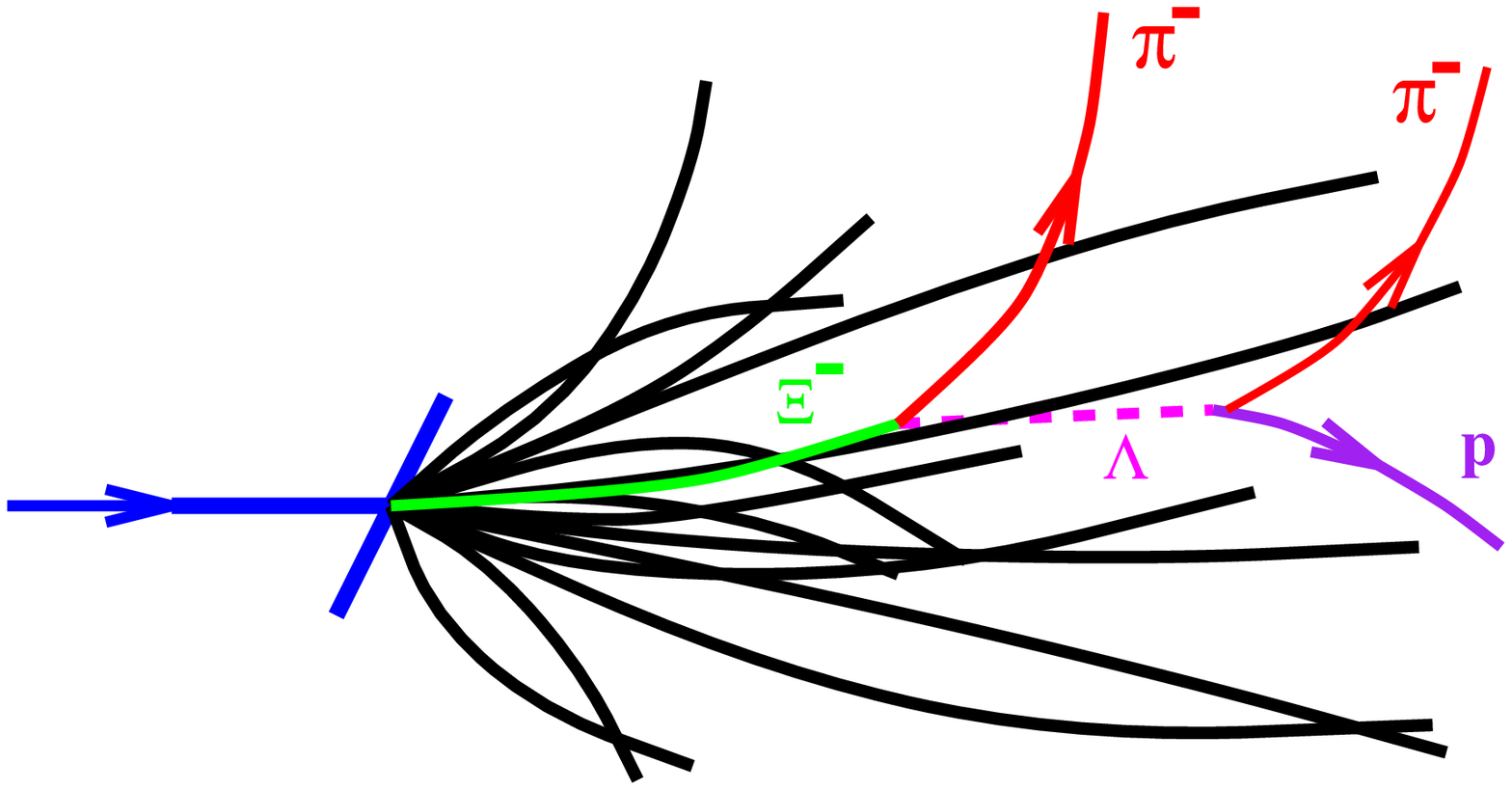,width=6cm}
  \caption{Left: Production of multi-strange hadrons through combination of strange quarks from a QGP.  \label{strange_enh}
Right: Detection of strange particles from a heavy ion collision due to decay topology. }
\end{figure}

\cite{strange_rafelski1,strange_rafelski4} has shown (Fig.~\ref{strangeness_formation}) that in a thermalized gluon-rich perturbative
QGP the $s \overline{s}$ production rate is as much as an order of magnitude
greater than the strangeness production rate in a hadron gas at a similar temperature, with gluon-gluon fusion reactions accounting for a large
majority of the production.
Due to the faster equilibration time and greater equilibrium strange quark density, therefore,  a system which has undergone a QGP phase transition will exhibit
an enhancement of strange quarks with respect to a system which has not. 
This enhancement will translate into an even greater enhancement of strange hadrons after hadronization.
Multi-strange hadrons ($\phi,\Xi,\Omega$) and their anti-particles will be particularly enhanced, since
their production in a hadron gas has a particularly high energy threshold
($p \overline{p} \rightarrow \Omega \overline{\Omega}$, requiring protons with momentum of $700 {\rm MeV}$ each) or a sequence of many reactions (e.g. $\pi \overline{p} \rightarrow K \overline{\Lambda},\pi \overline{\Lambda} \rightarrow K \overline{\Xi},\pi \overline{\Xi} \rightarrow K \overline{\Omega} $).   In a hadronizing strangeness-rich QGP, on the other hand, multi-strange baryons form by recombination of strange quarks, and should not be as suppressed according to most hadronization models (Fig.~\ref{strange_enh} left).

Hence, to test for QGP one has to look for strange particles
in a nucleus-nucleus collision and compare with p-A or p-p.
A clear enhancement, significantly raising with particle strangeness content, 
would constitute evidence for deconfinement.
Experimentally, this is facilitated by the fact that strange particles
decay weakly, with a lifetime comparable to the time of flight.  Hence, they can be reconstructed through decay topology (Fig.~\ref{strange_enh} right).

\begin{figure}[h]
\centering
  \psfig{figure=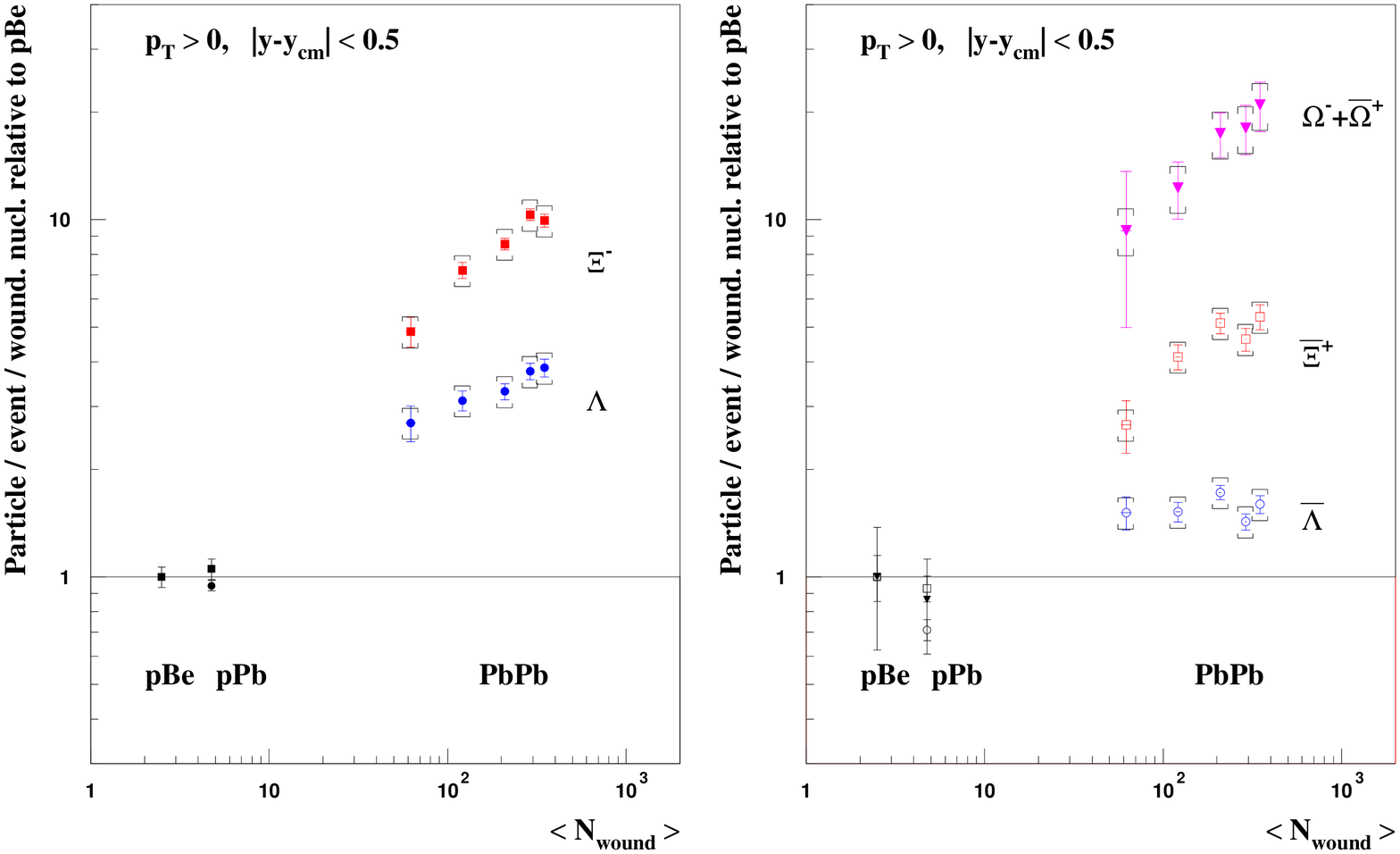,width=12cm}
  \caption{The nuclear modification factor (see section \ref{comparison}) for the production of hyperons as measured by the NA57 experiment, normalized by the number of participants.  Particle yields for p-p and Pb-Pb collisions are compared to p-Be, plotted against the number of participants, calculated within the Wounded nucleon model (see section \ref{comparison}) \label{exp_enhance}.}\end{figure}

As Fig.~\ref{exp_enhance} shows, a very clear experimental enhancement
has indeed been detected by the WA97 and NA57 collaborations \cite{wa97_enh}
(whose narrow-acceptance telescope is particularly suited to look for multi-strange particles).    The enhancement in all strange particles, and its rise
with strangeness content, constituted a crucial part of the supporting evidence
for CERN's claim to have produced deconfined matter \cite{cern_evidence}.
Microscopic nuclear interaction models have so far been unable to
even come close to reproducing the observed enhancement
\cite{urqmd_enhancement}.
\begin{figure}[h]
\centering
  \psfig{figure=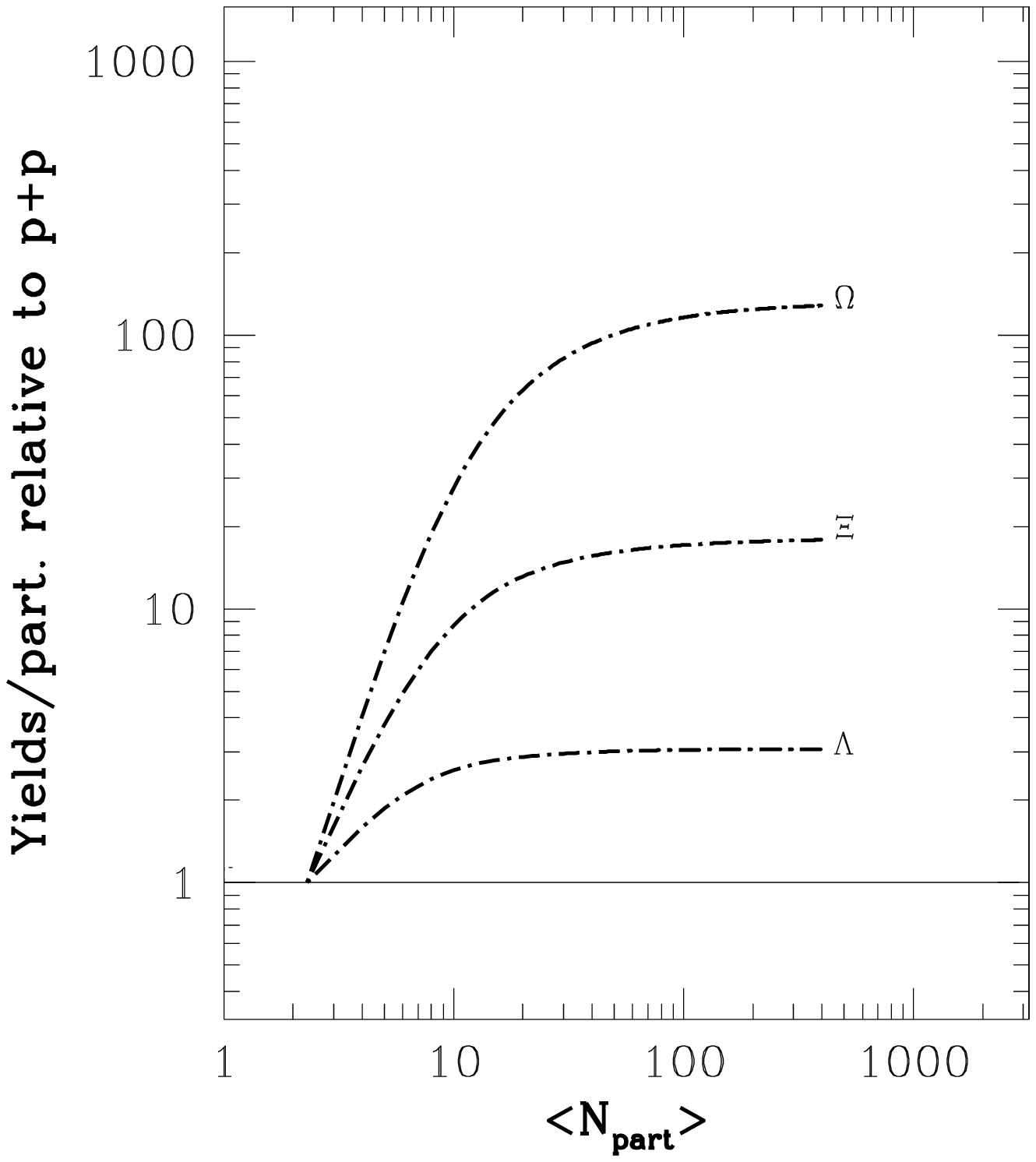,width=7cm}
  \psfig{figure=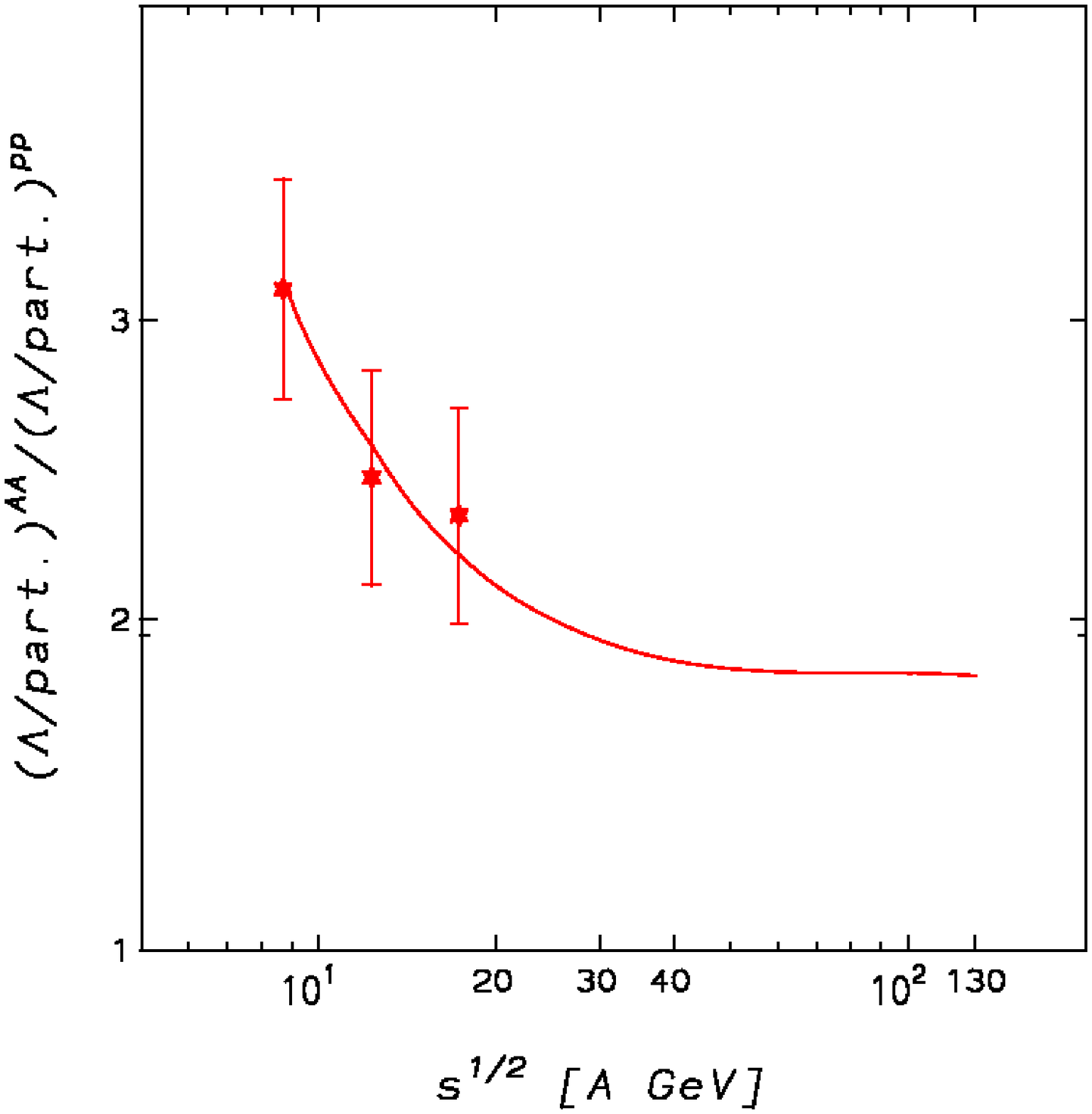,width=5cm}
  \caption{Nuclear modification factor for $\Lambda$ (right) and $\Lambda,\Xi,\Omega$ (left) production
calculated within a canonical statistical hadronization model.
The result is plotted against the number of participants $N_{part}$ (left) and center of mass collision energy per nucleon (right).
 \label{can_enhance} }
\end{figure}

One alternative explanation which would account for the observed enhancement
pattern is canonical suppression  \cite{canonical}.
Postulating that collisions in p-p and p-A systems achieve chemical
equilibration, the small volume of these systems necessitates that the canonical (rather than grand-canonical) ensemble be used to calculate strange
particle yields.   This means that, rather than introducing a chemical
potential for strangeness, only outgoing states with exact strangeness
conservation are counted in the statistical averaging.
As Fig.~\ref{can_enhance} shows, this leads to a sharp drop in strangeness
as system volume decreases below a critical value, with the effect 
being greater as strangeness content increases.

This explanation, however, has several problems.
For a start, it is not at all clear that equilibrium statistical mechanics can be
used to describe a p-p and p-A system. 
NA57 p-p and p-Pb data are well described through the
uRQMD nuclear microscopic model \cite{urqmd_enhancement}, where chemical equilibration is not assumed
or indeed predicted.

In addition, as Fig.~\ref{can_enhance} (left) shows, the pattern of enhancement is not predicted correctly by the canonical picture:  the canonical picture
has a sharp suppression at small volume and a long plateau as volume
becomes large and the canonical and grand-canonical pictures coincide.
Experimentally, however, strangeness scales linearly with number of participants in all (p-p,p-Be,p-Pb,Pb-Pb) collisions.
The scaling in the Pb-Pb system is steeper, but still linear with centrality, in accordance with a higher
$s \overline{s}$ production per unit volume per unit time. This is expected from QGP-based kinetic production \cite{rafcan}.

A final distinguishing test between canonical suppression and QGP production is to lower
the collision energy.  As Fig.~\ref{can_enhance} (right) shows, the canonical
model predicts an increase of enhancement (going to $\infty$ as the collision
energy approaches the $\Xi$ and $\Omega$ threshold).   The QGP model predicts
a decrease, with a sharp discontinuity at the energy where the QGP is not formed anymore.   Lower energy SPS runs will shortly measure the energy dependence of
enhancement and definitely clarify the situation.
\subsection{Collectivity}
The argument used in the derivation of strangeness enhancement can also be
applied to argue that in a quark-gluon plasma system local thermal equilibrium will
be reached on a much shorter timescale than in a system with hadronic degrees
of freedom.
Hence, the fact that most soft observables can be described through collective
dynamics (thermo and hydrodynamics) is a good indication that QGP has 
formed.   Are the particle yields described by a temperature and chemical
potential?    Are particle momentum distributions characterized
by one temperature and collective flow?   How well does a hydrodynamic approach
describe the system?
A good part of the subsequent chapters will be devoted to addressing these topics, so we will not dwell on them in detail here beyond mentioning
some motivational issues.

Any system, including a hadronic gas, will evolve collectively if given enough time.
However, a quark-gluon plasma, with its light colored degrees of freedom, should be much more
efficient as an entropy generator than a hadron gas.
Hence, we should expect a much stronger collective signal in a QGP than in a collection of
hadrons.

\begin{figure}[h]
{
\hbox to\hsize{\hss
  \includegraphics[width=0.5\hsize]{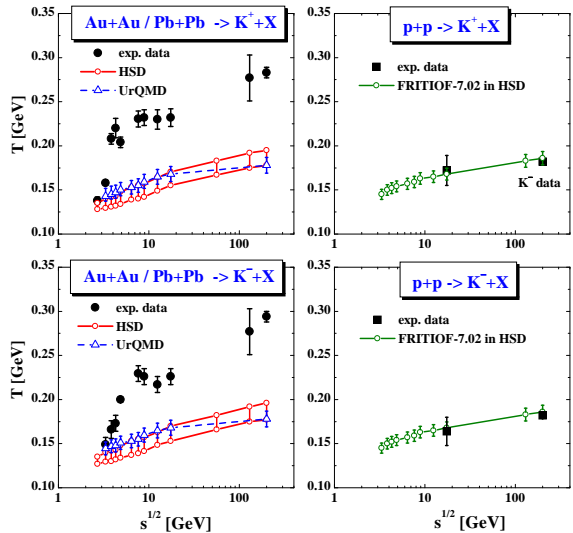}\hss}}
  \caption{Mass dependence of the apparent temperature (inverse slope in a logarithmic graph) as calculated by uRQMD and HSD models (see section \ref{comparison} for references).
 \label{uRQMDflow} }
\end{figure}

In this respect, we shall mention that microscopic hadronic simulations \cite{uRQMDflow} show that hadronic systems at SPS and RHIC energies do not have the time to develop a significant amount of collective flow, and hence mass dependence of the apparent temperature (inverse slope of the logarithm of the
transverse mass distribution) should not be large.
This is true for p-p but not A-A data (see Fig.~\ref{uRQMDflow}).
\begin{figure}
{
\hbox to\hsize{\hss
\includegraphics[width=0.25\hsize]{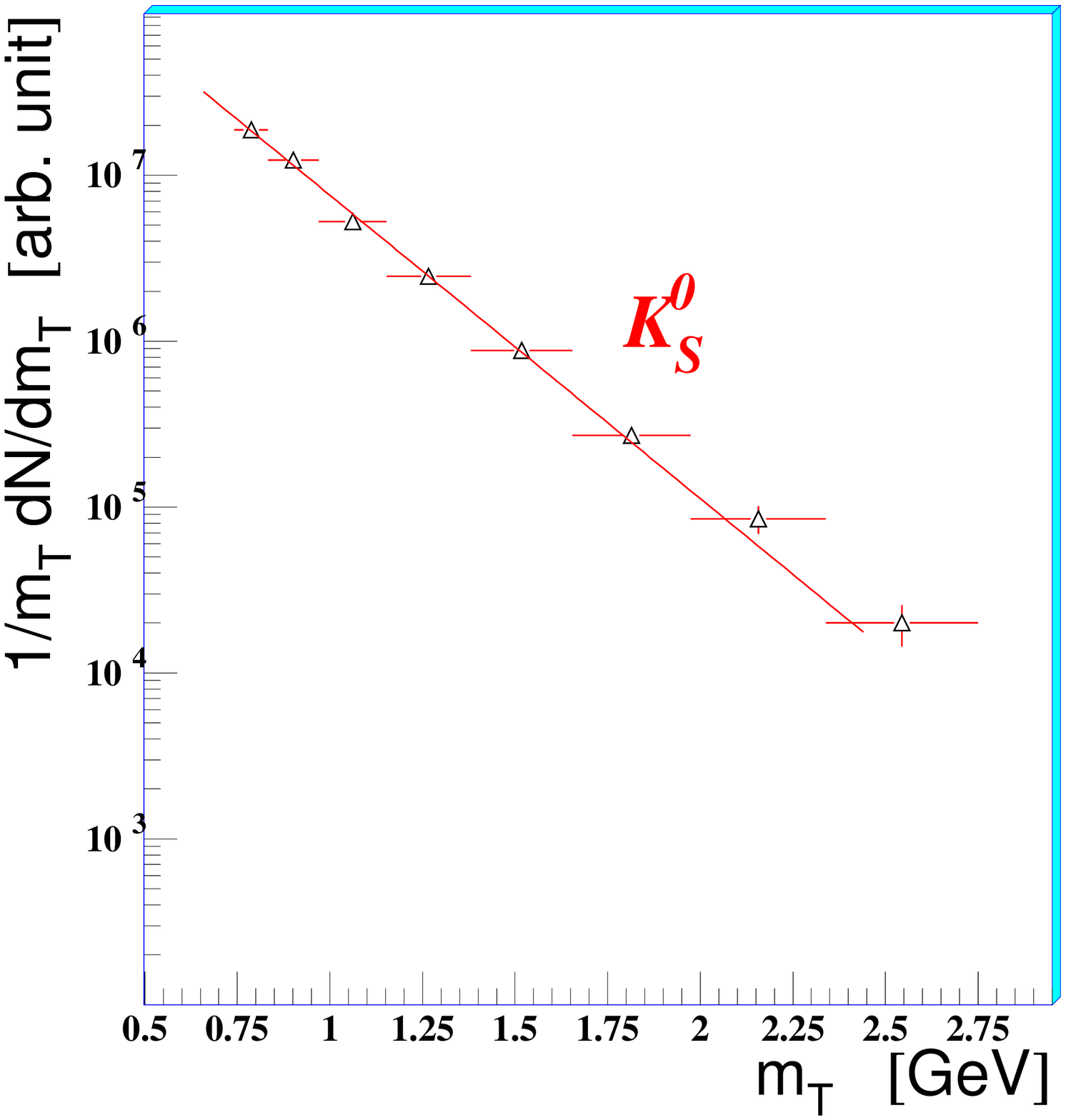}
\includegraphics[width=0.25\hsize]{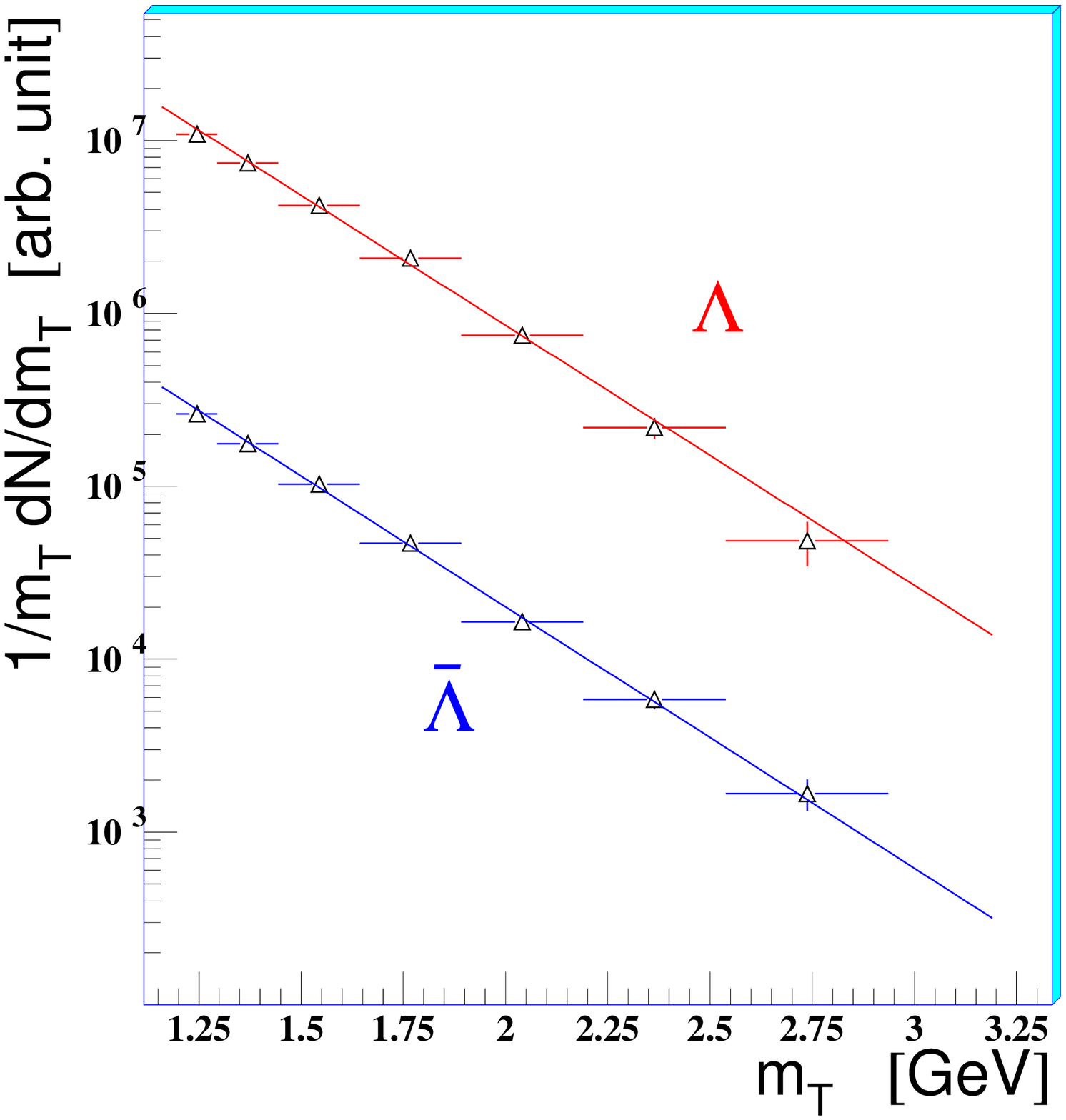}
\hss}}
{
\hbox to\hsize{\hss
\includegraphics[width=0.25\hsize]{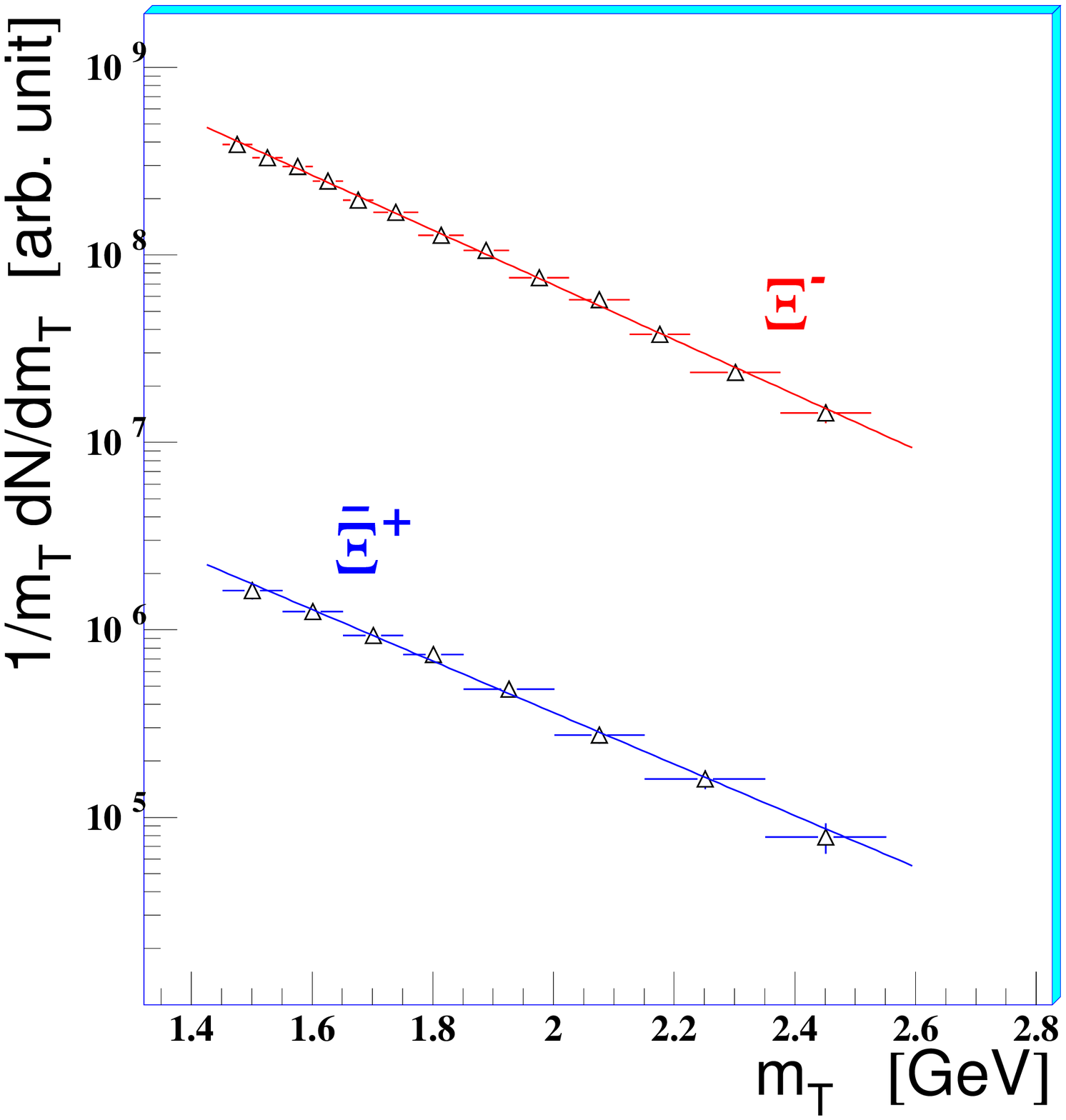}
\includegraphics[width=0.25\hsize]{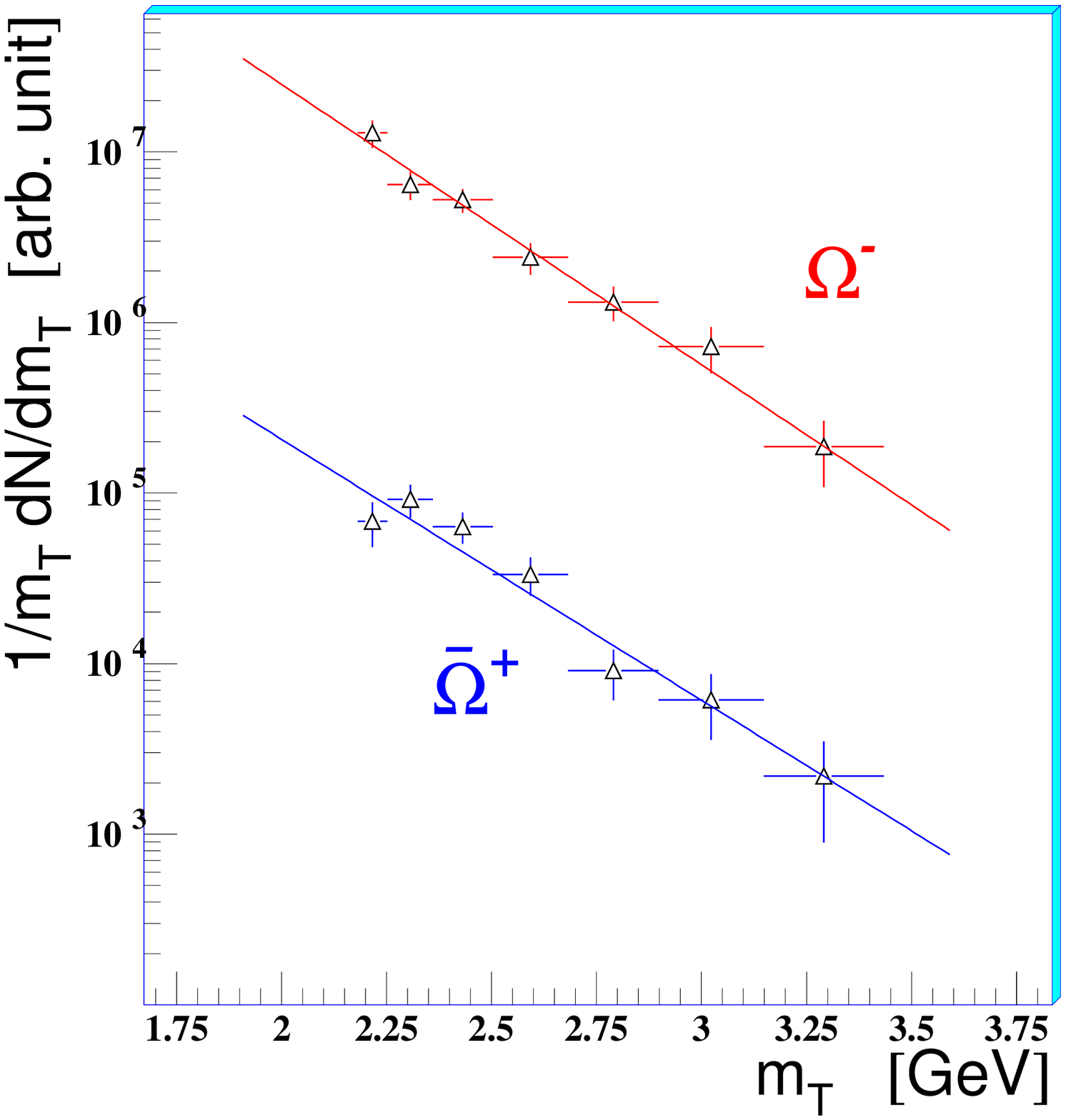} 
\hss} }
  \caption{Transverse mass distributions for particle and anti-particles as measured
by the WA97 experiment \label{tslopes} .}
\end{figure}
This is even more true in the case of anisotropic flow (discussed in detail in chapter 6).
The evolution of anisotropy is described  so well by the hydrodynamic picture that a very
early system thermalization is required \cite{HBTpuzzle}.   It is very difficult to see
how a hadronic model could achieve such a short thermalization timescale.

It should also be mentioned 
 that statistical production from an entropy-rich QGP is likely to look quite different from freeze-out of an interacting
hadron gas.
In case of a first-order phase transition or sharp cross-over, the system might evolve out of equilibrium at the transition, and
post-phase transition interactions can potentially destroy evidence for the earlier equilibration.
Given this, it is remarkable that several experiments found particles and
antiparticles to have exactly the same inverse slopes \cite{WA97spectra} (see Fig.~\ref{tslopes}).
In an interacting hadron gas, anti-particles should annihilate at a strongly
$p_T$ dependent cross-section.

\begin{figure}[h]
\centering
  \psfig{figure=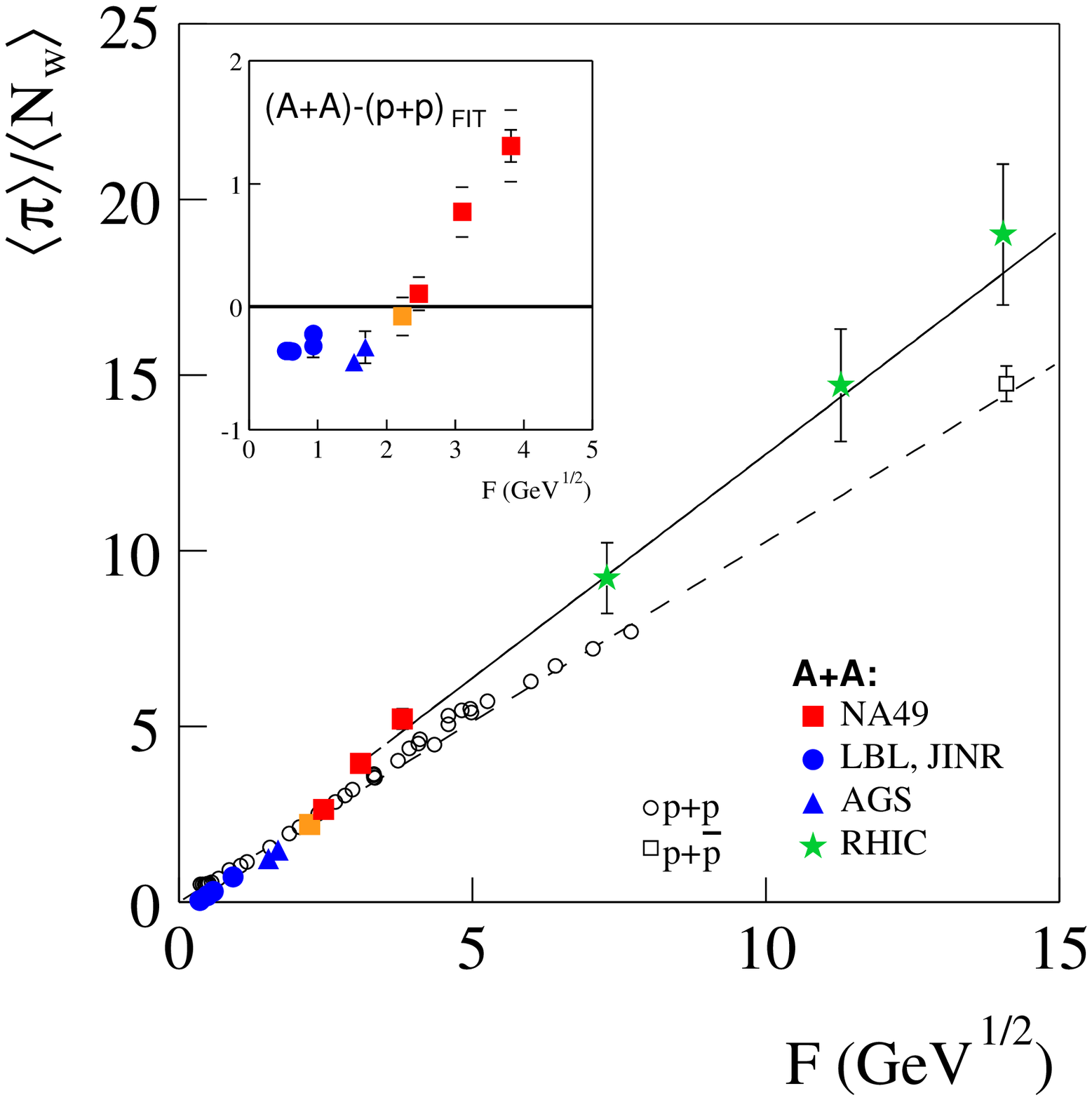,width=7cm}
  \psfig{figure=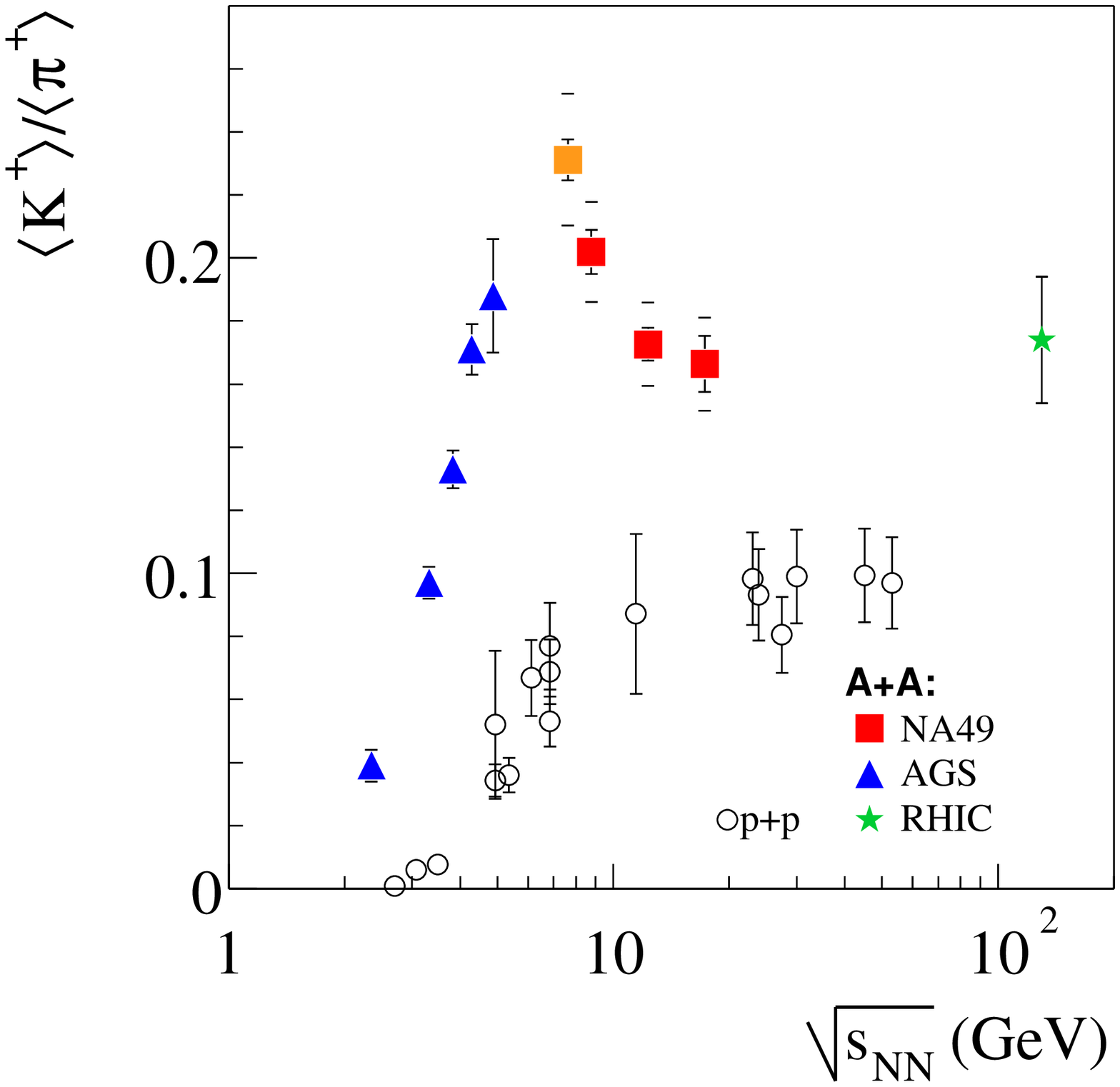,width=7cm}
  \caption{Left: $\pi$ production per wounded-nucleus a function of energy for A-A vs p-p collisions, plotted against Fermi's energy measure 
$F=(\sqrt{s_{NN}}-2m_N)^{3/4}/\sqrt{s_{NN}}^{1/4}$, where $\sqrt{s_{NN}}$ is the average of the center of mass energy per nucleon-nucleon collision.  
Right: K/$\pi$ ratio in nucleus-nucleus collisions collisions as a function of $\sqrt{s_{NN}}$. \label{gazdzicki} }
\end{figure}

Because of the large amount of entropy expected to be generated through QGP equilibration,
 it has also been suggested that to find the
phase transition we should look for a jump in the entropy per strangeness,
entropy per baryon number, or entropy per energy \cite{raf_s_bar}.
A recent systematic experimental study \cite{gazdzicki} uses this reasoning to cite the kink observed in $\pi$ ($\sim$ entropy) production at $\sim 4 {\rm GeV}$ (see Fig.~\ref{gazdzicki} (Left)) as well
as the flattening of the $K/\pi$ ratio (see Fig.~\ref{gazdzicki} (right))
as evidence for deconfinement.
\subsection{QGP evolution}
Concluding this overview of signatures, it is necessary to underline one of the most remarkable, and problematic aspects of the study of QGP in heavy ion collisions:  the range of physical approaches which needs
to be implemented to study the system.
At each stage of the system's evolution, the relevant physics and useful
approximations are in general completely different from that of the other phases.    In addition, different signatures probe different stages of this
evolution.

We shall give a chronological layout of the possible evolution of a QGP fireball, together with the methods which can be used to describe it.
\begin{description}
\item[Initial collision]  The short and energetic initial collision is usually
approximated through a parton cascade model \cite{parton_cascade}.
In this picture, the colliding partons are described as flat ``pancakes'' of
quarks and gluons, with the structure functions extrapolated from nucleus
deep inelastic scattering.   The collision dynamics is then governed by pQCD.

Recently, an alternative/complementary approach has emerged:  That
of the Color Glass Condensate \cite{cgc}.   From saturation physics one can
argue that at high energy heavy ion collisions the gluon occupation number
is significantly larger than one.  Hence, the colliding system can be modeled
as an incoherent classical $SU(3)$ field.
\item[Hydrodynamic evolution]  Somehow (the details are not entirely
clear), the initial system reaches local thermal equilibrium in a (hopefully!)
deconfined phase.    In this case, thermalization times will be very
fast \cite{strange_rafelski1} and hence the mean free path very small.
The bulk evolution of the system can then be optimally described by relativistic
hydrodynamics.

As we shall see later in the thesis, however, thermal and chemical equilibrium
can be two very different things.  Chemical equilibration is therefore
not necessarily as fast, and in fact does not necessarily occur for all quantities. This aspect of
QGP evolution is best modeled by transport theory, with the
collision terms given by quantum field theory \cite{strange_rafelski1},\cite{kapusta1},\cite{gyulassy1}.
\item[Hadronization]   This is the big unknown.  Quite simply, there is no
rigorous physical way to approach it.  It is non-perturbative, so Feynman
diagrams will not work.   It is also likely to be a far from equilibrium process,
so it is doubtful that lattice QCD will work as more than a qualitative indication of what's going on.

All we are left are effective models, based on basic physics such as
thermodynamics, relativity and conservation laws.
One  way to make quantitative predictions is entropy maximization.
This leads to statistical hadronization, the subject of this thesis.
Another recently proposed ansatz is the coalescence/recombination of partons \cite{coalescence}.   This picture has been proven to explain a range of phenomena
at intermediate $p_T$.  However, it cannot as yet be formulated self-consistently in a soft regime, where particle mass is non-negligible, as it violates energy-momentum conservation
(as well as entropy non-decrease).   Perhaps, in the future,
it will be possible to combine this approach with statistical hadronization \cite{biro}.

It should be noted that some observables described in this section
do not have to undergo hadronization.
Photons and dileptons leave the QGP with no further interactions.
The impact of hadronization on jets is also dynamically not relevant.
Since hadronization is a soft process, it will not change the momentum
distribution of the jet significantly, but will simply ``dress'' the outgoing
quark in a hadronic coating.

Soft probes (strangeness, charm, collectivity), on the other hand, will
undergo a potentially non-trivial hadronization before reaching the observer.
This means understanding hadronization is a crucial step in understanding
soft physics.   It also means, however, that soft probes are tools in understanding the phase transition from a quark-gluon plasma to a hadron gas.
\item[Post-hadronization evolution]   It is far from obvious that hadrons
stop interacting after they form.   In fact, one of the most important
questions one can ask about hadronization is the extent of interactions
which follow it.

One thing which is certain is that hadronic phase space is very different
from quark-gluon plasma phase space.   Hence, an equilibrated quark-gluon
plasma will probably produce an out-of-equilibrium hadron gas.
Equilibration times for a hadron gas are much slower than for a QGP, since
the lightest color-neutral boson has a mass of 137 ${\rm MeV}$.
Hence, the best way to analyze such a system is through transport theory.
Collision terms are measured
experimentally in elementary collisions or calculated through effective field theory techniques.
Quantum molecular dynamics models \cite{uRQMD,hsd} are based on this approach.
\item[Final freeze-out]  At this stage, particles decouple from the system and reach the observer.   It is sometimes fashionable to talk about a ``chemical
freeze-out'' (when inelastic collisions cease) and ``thermal freeze-out''
(when particles stop interacting altogether).
A look at particle reaction data shows that this description is inappropriate:
Inelastic reactions, such as strange quark exchange, can be just as low-threshold as elastic hadronic reactions.   It is better to talk about hadronization
(when hadrons become appropriate degrees of freedom) and freeze-out.
\end{description}
\section{The scope of this thesis}
This thesis, therefore, explores the effect of statistical hadronization
on the observed soft particle abundances and spectra in heavy ion collision.
We shall concentrate on hyperons, given their relevance to strangeness
enhancement.   However, non-strange hadrons shall also be examined.

We shall use experimental data (from SPS and RHIC) to constrain the statistical
hadronization picture, and differentiate between freeze-out scenarios.
We will test these scenario's ability to constrain both hadron yields and spectra, as well as direct detection
of unstable short-lived resonances.

Some of the questions which we would like to ask are:
\begin{itemize}
\item What are the temperature and chemical potentials characterizing the system?  
\item Is there evidence of transverse flow and other collective effects?
\item Does statistical hadronization happen in chemical equilibrium, or
are non-equilibrium effects important?
\item What is the dynamics of statistical hadronization?  How does particle emission proceed in spacetime?
\item What is the significance of the post-hadronization interacting
hadron gas phase?   What is its impact on observables?
\item What is the experimental data's sensitivity to these observables?
What is the significance of fit results, and how can they be further confirmed?
\end{itemize}
In the second chapter, we shall review relativistic statistical mechanics and how it
applies to hadronization.   We shall also describe how a variety of different freeze-out
scenarios arise out of statistical hadronization.
In chapters 3 and 4 we will use statistical hadronization to describe particle
spectra measured, respectively, in SPS and RHIC experiments.
In chapter 5, we will use yields of short-lived resonances as a way to differentiate between the
different freeze-out models and to constrain freeze-out dynamics.   
In chapter 6 we shall use azimuthal asymmetry in less than central collisions as a tool
to study freeze-out.   Finally, we will give a few conclusions and an outlook on
unresolved issues.

\setcounter{figure}{0}
\setcounter{equation}{0}
\setcounter{table}{0}
\chapter{Statistical hadronization: an overview}
\label{cha:stathad}
\section{Introduction}
The use of statistical mechanics to describe production of strongly interacting particles was pioneered by Fermi \cite{Fer50} and developed by 
Pomeranchuk \cite{Pom51}, Landau \cite{Lan53} and Hagedorn \cite{Hag65}.
Indeed, Hagedorn's observation \cite{Hag65} that an infinite number of hadrons of increasing mass leads to an exponential density of states, and hence
to a critical temperature where the canonical partition function diverges, provided a piece of evidence for a QCD phase transition before even the discovery of quarks.

The basic idea of this approach is that in a high-energy process driven by strong interaction a large number of particles is likely to be produced \cite{Fer50}.
In these circumstances, the dynamic part of the reaction cross-section will average out, and be considered a ``volume'' constant.
The distribution of each particle (1...N) will then be given by the corresponding phase space weight \footnote{See appendix C}
\begin{equation}
\label{ferphasespace}
d N_{1} \propto \prod_{i=2}^N \int \frac{d^3 p_i}{E_i} \delta(p^{\mu}_{total} - \sum_{i=1}^N p^{\mu}_i)
\end{equation}
It can be shown \cite{Fer50,Hag65,expcollective} that in the many-body infinite energy limit this distribution approaches an exponential, with the slope
related to the energy per particle (canonical limit).     If the phase space in Eq.~(\ref{ferphasespace}) also provides for quantum number conservation, it becomes possible to derive a grand-canonical limit.

The statistical model has been applied to every experimentally studied strongly interacting system, from $e^+ e^- \rightarrow hadrons$ to heavy
ions \cite{becattini}.  
It is not obvious, however, that particles described by such a statistical model are effectively thermalized.
In particular, a quark-gluon plasma, should exhibit a further degree of equilibration:  The fact that its composed of colored, massless
degrees of freedom means it should achieve local thermalization in a timescale considerably shorter than its evolution.
It is therefore expected that the QGP evolves as a continuous fluid.

\section{Thermodynamics and hydrodynamics from Kinetic theory}
Thermodynamics and hydrodynamics also arise as a $\tau \rightarrow \infty$ limit to the Boltzmann transport equation governing many-particle distributions
\begin{equation}
\label{boltzmann}
\frac{d f(x,p)}{d \tau} =\left( \frac{1}{m} p^{\mu} \frac{\partial}{\partial x^{\mu}} + F^{\mu} \frac{\partial}{\partial p^{\mu}} \right) f(x,p) = C[f](x,p)
\end{equation}
Where $C[f]$ is the Entropy-generating particle collision term.
We can describe the collision term semi-classically, by taking the cross-sections calculated in quantum field theory $\sigma (P,P' \rightarrow p,p')$ but
assuming that the distributions of the particles are uncorrelated beyond Quantum
statistics requirements.   For a theory invariant under time reversal, and considering just two body interactions\footnote{In general, the frequency of many-body interactions goes as a power of the mean-free-path, $<\sigma>_{n_{body}} \left(<N/V>\right)^{-n_{body}}$.   Hence, $C[f]$ is actually a perturbative series
and Eq.~(\ref{collision_term}) is the first term.  Both a QGP \cite{strange_rafelski1} and a hadron
gas \cite{uRQMD} have been described using just the two-body term.  However, this approach will break down in the non-perturbative limit, or equivalently when long-range correlations become too large for the semi-classical approximation.   A QGP hadronization can not therefore be described this way.}
\begin{equation}
\label{collision_term}
c[f](x,p)=\int d^3 [x',X,X',p,P,P']  \delta^4 (P+P' - p -p') \sigma(P,P' \Leftrightarrow p,p') 
\end{equation}
\[\ \left\{ \left[ f (X,P) f(X',P') + F f (X,P') f(X',P) \right] -   \left[ f (x,p) f(x',p') + F f (x,p') f(x',p) \right] \right\} \]
where F ensures the Fermion-Boson anti-symmetrization requirement
\begin{equation}
F=\left\{ \begin{array}{ll}  -1 & Fermions  \\  1 & Bosons  \\  0  & Boltzmann (Distinguishable)   \end{array}  \right.
\end{equation}
The Boltzmann H-theorem, generalized to quantum statistics \cite{huang} states that Entropy  
\begin{equation}
S= \int d^3 [x',X,X',p,P,P'] \left( -f(x,p) \ln[f(x,p)] - F (1 + F f(x,p)) \ln  [1 + F f(x,p)]   \right)
\end{equation}
will always monotonically increase, until detailed balance is reached, $C[f]$ will go do 0 and f reduces
to the entropy-maximizing Fermi-Dirac, Bose Einstein or Maxwell-Boltzmann distribution
\begin{equation}
\label{fd_be}
f_0 (p^{\mu}, x) = \frac{1}{\lambda_q^{-1} e^{u_{\mu}(x) p^{\mu} /T} + F} =
\sum_{n=1}^{\infty} F^{n-1} \lambda_q^n e^{n u_{\mu} p^{\mu} /T}  
\end{equation}
This will happen after ``many'' collisions will have taken place
\begin{equation}
\tau_{\infty} >>  \frac{1}{<\sigma><\frac{N}{V}>}
\end{equation}
($\tau_{\infty}$ is actually reached quite quickly in a strongly interacting system with high
particle density)

$f_0$ in Eq.~(\ref{fd_be}) is characterized by  temperature $T$ and fugacity $\lambda$, as well as $u^{\mu}(x)= \gamma (1,\vec{v})$,  the 4-velocity vector representing the motion of the volume element.
If the interactions will be strong enough that the mean free path of the particles
is negligible with respect to the collective motion motion of the system
\begin{equation}
\label{small_mfp}
\frac{d f(x,p)}{d \tau} >> \frac{1}{\tau_{\infty}}
\end{equation}
the system will always have local thermal equilibrium and evolve as
a continuous fluid
\begin{eqnarray}
\partial_{\mu} T^{\mu \nu} =0 \\
\partial_{\mu} n^{\mu} =0 
\end{eqnarray}
where $T^{\mu \nu}$ is the energy-momentum tensor and $\partial_{\mu} n^{\mu}$ is the number current (for any conserved current such as isospin, strangeness, etc.)
We can express these quantities in terms of pressure (P), energy density ($\epsilon$) and number density
\begin{eqnarray}
T^{\mu \nu}= \int \frac{p^{\mu} p^{\nu}}{E} f_0 (p^{\mu}, x) = P g^{\mu \nu} +(P+\epsilon) u^{\mu} u^{\nu}\\
\noindent n^{\mu} = \int n \frac{p^{\mu}}{E} f_0 (p^{\mu}, x)
\end{eqnarray}
Hence, if we know the equation of state of our fluid we will be able to close our system of equations
and evolve the system from any set of initial conditions.
\section{Statistical hadronization and the Cooper-Frye formula}
QGP, with its high number density of strongly interacting particles, is
a good candidate for an equilibrated fluid.   However, this situation
will change in the QGP-HG transition, in which the 10 colored massless
partons become over 200 color-neutral hadrons.

The Boltzmann formalism of Eq.~(\ref{boltzmann}) can not, as far as we can see, be applied to 
hadronization, since the assumptions that go in it (weakly correlated particles scattering locally) do
not apply to a non-perturbative quantum phase transition.
However, we know from lattice QCD that hadronization is a relatively fast process, and obviously
hadronization of a large thermally equilibrated system can not decrease entropy.
For this reason, and the fact that color-neutral hadrons are more weakly interacting than quarks, it is 
very likely that an equilibrated entropy-rich QGP
will emit hadrons according 
to phase space as described in section 2.1.

If the system is large, energy and quantum number conservation can be accomplished
by Lagrange multipliers, so, in the rest-frame with respect to the collective
flow the hadrons will be distributed according to $f_0$ as given in Eq.~(\ref{fd_be}). 

the post hadronization hadron current will be given by 
\begin{equation}
\label{cf_current}
j^{\mu} = \int d^3 p \frac{p^{\mu}}{E} f_0 (u_{\mu} p^{\mu}_{hadron}, T, \lambda_{hadron})
\end{equation}
where the Hadron fugacity is given by the product of the fugacities of its constituent
quarks quantum  (not necessarily the QGP's chemical potentials, as we'll see later)
\begin{equation}
\lambda_{hadron} = \prod_{q} \lambda_{q}
\end{equation}
Using the ``fast hadronization'' assumption again, we can define a ``Freeze-out
hypersurface'' defining a locus in space-time from which these
statistically hadronizing particles are emitted, labeled by a
4-vector $\Sigma^{\mu}$.
Since $\Sigma^{\mu}$ is a 3-surface in Minkowski 4-space, it can
always be expressed as a function of three parameters ($u,v,w$).
Its element can then be given in in a Lorentz-covariant way using
Stokes's theorem
\begin{equation}
\label{dsigma}
d^3 \Sigma_{\mu} =  \epsilon_{\mu \nu \alpha \beta} \frac{\partial \Sigma^{\nu}}{\partial u}  \frac{\partial \Sigma^{\alpha}}{\partial v}  \frac{\partial \Sigma^{\beta}}{\partial w}
\end{equation}
where $ \epsilon_{\mu \nu \alpha \beta}$ is the Levi-Civita symbol.
the number of particles produced in such a volume element is
then Lorentz-invariant, and computable in the volume element's rest frame as
\begin{equation}
\label{cf_element}
j_{\mu} d^3 \Sigma^{\mu} = d N  
\end{equation}
combining Eq.~(\ref{cf_current}) and Eq.~(\ref{cf_element})
we obtain the famous Lorentz-Invariant Cooper-Frye formula \cite{cf}
\begin{equation}
\label{cf}
E \frac{ dN}{d^3 p} = \int d^3 \Sigma_{\mu} p^{\mu} f(p_{\mu} u^{\mu},T,\lambda)
\end{equation}
If hadronization of the full volume takes a lot of time, the emitted particle can find itself in the QGP again if $p^{\mu} \Sigma_{\mu} <0 $.
It is still unclear how to handle this in a rigorous way, but typically one just truncates the distribution to exclude this unphysical region (perhaps adjusting the limits of integration to conserve energy, entropy etc.) \cite{cf_trunc1},\cite{cf_trunc2}
\begin{equation}
\label{cut_cf}
E \frac{ dN}{d^3 p} = \int d^3 \Sigma_{\mu} p^{\mu} f(p_{\mu} u^{\mu},T,\lambda)\Theta(\Sigma_{\mu} p^{\mu})
\end{equation}
where $\Theta(x)$ is the step function.

Since all hadrons apart from the pion are considerably heavier than
the typical hadronization temperature ($p_{\mu} u^{\mu} > m>>T$), the sum in Eq.~(\ref{fd_be}) can 
be truncated at n=0, corresponding to the Boltzmann approximation
\begin{equation}
\label{maxwell-boltzmann}
f(p_{\mu}) = \lambda e^{-p_{\mu} u^{\mu}/T}
\end{equation}

\subsection{Total particle yields}

If the thermodynamic parameters $T,\lambda$ do not vary within the hadronizing volume,
the total number of particles will be independent of $\Sigma^{\mu}$ and
$u^{\mu}$.
To see this, we integrate the Cooper-Frye formula over momentum space
\begin{equation}
N= \int dN = \int  \frac{d^3 p}{E} p^{\mu} d^3 \Sigma_{\mu} \lambda e^{p_{\mu} u^{\mu}/T} 
\end{equation}
and insert a $u_{\mu} u^{\mu}$ $(=1)$ in the integrand.
The two integrals then decouple
\begin{equation}
N = \left[ \int d^3 \Sigma_{\mu} u^{\mu}  \right] \left[ \int d^3 p \frac{p^{\mu} u_{\mu}}{E}  e^{p_{\mu} u^{\mu}/T}  \right]
\end{equation}
The first integral is just a normalization constant.
Since $p^{\mu} u_{\mu}=E_{rest}$, we are left with
\begin{equation}
N = V g \int d^3 p \lambda e^{-\sqrt{p^2+m^2}/T}
\end{equation}
where g is the particle degeneracy  (For a colorless
particle, it will be equal to $2S+1$ where S is the spin) 
\footnote{It is perhaps not immediately intuitive how a non interacting gas dispersion relation can be
used to describe a strongly interacting system.  The ``trick'', here, is that we will generate all the strongly excited states (resonances) in the data book, and
the strong interaction excited states can be considered, to a good approximation, as narrow resonances.   As was explicitly shown \cite{dashen} if resonances
are narrow enough, an ideal gas of resonances is an appropriate description.  As this section shows, our approximation can be improved through finite width effects}.
Using the Bessel function definition \cite{gradshteyn}
\begin{equation}
K_n (x) = \frac{2^n n!}{(2n)!} x^{-n} \int_{0}^{\infty} \frac{dz}{\sqrt{z^2+x^2}} z^{2n} e^{-\sqrt{z^2+x^2}}
\end{equation}
we can find N analytically
\begin{equation}
\label{Nk}
N = g \frac{4 \pi}{(2 \pi)^3 } m^2 \lambda T K_2 \left( \frac{m}{T}  \right)
\end{equation}
The energy density in a rest frame with respect to the collective flow also
follows
\begin{equation}
\epsilon = \int d^3 \sqrt{p^2 + m^2 }  \lambda e^{-\sqrt{p^2+m^2}/T} = g \frac{4 \pi}{(2 \pi)^3 } m^3 \lambda T \left(\frac{3 T}{m} K_2 \left( \frac{m}{T} \right) + K_1 \left( \frac{m}{T} \right)  \right)
\end{equation}
Comparing Eq.~(\ref{maxwell-boltzmann}) and Eq.~(\ref{fd_be})
it can be seen that these formulae can be generalized to the
Fermi-Dirac and Bose-Einstein distribution by the substitution 
\begin{equation}
\lambda T  K_l \left( \frac{m}{T} \right) \rightarrow 
\sum_{n=1}^{\infty} (\pm 1)^{n+1} \frac{T}{n} \lambda^n   K_l \left( \frac{n m}{T} \right)
\end{equation}

If the particle has a finite width, the Bessel function will be further
integrated over the range of masses to take the mass spread into account.
For a resonance with width, we obtain
\begin{equation}
N_i (m_i) \rightarrow \frac{1}{N_0} \sum_{\forall i \rightarrow j} 
\int_{m_{threshold}}^{\infty} n_i (M) F_{\Gamma} (M,\Gamma_i (b_{i \rightarrow j},M)) dM 
\label{widthn}
\end{equation}
where
\begin{itemize}
\item $b_{i \rightarrow j}$ is the decays branching ratio
\item $m_{threshold}$ is the threshold mass for the decay, $\sum_j m_j$
\item $F_{\Gamma} (M,\Gamma_i)=\frac{\Gamma \Gamma_i}{(M-m_i)^2+\Gamma_i^2/4}$ is the Breit-Wigner
formula
\item $\Gamma_i (b_{i \rightarrow j},M)$ is the energy-dependent width for the decay under consideration.
In general, this is a non-trivial particle dependent function.
Here, we  only consider the dominant  energy dependence of the width, namely 
the decay threshold energy phase space factor.
The explicit form for decays with low relative angular momentum has been studied through corresponding
reverse production cross-sections, and found to be \cite{phase1,phase2,phase3} 
\begin{equation}
\label{energywidth}
\Gamma_i (b_{i \rightarrow j},M)= b_{i \rightarrow j} \left[ 1-\left(\frac{m_{threshold}}{m}\right)^2\right]^{l+\frac{1}{2}} \Gamma^*
\end{equation}
where $l$ is the relative angular momentum of the decay and $\Gamma^*$ is the energy-independent constant found in the particle data book \cite{pdg}
\end{itemize}
$N_0$ is the Breit-Wigner and phase space normalization
\begin{equation}
\label{norm}
N=\int_{m_{threshold}}^{\infty} F_{breit-wigner}(M,\Gamma_i (b_{i \rightarrow j},M) ) dM
\end{equation}
\[\
=\int_{m_{threshold}}^{m_1} F_{breit-wigner}(M,\Gamma_i(b_{i \rightarrow j},M))  dM+
\int^{1}_0 \frac{ m_1 \Gamma_i \left(b_{i \rightarrow j},\frac{m_1}{z}\right)}{(m_1-m_i z)^2 + \frac{\Gamma^2\left(b_{i \rightarrow j},\frac{m_1}{z}\right)  z^2}{4}} dz
 \footnote{This change of variables, $\frac{m_1}{m}=z$, used within our programs, is absolutely essential for the numerical evaluation of the integral.   Otherwise, the
result will be plagued by $\Gamma$-dependent systematic errors, since N converges slowly as the upper limit of the numerical integration $\rightarrow \infty$.
On the other hand, the integral with changed variables is also prone to error if the width is too small, since the Gaussian integration might miss the peak.
The solution, which our calculations implement \cite{share}, is to split the integral in two parts:
$\int_{m_{threshold}}^{\infty} = \int_{m_{threshold}}^{m_0 + 2 \Gamma} + \int_{m_0 + 2 \Gamma}^{\infty}$.  The first part can be done through conventional integration, the second
through variable change}\]
\section{Treatment of resonance decays}
\begin{figure}[h]
\centering
  \psfig{figure=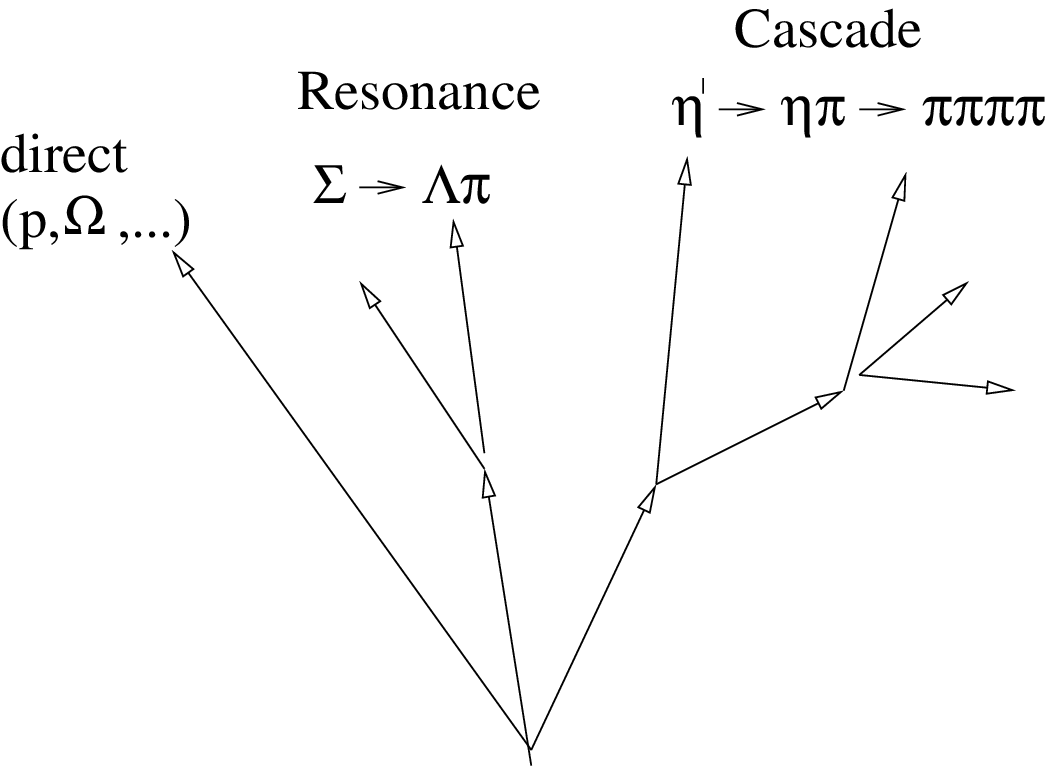,width=7cm}
  \psfig{figure=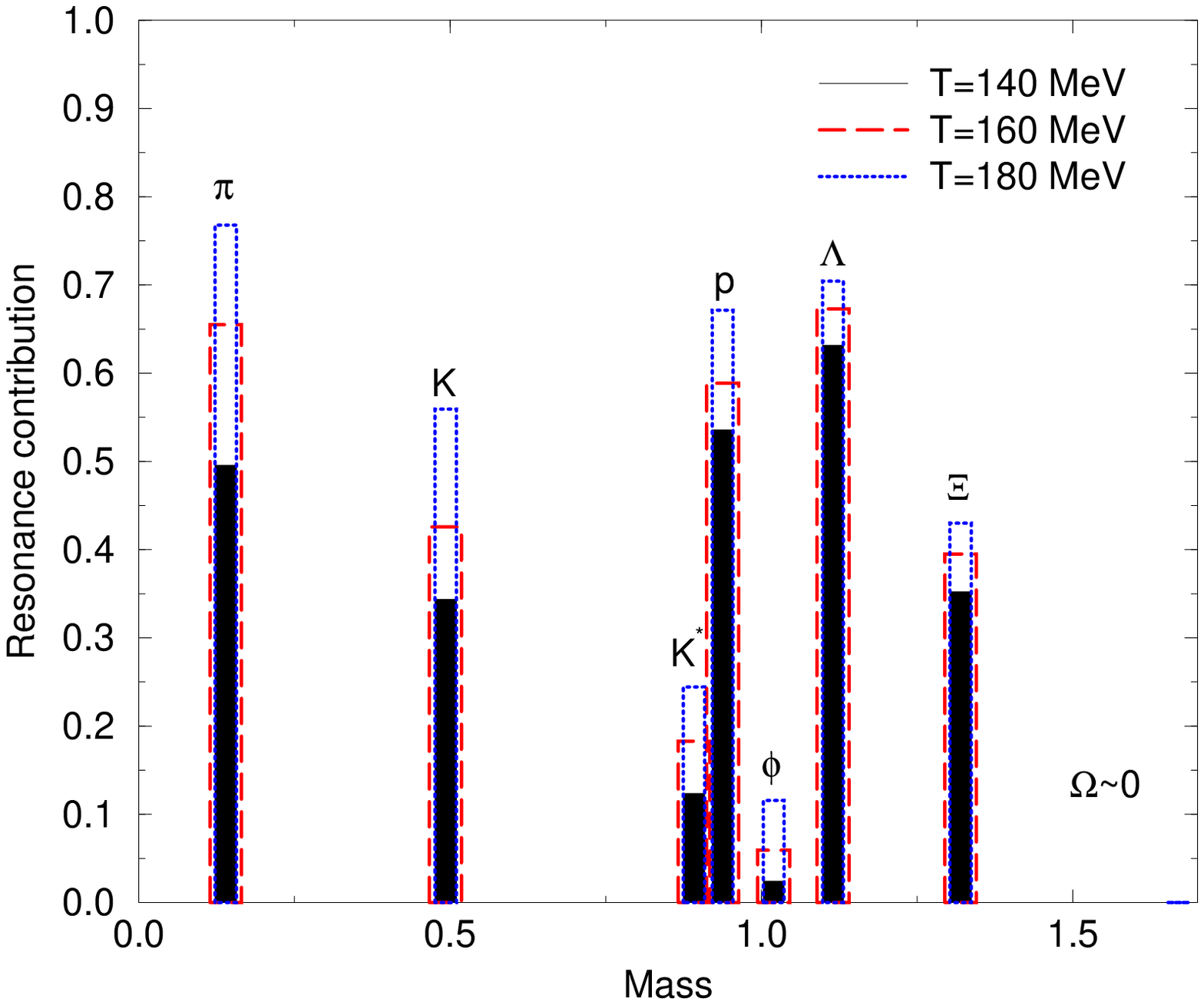,width=7cm}
  \caption{Left: Typical resonance decay patterns.  Right:  Resonance contribution to observed particle
yield  as a function of freeze-out temperature \label{rsyield}.}
\end{figure}

Most of the 200 particles produced in statistical hadronization will be short-lived resonances, whose decay is undetectable through an analysis of the particle trajectories as it happens  after
the particle was produced.   More complicated cascades, with sequential decays, are also possible.
While the number of particles made from these decays is suppressed by the high mass of the resonance, they
are also enhanced by resonance's typically high spin degeneracy and the fact that many particles are produced in a typical decay  (Fig.~(\ref{rsyield}) left).

The expected yield of observed ``light'' particles will then have to include the products of
all resonance decays of short-lived products.
As Fig.~\ref{rsyield} (right) shows, there is no way that this contribution can be neglected.
However, on the bright side, resonances are an important tool for QGP diagnostics:  Since their quark
numbers are the same as the base particles, their relative abundance is controlled by the temperature only.  Their abundance is also sensitive to post-hadronization dynamics (see chapter 5).   In short, resonances will be a very sensitive freeze-out probe, one which will be explored from
different angles in the next chapters of this thesis.

To include resonances in particle yields and ratios, it is enough to sum the resonance decay products
to the base yields
\begin{equation}
\label{Nres}
N_{i} \rightarrow N_i+\sum_{\forall j\rightarrow i+...} b_{j\rightarrow i+...} N_j
\end{equation}
where $N_i$ is given as in Eq.~(\ref{Nk}) and $ b_{j\rightarrow i+...}$ is the branching ratio.
The only technical difficulty in this calculation is that decay products in the particle data
book \cite{pdg} are given up to Clebsh-Gordan coefficient factor of Isospin.
In case of two-body decays, we calculated C-G coefficients automatically using the algorithms
in CERNLIB \cite{cernlib}.
Three body decays, are a bit more problematic, since there are several possible ways of combining 
partial isospin sums.         We averaged over these, giving ($J,m$ a and $j_{i},m_{i}$ refer to the total
and z isospin components of the resonance and decay products respectively)  
\begin{eqnarray}
b_{j\rightarrow 1 2}&=&b_{0\rightarrow 1 2}^0 (<J_0 m_0 | j_1 j_2 m_1 m_2>)^2 \\
b_{j\rightarrow  1 2 3}&=&b_{0\rightarrow 1 2 3}^0 \frac{1}{3}
[
\sum_{j_{12}}  (<J_{12} m_{12} | j_1 j_2 m_1 m_2>   <J_0 m_0 | j_{12} j_{3} m_{12} m_3>)^2   \\ \nonumber
&+&\sum_{j_{13}}  (<J_{13} m_{13} | j_1 j_3 m_1 m_3>   <J_0 m_0 | j_{13} j_{2} m_{13} m_2>)^2  \\ \nonumber
&+&\sum_{j_{23}}  (<J_{23} m_{23} | j_2 j_3 m_2 m_3>   <J_0 m_0 | j_{23} j_{1} m_{23} m_1>)^2
]
\end{eqnarray}
\subsection{Spectra}
The contribution of resonances to particle spectra is non-trivial, since the full kinematics of the decay has to be understood.
However, the resonance contribution to spectra is where the data is really sensitive to freeze-out
dynamics since post-hadronization  interactions might re-thermalize resonance decay products.
Hence, searching for a bona-fide resonance contribution which could be reconstructed by invariant
mass is a stringent test of fast statistical freeze-out.

In analogy with Eq.~(\ref{Nres}) we assume that 
\begin{equation}
\label{respectra}
E_1 \frac{dN_1}{d^3 p_1} = \left( E_1 \frac{dN_1}{d^3 p_1} \right)_{direct} + \sum_{\forall j \rightarrow 1 2..n}  \left( E_1 \frac{dN_1}{d^3 p_1} \right)_{j \rightarrow 1 2..n}
\end{equation}
where the first term is given by the Cooper-Frye formula and the second gives the distribution
of decay products in a cascade.
Assuming $j$ is also produced through statistical hadronization, all is left is to find $E_1 \frac{dN_1}{d^3 p_1}$ in terms of $E_j \frac{dN_j}{d^3 p_j}$ (which will again be given by the Cooper-Frye formula.  The relations we will develop will themselves be recursive, to handle the case when j has a cascade component).

We shall assume that all the short-lived resonances have decayed.   We shall also assume that lots of particles are emitted in all directions, thereby averaging over any matrix-element dependence on factors such as spin.
The distribution of the resonance decay product momenta will then be given simply by the integral over
the available region of the Lorentz-invariant phase space \footnote{See appendix C}.
\begin{equation}
\label{decayphase}
 \left( E_1 \frac{d^3 N_1}{d^3 p_{1}} \right)_{j \rightarrow 12..n}= B \int \frac{d^3 p_j}{2 E_j} \int T_n \left( E_j \frac{d^3 N_j}{d^3 p_{j}} \right)
\end{equation}
\begin{equation}
\label{Tn}
T_n=\prod_{i=2}^{n} \frac{d^3 p_i}{2
E_i} \delta(\sum_{i=2}^N p_i-p_1) \delta(\sum_{i=2}^N E_i-E_{1}-E_j)
\end{equation}
B is a normalization factor to make sure that
\begin{equation}
\frac{N_{j\rightarrow12..n}}{N_j}=b_{j \rightarrow 12..n} 
\end{equation}
\subsubsection{Two body decay}
For a two-body decay, $j \rightarrow 1 2$, we can go to the resonance's rest frame (denoted by the $*$)where
\begin{eqnarray}
E^*_{1}=\frac{1}{2 m_1} (m_1^2+m_j^2-m_2^2)  \label{stare} \\
p^*_{1}=\sqrt{E^*_1-m_1^2}=-p^*_2  \label{starp}
\end{eqnarray}
and take advantage of energy conservation and Lorentz-invariance
\begin{equation}
 p_{j \mu} p_{1}^{\mu} = p_{j \mu}^{*} p_{1}^{\mu *} = E_1^* m_j 
\end{equation}
hence, the integral in Eq.~(\ref{decayphase}) for just two bodies reduces to 
\begin{equation}
E_1 \frac{dN_1}{d^3 p_1}=B \int  \frac{d^3 p_j}{E_j} \delta \left( \frac{p_{j \mu} p_{1}^{\mu}}{m_{j}}-E^{*}_{1} \right)
\left( E_j \frac{dN_j}{d^3 p_j} \right)
\label{reso1}
\end{equation}
After expanding  the delta function argument explicitly we get a condition in terms of the transverse mass and momentum ($m_T,p_T$) and rapidity ($y$)\footnote{See appendix A}
\begin{equation}
\frac{m_{T1} m_{Tj} \cosh(y_1-y_j) - p_{T1} p_{Tj} \cos(\theta_1 - \theta_j)}{m_j}-E^*_1=0
\label{p1p2_explicit}
\end{equation}
Rearranging this becomes
\begin{eqnarray}
 (e^{y_j-y_1})^2-2 A e^{y_j-y_1} +1=0\\
 A=\frac{m_j E^*_1 +p_{T1} p_{Tj} \cos(\theta_1 - \theta_j)}{m_{T1} m_{Tj}} \label{coshres} 
\end{eqnarray}
Which introduces constraints linking rapidity, transverse mass and relative angle
\begin{equation}
y_j=y_1+\ln(A \pm \sqrt{A^2-1})  \; \;  A>1 
\label{Asol}
\end{equation}
Eq.~(\ref{Asol}) can be used to eliminate rapidity from the integration and to
introduce integration constraints on $m_T$:
\begin{equation}
\frac{m_j E^*_1 +p_{T1} p_{Tj} \cos(\theta_1 - \theta_j)}{m_{T1} m_{Tj}}>1
\label{intlimits1}
\end{equation}
together with $ |\cos(\theta_1 - \theta_j)| \leq 1$ leads to
\begin{eqnarray}
m_j E^*_1 - p_{T1} p_{Tj} >  m_{T1} m_{Tj} \label{intup1}
\end{eqnarray}
with the limiting cases
\begin{equation}
m_j E^*_1 - p_{T1} \sqrt{m_{Tj}^2-m_j^2}= m_{T1} m_{Tj}
\label{intlimits2}
\end{equation}
Rearranging and squaring, we get
\begin{equation}
p_{T1}^2 (m_{Tj}^2-m_j^2)= (m_j E^*_1 -  m_{T1} m_{Tj})^2
\label{intlimits3}
\end{equation}
putting this in the quadratic form and remembering that $m_{T1}^2-p_{T1}^2=m_1^2$
\begin{equation}
m_1^2 m_{Tj}^2 - 2 m_{Tj} m_{T1} m_j E^*_1 + m_j^2 (p_{T1}^2+E^{*2}_{1})=0
\label{intlimits4}
\end{equation}
Using the quadratic formula, we get two solutions (a factor of 2 cancels out
from top and bottom)
\begin{equation}
m_{Tj}= \frac{ m_{T1} m_j  E^*_1 \pm \sqrt{( m_{T1} m_j )^2-m_1^2  m_j^2 ( E^*_1+p_{T1}^2) }}{m_1^2}
\label{intlimits5}
\end{equation}
which simplify since
\begin{equation}
 \sqrt{ m_{T1}^2 m_j^2 E^{*2}_{1}-m_1^2  m_j^2 p_{T1}^2-m_1^2  m_j^2 E^{*2}_{1} }= 
m_j \sqrt{ m_{T1}^2 E^{*2}_{1} - m_1^2  p_{T1}^2 - m_1^2 E^{*2}_{1}}  \nonumber
\end{equation}
\begin{equation}
 =m_j \sqrt{p_{T1}^2 E^{*2}_{1}- p_{T1}^2 m_1^2}=m_j p_{T1} p^*_1
\label{discriminant}
\end{equation}
If one puts this into Eq.~(\ref{intlimits5}) one gets
\begin{equation}
m_{Tj}^{\pm}=\frac{m_j}{m_1^2} (m_{T1} E^*_1 \pm p_{T1} p^*_1)
\label{intlimits}
\end{equation}
reassuringly, this is always physical since $m_{T1} E^*_1 > p_{T1} p^*_1$

In a similar vein, Eq.~(\ref{Asol}) yields the constraints for the relative angle $\theta_j$.
Putting in the form of A (Eq.~\ref{coshres}) into Eq.~(\ref{Asol}) and manipulating one gets
\begin{equation}
 \frac{-m_j E^*_1 + m_{T1} m_{Tj}}{p_{T1} p_{Tj}}<\cos(\theta_j-\theta_1)
\label{thetalimits1} 
\end{equation}
\begin{equation}
 \theta_1- \arccos \left(\frac{-m_j E^*_1 + m_{T1} m_{Tj}}{p_{T1} p_{Tj}} \right) < \theta_j < \theta_1+ \arccos \left(\frac{-m_j E^*_1 + m_{T1} m_{Tj}}{p_{T1} p_{Tj}} \right)
\end{equation}
Putting everything together, we finally have an analytically solvable
expression for the decay products distribution in terms of the resonance distribution
\begin{equation}
 E_1 \frac{dN_1}{d^3 p_1}=B  \int^{m_T^{+}}_{m_T^{-}} m_{Tj} d m_{Tj} \int^{\theta^{+}}_{\theta^{-}}  d \theta_j \int dy_j 
\delta(y_j=y_1+\ln(A \pm \sqrt{A^2-1})) \left(  E_j \frac{dN_j}{d^3 p_j} \right)
 \label{resofinal} 
\end{equation}
\begin{eqnarray}
 A=\frac{m_j E^*_1 +p_{T1} p_{Tj} \cos(\theta_1 - \theta_j)}{m_{T1} m_{Tj}}   \label{resofinaly} \\
 \theta^{\pm}=\theta_1\pm \arccos \left(\frac{-m_j E^*_1 + m_{T1} m_{Tj}}{p_{T1} p_{Tj}} \right) \label{thetahigh}
\end{eqnarray} 
Integrating over Eq.~(2.56) to find $N_1/N_j$ we discover that the normalization has to be
\begin{equation}
B=\frac{b_{j\rightarrow 1}}{4 \pi p^{*}}
\end{equation}
An important simplification arises in the rapidity-invariant limit \cite{bjorken_boost}:\\
In this case, the integral over the $\delta$ function in Eq.~(\ref{resofinal}) has no effect, and the different Fourier components 
of the distributions
\begin{eqnarray}
\label{fourier}
E_1 \frac{dN_1}{d^3 p_1}=v_{10}+v_{11} \cos(\theta)+v_{12} \cos(2 \theta)...\\
E_j \frac{dN_j}{d^3 p_j}=v_{j0}+v_{j1} \cos(\theta)+v_{j2} \cos(2 \theta)...
\end{eqnarray}
do not mix (ie $v_{Jo}$ has no effect on $v_{1 m \ne i}$) \cite{flork_priv}

If the original distribution function is azimuthally invariant (no $\theta$ dependence), Eq.~(\ref{resofinal}) simplifies to
\cite{heinzreso}
\begin{eqnarray}
\label{resofinal_azim}
\frac{dN}{d {m^2_{T1}} d y_1 } &=&
\frac{ b}{4 \pi p^{*}_1}
\int_{y_{-}}^{y_{+}} dy_j
\int_{m_{T}-}^{m_{T}+} dm_{Tj}^{2} J 
\frac{d^2 N_{j}}{dm_{Tj}^{2} dy_j} , \\
\noindent J&=&\frac{m_j}{\sqrt{p_{Tj}^2 p_{T1}^2 -(m_R E^{*}_1 - m_{Tj} m_{T1} \cosh (y_j-y_1)^2) }},\\
 y_{\pm}&=&y_1 \pm \sinh^{-1} \left( \frac{p^{*}_1}{m_{T1}} \right) \\
m_{T}^{\pm}&=&m_j 
\frac{E^{*}_1 m_{T1} \cosh(y_j-y_1) \pm p_{T1} \sqrt{p^{*2}_1-m_{T1}^{2} \sinh^{2} (y_j-y_1)}}
{m_{T1}^{2} \sinh^{2} (y_j-y_1)+m^{2}_1}
\end{eqnarray}
\subsubsection{Decays of three bodies and more}
The strategy to pursue if there are more than 3 bodies in the decay is to change
variables from $p_1,p_2,...,p_n$ to $p_1,p_2,...,p_{n-1},s_{2..n},p_{2..n}$ where
\begin{equation} 
s_{2..n}=(\sum_{i=2}^{n} E_i)^2 -(\sum_{i=2}^{n} p_i)^2 
\end{equation}
is the invariant mass of ``all the other'' decay products \cite{phasespace} and $p_{2..n}$ is
their combined momentum.
The invariant phase-space in Eq.~(\ref{decayphase}) becomes, applying the $\delta$ function in
Eq.~(\ref{reso1})
\begin{eqnarray}
\label{manybody}
E_1 \frac{d n_1}{d^3 p_1} = B 
\int^{s_+}_{s_-} ds_{2..n} \left| \frac{\partial (p_1,p_2,...,p_{n-1},s_{2..n},p_{2..n} ) }{\partial (p_1,...,p_{n})} \right| \\ \nonumber \int \frac{d^3 p_{j}}{E_{j}} \delta \left( \frac{p_{j \mu} p_{1}^{\mu}}{m_{j}}-E^*_{1}\right)  \int T_{n-1}  \left( E_j \frac{d^3 N_j}{d^3 p_{j}} \right)  \\ \nonumber
E_{2..n}^2=\sqrt{p_{2..n}^2+s_{2..n}}\\
E_{1}^2=\sqrt{p_{2..n}^2+m_1}
\end{eqnarray}
here $T_n$ is given as in Eq.~(\ref{decayphase}), the Jacobian is
\begin{equation}
\left| \frac{\partial (p_1,p_2,...,p_{n-1},s_{2..n},p_{2..n} ) }{\partial (p_1,...,p_{n})} \right| =  \frac{\sqrt{m_j-(s_{2..n}+m_1)^2} \sqrt{m_j-(s_{2..n}-m_1)^2}}{m_j}
\end{equation}
 and the limits of integration are given by energy
conservation
\begin{equation}
s_{-}=(\sum_{i=2}^n )^2 \; \;
s_{+}=(m_j-m_1)^2
\end{equation}
Thus, the n-body decay can be expressed as a convolution of a 2-body and an n-1 body decay, and further
convoluted to n 2-body decays.

Unsurprisingly, the decay products of the cascades such as 
\begin{equation}
\label{eta'}
\eta' \rightarrow \eta \pi \rightarrow \pi\pi\pi\pi
\end{equation}
(Fig.~\ref{rsyield} left) will have a very similar distribution:
$E_j \frac{dN_j}{d^3 p_j}$ of the intermediate state ($\eta$ in Eq.~(\ref{eta'}) ) will be fed 
back into  a formula such as Eq.~(\ref{decayphase}), and the only difference from Eq.~(\ref{manybody}) 
is that the intermediate mass is a constant
\begin{equation}
\left| \frac{\partial (p_1,p_2,...,p_{n-1},s_{2..n},p_{2..n} ) }{\partial (p_1,...,p_{n})} \right| \rightarrow \delta(s-m_{intermediate})
\end{equation}
or a Lorentzian if $\Gamma \ne 0$.

If the resonance distribution is azimuthally invariant Eq.~(\ref{manybody}) for the 3-body case
simplifies to \cite{heinzreso}
\begin{equation}
\label{3bodyazim}
\frac{dN_1}{d^3 p_1} =B  \frac{\sqrt{m_j-(s_{2..n}+m_1)^2} \sqrt{m_j-(s_{2..n}-m_1)^2}}{m_j} 
\left( E_{s \rightarrow 23} \frac{dN_{s \rightarrow 23}}{d^3 p_{s \rightarrow 23}} \right) 
\end{equation}
\begin{eqnarray}
B&=&\frac{m_j b_{j\rightarrow 123}}{2 \pi Q[(m_j+m_1)^2,(m_j-m_1)^2,(m_2-m_3)^2,(m_2+m_3)^2]}\\
Q[a,b,c,d]&=& \int^{b}_{c} \frac{dx}{x} \sqrt{(a-x)(b-x)(x-c)(x-d)}
\end{eqnarray}
where $\left( E_{s \rightarrow 23} \frac{dN_{s \rightarrow 23}}{d^3 p_{s \rightarrow 23}} \right)$ is
given by Eq.~(\ref{resofinal_azim}).

From the above discussion, it is clear that in general an n-body decay will involve $2 n+1$ numerical
integrals.    Cascades of n steps with a $n_i$ decay at each step will require a total of $\sum_{i=1}^n n_i$ integrals to find the final decay products from the resonance distribution.

It is clear that in general Monte-Carlo integration becomes the only way to evolve decays beyond
a simple topology.     Algorithms such as MAMBO \cite{mambo} can generate points with a uniform density
in n-body phase space.   In the subsequent chapters we will use MAMBO to perform integrals within the
Monte-Carlo program, and multi-dimensional Gaussian integration \cite{cernlib} in our fits.
The latter, therefore, can only accommodate a restricted number of resonances
where the decay topology is not too complicated.
\section{What kind of statistical hadronization?}
The formation of QGP should be accompanied by statistical hadronization at some point:
This, in a sense, true by definition, since phase transitions characterize equilibrated systems.
However, by itself, statistical hadronization is not a proof of quark gluon plasma formation, since
any system will thermalize after enough time as discussed in section 2.2.

While, as discussed in sections 1.5.5 and 1.5.6, the equilibration time of a system of quarks
should be larger by an order of magnitude than that of a system of hadrons, hence evidence of early thermalization
can be construed as a QGP signature, some thought should be put into whether statistical hadronization of
a QGP can be distinguished from freeze-out of an equilibrated hadronic system.

An obvious signal to look for is whether the hadron formation temperature approaches the temperature predicted
by lattice for deconfinement.  However, as explained in chapter 1, lattice results have still to converge
on a definite value for the phase transition temperature.   In addition, if the phase transition is
sharp enough, phenomena such as super-cooling can lower the temperature at which hadrons are emitted.

A phase transition from a QGP to a hadron gas should also considerably increase the system's mean free path.
Due to this, the Cooper-Frye prescription should be particularly appropriate.
While this has certain consequences, which the next two chapters will explore, it is also hardly a definite
distinction, due to the Cooper-Frye formula's flexibility in the choice of $d^3 \Sigma$ and $u^{\mu}$ (see chapter 4)

Finally, it should be mentioned that the phase space available to QGP is very different to that available to a HG:
The lightest degree of freedom in a (perturbative) QGP is massless, with a degeneracy of 16.   The lightest
degree of freedom in a Hadron Gas, on the other hand, has a mass comparable to the hadronization temperature
and a degeneracy of 3.   This suggests that, if the phase transition is sharp, statistical emission of hadrons
from a QGP will not be based on chemical equilibration at all, but rather on production of hadrons
from quarks with thermal weights.  Section 2.4.3 describes a statistical hadronization model based on these
assumptions.   

These considerations suggest that ``statistical hadronization'' actually encompasses a variety of models, different in both
physical interpretation and observational consequences.   This is indeed the case.    The following two sub-sections describe
the two ``camps'' into which statistical models can be categorized.
\subsection{Long freeze-out}
\begin{figure}[h]
\centering
  \psfig{figure=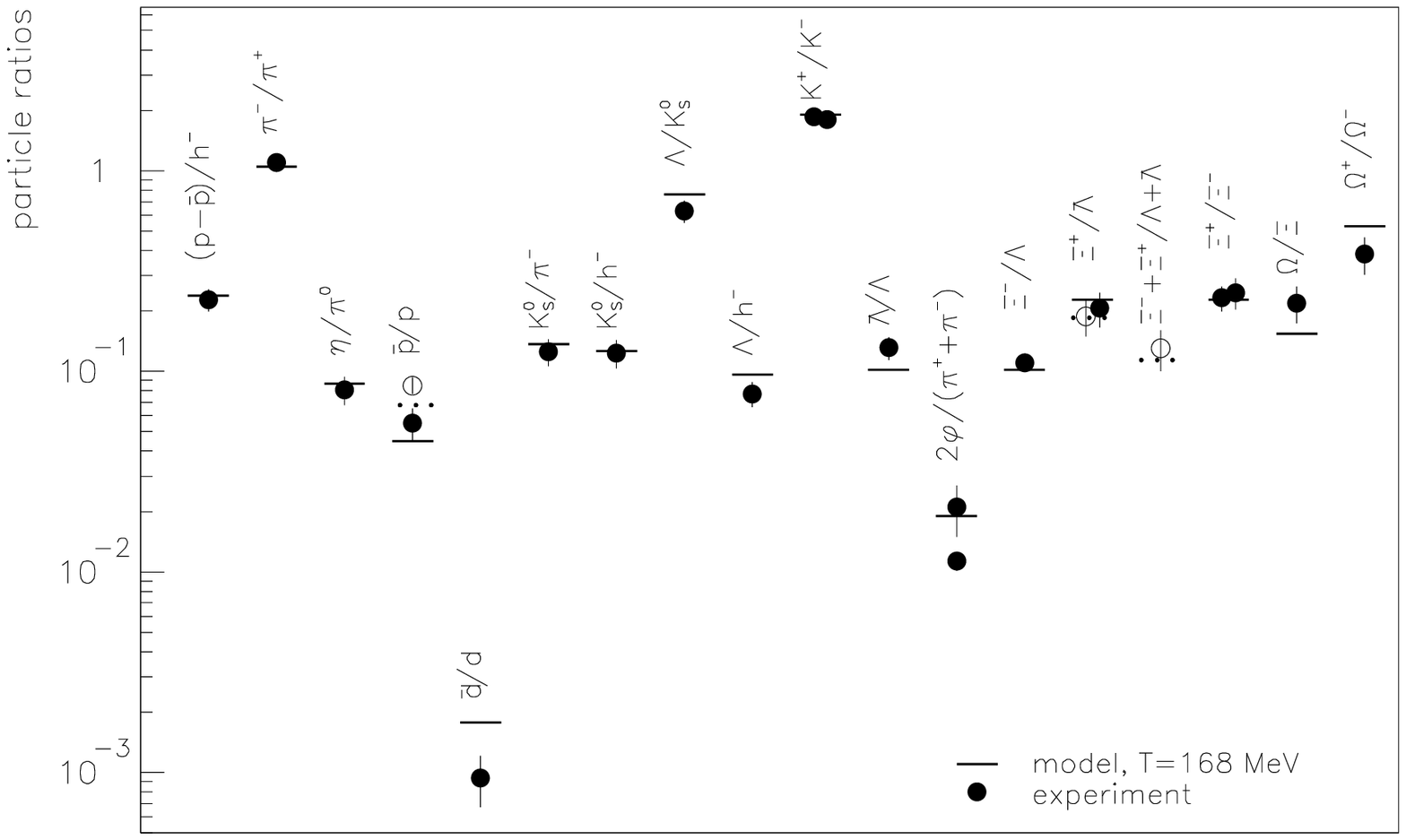,width=7cm}
  \psfig{figure=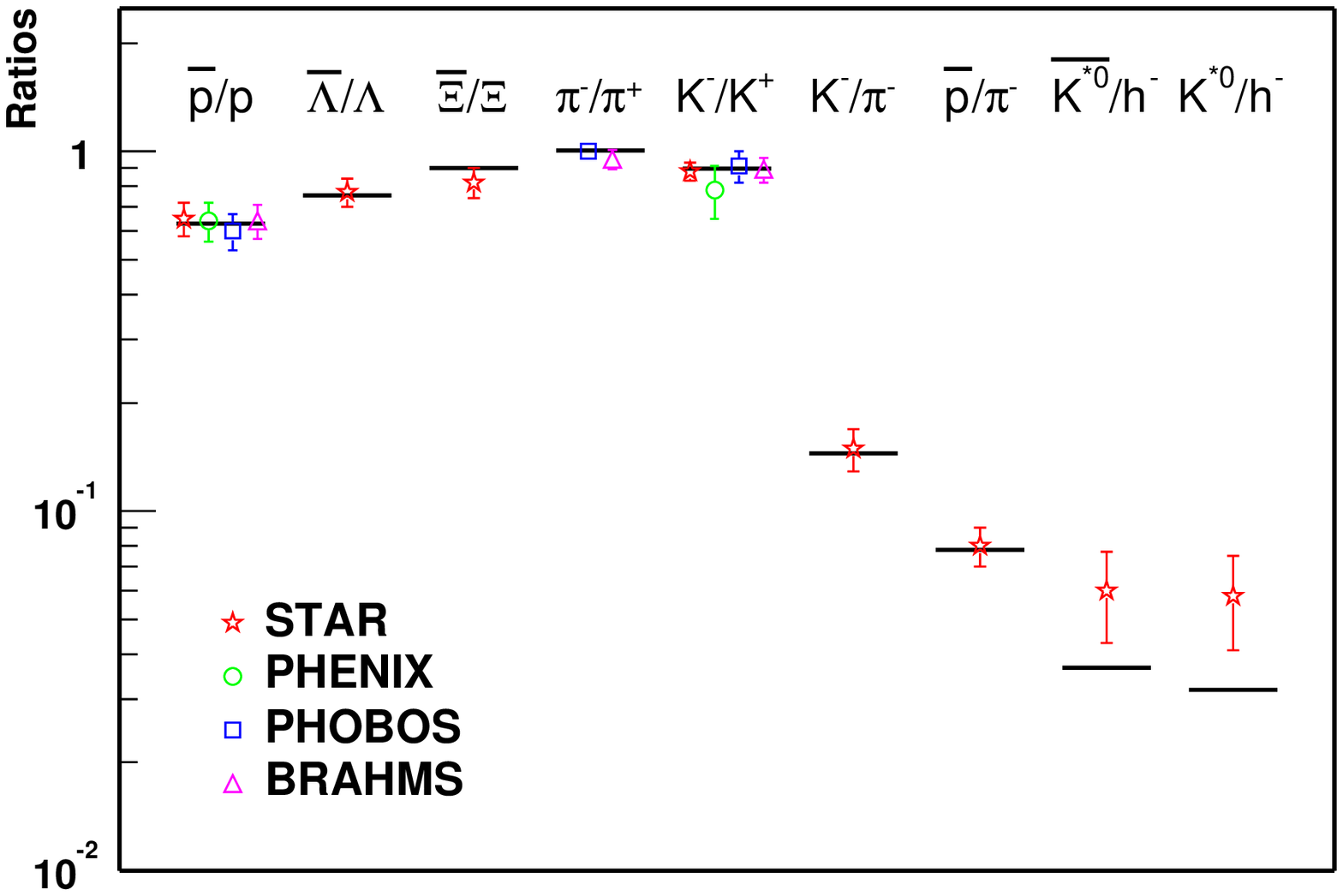,width=7cm}
  \caption{Equilibration fits for SPS (left) and RHIC (right) energies \label{bdm}.  The $\chi^2/\mathrm{DoF}=38/16$ at the SPS and $5.7/7$ at RHIC.}
\end{figure}

The most naive model for freeze-out assumes particle formation at the deconfinement temperature, a long-interacting
hadron gas phase, and thermal freeze-out at $\sim 100$ ${\rm MeV}$.   Superficially, such a picture has
experimental support.
It is possible to fit hadrons produced at RHIC \cite{bdmrhic} and SPS \cite{bdmsps} (Fig.~(\ref{bdm}))
using a model based on chemical equilibration $\lambda_q=\lambda_{\overline{q}}^{-1}$ and Eqns.~(\ref{Nk}) and~(\ref{Nres}) , and get a
temperature of the order of $\sim 170 {\rm MeV}$.  The chemical potential decreases with higher energy, as expected (see Chapter 1).
In this model hadrons will continue interacting, so 
particle spectra will have to be modeled with a different set of thermodynamic
parameters. Experiments have done these fits, neglecting resonance abundances, and have found a spectra freeze-out temperature of 100${\rm MeV}$ for all particles except $\Xi$ s and $\Omega$, consistent
with a staged thermal freeze-out \cite{castillo},\cite{van-leuween} where particles with fewer
interactions with pions (particularly the $\Omega$, as there is no $\Omega \pi$ resonance) freeze-out earlier.

$\Sigma^{\mu}$ in the Cooper-Frye formula will then not be a good physical description of
particle emission, since particles will not be emitted from a definite point in spacetime, but will
emerge from a ``freeze-out hyper-volume'' (a continuous limit of fewer and fewer interactions).
However, the Cooper-Frye description can still be used as a prescription to switch from a continuous medium
to a gas of hadrons (not necessarily at a phase transition, but simply the space-time region the mean free path
becomes ``large'') \cite{hydro_shuryak} 

This picture, however, has problems both at a fundamental level and in terms of experimental agreement.
The fits in Fig.~\ref{bdm} actually have a very small statistical significance.
The experimental fits performed in \cite{castillo},\cite{van-leuween} are demonstrably meaningless,
since they do not include resonances which have already been experimentally detected (see chapter 5).
On a fundamental level,   It is hard to see how entropy gets conserved
in a QGP-HG chemically equilibrated phase transition unless the transition is 
slow enough to permit the system's
volume to increase significantly (something not corroborated by the lattice).

It is also difficult to see why particle spectra should be described by a hydrodynamical model at all,
given that an interacting hadron gas at $T=120-170$ should include many out-of-equilibrium inelastic
interactions.   In particular, the possibility for baryon-antibaryon annihilation at any cross-section
makes it puzzling that the $m_T$ slopes for particle and anti-particle as virtually identical, as shown
in chapter one.  (Fig.~\ref{uRQMDflow}, \cite{WA97spectra}).
This feature is predicted by hydrodynamic models with a few simple reactions
(quark-level kinetics, $gg \leftrightarrow q\overline{q}$), but certainly not in an interacting hadrons gas picture with many particle-specific channels
($\Lambda + \overline{\Lambda} \leftrightarrow$ pions,$p+\overline{\Lambda} \leftrightarrow$ Pions and Kaons,...) .

Another failure of the low temperature scenario is Hanbury-Brown-Twiss interferometry (HBT) applied to $\pi$ and K.   This technique, which  relies on
bosonic two-particle correlations assuming no quantum mechanics before emission, has been used to estimate the spacetime shape of the system when interactions stop \cite{HBTreview}.
It has been found that ``naive'' hydrodynamics with early freeze-out fails to explain the observed correlations \cite{HBTpuzzle} unless fast emission and a small freeze-out radius (which contradict the model)
are introduced by hand.   

While the effectiveness of HBT as a description has been questioned in light
of its assumptions  (For instance, how to properly take  short lived resonances \cite{HBTreso} and final state interactions \cite{HBTrqmd,HBTint} into account?), the need of hydrodynamic models to introduce fast freeze-out Ad Hoc to
even qualitatively explain the data, together with its other failings,  is a strong motivation to look for a scenario where fast freeze-out
is part of the model, and particles stop interacting soon after formation.
\subsection{Explosive freeze-out}
\begin{figure}[h]
\centering
  \psfig{figure=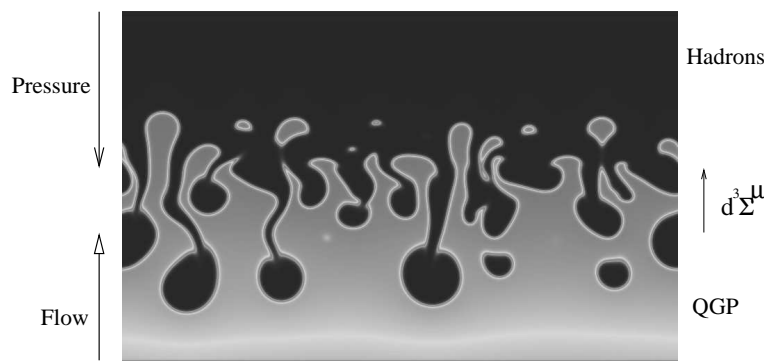,width=10cm}
  \caption{How expansion plus vacuum pressure leads to viscous fingering \label{fingering}.}
\end{figure}

One framework which would explain the rapid freeze-out and lack of post-hadronization interactions
is explosive hadronization of a super-cooled plasma \cite{sudden1},\cite{sudden2}.
If the QGP-HG phase transition is first order, at a certain critical temperature/density the two phases
will coexist.  We can estimate this critical density through the Bag model, described in the previous
chapter (section 1.4.1)
\begin{equation}
T^{\mu \nu}_{QGP} = T^{\mu \nu}_{HG} + B g^{\mu \nu}
\end{equation}
where $B$ is the vacuum pressure/bag constant.
If the Quark gluon plasma exhibits strong collective expansion, ie
\begin{equation}
T^{\mu \nu}_{QGP} = P g^{\mu \nu} + (P+\epsilon) u^{\mu} u^{\nu}
\end{equation}
it will expand past the phase of coexistence.   The QGP will then experience negative pressure, from
the outside vacuum.   What happens in this situation is described in hydrodynamics as ``viscous fingering'' \cite{fingering}:  A mechanical instability develops on the surface where the two phases meet, and the lower pressure
QGP ``fingers'' into the higher-pressure vacuum.   Hadron emission will then occur through ``bubbling''
at that surface, as the mechanical instability tears the fireball apart (Fig.~\ref{fingering}).
The hadronization freeze-out surface will then acquire a physical meaning as the surface where the
two vacua come into contact and viscous fingers develop.   Indeed, the whole freeze-out dynamics is
modeled through the choice of $\Sigma^{\mu}$ and the Matching conditions \cite{sudden3}
\begin{eqnarray}
d^3 \Sigma_{\mu} (T^{\mu \nu}_{QGP}-T^{\mu \nu}_{HG}) = 0 \\
d^3 \Sigma_{\mu} (n^{\mu}_{QGP}-n^{\mu}_{HG})=0\\
d^3 \Sigma_{\mu} (S^{\mu}_{HG}-S^{\mu}_{QGP}) \geq 0
\end{eqnarray}
Where $n^{\mu}$ refers to any conserved quantum number current (flavor, strangeness, charge etc.).
It becomes clear that the last equation is the most problematic.  While the first two equations
can generally be satisfied through a readjustment of the thermodynamic parameters at freeze-out
\cite{sudden4}, the entropy of the hadron gas will in general tend to always be much lower than that
of the QGP due to the fact that HG degrees of freedom are much more massive.
In a slow freeze-out scenario entropy will be generated through flow.
If freeze-out is sudden, however, entropy conservation across the freeze-out surface in equilibrium
is only possible by either choosing carefully the freeze-out parameters or  post-hadronization
reheating (release of latent heat) \cite{sudden4}.

A more general freeze-out scenario is provided by dropping the assumption that hadronization happens under flavor 
chemical equilibrium.
Chemical equilibrium requires that particle and antiparticle fugacities
are inverse of each other
\begin{equation}
\lambda_q = \lambda_{\overline{q}}^{-1}
\end{equation}
which, by the law of mass action, corresponds to the equilibration of creation and annihilation
processes
\begin{equation}
q \overline{q} \rightarrow gg \Leftrightarrow gg \rightarrow q \overline{q}
\end{equation}
According to the Boltzmann equation Eq.~(\ref{boltzmann}) this condition will be reached
for all quark flavors after infinite time.   To model the approach of this condition, 
one defines the phase space occupancy $\gamma_q$ so that
\begin{equation}
\lambda_q \rightarrow \lambda_{q} \gamma_q  \; \; \; \gamma_q=\gamma_{\overline{q}} 
\end{equation}
Equilibrium will then be reached  when $\gamma_q=1$.
Several authors have allowed the possibility that strangeness is not yet in chemical equilibration
when freeze-out happens \cite{becattini}, but it is still generally assumed that light quarks
are equilibrated.   It is not immediately clear, however, why any quantum number should be equilibrated at all in a phase transition during dynamic
non-equilibrium.

\begin{figure}[h]
\centering
  \psfig{figure=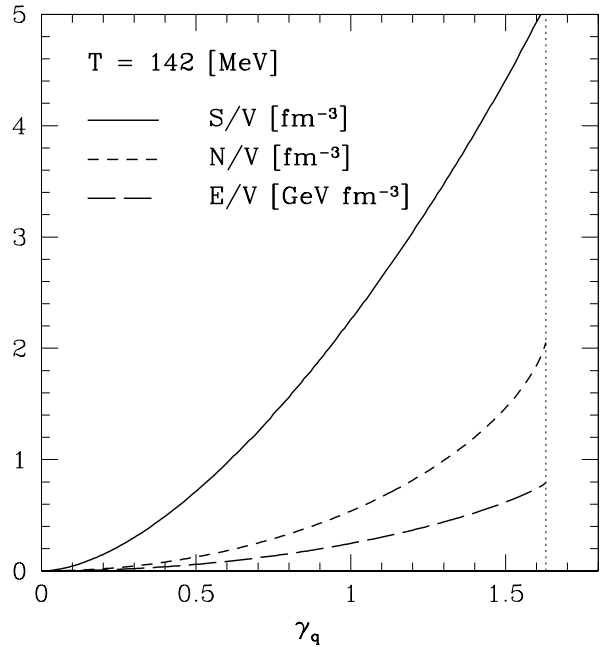,width=9cm}
  \caption{Entropy, Energy and number density as $\gamma_q>1$ increases  \label{gammaq}.}
\end{figure}

It turns out that if one allows light quark phase space to be overpopulated, i.e. grow above equilibrium
($\gamma_q > 1$), the entropy of the hadron gas starts growing rapidly with $\gamma_q$
\cite{rafgammaq} (Fig.~\ref{gammaq}).
This region is phase space is kinetically not reachable through a hadron gas evolution such as that described in 
Eq.~(\ref{boltzmann}).   However, if the particles are emitted from a rapidly non-perturbatively hadronizing 
$q \overline{q}$ rich QGP, over-saturation might occur.   This is the only way found as yet to contain the entropy
created during the QGP stage without reheating or expansion (which requires slow hadronization)
The value of $\gamma_q$ is bounded above by the Bose-Einstein condensation condition for pions
\begin{equation}
\frac{1}{\gamma_q^{-2} e^{m_{\pi}/T}-1} < \infty,   \; \; \; \gamma_q < e^{m_{\pi}/2T}
\end{equation}
above which a Bose-Einstein Condensate (BEC) of pions forms, which consumes energy while carrying no entropy.
The BEC constraint works well with the super-cooling hypothesis, since $\gamma_q$ has more freedom
to rise to the point at which $S_{QGP}<S_{HG}$ only as the critical temperature becomes
fairly small $\sim 140 {\rm MeV}$.

This picture is phenomenologically in accord with the Gribov hadronization scenario described
in section 1.3.1.     As energy density reaches a critical value, the color potential between
quarks and gluons 
grows  up to the point where it is energetically convenient to fragment gluons into $q \overline{q}$
pairs to maintain large-scale color neutrality.
The $q \overline{q}$ phase space is then overpopulated as the gluons are depleted by $g \rightarrow q \overline{q}$ reactions.
\begin{table}[bt]
\begin{center}
\caption{
\label{table_raf1sps} The chemical freeze-out  statistical parameters found for 
 nonequilibrium (left) and semi equilibrium (right) 
fits to SPS results. We show $\sqrt{s_{NN}}$,  
the  temperature $T$,  light quark fugacity $\lambda_q$, 
strange quark fugacity $\lambda_s$,
the quark occupancy parameters $\gamma_q$ and
$\gamma_s/\gamma_q$. Bottom line presents
the statistical significance of the fit. 
The star  (*) indicates for $\lambda_s$ that it is a value resulting from
strangeness conservation constraint. For $\gamma_q$ that there is 
an upper limit  to which the value converged,
$\gamma_q^2<e^{m_\pi/T}$ (on left), or that the value of $\gamma_q=1$
is set (on right).
}\vspace*{0.2cm}
\begin{tabular}{l|ccc|ccc}
\hline\hline
$\sqrt{s_{NN}}$\,[${\rm GeV}$]    &17.2   & 12.3  &8.75    & 17.2   & 12.3  &8.75 \\[0.1cm]
\hline
$T$\,[${\rm MeV}$]       &$135\pm3$    & $135\pm3$      &$133\pm2$       & $157\pm4$     &$156\pm4$         & $154\pm3$\\[0.1cm]
\hline
$\lambda_q$      &$1.69(5)$ &$1.98(6)$ &$2.56(6)$ &$1.74(5)$ &$2.03(7)$   &$2.69(8)$\\[0.1cm]
$\lambda_s$      &$1.23^*$      &$1.27^*$       &$1.31^*$      &$ 1.20^*$       &$1.24^*$        &$ 1.24^*$   \\[0.1cm]
\hline
$\gamma_q$          &$1.68^*$     &$1.68^*$        & $1.69^*$    & $1^*$    & $1^*$           & $1^*$ \\[0.1cm]
$\gamma_s/\gamma_q$  &$0.91(6)$ & $0.83(4)$  &$0.85(6)$  &$0.66(4)$    &$0.60(4)$  &$0.67(5)$ \\[0.1cm]
\hline
$\chi^2/$dof      &11.4/6  & 4.3/2      &2.3/4  & 23/7   &8.9/3    & 4.0/5  \\[0.1cm]
\hline\hline
\end{tabular}
\end{center}
 \end{table}
\begin{table}[htb]
\caption{
\label{table_raf1} freeze-out  statistical parameters found for 
 non-equilibrium (left) and $\gamma_q=1$ ,$\gamma_s$ fitted (right) 
fits to RHIC results. We show $\sqrt{s_{NN}}$,  
the  temperature $T$,  baryochemical potential $\mu_b$, 
strange quark chemical potential $\mu_s$,
strangeness chemical potential $\mu_{\rm S}$, 
the quark occupancy parameters $\gamma_q$ and
$\gamma_s/\gamma_q$, and in the bottom line 
the statistical significance of the fit. 
The star  (*) indicates that there is an upper limit on the value of 
$\gamma_q^2<e^{m_\pi/T}$ (on left), and/or that the value is set (on right).
}\vspace*{0.2cm}
\begin{center}
\begin{tabular}{l|cc|cc}
\hline\hline
$\sqrt{s_{NN}}$\,[${\rm GeV}$]    &200          & 130        &200          & 130    \\[0.1cm]
\hline
$T$\,[${\rm MeV}$]                &$143\pm7$    & $144\pm3$  &$160\pm8$    & $160\pm4$\\[0.1cm]
$\mu_b$\,[${\rm MeV}$]            &$21.5\pm31$&$29.2\pm4.5$   &$24.5\pm3$   &$31.4\pm4.5$ \\[0.1cm]
$\mu_s$\,[${\rm MeV}$]            &$2.5\pm0.2$ &$ 3.1\pm0.2$ &$2.9\pm0.2$ &$ 3.6\pm0.2$ \\[0.1cm]
$\mu_{\rm S}$\,[${\rm MeV}$]            &$4.7\pm0.4$ &$ 6.6\pm0.4$ &$5.3\pm0.4$ &$ 6.9\pm0.4$ \\[0.1cm]
\hline
$\gamma_q$                &$1.6\pm0.3^*$&$1.6\pm0.2^*$& $1^*$ & $1^*$ \\[0.1cm]
$\gamma_s/\gamma_q$       &$1.2\pm0.15$ & $1.3\pm0.1$ &$1.0\pm0.1$&$1.13\pm0.06$ \\[0.1cm]
$\chi^2/$dof               &2.9/6        & 15.8/24    &4.5/7       & 32.2/25  \\[0.1cm]
$P_{\rm true}$              &90\%+        & 95\%+    &65\%       & 15\%  \\[-0.7cm]
\end{tabular}
\end{center}
 \end{table}
Tables~\ref{table_raf1sps} and~\ref{table_raf1} \cite{strange_rafelski2} and \cite{strange_rafelski3} make
it clear that fits where $\gamma_q$ is a variable parameter consistently prefer values of $\gamma_q$
just below the BEC limit.    In this minimum, $\mu_s=T \ln(\lambda_s)$ goes to a value very close to 0, in
accordance to a scenario where strangeness is produced through annihilation and gluon fusion
reactions. These fit's statistical significance is markedly better
than fits based on the equilibrium hypothesis.
\section{How to falsify freeze-out models}
Sudden non-equilibrium freeze-out has therefore passed the elementary comparison with experimental
data.   It can fit observed hadron yields to an acceptable statistical significance.
It also seems that super-cooling can explain the small volume and fast emission times inferred
by HBT observations \cite{sudden3}, though this has yet to undergo a quantitative corroboration.

However, the sudden freeze-out picture allows for much more stringent falsifiability tests to be made, 
since it demands that all soft variables be described by the same set of statistical mechanics
parameters.    The rest of this thesis Performs some of these falsifiability tests, which can
be broadly grouped in three categories.
\begin{description}
\item[Particle spectra]
In the sudden hadronization picture, particles form from a transversely expanding system and
undergo little or no re-interaction after formation.   Therefore, the same temperature which describes
hadron abundances should, with the addition of transverse flow, also be able to describe the shape
of hadron spectra.   Moreover, because of the absence of the hadron gas phase, the number of
hadrons in each particle species is unchanged since formation, except for the decay of short-lived
resonances.

Therefore, spectra should be normalized through hadronic chemical potentials, and the resonance
admixture into the spectra should be prominently present.  The non-equilibrium
parameters $\gamma_q$ fitted in table~\ref{table_raf1} should also be
necessary to describe particle spectra.
These predictions are tested in detail within the next two chapters.
\item[Short-lived resonances]
Short lived resonances (strong excited states) provide a direct test of post-hadronization interactions,
since their final yield will be affected by them.   The sudden freeze-out model demands, therefore,
that their abundance be determined by the same temperature and chemical potentials that determine
long-lived particles.   Any deviation from this can be attributed to post-hadronization dynamics.
This is explored quantitatively in chapter 5.
\item[Exotica (Pentaquarks etc.)]  Pentaquarks should be strongly enhanced in a non-equilibrium model:
Their mass seems to be similar to that of other short-lived resonances \cite{Dia97}, but, of course,
they have a different number of quarks (5 in the pentaquark case).
Hence, in non-equilibrium models they should be enhanced with respect to similar baryons by a
factor of $\sim \gamma_q^2$.  This is a large difference for $\gamma_q \sim 1.5$.
This is also explored in chapter 5.
\item[Freeze-out dynamics]
If hadronization happens through a vacuum instability at the QGP-HG contact surface, the freeze-out
hypersurface $d^3 \Sigma^{\mu}$ acquires a physical significance connected to the confining phase transition.
Therefore 
\begin{itemize}
\item Experimental data should strongly prefer a particular choice of $d^3 \Sigma^{\mu}$
\item The preferred $d^3 \Sigma^{\mu}$ should be localized on a surface moving with time across
the fireball.
\end{itemize}
Chapter 4 will explore whether the $d^3 \Sigma^{\mu}$ choice can be made through an analysis
of particle spectra alone.   Chapters 5 and 6 will provide ways to constrain  $d^3 \Sigma^{\mu}$ further.
\end{description}

\setcounter{figure}{0}
\setcounter{equation}{0}
\setcounter{table}{0}
\chapter{Particle spectra at SPS: a spherical ansatz}
\label{cha:spectra_sps}

\section{Sudden freeze-out spectra}
The WA97 and NA57 experiments have, for the first time, measured the
normalized spectra of a wide range of strange hadrons, ranging
in mass from the $K_S$ to the $\Omega$ \cite{WA97spectra}.
The relative yield of these particles has been one of the
main pieces of evidence pointing to the formation of quark gluon
plasma at SPS energies \cite{cern_evidence}.
Correspondingly the spectra's variety can provide a very stringent 
constraint on the system's freeze-out conditions.

In particular, when sudden QGP breakup occurs, 
the  spectra of hadrons are not formed at a range of 
stages in fireball  evolution, but arise rather
suddenly. Most importantly, particles of very different 
properties are produced by the same mechanism and thus are 
expected to have similar
$m_\bot$-spectra as is indeed observed \cite{WA97spectra}.  
The reported symmetry of the strange
baryon and antibaryon spectra is strongly suggesting that 
the same reaction mechanism produces $\Lambda$ and $\overline\Lambda$ 
and $\Xi$ and $\overline\Xi$. 
This is a surprising, but rather clean
experimental fact, which we will 
interpret quantitatively in this paper. 

When the momentum distributions of final state particles 
stop evolving during the fireball evolution, we speak of thermal
freeze-out. Because a spectrum of strange hadrons includes
directly produced, and heavy resonance decay products,
one can determine the freeze-out temperature and 
dynamical velocities of fireball evolution 
solely from the study of  precisely known 
shape of the particle spectra. We demonstrate
this in some detail in section 
\ref{chidata}. The physical mechanism is that 
the freeze-out temperature  determines the 
relative contribution  of each decaying resonance
while the shape of each  decay contribution differs 
from the thermal shape, see section~\ref{Thermal}. 

We note that we 
make in our analysis the tacit assumption that 
practically all decay products of resonances 
are thermally not re-equilibrated, which is equivalent 
to the assumption of sudden freeze out. This is consistent 
with our finding that  the $m_\bot$ strange baryon 
and antibaryon  distributions of $\Lambda, 
\overline\Lambda, \Xi, \overline\Xi$ froze out 
near to the condition at which the chemical
 particle yields were  established. 

One of the key objectives of this  work is to present 
a comparison between thermal 
and chemical freeze-out analysis results for temperature,
(explosion) collective velocity and other chemical 
and dynamical parameters. It is important to realize that 
particle spectra and yields are sensitive to magnitude of 
collective matter flow, in which 
produced particles are born, for somewhat different 
reasons:
1) in thermal analysis the collective flow combines with
thermal freeze-out temperature to fix the shape of each particle 
spectrum, and temperature is also controlling the relative yield of 
contributing resonances -- see previous paragraph for 
a here relevant tacit assumption -- thus both $T$ and $v$ 
are fixed by  the shape of $m_\bot$ data; 
2) in chemical analysis the  particle yields required are obtained
integrating spectral yields, with experimental 
acceptance in $p_\bot,y$ implemented. Since many
particles have a too small  particle momentum to be
usually observed,
the acceptance-cut yields used in chemical analysis 
depend quite sensitively
on parameters which   deform the soft part of the 
spectra without changing
the number of produced particles, such as is the flow velocity.
For this reason precise particle spectra and yields
are allowing to draw conclusions about the proximity of 
thermal and chemical 
freeze-out conditions. 
 
\section{Spherically symmetric freeze-out}

Collisions at the SPS are characterized by a high degree of stopping power:
Most of the primordial particles lose their longitudinal momentum in the initial collision, and
the mid-rapidity region has a high particle/antiparticle imbalance \cite{NA49stop} and hence
baryochemical potential.
Hence, it is reasonable to assume that the system will not have a longitudinal rapidity
structure at freeze-out.   
One such emission surface which has the advantage of analytical simplicity is spherically
symmetric freeze-out.   Such an ansatz is also favoured by hydrodynamic evolution, since the
hydrostatic force per unit volume is minimized.

We proceed to construct a spherically symmetric freeze-out hypersurface, in which the freeze-out
time depends on the radius alone
\begin{equation}
\label{sigma_sph}
\Sigma^{\mu}= \left( \begin{array}{c} t_f (r) \\ r \sin (\theta) \cos(\phi) \\  r \sin (\theta) \sin(\phi) \\r \cos(\theta) \end{array} \right)
\end{equation}
Parameterizing this surface in terms of $r,\theta,\phi$ we get, according to Eq.~(\ref{dsigma}), an emission
element of this form:
\begin{equation}
\label{dsigma_sph}
d^3 \Sigma^{\mu}= r dr d\theta \left( \begin{array}{c} 1 \\ \frac{\partial t_f}{\partial r} \sin (\theta) \cos(\phi) \\  \frac{\partial t_f}{\partial r} \sin (\theta) \sin(\phi)\\ \frac{\partial t_f}{\partial r} \cos(\theta) \end{array} \right)
\end{equation}
combining this emission shape with a spherically symmetric flow we get an explicit form from the Cooper-Frye formula (where $\theta$ is the configuration space emission coordinate and $\phi$ the particle momentum
coordinate).   
\begin{equation}
\label{f_sph}
E \frac{dN}{d^3 p} = N  \prod \lambda_i \gamma_i \int r^2 dr \int \sin(\theta) d \theta \left( E-p_T  \frac{\partial t_f}{\partial r} \cos(\theta - \phi) \right) e^{-\frac{\gamma}{T} (E-v p_T \cos(\theta-\phi))}
\end{equation}
Using modified Bessel functions, the integral over $\theta$ can be done analytically
\begin{equation}
\label{f_sph2}
E \frac{dN}{d^3 p} = N  \int r^2 dr \sqrt{\frac{T}{\gamma v p_T}} \left( E I_{1/2} \left( \frac{\gamma v p_T}{T} \right)-p_T  \frac{\partial t_f}{\partial r}  I_{1/2} \left(\frac{\gamma v p_T}{T} \right) \right)
\end{equation}
and the Normalization N includes the degeneracy and chemical potentials
\begin{equation}
N= g_i \prod \lambda_i \gamma_i
\end{equation}
This approach can be generalized to B-E and F-D statistics, relevant
for $\pi$, as, for realistic chemical potentials, these distributions can be represented as a converging series of Boltzmann-like therms
\begin{equation}
\lambda e^{-\frac{p_{\mu} u^{\mu}}{T}} \rightarrow \sum_{n=0}^{\infty} (\pm 1)^n_{BE/FD} \lambda^{n+1} e^{-(n+1)\frac{p_{\mu} u^{\mu}}{T}} 
\label{boltz_fdbe}
\end{equation}
For a real hydrodynamical system the integration over r will not be trivial, since flow will be a non-trivial, equation of state dependent function of r.
For our fit, we shall average the system using one flow.
\begin{equation}
E \frac{dN}{d^3 p} = N V E \frac{dN}{d^3 p} \left( <v>,<\gamma> \right)
\end{equation}
The consequences of such averaging will be explored in the next chapter in detail
\section{Thermal freeze-out analysis of SPS data}\label{Thermal}
In recent months experiment WA97 
determined the absolute normalization of
 the published 
$m_\bot$ distribution \cite{WA97spectra},  and we
took the opportunity to perform the spectral shape analysis
and will compare  our results to those obtained in chemical
yield analysis \cite{strange_rafelski3} in order to check if the 
thermal and chemical freeze-out conditions are the same.
Our analysis continues and this report gives its current status.

We report here a simultaneous analysis of absolute yield and shape of 
WA97 results of six $m_\bot$-spectra of $\Lambda,\,\overline\Lambda, \,\Xi,\,
\overline\Xi,\, \Omega+ \overline\Omega, \, K_s=(K^0+\overline{K^0})/2$
in four centrality bins.
If thermal and chemical freeze-outs are identical, our
present results  must  be consistent with 
earlier chemical analysis of hadron yields. Since
the experimental data we here study 
is dominated by the shape of $m_\bot$-spectra
and not by relative particle yields, our analysis is de facto 
comparing thermal and chemical freeze-outs. 

We have found, as is generally believed and expected, 
 that all hadron $m_\bot$-spectra are strongly influenced by 
resonance decays. Thus we apply standard procedure to 
allow for this effect \cite{heinzreso} and described in the previous chapter.

Since particle spectra we consider have a good relative
normalization, only one parameter is required for 
each centrality in order to describe the absolute normalization
of all six hadron spectra. This is for two reasons important:\\
a) 
we can check if the volume from which strange hadrons 
are emitted grows with centrality of the collision as
we expect;\\
b) 
we can determine which region in $m_\bot$ 
produces the excess of $\Omega$  noted in the
chemical fit \cite{strange_rafelski3} is coming from.

However, since the normalization $V_{\scriptsize\rm QGP}$ common for
all particles at given centrality comprises additional  
experimental acceptance normalization, we have  not 
determined the value of the fireball emission volume at each centrality. 
Hence we will be presenting the volume parameter as
function of centrality  in arbitrary units.

The best thermal and chemical  parameters are found by minimizing the total relative
error $\chi^2$ as defined in Appendix B, where the data points are given by the
 WA97 \cite{WA97spectra} spectra for
 K$^{0}$, $\Lambda$, $\overline{\Lambda}$, $\Xi$, $\overline{\Xi}$,
$\Omega + \overline{\Omega}$. We have checked the validity of the 
statistical analysis by the usual method, i.e. omission of 
some data in the fit. 

Only in case of Kaons we find any impact of such a
procedure. Noting that the statistical error of kaon spectra 
is the smallest, we have established how a a systematic
error which could be for Kaons greater than statistical 
error would influence our result.  For this purpose 
we assign to   K$^0$ experimental results in most of our analysis  
an `error'  which  we arbitrarily  have chosen to be 5 times greater 
than the statistical error. In this way the weight of the kaon spectra 
in the analysis is greatly reduced. In the first result figure below 
(Fig.~\ref{TdT}) we present both results, standard K$^0$ error and enlarged 
error. We see that while in individual result some change can occur,
overall the physical result of both analysis are consistent. Thus we 
can trust in the combined study of hyperon and kaon data. This conclusion
is reaffirmed in section~\ref{chidata}, where we will see that the 
minimization of $\chi^2$ involve more or less pronounced 
minima, depending on the error size of the kaon spectra,
see Fig.~\ref{TchiT2}. In most calculations we present in this
paper we will be using, unless otherwise said, hyperon results combined
with the Kaon data with 5 times enlarged statistical error. We believe
that in this way we will err on the conservative side in our 
physical conclusions.

\section{Overview of the results}
We show here a slate of results obtained within the 
approach outlined above. First we address the parameters
determining the shape of the $m_\bot$ distributions,
that is $T,v,\frac{\partial t_f}{\partial r}$.

\begin{figure}[tb]
\vspace*{-1.3cm}
\centerline{
\psfig{width=9.cm,clip=,angle=-90,figure=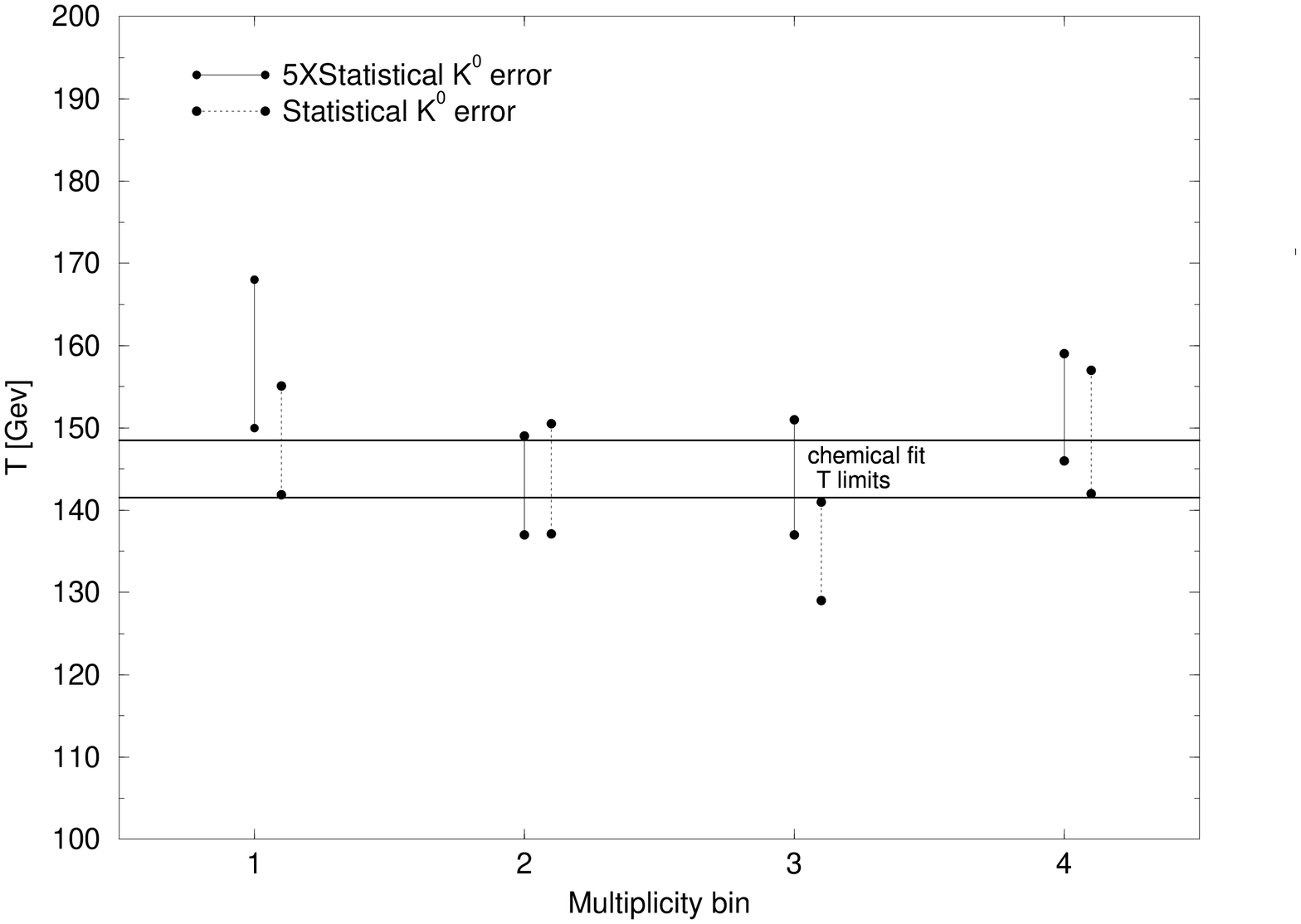}
}
\vspace*{-0.6cm}
\caption{(color on-line) 
Thermal freeze-out temperature $T$ 
for different centrality bins compared to 
chemical freeze-out analysis shown by
horizontal solid lines. Original statistical error is used in the 
dotted results, five times statistical error for  kaon data is used 
in solid vertical lines. 
\label{TdT}
}
\end{figure}

As function of the centrality bin we show in Fig.~\ref{TdT}
the freeze-out temperature $T$ 
of the $m_\bot$ spectra. The horizontal lines 
delineate range of result of the most recent  chemical 
 freeze-out analysis , see  Ref. \cite{strange_rafelski3}.
It is reassuring that we find a result
consistent with the purely chemical analysis of data
that included non-strange hadrons \cite{strange_rafelski3}. 
There is 
no indication of a significant or systematic change of $T$ with centrality.
This is consistent with the believe that the formation of the new state of 
matter at CERN is occurring in all centrality bins explored by the 
experiment WA97. Only most peripheral interactions  
produce a change in the pattern of strange hadron production \cite{Kab99}. 
The (unweighted) average of all results shown in Fig.~\ref{TdT}
produces a freeze-out temperature at the upper boundary of the 
the pure chemical freeze-out analysis result, $T\simeq 145$\,${\rm MeV}$. 
It should be noted that in chemical analysis 
$\frac{\partial t_f}{\partial r}=v$ \cite{strange_rafelski3}, which may be the cause of this 
slight difference between current analysis average and the earlier 
purely chemical analysis result. 

\begin{figure}[tb]
\vspace*{-1.3cm}
\centerline{\hskip -0.9cm
\epsfig{width=10.cm,clip=,figure=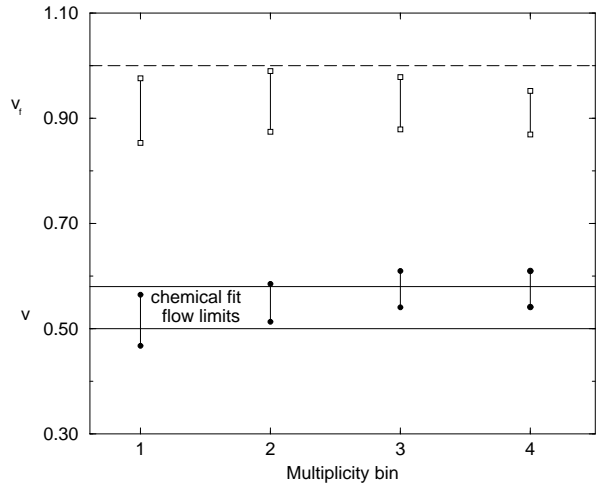}
}
\vspace*{-0.9cm}
\caption{(color on-line) 
Thermal freeze-out flow velocity $v$ (top) and 
break up (hadronization) velocity $\frac{\partial t_f}{\partial r}$ 
for different centrality bins. Upper limit $\frac{\partial t_f}{\partial r}=1$ (dashed line) and 
chemical freeze-out analysis limits for $v$ (solid lines) are also shown.
\label{Tdv1v2}
}
\end{figure}

The magnitudes of the 
collective expansion velocity $v$  and the  break-up (hadronization) speed 
parameter $\frac{\partial t_f}{\partial r}$ are presented in Fig.~\ref{Tdv1v2}.
For $v$ (lower part of the figure) 
we again see consistency with earlier chemical freeze-out
analysis results, and there is no confirmed systematic trend
in the behavior of this parameter as function of centrality. 

Though within the experimental error, one could argue 
inspecting  Fig.~\ref{Tdv1v2} that there is 
systematic increase in transverse flow velocity $v$ with centrality and thus 
size of the system. This is expected, since the more central events comprise 
greater volume of matter, which allows more time for development
of the flow.  Interestingly, it is in $v$ and not $T$ that we find the 
slight change of spectral slopes noted in the presentation of the 
experimental data \cite{WA97spectra}. 

The value of the break-up (hadronization) speed 
parameter $\frac{\partial t_f}{\partial r} [=1/(\partial r_f/\partial t_f)]$ 
shown in the top portion of Fig.~\ref{Tdv1v2} is near to 
velocity of light which is highly  consistent with the picture of a 
sudden breakup of the  fireball. This 
hadronization surface velocity $\frac{\partial t_f}{\partial r}$ was in the earlier chemical
fit set to be equal to $v$, as there was not enough sensitivity in
purely chemical fit to  determine the value of $\frac{\partial t_f}{\partial r}$. 

\begin{figure}[tb]
\vspace*{-1.3cm}
\centerline{\hskip 0.5cm
\epsfig{width=9.5cm,clip=,angle=-90,figure=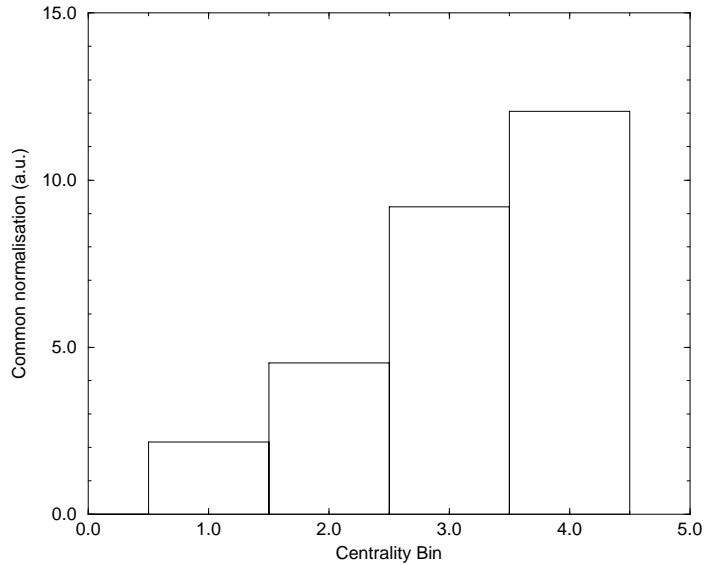}
}
\vspace*{-0.2cm}
\caption{ 
Hadronization volume (arbitrary units)
 for different centrality bins. 
\label{Tdiagnorm}
}
\end{figure}

Unlike the temperature and two velocities, the overall normalization
of hadron yields, $V^h$ must be, and is strongly centrality dependent, as is
seen in Fig.~\ref{Tdiagnorm}. This confirms in quantitative way the
believe that the entire available fireball volume is available for 
hadron production. The strong increase in the volume
by factor 6 is qualitatively 
consistent with a geometric interpretation of the 
collision centrality effect. Not shown is the error propagating from the
experimental data which is strongly correlated to the chemical
parameters discussed next. This systematic uncertainty is another reason 
we do not attempt an absolute unit volume normalization. 

The 4 chemical parameters $\lambda_q,\lambda_s, \gamma_q, \gamma_s/\gamma_q$ are
shown in the following Figures~\ref{Tdlqls},\ref{Tdgamqgams}.
These parameters determine along with $V^h$ the final particle
yield. Since we have 5 parameters 
determining normalization of  6 strange hadron spectra, and as discussed we 
reduce the statistical weight of Kaons, there
is obviously a lot of correlation between these 4 quantities, and thus 
the error bar which reflects this correlation,  is significant. 

The chemical fugacities $\lambda_q$ and $\lambda_s$ shown in 
Fig.~\ref{Tdlqls} do not exhibit
a systematic centrality dependence. This is consistent with the result
we found for $T$ in that the freeze-out properties of the fireball are
seen to be for the temperature and chemical potential values independent
of the size of the fireball. Comparing to the earlier chemical
freeze-out result in Fig.~\ref{Tdlqls} one may argue that 
there is a systematic downward deviation in $\lambda_q$. However,
 this could be  caused by the fact that the 
chemical freeze-out analysis allowed for
isospin-asymmetric $\Xi^-(dss)$ yield \cite{strange_rafelski3}, 
while out present analysis is not
yet distinguishing light quarks.

\begin{figure}[tb]
\vspace*{-2.9cm}
\centerline{\hskip -2.5cm
\epsfig{width=11.cm,clip=,angle=-90,figure=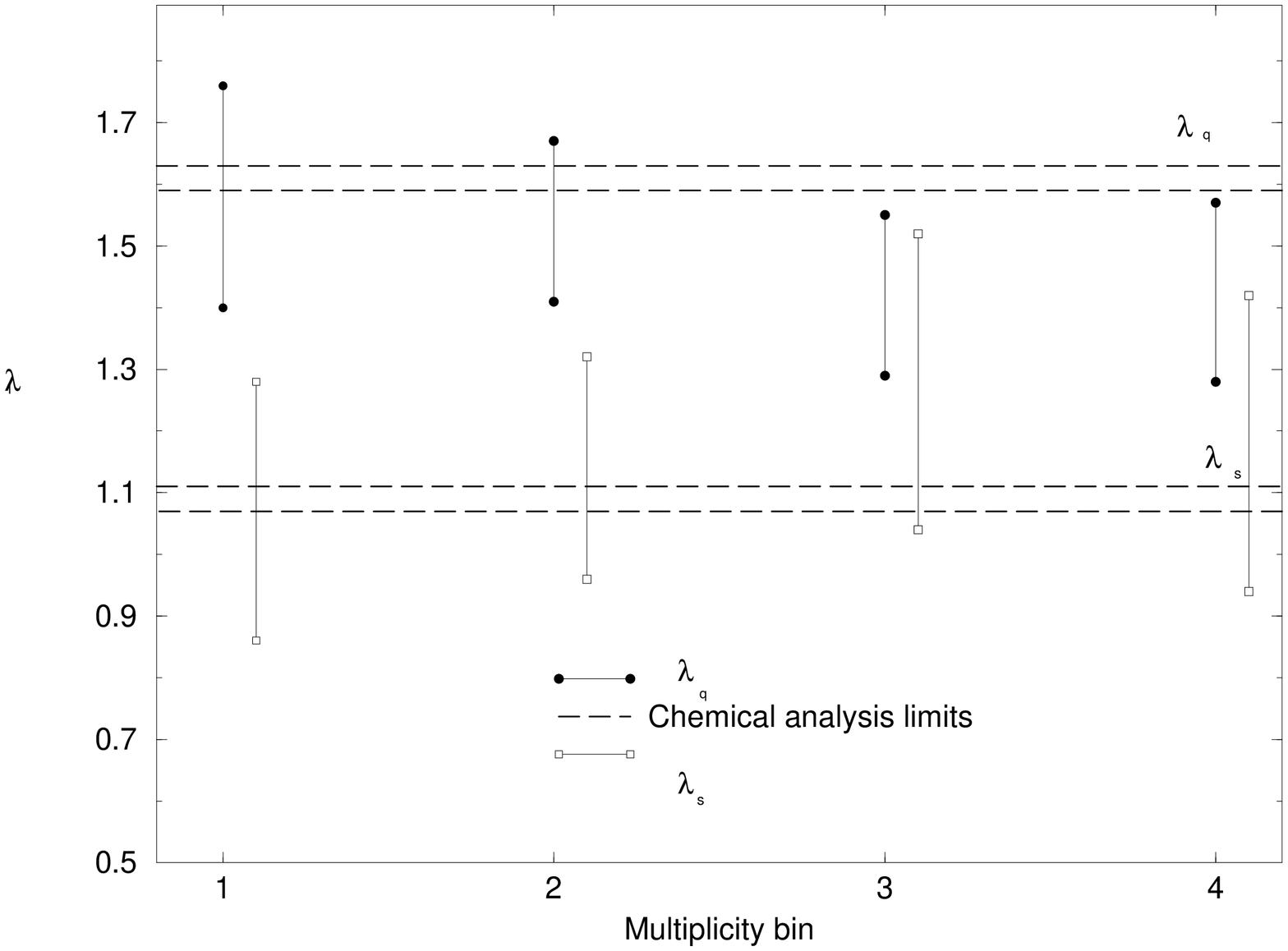}
}
\vspace*{-0.2cm}
\caption{ 
Thermal analysis chemical quark fugacity $\lambda_q$ (top) and 
strange quark fugacity $\lambda_s$ (bottom)
 for different centrality bins compared to 
the chemical freeze-out analysis results. 
\label{Tdlqls}
}
\end{figure}

The ratio $\gamma_s/\gamma_q$  shown in bottom portion of  Fig.~\ref{Tdgamqgams}
 is systematically smaller than unity, consistent with many years of prior analysis:
when $\gamma_q=1$ is tacitly chosen, this ratio is the value of $\gamma_s$
in analysis of strange baryons.
We have not imposed a constrain on the range of $\gamma_q$ 
(top of Fig.~\ref{Tdgamqgams}) and thus 
values greater than the pion condensation point 
$\gamma_q^*=e^{m_\pi/2T}\simeq 1.65$ (thick line) can be expected, but in fact 
do not arise.

It is important to explicitly check
how well the particle $m_\bot$-spectra are
reproduced. We group all bins in one figure and show 
in Figs.~\ref{TdLamtot},~\ref{TdALamtot},~\ref{TdXitot},~\ref{TdAXitot} 
in sequence  $\Lambda,\,\overline\Lambda,\,\Xi,\,\overline\Xi$. 
It is important to note that there are some significant 
deviations which appear to be falling outside of the trend set by the 
other measurements. --this  occurs for $\Lambda$ as well 
but remains invisible in the figure due to the smallness of the 
experimental error bar. Overall, the description of the shape of the spectra 
is very satisfactory.

\begin{figure}[tb]
\vspace*{-2.5cm}
\centerline{\hskip -1.cm
\epsfig{width=9.5cm,height=9.5cm,clip=,angle=-90,figure=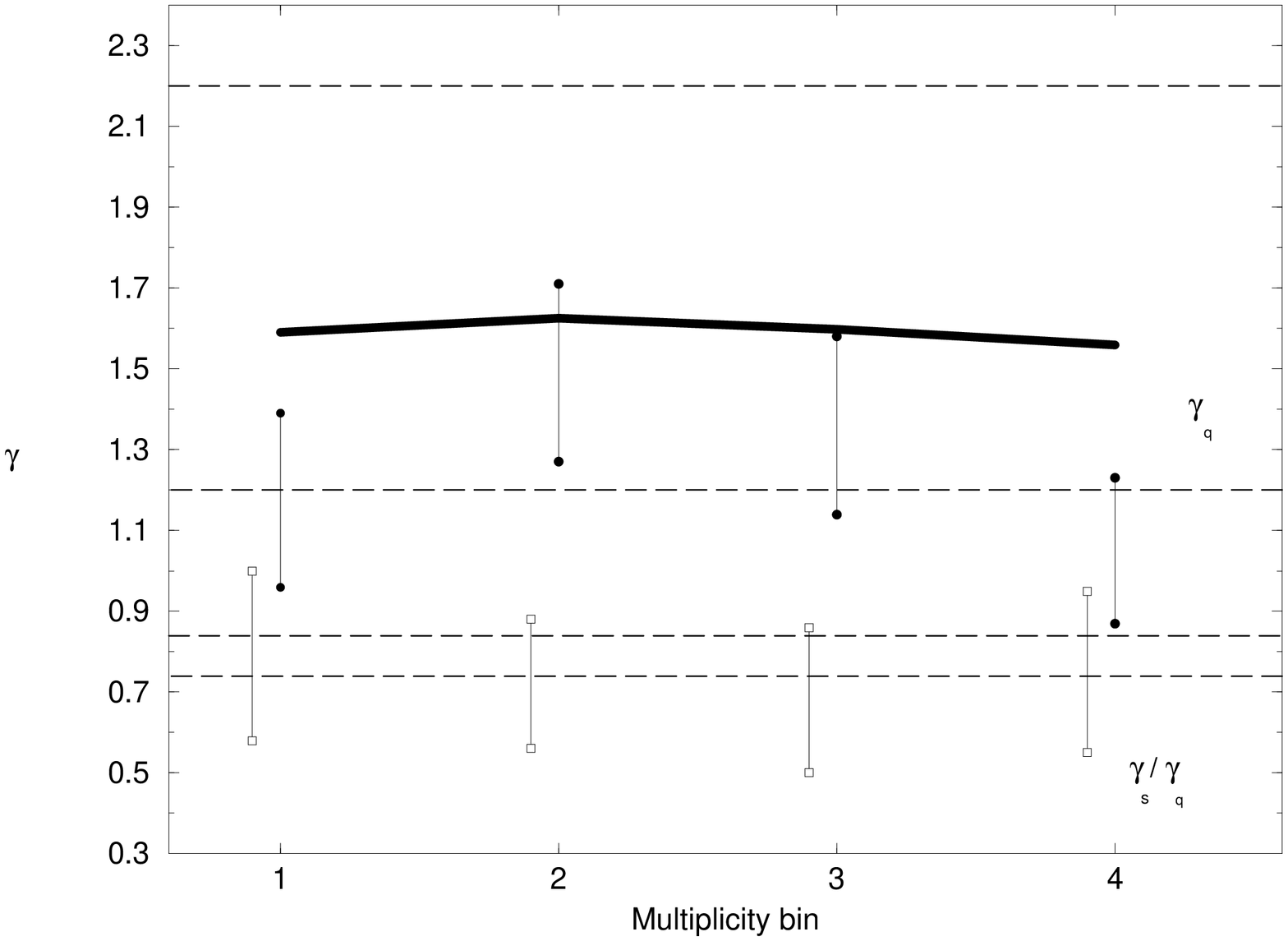}
}
\vspace*{.7cm}
\caption{ 
Thermal analysis chemical quark abundance parameter $\gamma_q$ (top)
and $\gamma_s/\gamma_q$ (bottom)  for different centrality bins compared to 
the chemical freeze-out analysis. Thick line: upper limit due to pion
condensation.
\label{Tdgamqgams}
}
\end{figure}

\begin{figure}[tb]
\vspace*{-1.8cm}
\centerline{\hskip 0.5cm
\epsfig{width=10.cm,clip=,angle=-90,figure=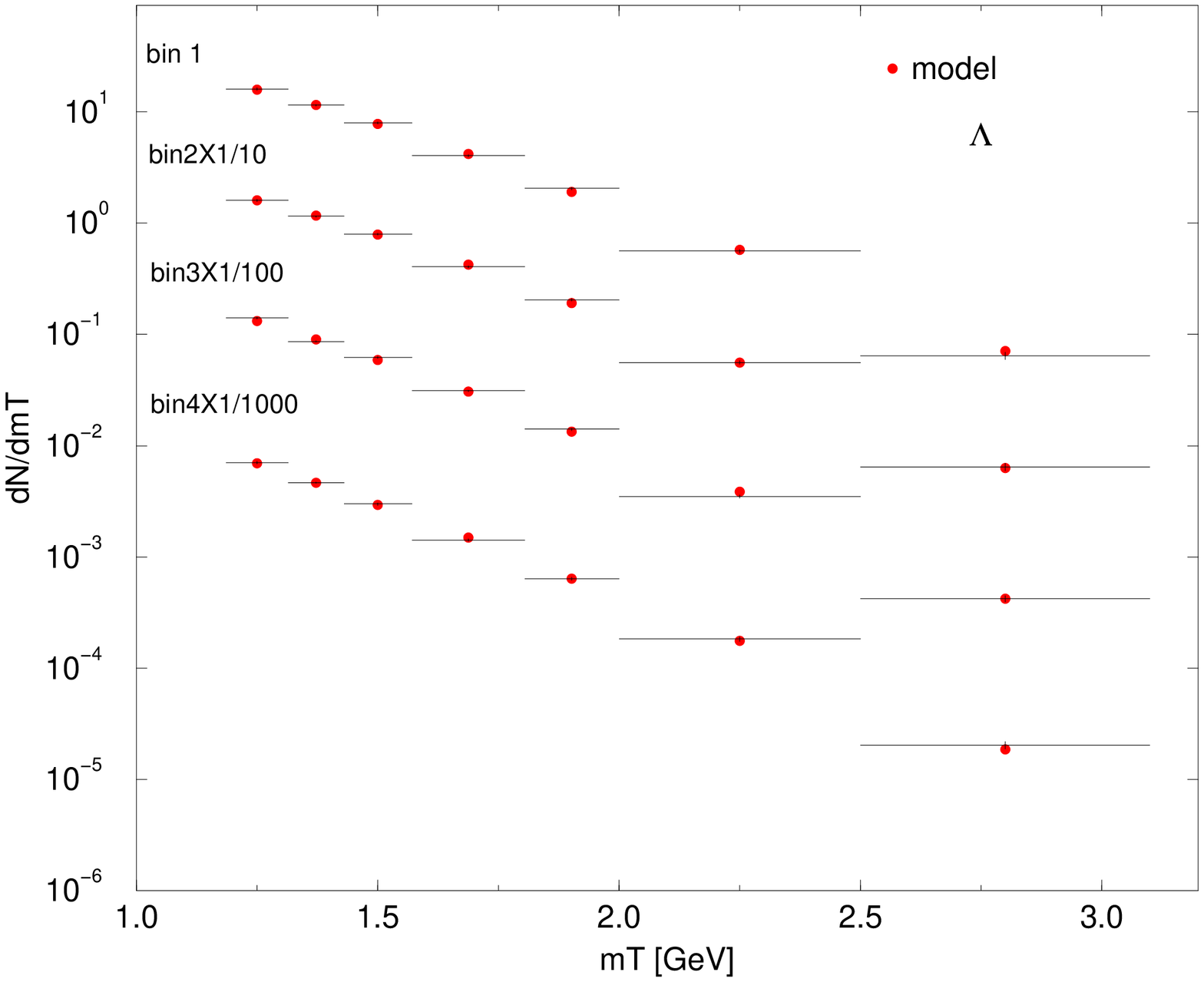}
}
\vspace*{-0.2cm}
\caption{(color online) 
Thermal analysis $m_T$ spectra: $\Lambda$ 
\label{TdLamtot}.
}
\end{figure}

We also describe the K$_0$data extremely well,
especially 
in the $m_\bot$ range which is the same as that for
hyperons considered earlier,  as is seen in Fig.~\ref{TdK0All}.
We recall that these results were obtained reducing the 
statistical significance of Kaon data, and thus the conclusion
is that hyperons predict both the abundance and shape of 
kaon spectra. Moreover, all the strange hadron spectra can be
well described within the model we have adopted.  

\begin{figure}[tb]
\vspace*{-1.5cm}
\centerline{\hskip 0.5cm
\epsfig{width=10.cm,clip=,angle=-90,figure=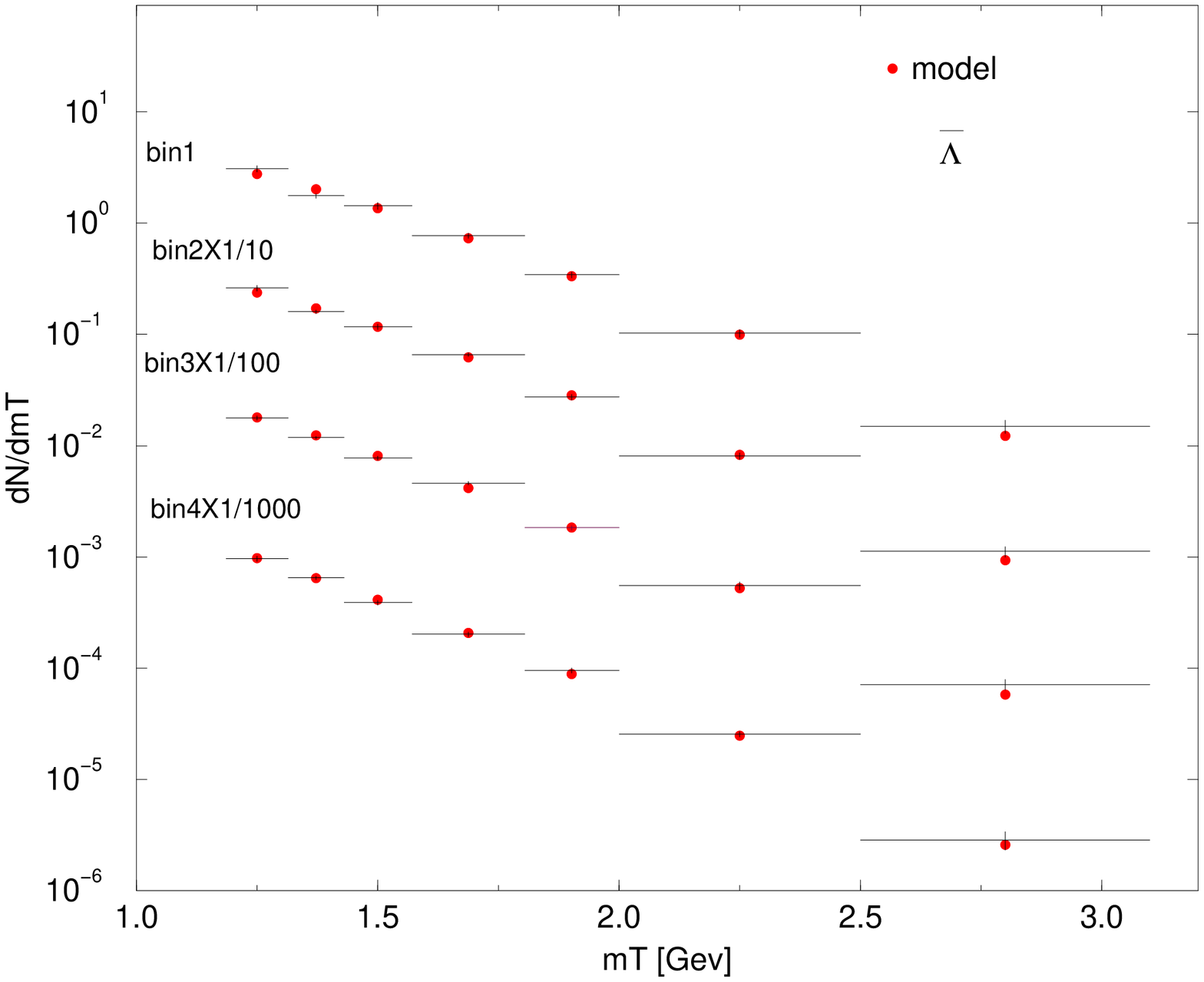}
}
\vspace*{-0.2cm}
\caption{(color online) 
Thermal analysis $m_T$ spectra: $\overline\Lambda$. 
\label{TdALamtot}
}
\end{figure}

\begin{figure}[tb]
\vspace*{-1.5cm}
\centerline{\hskip 0.5cm
\epsfig{width=10.cm,clip=,angle=-90,figure=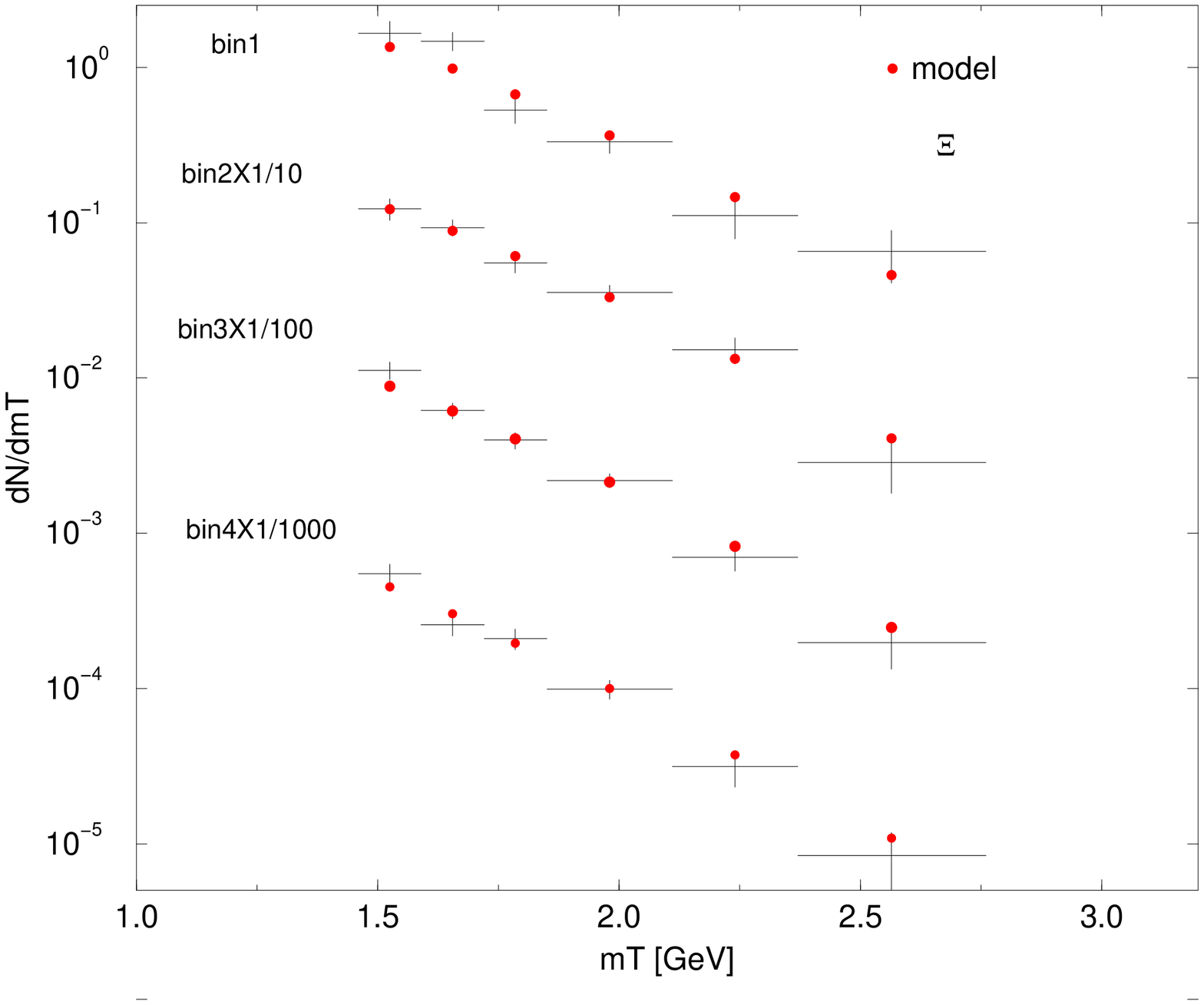}
}
\vspace*{-0.2cm}
\caption{(color online) 
Thermal analysis $m_T$ spectra: $\Xi$.
\label{TdXitot}
}
\end{figure}

\begin{figure}[tb]
\vspace*{-1.5cm}
\centerline{\hspace*{0.70cm}
\epsfig{width=10cm,clip=,angle=-90,figure=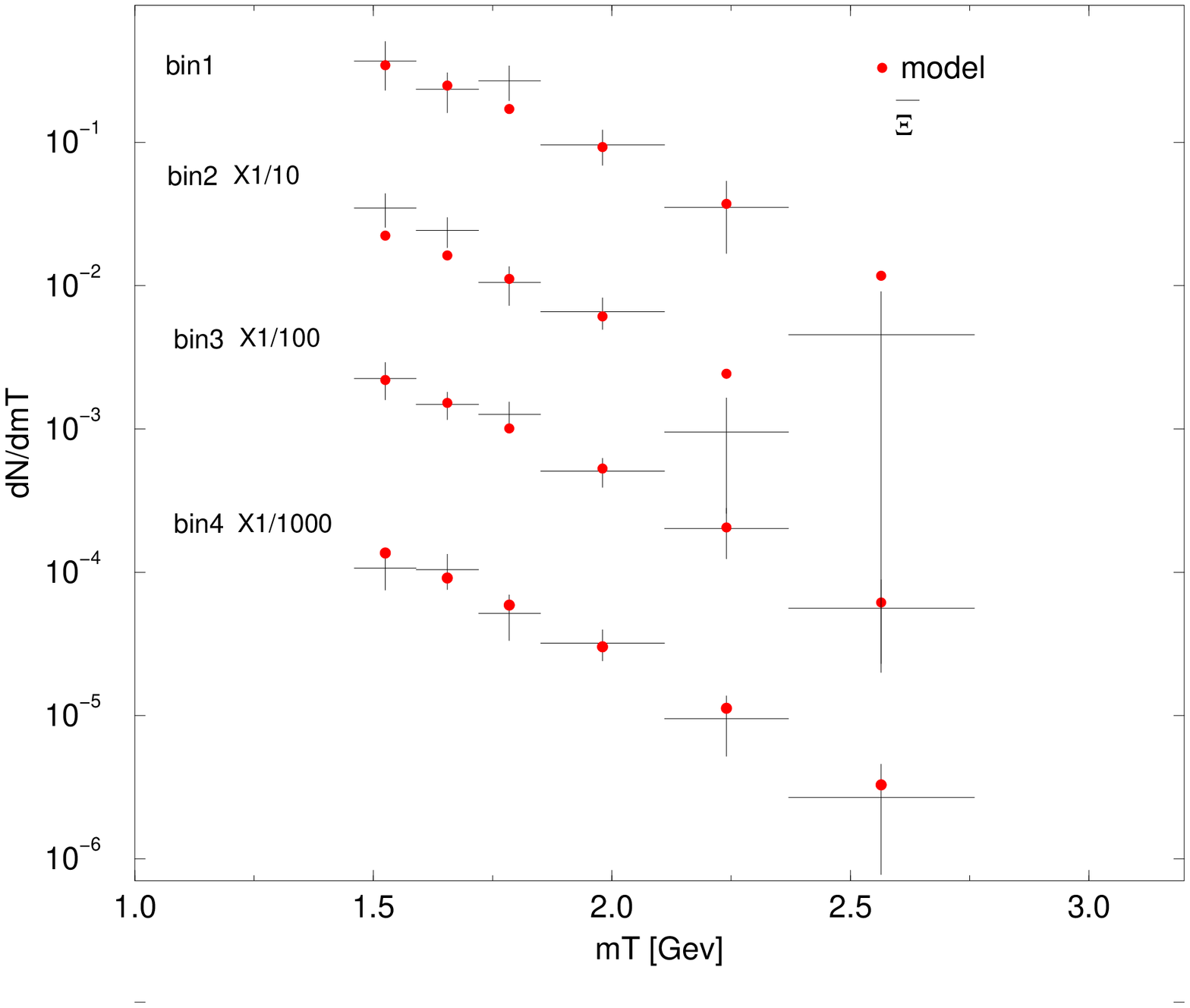}
}
\vspace*{-0.2cm}
\caption{(color online) 
Thermal analysis $m_T$ spectra: $\overline\Xi$.
\label{TdAXitot}
}
\end{figure}

\begin{figure}[h]
\centerline{\hskip 0.5cm
\epsfig{width=10.cm,clip=,angle=-90,figure=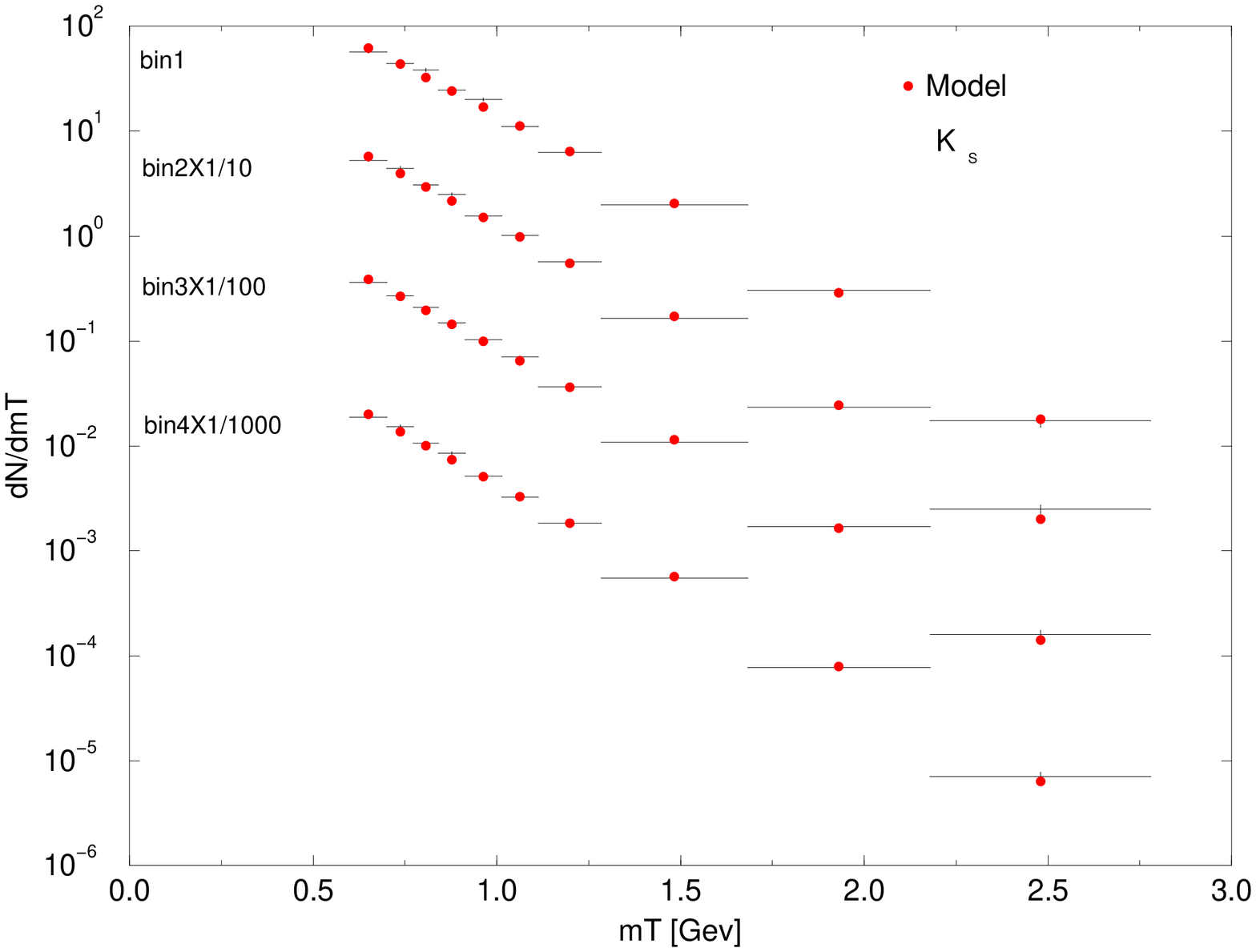}
}
\caption{(color online) 
Thermal analysis $m_T$ spectra: $K_s$.
\label{TdK0All}
}
\end{figure}

\section{Statistical significance of the 
results presented}\label{chidata}
We have analysed the validity and consistency 
of our data analysis by exploring $\chi^2/\mathrm{DoF}$ 
profiles. These are obtained by fixing the 
value of one of the parameters (we consider $T_f,
v, \partial t_f/\partial r_f=1/\frac{\partial r_f}{\partial t_f}$), and computing 
the related  $\chi^2_{\rm T}/\mathrm{DoF}$, the total error 
divided by degrees of freedom. These are the number 
of measurements minus number of parameters, $\mathrm{DoF}$ is
typically 33 in this data analysis.  All curves must have 
the same $\chi^2_{\rm T}/\mathrm{DoF}$ at the minimum and this 
minimum must point to the value of parameters we report. 
For the temperature $T$ we produced two results 
shown in Fig.~\ref{TchiT2}, in the bottom section
for experimental (statistical) $K^0$ measurement error,
and in the top part for the five times enlarged error.
We recall that both results are presented in Fig.~\ref{TdT}.
We note that there is a pronounced $\chi^2_{\rm T}/\mathrm{DoF}$ 
minimum shown on logarithmic scale) 
for all 8  results of which the average value is at 
$T=145$ ${\rm MeV}$. 
\begin{figure}[h]
\centerline{
\epsfig{width=9.5cm,clip=,figure=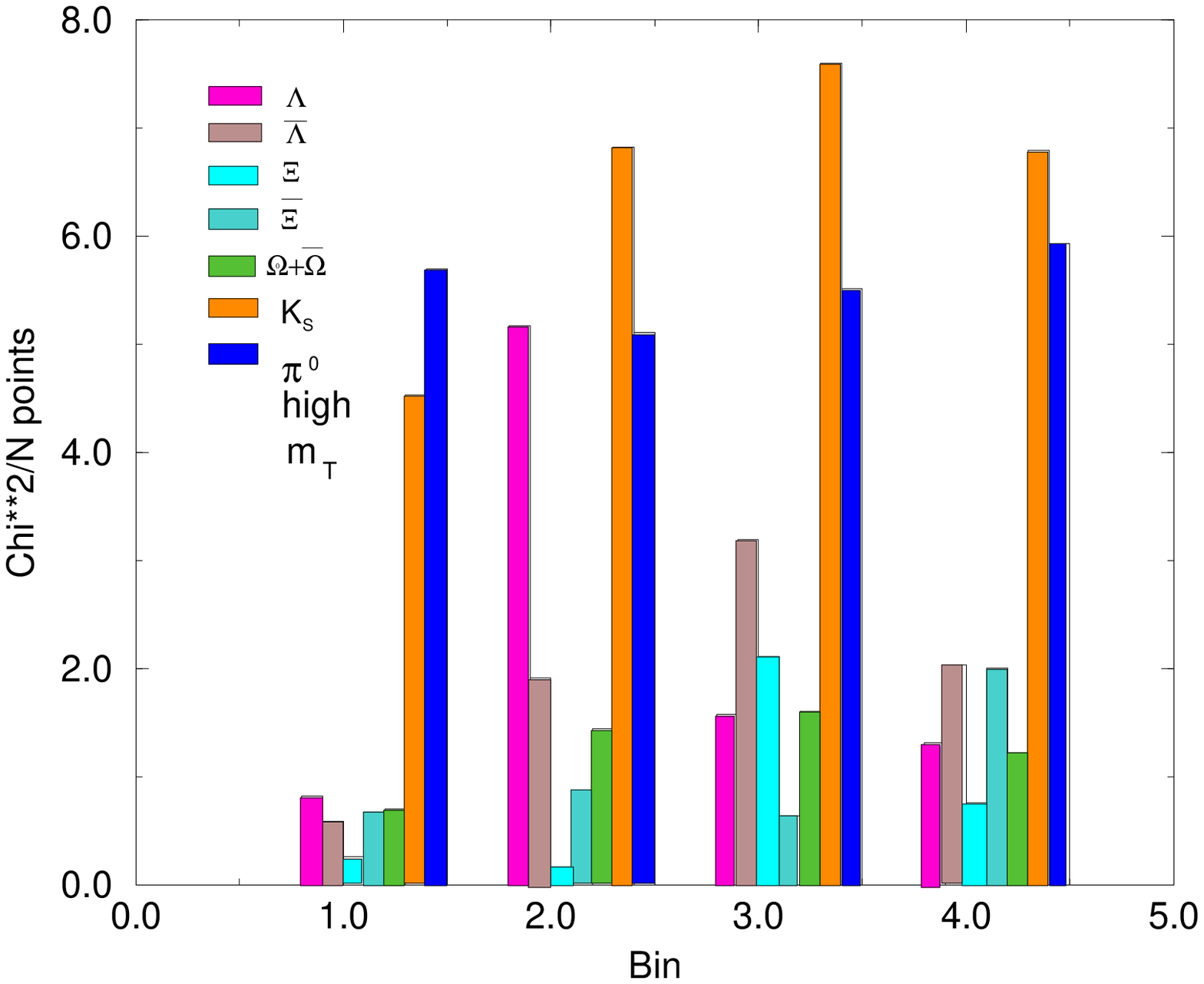}}
\caption{(color online) 
$\chi^2$/Dof per particle data point.   As can be seen, $K_s$ (error given by the experiment) gives the dominant
discrepancy
\label{diagchi}.
}
\end{figure}
We show in Fig.~\ref{Tchivv2} the 
profile of $\chi^2_{\rm T}/\mathrm{DoF}$  for the collective flow velocity
v (top) and the freeze-out surface 
$\partial t_f/\partial r_f=1/\frac{\partial r_f}{\partial t_f}$ motion (bottom) being 
fixed. These minima can be shown on linear scale. We note
a mild secondary minimum in the region 
$v\simeq 0.25$--0.35. However, the
minima we find at $v=0.5$--0.58 are by far more significant.
$\partial t_f/\partial r_f=1/\frac{\partial r_f}{\partial t_f}$ is converging to a sharp
minimum seen in bottom portion of  Fig.~\ref{Tchivv2}
at at a value consistent with the sudden breakup scenario.
It is necessary to include $\partial t_f/\partial r_f=1/\frac{\partial r_f}{\partial t_f}$
along with $v$ in the analysis to find this result, 
which was not always done in other studies of particle spectra.

\begin{figure}[h]
\epsfig{width=6.5cm,clip=,figure=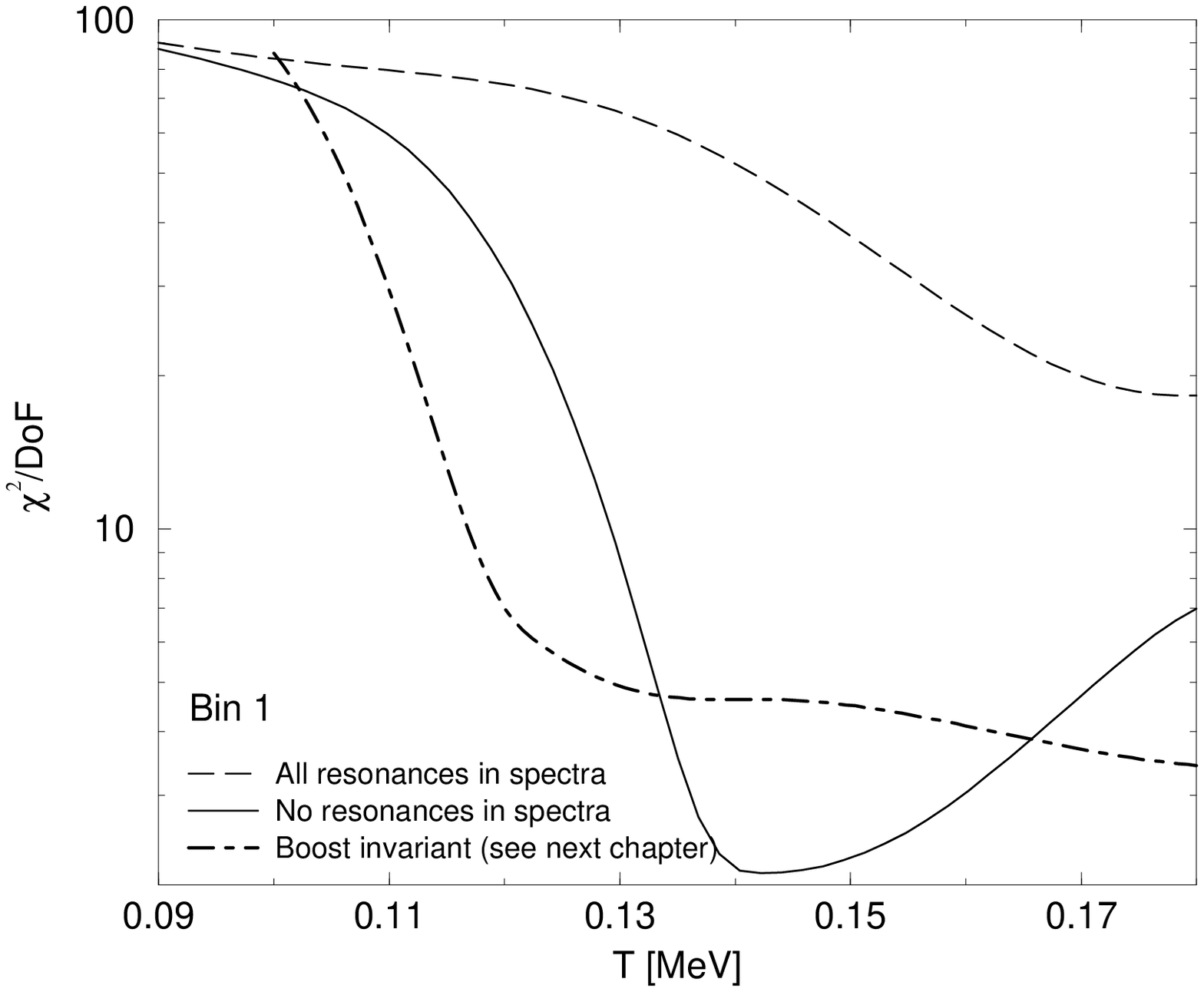}
\epsfig{width=6.5cm,clip=,figure=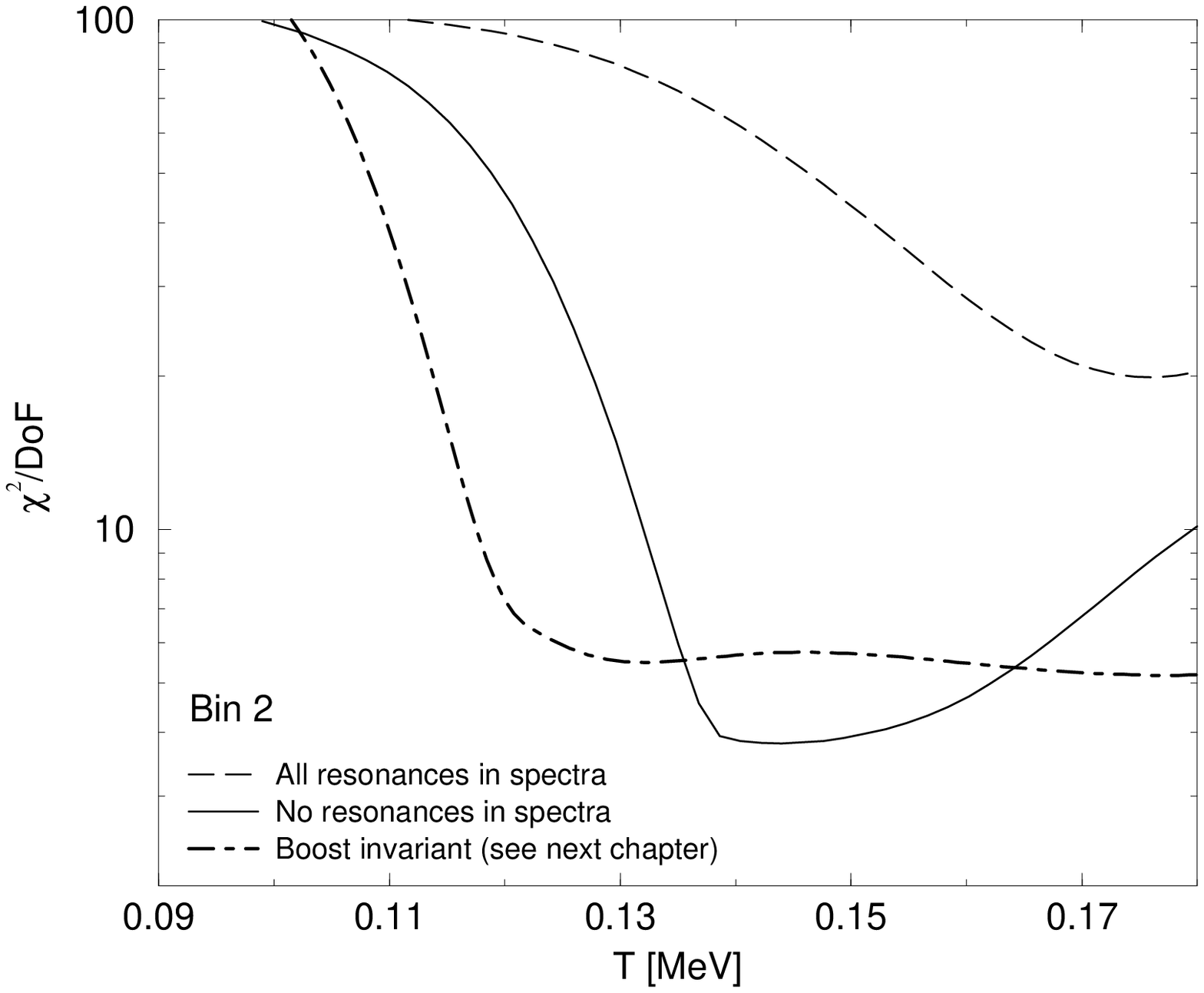}\\
\epsfig{width=6.5cm,clip=,figure=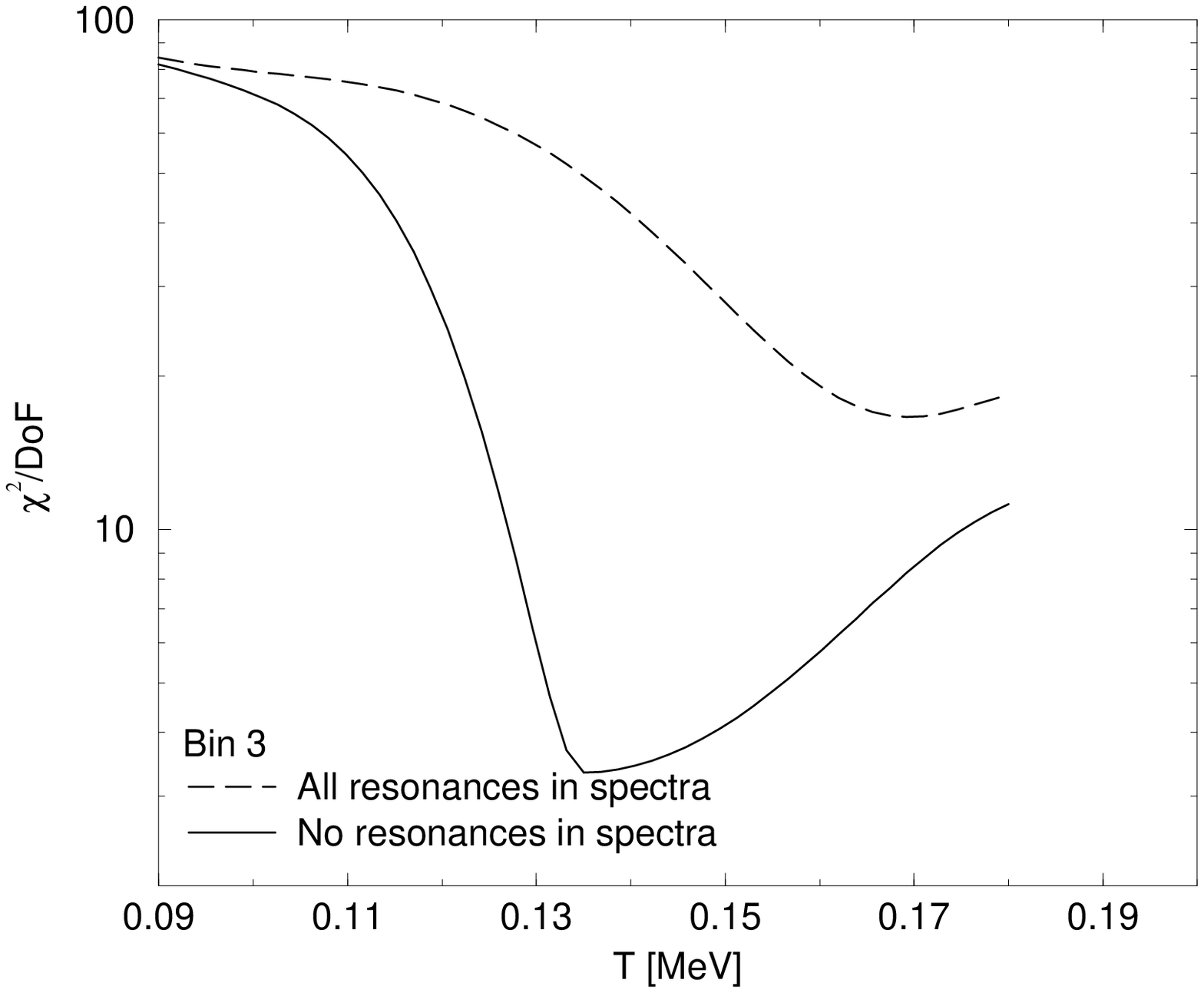}
\epsfig{width=6.5cm,clip=,figure=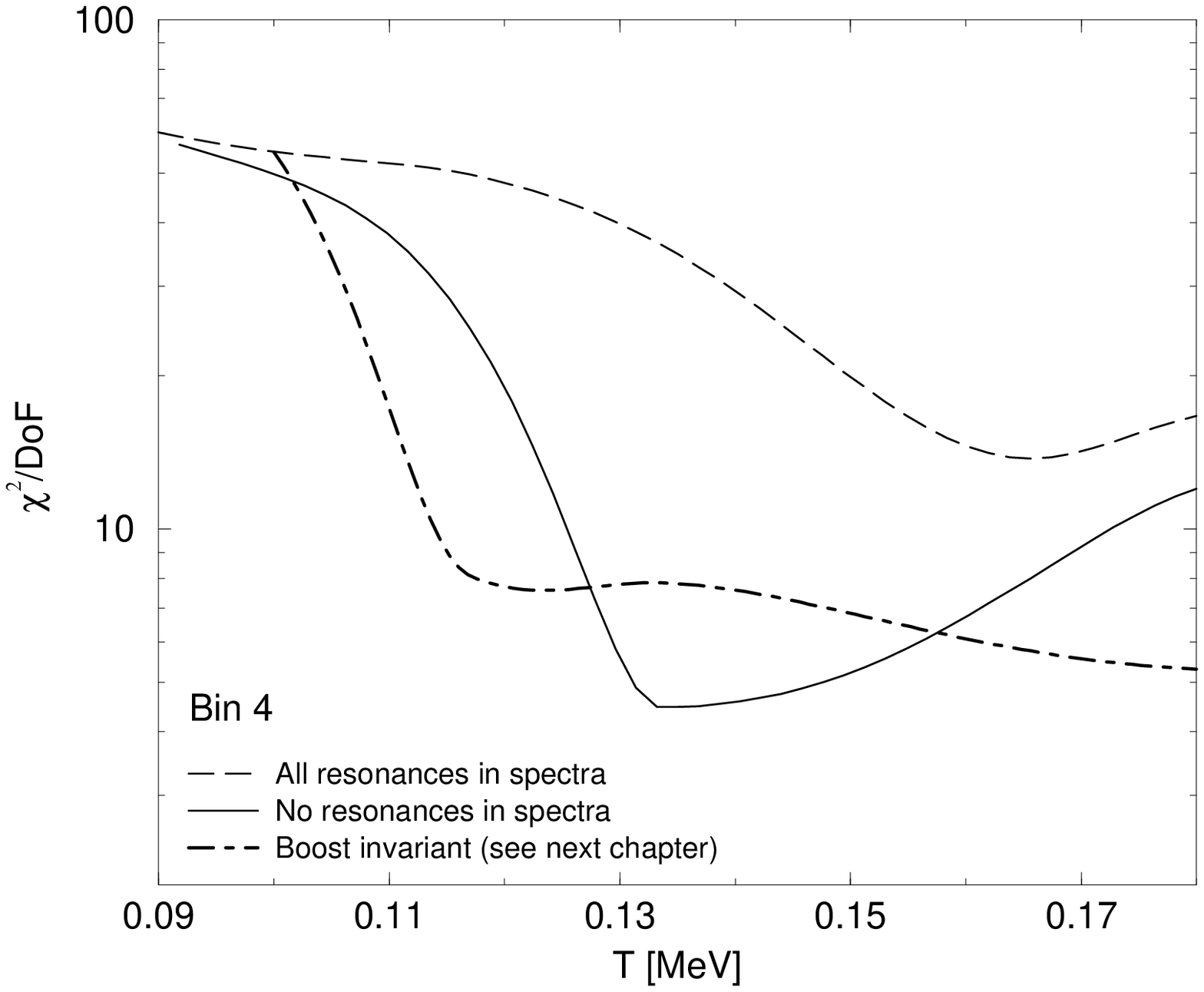}
\caption{ 
The total error divided by degrees of freedom for different
 centrality bins, shown as function of (fixed) 
freeze-out temperature $T$, bottom for the  experimental
value of the (statistical) $K^0$ error, top for 
the 5 times enlarged kaon data statistical errors.
\label{TchiT2}
}
\end{figure}

\section{Omega spectra}
In Fig.~\ref{TdOAOtot}
all four centrality bins for the sum $\Omega+\overline\Omega$ are
shown. We see that we systematically  under predict the two lowest
$m_\bot$ data points. Some deviation at high $m_\bot$ may be attributable
to acceptance uncertainties, also seen in the the 
$\Xi$ result presented earlier in Fig.~\ref{TdAXitot}. 
 We recall that there is a disagreement with the 
Omega yields in the chemical analysis, which thus does
not include in the analysis the production of $\Omega$. In the here presented 
analysis we see that this disagreement is arising at
low momentum.

The low-$m_\bot$ anomaly also explains why the 
inverse $m_\bot$ slopes for $\Omega,\overline\Omega$ 
are smaller than the values seen in all other strange (anti)hyperons.
One can presently only speculate about the processes 
which contribute to this anomaly. 
We note that the 1--2s.d. deviations in the 
low  $m_\bot$-bins of the $\Omega+\overline\Omega$ 
spectrum translates into 3s.d. 
deviations from the prediction of the chemical analysis. 
This is mainly a consequence of the fact that after summing
over centrality and $m_\bot$, the statistical error which dominates
$\Omega+\overline\Omega$ spectra becomes relatively small, and
as can be seen the low $m_\bot$ excess practically doubles
the $\Omega$ yield. 

\begin{figure}[tb]
\vspace*{-1.3cm}
\centerline{
\epsfig{width=9.5cm,clip=,angle=-90,figure=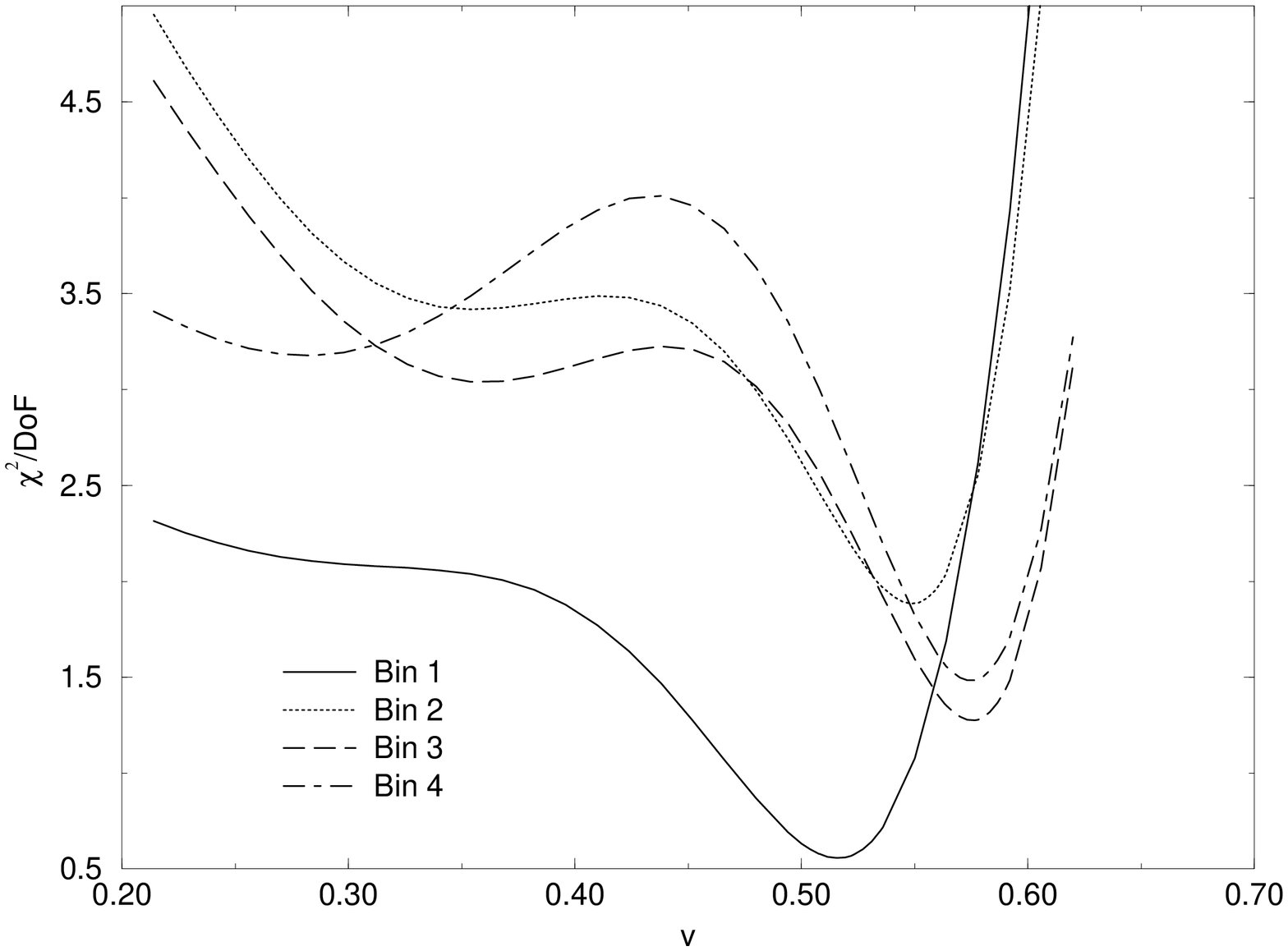}
}
\vspace*{-1.7cm}
\centerline{
\epsfig{width=9.5cm,clip=,angle=-90,figure=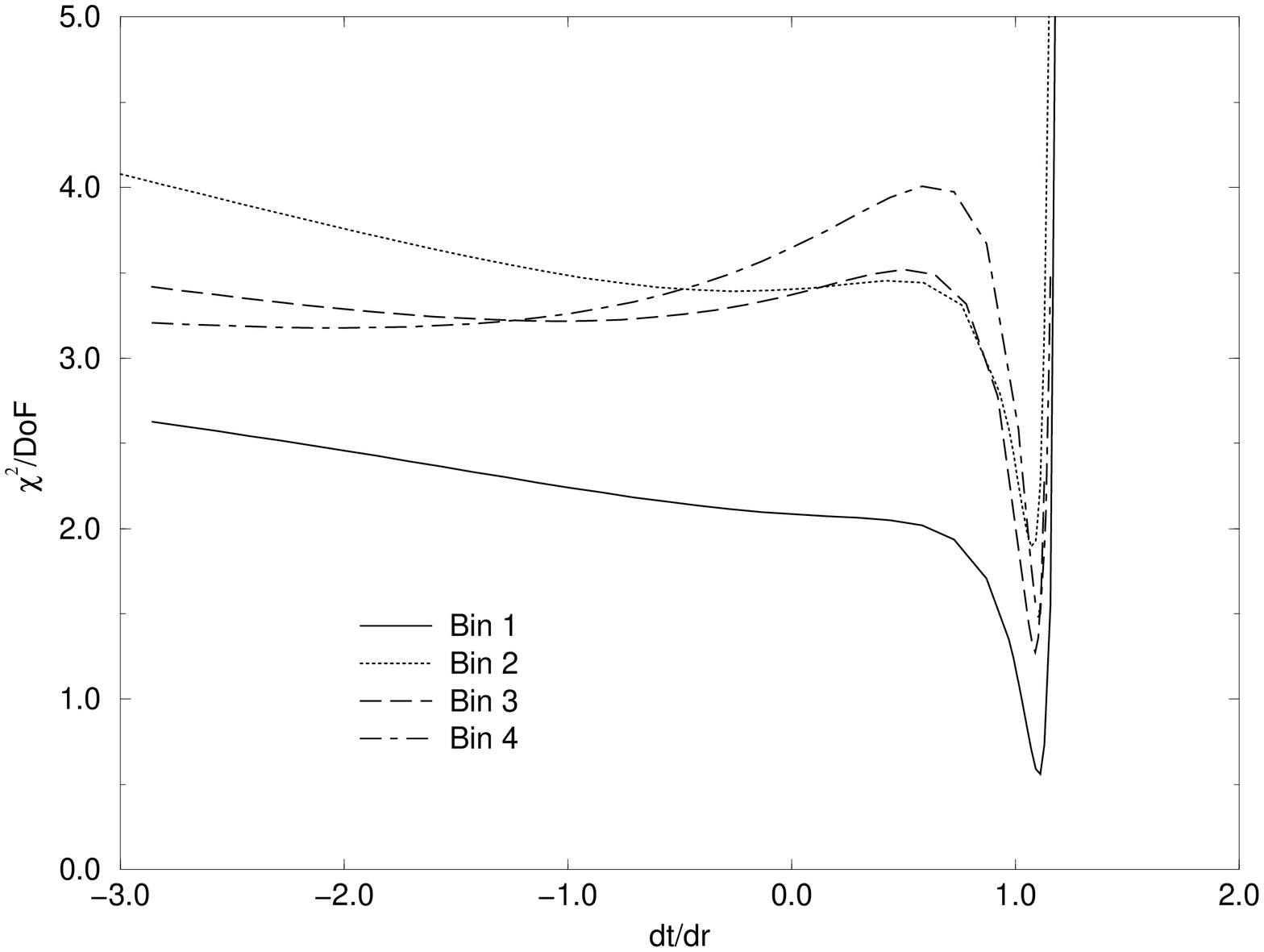}
}
\vspace*{-.3cm}
\caption{ 
The total error divided by degrees of freedom  for different
 centrality bins, shown as function of (fixed) flow velocity  
 $v$ on top and for (fixed) freeze-out surface 
$\partial t_f/\partial r_f=1/\frac{\partial r_f}{\partial t_f}$ dynamics on the bottom. 
\label{Tchivv2}
}
\end{figure}

\begin{figure}[tb]
\vspace*{-1.5cm}
\centerline{
\epsfig{width=9.5cm,clip=,angle=-90,figure=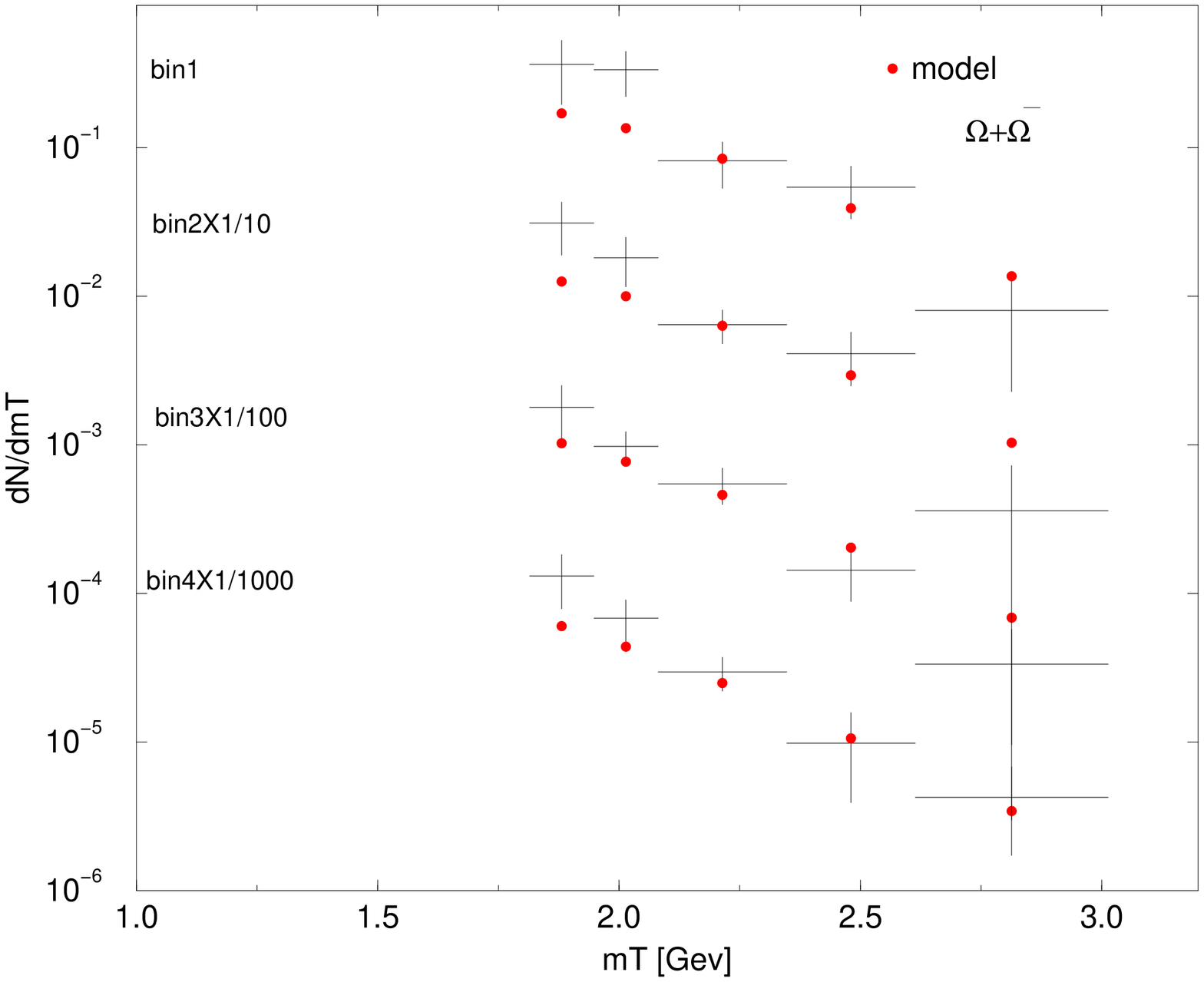}
}
\vspace*{-0.6cm}
\caption{(color online) 
Thermal analysis $m_T$ spectra: $\Omega+\overline\Omega$.
\label{TdOAOtot}
}
\end{figure}

\section{Discussion of SPS spectra}
Our thermal freeze-out analysis confirms that
CERN-SPS  results decisively show  interesting and new 
physics, and confirms the reaction picture of
a suddenly hadronizing QGP-fireball with both
chemical and thermal freeze-out being the same. 
In our view driven
by internal pressure, a  quark-gluon fireball expands and 
ultimately  a sudden breakup (hadronization) 
into final state  particles occurs which reach detectors 
without much, if any, further rescattering. The required
sudden fireball breakup  arises as the fireball super-cools, and in this 
state encounters a strong mechanical instability \cite{strange_rafelski3}. Note that
deep super cooling requires a  first order phase transition.

The remarkable similarity of 
$m_\bot$ spectra reported by  the  WA97 experiment
is interpreted by a set of freeze out parameters,
and we see that production mechanism of 
$\Lambda$, $\overline{\Lambda}$, and  $\Xi$, 
$\overline\Xi$ is the same. This symmetry, including matter--antimatter 
production is an important cornerstone of the claim that the strange
antibaryon data can only be interpreted in terms of direct particle emission
from a deconfined phase.

The reader must remember that  in presence of conventional 
hadron collision based physics, the production mechanism 
of antibaryons is quite different from that of baryons 
and a similarity of the $m_\bot$ spectra is not 
expected. Moreover, even if QGP is formed, but a equal
phase of confined particles is present, the annihilation of 
antibaryons in the baryon rich medium created at CERN-SPS energy
would deplete more strongly antibaryon yields, in
particular so  at small particle momentum, 
with the more abundant baryons remaining less influenced. 
This effect is not observed \cite{WA97spectra}. 

Similarity of $m_\bot$-spectra 
does not at all imply in our argument a similarity of particle rapidity spectra.
As hyperon are formed at the fireball breakup, any remaining longitudinal
flow present among fireball constituents 
will be imposed on the product particle, thus $\Lambda$-spectra
 containing potentially two original valence quarks are stretched in
$y$, which  $\overline\Lambda$-$y$-spectra are not, as they are made from newly 
formed particles. All told, one would expect that anti-hyperons can
 appear with a thermal
rapidity distribution, but hyperons will not. But both have the same 
thermal-explosive collective flow controlled shape of $m_\bot$-spectra.

We have shown  that  thermal freeze-out 
condition for strange hadrons (K$^0_s, \Lambda, \overline\Lambda,
\Xi, \overline\Xi$)  agrees within error with chemical freeze-out
and we have confirmed the  freeze-out temperature $T\simeq 145$\,${\rm MeV}$.
These findings about the similarity of thermal and chemical
freeze-out were controversial, when   the experimental single 
particle spectra  were lacking precision, since pion spectra and
two particle correlation analysis did not yield this result. 
However, this paper
studies the  precise hyperon and kaon  $m_\bot$ spectra
which reach to relatively low $p_\bot$ and compares 
with definitive chemical analysis of SPS data. The two particle
correlation analysis involves pions, which unlike strange
hadrons here considered, are potentially witnesses to other 
physics than the properties of dense and hot quark-gluon phase.

We were able to determine the freeze-out surface
$1/\frac{\partial r_f}{\partial t_f}=\partial t_f/\partial r_f$
dynamics and have shown that the break-up velocity $\frac{\partial t_f}{\partial r}$ is nearly
the velocity of light, as would be expected in  a sudden breakup of a
QGP fireball.  A study with $\partial t_f/\partial r_f$  has not
been previously considered, and only collective flow is included in
the description of the particle source. In our analysis we 
find a slight increase of the transverse expansion 
velocity with the size of the fireball volume, 
but consistently $v\le 1/\sqrt{3}$. 

We have reproduced the strange particle spectra in all centrality bins. 
Our findings rely strongly on
results obtained by WA97 at smallest accessible  particle momentum, 
and this stresses the need to reach to smallest possible
$p_\bot$ in order to be able to explore the physics of particle
freeze-out from the deconfined region.  Moreover, we 
demonstrated that the experimental production data 
of $\Omega+\overline\Omega$ 
has a noticeable systematic low $p_\bot$ enhancement anomaly 
present in all centrality bins. This result shows that it is not a
different temperature of freeze-out  of $\Omega+\overline\Omega$ 
 that leads to more enhanced yield, but a soft momentum secondary source
which contributes almost equal number of soft $\Omega+\overline\Omega$
compared to the systematic yield predicted by the other strange hadrons. 

\setcounter{figure}{0}
\setcounter{equation}{0}
\setcounter{table}{0}
\chapter{Particle spectra at RHIC: A boost-invariant ansatz}
\label{cha:spectra_rhic}

\section{Boost-invariant statistical hadronization}
\label{amb}
\subsection{General remarks }
The advent of RHIC data has coincided with the consensus  that
a model based on statistical hadronization, combined
with transverse expansion, can explain both the abundancies
and the transverse momentum distributions of hadrons produced
in heavy ion collisions.
The fitted parameters, and in particular the temperature, however,
have varied considerably,
ranging from as low as 110 ${\rm MeV}$ \cite{burward-hoy,van-leuween}
to 140 {\rm MeV} \cite{rafelski2002}
to as high as 160 and 170 ${\rm MeV}$
\cite{florkowski,bdmrhic}.

Such discrepancies are not very surprising, since the models differ
considerably.
However, this means that before we can say that the freeze-out temperature has
been determined, we must understand precisely the origins
of these differences, and try to ascertain which model is ``the best'',
both in terms of physical reasonableness and as an ansatz to fit the data.
We shall proceed to give an overview of the ways in which these models
differ.

Firstly, spectra are normalized very differently in different models.
Some recent work fits the particle slopes
 only~\cite{burward-hoy,vanleeuwen,NA57spectra,castillo}, 
treating the normalization of each particle
as a free parameter. 
This approach can be argued for assuming a long-lived post-hadronization
``interacting hadron gas phase'' in which individual hadron abundances
subject to inelastic interactions evolve away from chemical equilibrium.
This particular  reaction picture clashes with e.g. the fact 
that  short-lived resonance ratios can be described 
within  the statistical hadronization 
 model using  the chemical (statistical hadronization) freeze-out 
 temperature obtained  in  stable particle studies ( \cite{bdmrhic} and
next chapter).
This implies
 that in principle  the relative normalization  of the particle spectra 
should be derived from a hadronization scenario involving 
flavor chemical potentials. In fact a study of RHIC spectra 
 finds that the  normalization can be accounted for~\cite{florkowski}, and 
 that the chemical equilibration temperature also describes
particle spectra well. This is suggesting that  any post-hadronization re-interaction
phase is short and has  minor influence on the particle yields.

The problem is that the different ways to derive hadronization
 particle distributions  have a profound effect on the resulting fitted temperature.
Temperature affects the absolute number of particles 
through several mechanisms  and anti-correlates 
with the  phase space occupancy parameters $\gamma_i, i=u,d,s$~\cite{strange_rafelski3}.
It has been found that the introduction of these parameters, motivated by the need to conserve entropy at hadronization~\cite{strange_rafelski3} decrease
the $\chi^2/$ per degree of freedom considerably and lowers the freeze-out temperature by ~30 ${\rm MeV}$
\cite{strange_rafelski3}.
Other workers assume
the light flavors are in chemical equilibrium~\cite{becattini,bdmsps,bdmrhic,florkowski}.

Additionally, when fitting the particle spectra, the system's
spatial shape and the way the freeze-out progresses in time have
a considerable effect on the form of particle distributions, and hence
on the fitted temperature and matter flow.  
The impact of freeze-out geometry and dynamics on particle spectra were examined well before  RHIC data became available 
\cite{heinzreso} and it was realized that
an understanding of freeze-out is essential for the statistical
analysis of the fireball \cite{HBTpuzzle}.  Even though this matter 
has been clearly recognized, a systematic
analysis  of how freeze-out geometry affects particle distributions
is for the first time attempted here. In fact, each of the models used in the study of particle
spectra \cite{vanleeuwen,castillo,burward-hoy,florkowski}
employs  a different choice
of freeze-out geometry, based on different, often tacitly assumed, 
hadronization scenarios. Thus an understanding for the influence of hadronization
mechanism is impossible to deduce from this diversity. 

\subsection{Freeze-out geometry}
At RHIC collision energies the measured 
$dN/d \eta$ ~\cite{phobos,brahms}
indicates that around mid-rapidity the system conditions can be approximated
by the Bjorken picture~\cite{bjorken_boost}.

To describe particle spectra measured around mid-rapidity, therefore, 
boost invariance 
becomes the dominant symmetry on which freeze-out geometry should be based.
We shall construct a general hadronization scenario, combining
the cut Cooper-Frye formula  (Eq.~(\ref{cut_cf})) with boost-invariance.

A particle's energy  at rest with respect to the collective flow is given by
combining the flow field (parametrized by the transverse and longitudinal
rapidities $y_{L,T}$ as well as the emission angle $\theta$) with
the observed particle's momentum (parametrized by the rapidity y, transverse
momentum $p_T$ and angle $\phi$) 
\begin{equation} 
u^{\mu} = 
\left( 
\begin{array}{l}
\cosh (y_L) \cosh(y_T) \\ 
\sinh(y_T) \cos(\theta) \\ 
\sinh(y_T) \sin(\theta) \\ 
\sinh (y_L) \cosh(y_T) 
\end{array} 
\right),\quad
p^{\mu} = 
\left( 
\begin{array}{c}
m_T \cosh (y)  \\ 
p_T \cos(\phi) \\ 
p_T \sin(\phi) \\ 
m_T \sinh (y)  
\end{array} 
\right)
\label{flowprof}
\end{equation}
The rest energy will then be
\begin{equation}
\label{goesas}
p_{\mu} u^{\mu} = m_T \cosh(y_T) \cosh(y-y_L) - \sinh(y_T)  \cos(\theta-\phi) p_T.
\end{equation}
The requirement for the Bjorken picture is that the emission volume element
has the same $y_L$ dependence:
\begin{equation}
p_{\mu} d^3 \Sigma^{\mu} \sim A \cosh(y-y_L)+B.
\end{equation}
This constrains the freeze-out hypersurface to  be
of the form
\begin{equation}
\label{bjorkfreeze}
\Sigma^{\mu}= (t_f \cosh(y_L),x,y,t_f \sinh(y_L)).
\end{equation}
Here $t_f$ is a parameter invariant under boosts in the \textit{z} direction,
whose physical significance depends on the model considered.

For central collisions, a further simplifying constraint
is provided by the cylindrical symmetry, which forces $t_f$,
 as well as $y_L$ and $y_T$, to be independent of the angles
$\theta$ and $\phi$.
The freeze-out hypersurface can be parametrized, in this case, as
\begin{equation}
\label{cylfreeze}
\Sigma^{\mu}=\left( t_f(r) \cosh(y_L),r \sin(\theta),r \cos(\theta),t_f (r) \sinh(y_L) \right),
\end{equation}
\begin{equation}
\label{blastsurf}
d^3 \Sigma^{\mu} = t_f r dr d \theta d y_L 
\left(
\cosh(y_L) 
 \frac{\partial t_f}{\partial r} \cos(\theta),
 \frac{\partial t_f}{\partial r} \sin(\theta),
\sinh(y_L)
\right)
\end{equation}
And the emission element takes the form
\begin{equation}
\label{blastemission}
p^{\mu} d^3 \Sigma_{\mu} = \left[ m_T \cosh(y-y_L)
 - p_T \frac{\partial t_f}{\partial r} \cos(\theta-\phi) \right] t_f r dr d \theta d y_L,
\end{equation}
with the same dependence on the angle as Eq.~(\ref{goesas}).

Putting Eq.~(\ref{goesas}) and Eq.~(\ref{blastemission}) together, the particle's momentum distribution function
in the Boltzmann approximation is given by
\begin{equation}
\label{before_int}
E \frac{dN}{d^3 p} \propto  \left[ m_T \cosh(y-y_L) 
- p_T \frac{\partial t_f}{\partial r} \cos(\theta-\phi) \right] 
\end{equation}
\[\ \exp \left[ -\frac{m_T \cosh(y_T) \cosh(y-y_L) - \sinh(y_T)  \cos(\theta-\phi) p_T}{T} ]\right] t_f r dr d \theta d y_L \]
(the quantum statistics case, relevant for $\pi$, can be obtained through the
prescription in Eq.~(\ref{boltz_fdbe})).
For an approximately boost-invariant cylindrical system we can use Bessel
functions to perform the integrals over $y_L$ and $\theta$ analytically
\begin{eqnarray}
\label{bessels}
K_n (z) = \int_{-\infty}^{\infty} \cosh(ny) e^{-\cosh(y)} dy\\
I_n (z) = \frac{1}{2 \pi} \int_{0}^{2 \pi} \cos(n \theta) e^{-\cos(\theta)} d\theta
\end{eqnarray}
and obtain 
\begin{equation}
\label{after_int}
E \frac{dN}{d^3 p} \propto 
 m_T  I_{0} (\alpha p_T)  K_1 (\beta m_T)
-
p_T \frac{\partial t_f}{\partial r} I_{1} (\alpha p_T)    K_0 (\beta m_T)
\end{equation}

What distinguishes the models currently considered is the time component
of the freeze-out surface.
The most general freeze-out hypersurface compatible with cylindrical 
symmetry is
provided by Eq.~(\ref{cylfreeze}). Generally, $t_f$ (a generic
function of \textit{r}) represents the time, in a frame co-moving with the
longitudinal flow, at which the
surface at distance \textit{r} freezes out.

The fits in Refs. ~\cite{burward-hoy,vanleuween,castillo} are based on a particular
case of such a freeze-out surface, in which $t_f$ is completely independent
of $r$ ($\partial t_f/\partial r=0$).
Such a picture's  physical reasonableness can be questioned, e.g., 
why should spatially distant volume
elements, presumably with different densities and moving at different 
transverse velocities, all freeze out simultaneously in a longitudinally
co-moving frame?. However, such a simple model  can perhaps serve as an 
 approximation.

More generally, the ``burning log'' model, discussed in the last chapter
for the spherically symmetric case,  assumes that the 
emission occurs through a
three-dimensional hadronization surface which is moving 
at a constant ``velocity'' ($\partial t_f/\partial r$ throughout
the fireball. Both boost-invariant and spherically symmetric versions 
of burning log model were considered. Even if the hadronization
velocity encompasses an extra parameter, it is worth
considering since it is based on a  physically motivated hadronization  picture.
Moreover, the burning log picture is a suitable framework in the study 
of sudden hadronization. Sudden hadronization occurs
when the fireball encounters a mechanical instability,
which combined with the fireball's high transverse
flow ensures that the emission surface spreads to the interior of the
fireball with $\partial t_f/\partial r \simeq c$.
All of the indications suggested for such a picture seem to be borne out
by both SPS and RHIC data~\cite{strange_rafelski3,rafelski2002}.

\begin{figure}
\centerline{\resizebox*{!}{0.3\textheight}
{
\includegraphics{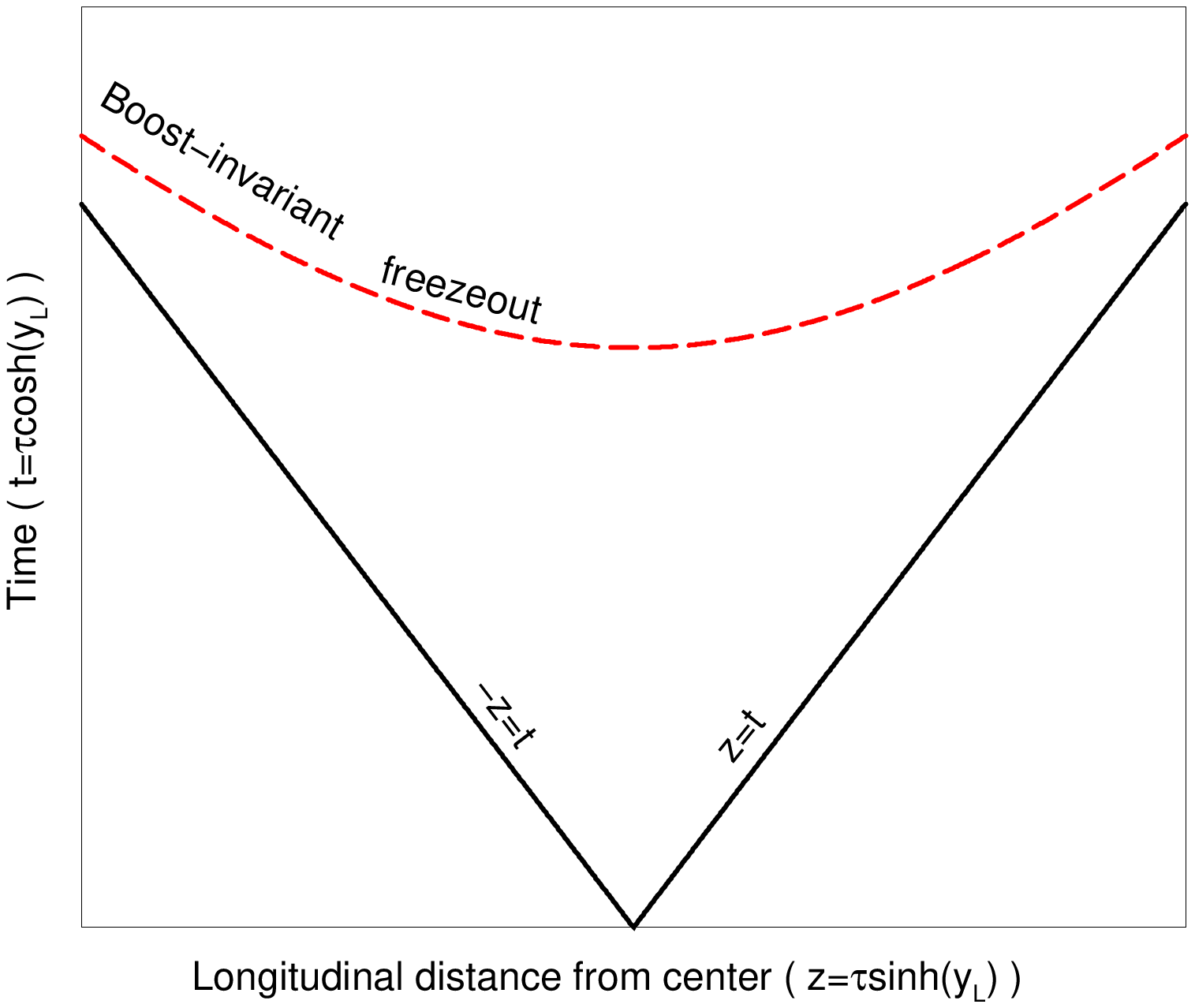}
}
\resizebox*{!}{0.3\textheight}
{
\includegraphics{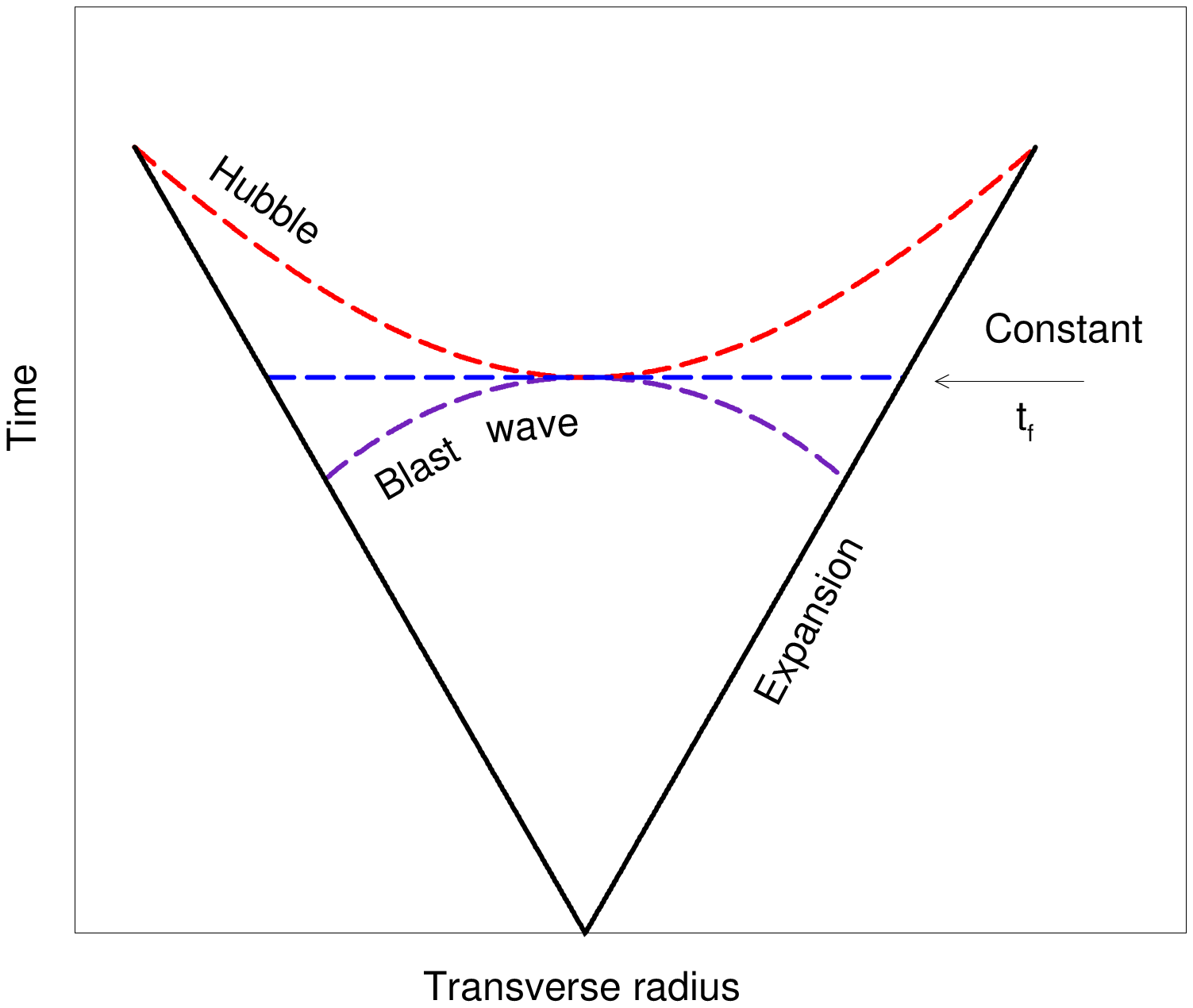}
}}
\caption{(Color on-line) While boost-invariance fixes the longitudinal
freeze-out structure (left), several scenarios exist for the transverse dependence of
 freeze-out (right).   For spherical freeze-out, only plot on the right applies 
\label{figsurf}.}
\end{figure}

\begin{table}
\caption{(Color on-line) Freeze-out hypersurfaces at contours of
  constant radii. \label{hypsurf}}
\vspace{0.2cm}
\begin{tabular}{lcc}
\hline
Surface 
& 
$\Sigma^{\mu}$ 
&  
$E \frac{dN}{dp^3}$ \footnote{$\beta=\cosh(y_T)/ T ,\alpha=\sinh(y_T)/T$}
\\ \hline
$\begin{array}{c}
\mbox{Constant\ } t_f \\
{\partial t_f}/{\partial r}=0
\end{array}
$
& 
$\left( \begin{array}{c}
t_f \\ \vec{r} \end{array} \right)$ 
& 
$m_T K_1 (\beta m_T)I_{0} (\alpha p_T)$
\\ \hline
$\begin{array}{c}
\mbox{Hubble}\\
(\mbox{constant}\ \tau_f) 
\end{array}
$
& 
$\tau_f 
\left( \begin{array}{l}\cosh(y_L) \cosh(y_T) \\ 
       \sinh(y_T) \cos(\theta) \\
        \sinh(y_T) \sin(\theta) \\ 
          \sinh(y_L) \cosh(y_T) 
\end{array} 
\right)
$ 
&
$\begin{array}{c}
m_T \cosh(y_T) I_{0} (\alpha p_T) K_1 (\beta m_T )-\\ 
p_T \sinh(y_T) I_{1}(\alpha p_T)  K_0 (\beta m_T)
\end{array}
$
\\ \hline
$\begin{array}{c}
\mbox{burning log}\\
\mbox{(boost invariant)}
\end{array}
$
& 
$\left( \begin{array}{c} t_f (r) \cosh(y_L) \\
                  r \cos(\theta)\\        
                  r \sin(\theta)\\ 
                  t_f (r) \sinh(y_L) \end{array} \right)$
& 
$\begin{array}{c}
m_T  I_{0} (\alpha p_T)  K_1 (\beta m_T)
-\\
p_T \frac{\partial t_f}{\partial r} I_{1} (\alpha p_T)    K_0 (\beta m_T)
\end{array}
$
\\ 
\hline 
\\
$\begin{array}{c}
\mbox{burning log}\\
(\mbox{spherical})
\end{array}
$
& 
$\left( \begin{array}{c} t_f \\ r \vec{e}_r \end{array} \right)$
&
$\begin{array}{c}
e^{-E/T} \sqrt{\frac{T}{p_T \sinh(y_T)}} 
(E I_{1/2} (\alpha p_T) -\\
 p_T \frac{\partial t}{\partial r} I_{3/2} (\alpha p_T))
\end{array}$
\\ \hline
\end{tabular}
\end{table}

An approach based on the hypothesis of initial state ``synchronization''
by the primary instant of collision and the following independent but 
equivalent evolution of all volume elements  assumes that each element 
of the system undergoes freeze-out at the same proper time $\tau$.
In this framework 
each fireball element expands and cools down
independently, hadronizing
when its temperature and density reach the critical value.
This model  was successfully used to describe 
RHIC $m_T$-spectra~\cite{florkowski}.
In this approach  $t_f$ in Eq.~(\ref{cylfreeze}) is equal to
$\tau \cosh(y_T)$ and the hadronization hypersurface in Eq.~(\ref{blastsurf})
becomes proportional to the flow vector:
\begin{eqnarray}
\label{hubblefreeze}
\Sigma^{\mu}&=& \tau u^{\mu}\\
d^3 \Sigma^{\mu} &=& \tau r dr d\theta d y_L u^{\mu}=dV u^{\mu}\\
r &=& \tau \sinh(y_T).
\end{eqnarray}
In this   hadronization model  the heavy ion fireball  behaves similarly
to the expanding Hubble universe. In the  
`Hubble' scenario, the Cooper-Frye formula
reduces to the Touscheck Covariant Boltzmann distribution
~\cite{jansbook,tous1,tous2}.
\begin{eqnarray}
\label{touscheck}
\frac{V_0d^3p}{(2\pi)^3} e^{-E/T}&\to& \frac{V_\mu p^\mu}{(2\pi)^3}
d^4p\,2\delta_0(p^2-m^2)e^{-p_\mu u^\mu/T}\, \\
V^{\mu}&=&V_0 u^{\mu}
\end{eqnarray}
(Where \textit{V} is the co moving fireball's volume element in the local
rest frame.)

To summarize and illustrate the diversity of distinct  hadronization 
geometries we present in Table~\ref{hypsurf} and Fig.~\ref{figsurf} the 
freeze-out scenarios examined here. As we shall see the 
 choice of freeze-out geometry produces in a fit of experimental 
data a non trivial
effect capable of altering significantly the understanding 
of statistical hadronization parameters.

\begin{figure}
\centerline{\resizebox*{!}{0.3\textheight}
{
\includegraphics{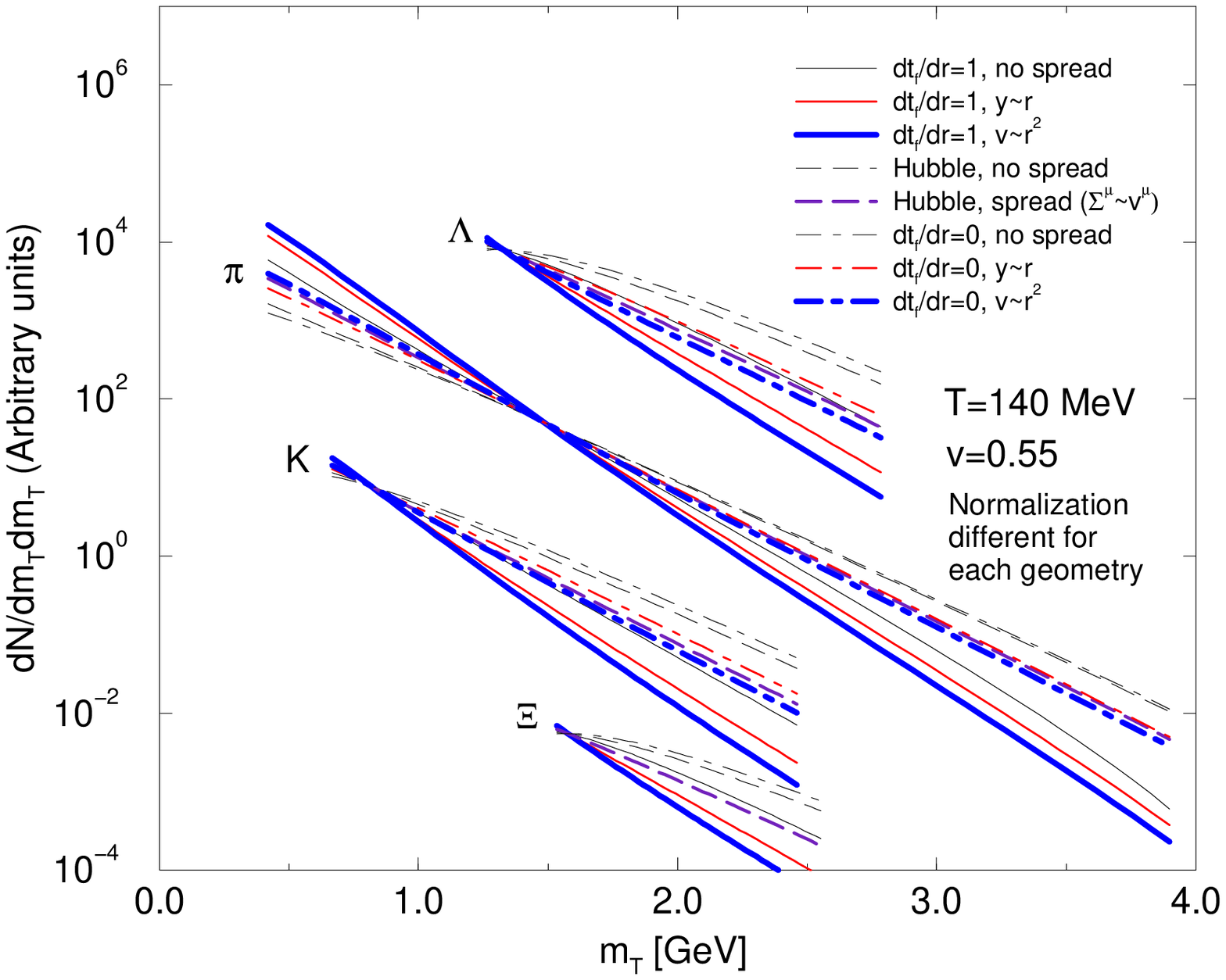}
}
\resizebox*{!}{0.3\textheight}{
\includegraphics{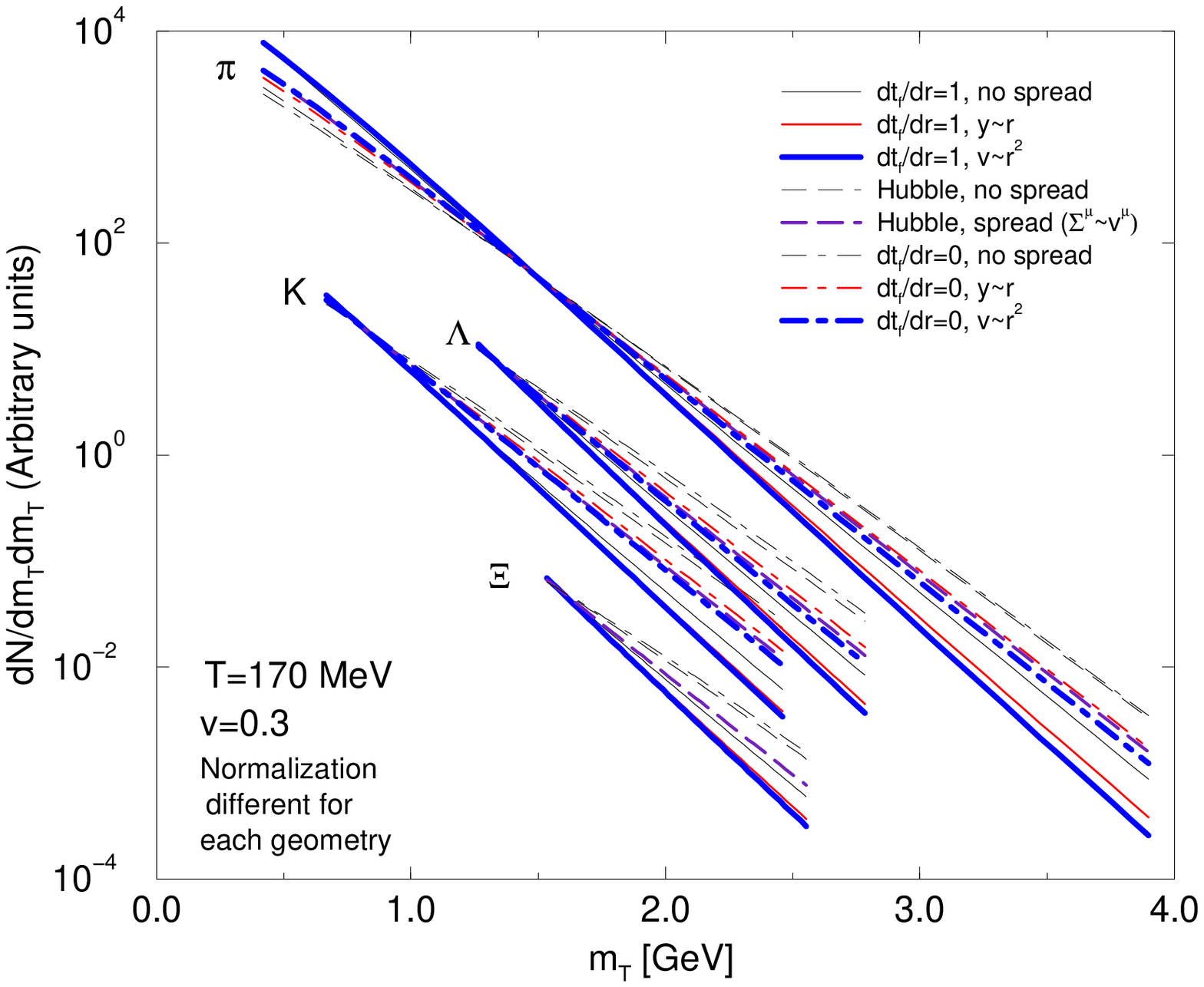}
}}
\caption{(Color on-line) $\pi$,$K$,$\Lambda$ and $\Xi$ $m_T$ distributions 
obtained with different freeze-out models and flow profiles.
For this and subsequent figures, a uniform density profile was assumed
\label{mcplots}.}
\end{figure}

\subsection{Flow profile}
Hydrodynamical expansion of the fireball implies in 
general that  each volume element
 will have a different density and transverse expansion rate.
For this reason, the integral over $d^3 \Sigma$ can  span a range
of flows, weighted by density. In first approximation one can  
fit data using just an ``average'' flow
velocity throughout the entire fireball~\cite{NA57spectra,torrieri_sps1}:
\begin{eqnarray}
E \frac{dN}{d^3 p}&=& \int r dr (E-p_T \frac{dt_f}{dr}) f(T,y_T (r),\lambda) 
\nonumber\\
&\propto& (E-p_T \frac{dt_f}{dr}) f(T,\langle y_T\rangle,\lambda).
\label{1flow}
\end{eqnarray}

However, if one wants to properly identify 
$\frac{dt_f}{dr}$, the  flow profile should be  taken into account.
Hydrodynamic simulations~\cite{hydro_shuryak} accompanied by assumption that
 freeze-out happens when a volume element reaches a critical
energy density indicate that the transverse rapidity will depend linearly
with the radius i.e. $v_T \sim \tanh(r)$.
This condition, however, is appropriate for a static freeze-out
and will not in general hold if the freeze-out is sudden.
Other flow profiles have been tried in the literature, arising 
from dynamical hypothesis. For example, 
the assumption that the  freeze-out occurs at the same time
$t_f$ results in a quadratic ($v \propto r^2$) flow profile~\cite{hydroheinz}, 
which has also  been used recently in fits to data~\cite{castillo}.
In the  Hubble fireball~\cite{florkowski} the  freeze-out conditions
will also result in a distinctive flow profile. Specifically with 
$\Sigma^{\mu} \propto u^{\mu}$, we have $\gamma v \propto r$.

Density profiles  also depend on the assumed initial condition
and the equation of state of the expanding QGP.
It has been shown~\cite{wa98} that different density choices have
a considerable effect on both the temperature and flow fits at SPS
energies.  

Fig.~\ref{mcplots} shows how the choice in
hadronization dynamics and flow profiles at same  given
freeze-out temperature and transverse flow can result in a range 
of inverse spectral slopes. Here the density profiles were assumed to be uniform.
It is  clear that the same freeze-out parameters  give
rise to a variety of substantially different particle
spectra. Conversely, fits to experimental data will only produce 
reliable information on the freeze-out conditions if and when we
have a prior knowledge of the hadronization geometry and dynamics. 
Therefore, conclusions about statistical model fits,
as well as arguments whether freeze-out occurs  simultaneously for different particles
or not, cannot be answered  while the models used to fit the data are plagued by
such uncertainties.

We now proceed to use a Monte-Carlo method to calculate the effect these uncertainties
have on fitted data.    Afterwards, we will apply a range of fits to RHIC models to see if the
``right'' hadronization model can be isolated from particle spectra alone.  
\begin{figure}[tb]
\begin{center}
\epsfig{width=6.5cm,clip=,figure=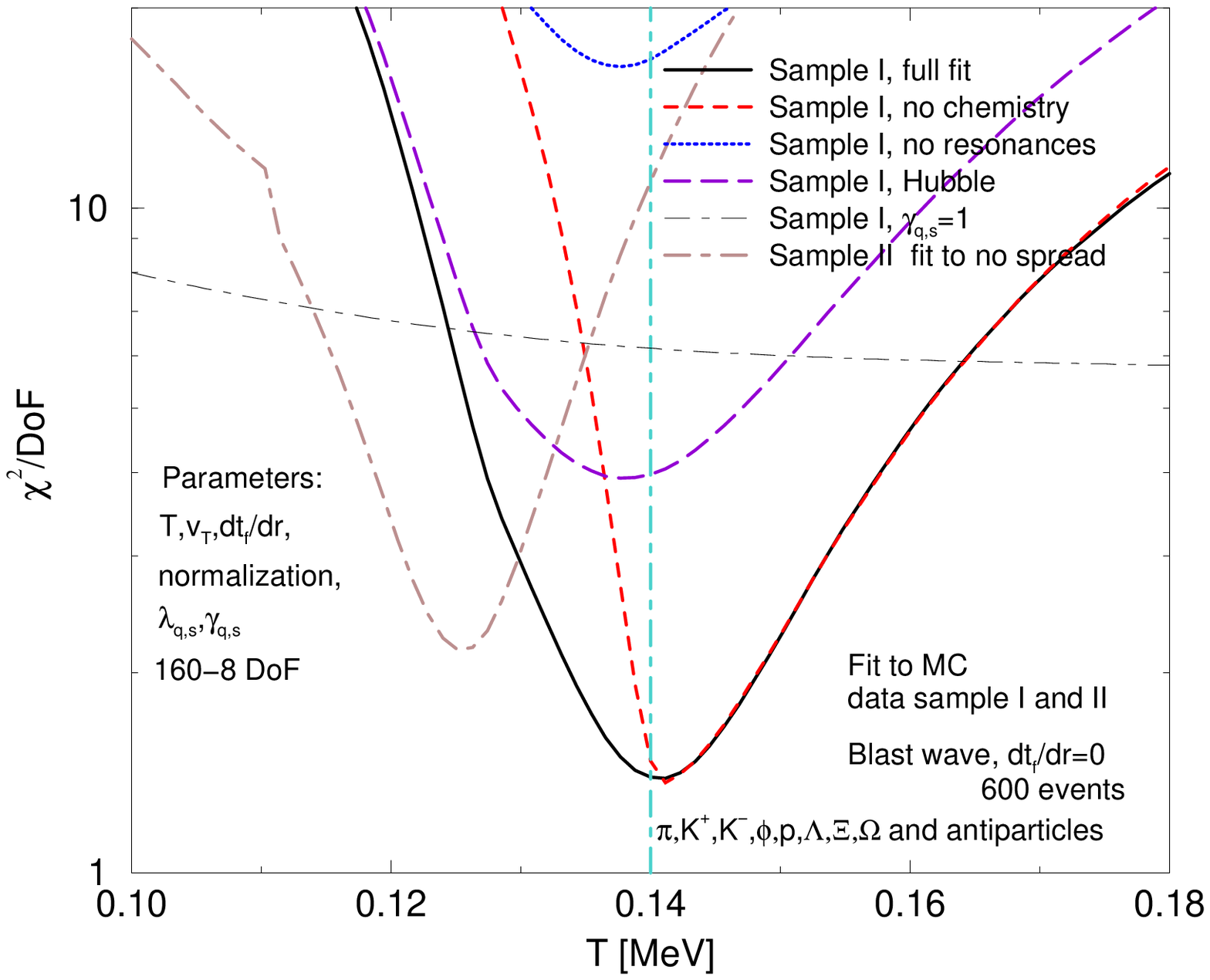}
\epsfig{width=6.5cm,clip=,figure=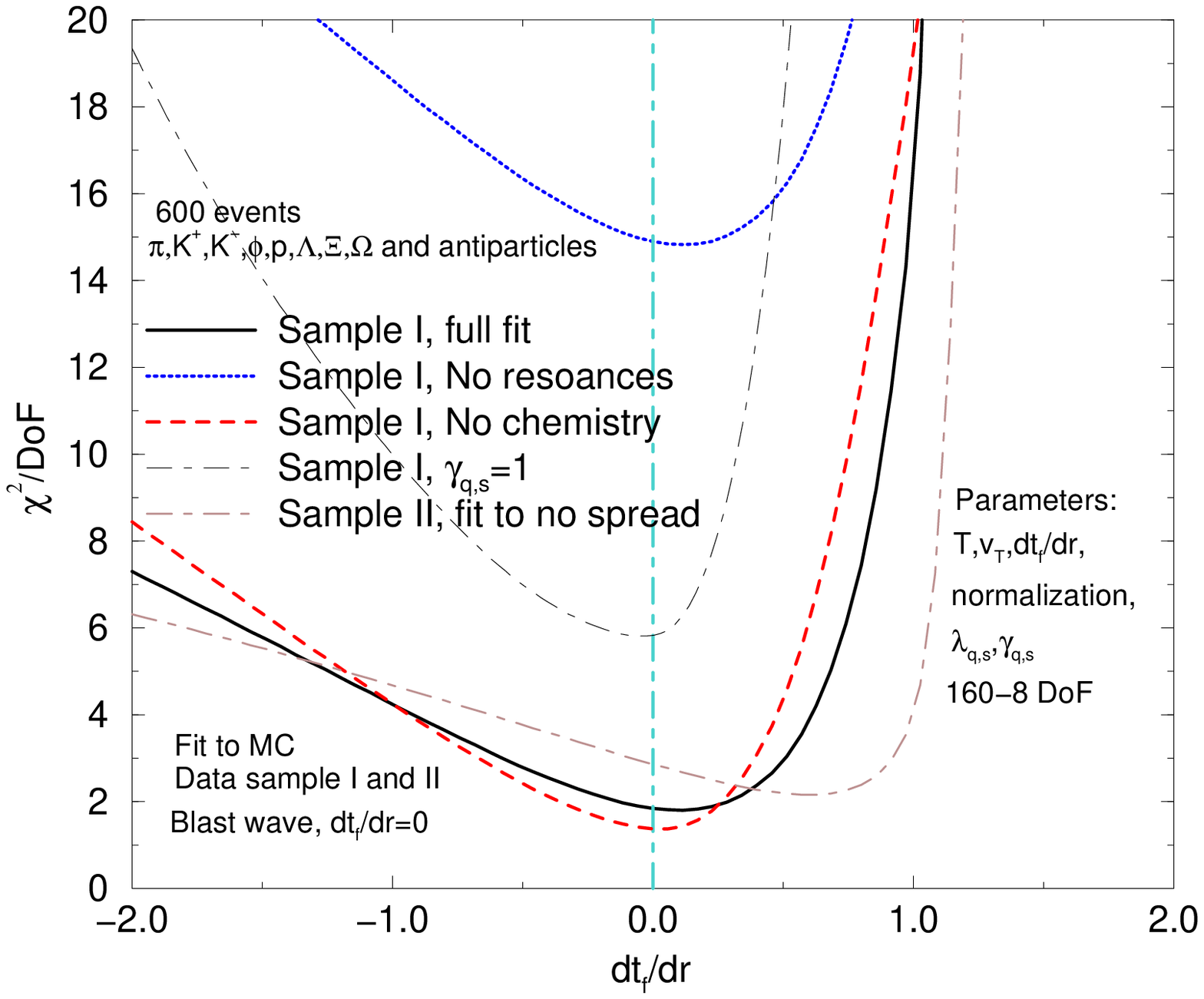}
\end{center}
 \label{prof0}
\caption{(color on-line)
Results of fits to Monte-Carlo generated data samples I and II ($\frac{dt_f}{dr}=0$):
$\chi^2$ profiles for temperature (left) and fitted $\partial t_f/\partial r$ (right).   The models used in the fits are described in sections 1.1 and 1.2
Full fit includes fitted chemical potentials and $\gamma_{q,s}$, resonances and fitted $\partial t_f/\partial r$.
Last profile shows effect of fitting sample II using Eq.~(\ref{1flow})  \label{prof0}. }
\end{figure}
\section{Sensitivity to model choice}
The ambiguities presented in the previous section mean that it is important
to study their effect on the statistical model's fitted parameters.
One way to do this is to use a Monte-Carlo to generate data according
to a particular freeze-out model, and to see what happens if the ``wrong'' model is used to perform the fit.
We have written a Monte-Carlo program which can be used for this
purpose.
An acceptance/rejection algorithm is used to generate particles
in a statistical distribution in the volume element's rest frame.
The accepted particles are then Lorentz transformed to the lab frame.
(Any flow and density profile, as well as any freeze-out surface can 
be accommodated).
Resonance decays are handled through Eq.~(\ref{decayphase}), using the MAMBO algorithm \cite{mambo} to generate points in phase space.
Output can be used to generate spectra or fed into a microscopic model such as uRQMD \cite{uRQMD}.

\begin{figure}
\begin{center}
\epsfig{figure=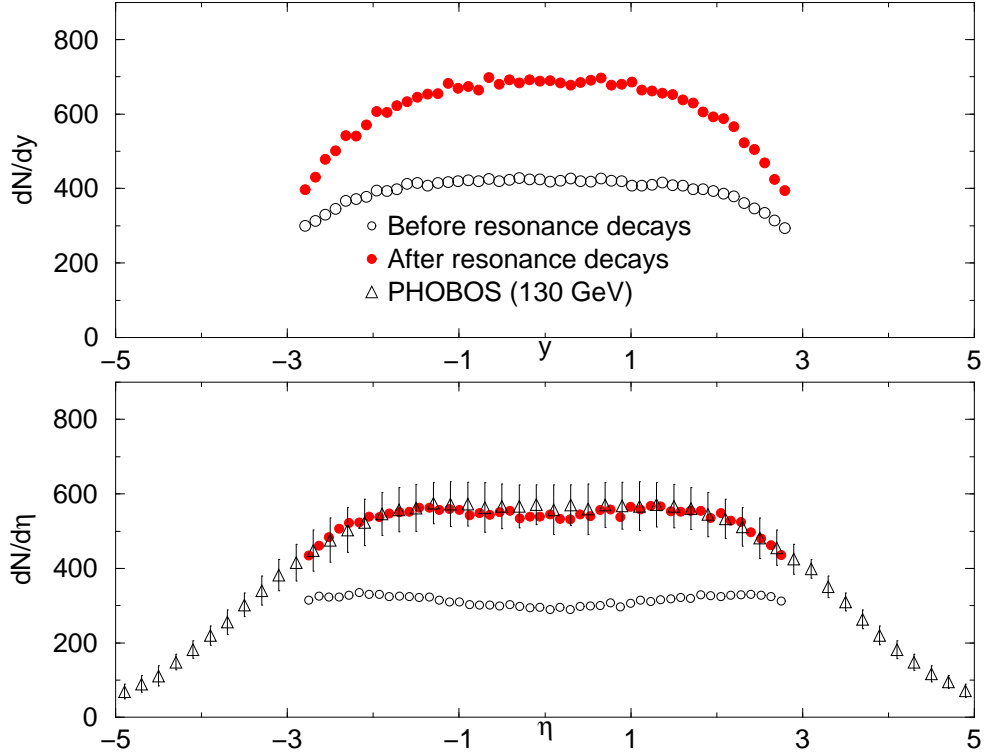,width=13cm}
\caption{(color on-line)  $dN_{tot}/d(y,\eta)$ arising from freeze-out of a Bjorken fluid, before (open circles) and after (solid circles) resonance decays.   
The normalization is arbitrary, chosen to coincide with PHOBOS  data (triangles)
\label{globals}. }
\end{center}
\end{figure}

The Monte-Carlo output was used to produce the data points in Fig.~\ref{globals}.
It can be seen that a boost-invariant statistical hadronization
can explain the global properties of the system such as $dN_{tot}/d\eta$.   It can also be verified that
the role of resonances is absolutely crucial.
We proceeded to generate three datasets of particles.
Each data set had a temperature of 140 ${\rm MeV}$, a maximum transverse
flow of 0.55, and out of equilibrium chemistry ($\gamma_q=1.4,\gamma_s/\gamma_q=0.8$).
Generated particles include $\pi,K^{+},K^{-},p,\overline{p},\Lambda,\Xi,\Omega$ and their resonances.
The three samples differ in their choice of freeze-out geometry (specifically $\partial t_f/\partial r$) and flow profile:
\begin{description}
\item[Sample I]  $\partial t_f/\partial r=0$ and no flow profile, as fitted in \cite{NA57spectra}
\item[Sample II] $\partial t_f/\partial r=0$ and a quadratic
flow profile, as fitted in \cite{burward-hoy,vanleeuwen,castillo}
\item[Sample III]  $\partial t_f/\partial r=1$, the boost-invariant analogue of \cite{torrieri_sps1}.
\end{description}
\begin{figure}[h]
\begin{center}
\epsfig{width=6.5cm,clip=,figure=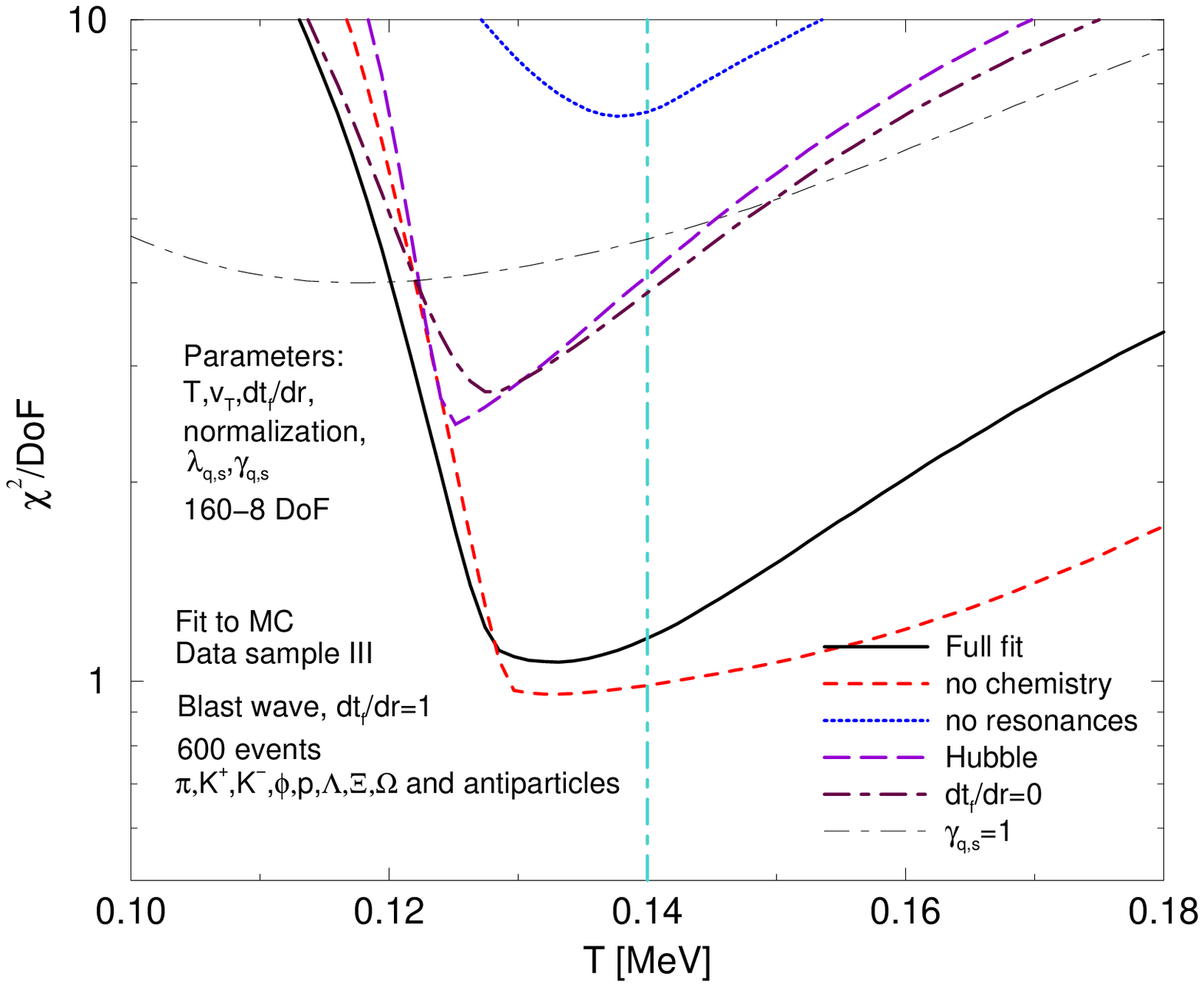}
\epsfig{width=6.5cm,clip=,figure=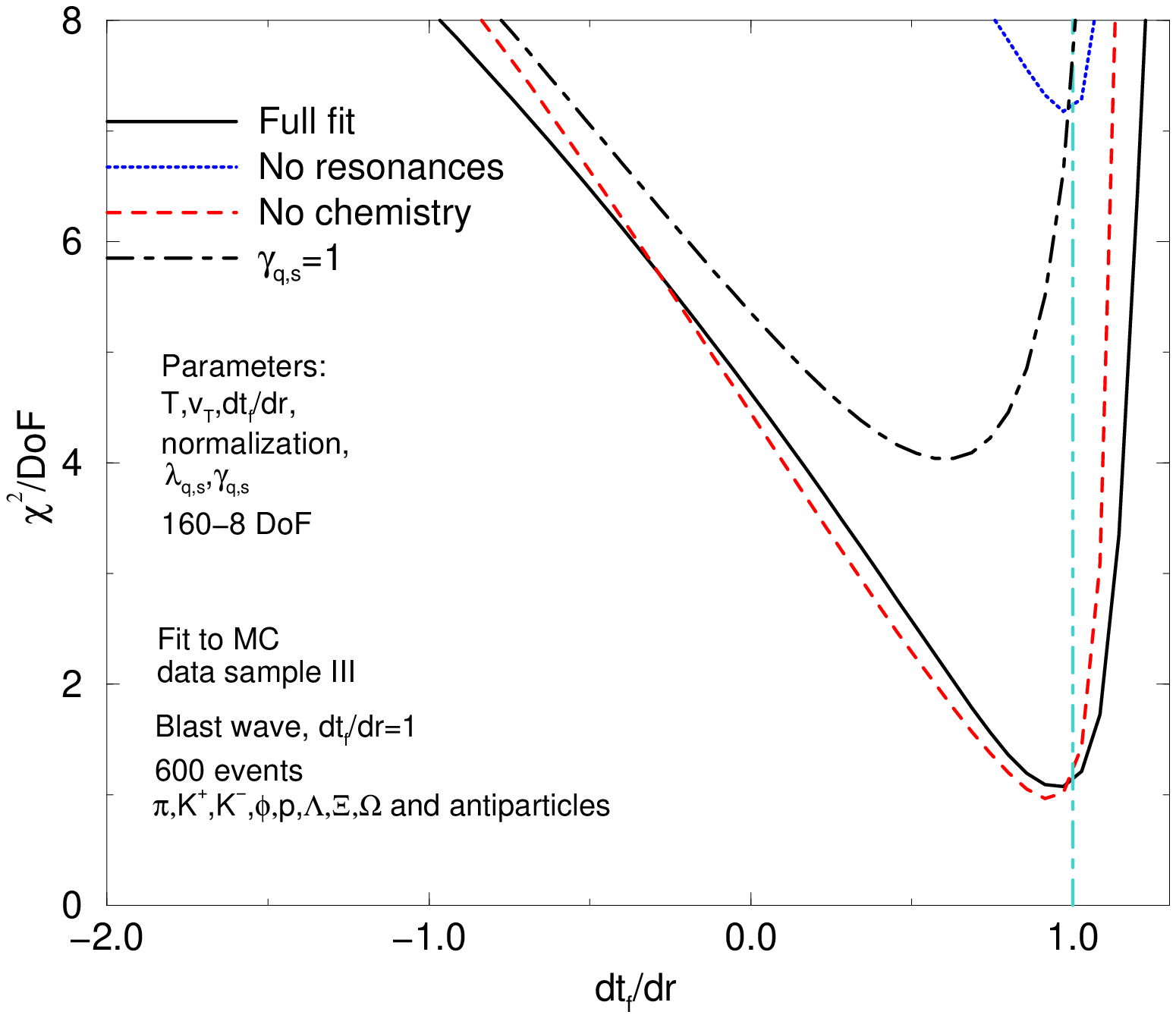}
\end{center}
\caption{(color on-line)
Results of fits to Monte-Carlo generated data samples III ($\frac{dt_f}{dr}=1$):
$\chi^2$ profiles for temperature (left) and fitted $\partial t_f/\partial r$ (right).   The models used in the fits are described in sections 1.1 and 1.2
Full fit includes fitted chemical potentials and $\gamma_{q,s}$, resonances and fitted $\partial t_f/\partial r$.
\label{prof1} }
\end{figure}
We have fitted the three samples to a variety of models, producing
$\chi^2/\mathrm{DoF}$ profiles for freeze-out temperature and the $\partial t_f/\partial r$ parameter.
Fig.~\ref{prof1} shows the profiles resulting in the fit to sample III while samples I and II are shown in Fig.~\ref{prof0}.
A full chemistry model with resonances  (solid black line) seems to be equivalent (as far
as the position of the $\chi^2$ minimum and the value of $\chi^2/\mathrm{DoF}$) 
to a fit in which normalization is particle-specific (dashed red line).
However, the chemical potential fit has greater statistical
significance since considerably more
degrees of freedom are required for arbitrary normalization.

If chemical potentials are included resonances
become essential since a fit with chemical potentials but no resonances
(dotted blue line) loses all statistical significance.
Similarly, the physical presence of non-equilibrium
($\gamma_{q,s} \neq 1$) means chemical potentials have to include
the non-equilibrium parameter for the fit to be meaningful (black dot-dashed line).
The freeze-out geometry does not seem to impact the temperature minimum that much.
However,  the correct freeze-out geometry can
be picked out by a comparison of fits to different models by choosing
the model with the lower $\chi^2/\mathrm{DoF}$.
Moreover, data sensitivity to temperature is strongly affected by freeze-out
geometry:
Comparing the temperature profiles for different choices of $\partial t_f/\partial r$ (Figs.~\ref{prof0} and~\ref{prof1}) it is
apparent that the temperature $\chi^2$ minimum is more definite in
the $\partial t_f/\partial r=0$ case.
In the case of explosive freeze-out, the correlation
between temperature and other parameters in the fit (notably flow) increases, resulting in a shallow $\chi^2$ increase at larger than minimum temperatures.

Finally it should be noted that flow profile, freeze-out geometry and temperature appear to be
strongly correlated.
If data sample II is fitted with a distribution with no flow profile (such as Sample I) there is a non-negligible shift in both the fitted temperature and
$\partial t_f/\partial r$, and a small rise in $\chi^2/\mathrm{DoF}$ (Fig.~\ref{prof0}, brown dot-dashed line).

As fits including both flow profile and resonances are computationally very intensive, we shall have to limit ourselves to the observation
that the uncertainty in flow profile is a source of systematic error here.  In the following chapter, a probe capable of disentangling
flow profile from freeze-out will be developed.

\section{Fit to RHIC data ($\sqrt{s_{NN}}=130 {\rm GeV}$)}
\begin{figure}[h]
\begin{center}
\epsfig{width=6.5cm,clip=,figure=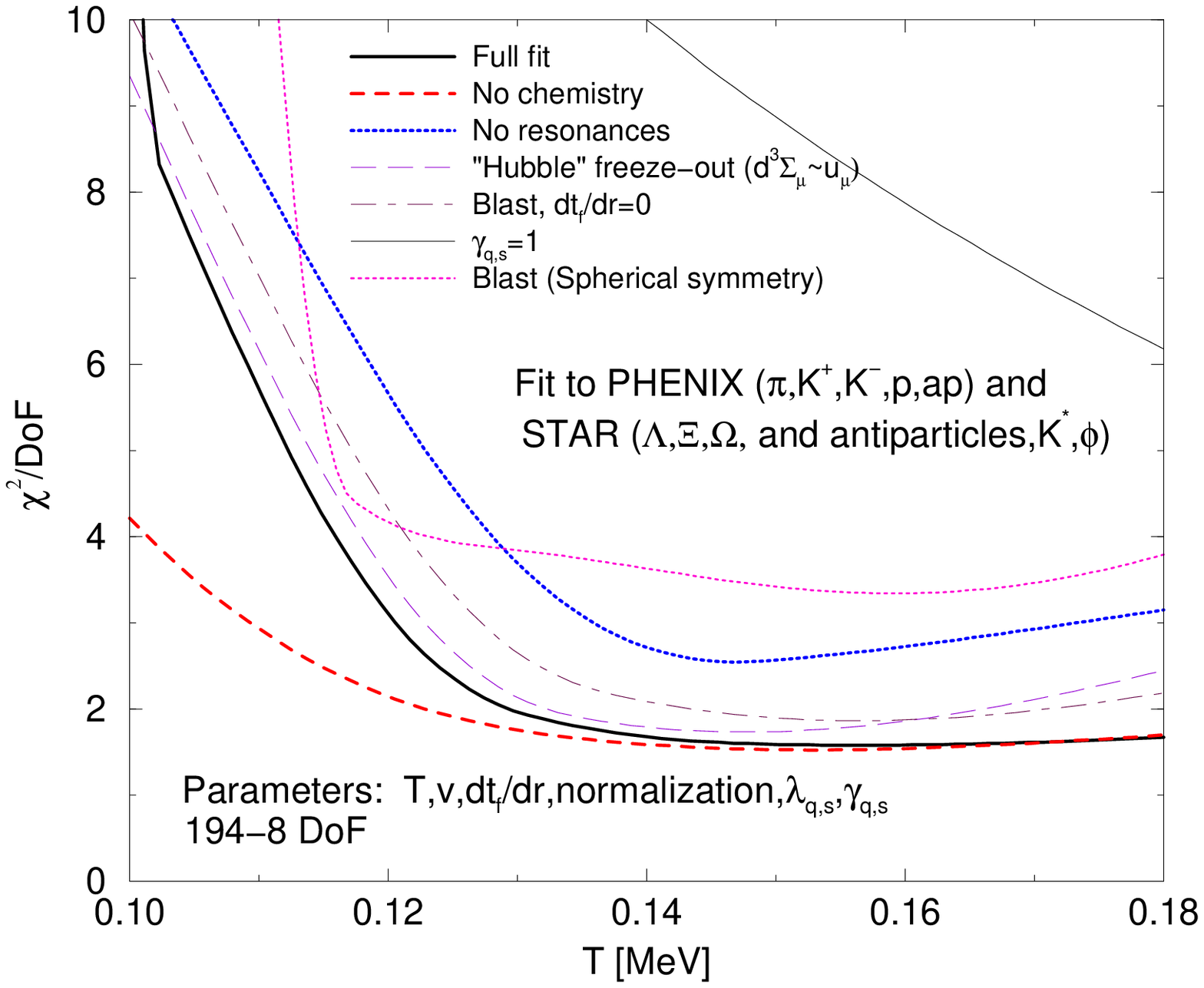}
\epsfig{width=6.5cm,clip=,figure=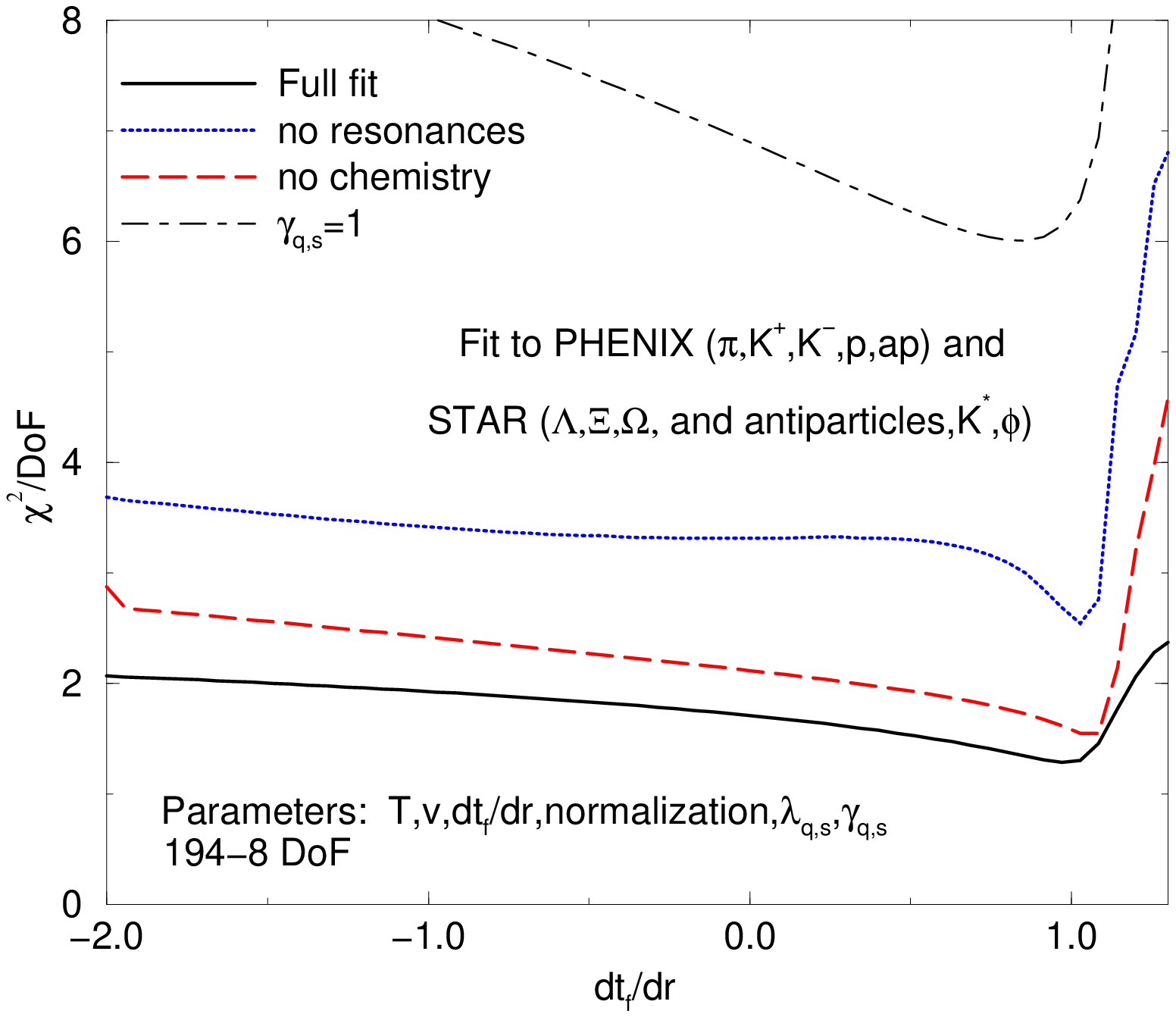}
\end{center}
\caption{(color on-line)Fit to RHIC data
profiles for temperature (left) and fitted $\partial t_f/\partial r$ (right).   The models used in the fits are described in sections 1.1 and 1.2
Full fit includes fitted chemical potentials and $\gamma_{q,s}$, resonances and fitted $\partial t_f/\partial r$.
\label{fitdata}. }
\end{figure}
Finally, we have performed a fit to the available RHIC data.
The data sample we used is the same as the one used for the Monte-Carlo, but,
since the STAR and PHENIX spectra had different trigger requirements (notably
centrality) and acceptance regions, we have used two different
system volumes, one for STAR particles and another one for PHENIX.

The results are similar to the Monte Carlo data in many ways.
The $\chi^2/\mathrm{DoF}$ was only slightly larger.\\
A fit  with particle-specific normalization gives a very similar $\chi^2$ and
fitted temperature to a fit including chemistry and resonances.
If chemistry is included in particle spectra analysis, resonances and
non-equilibrium are essential.
The fit to freeze-out geometry weakly points to
$\partial t_f/\partial r=1$, a picture that is supported by the
temperature $\chi^2$ profile, virtually identical to data sample III.
However, we can not claim our study to be complete in this respect, since we have not yet
investigated the effect of including
flow profile in the models.
As Monte-Carlo simulations have shown, the conclusion can differ
once these are taken into account.

\begin{figure}[h]
\begin{center}
\epsfig{width=6.5cm,clip=,figure=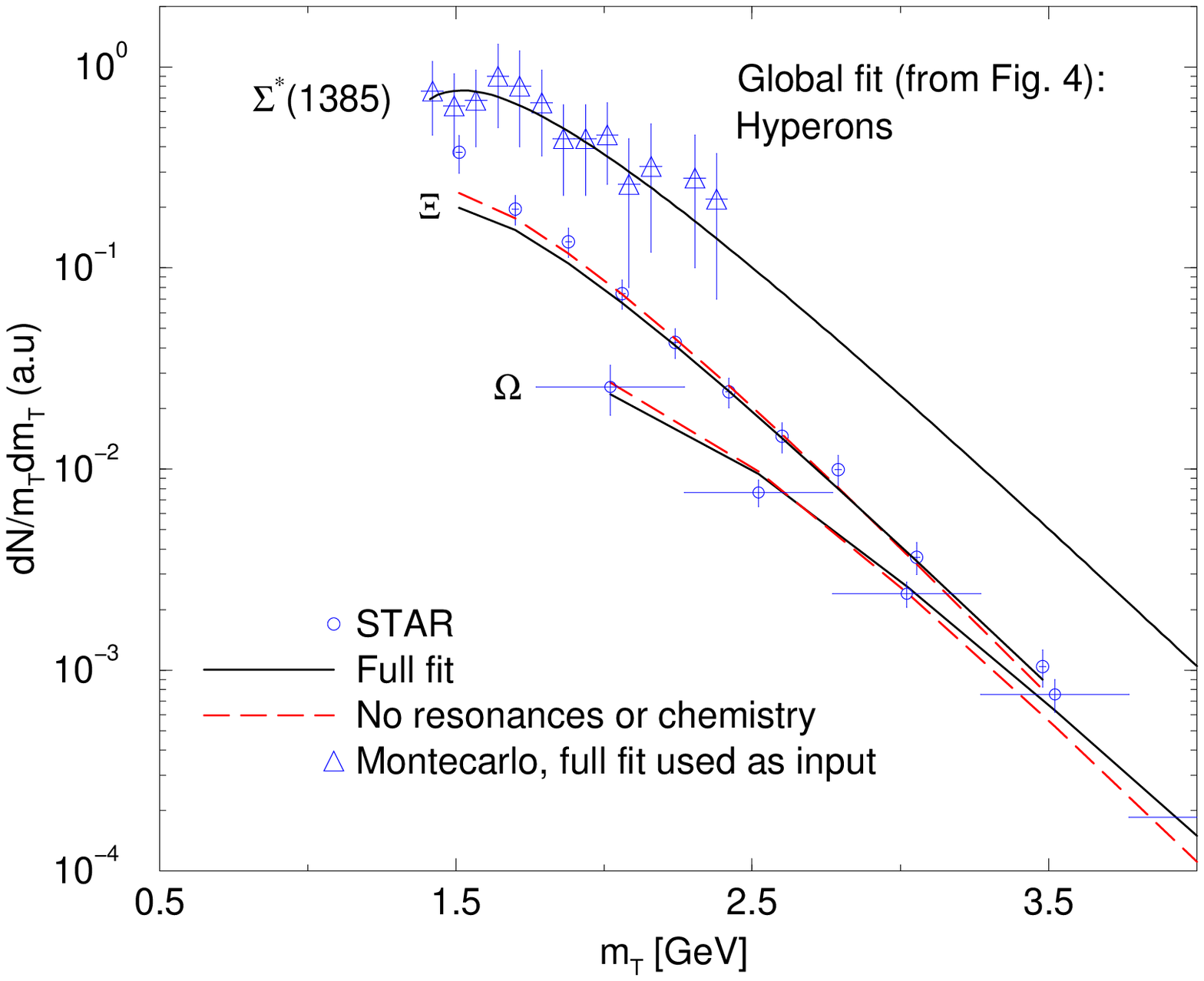}
\epsfig{width=6.5cm,clip=,figure=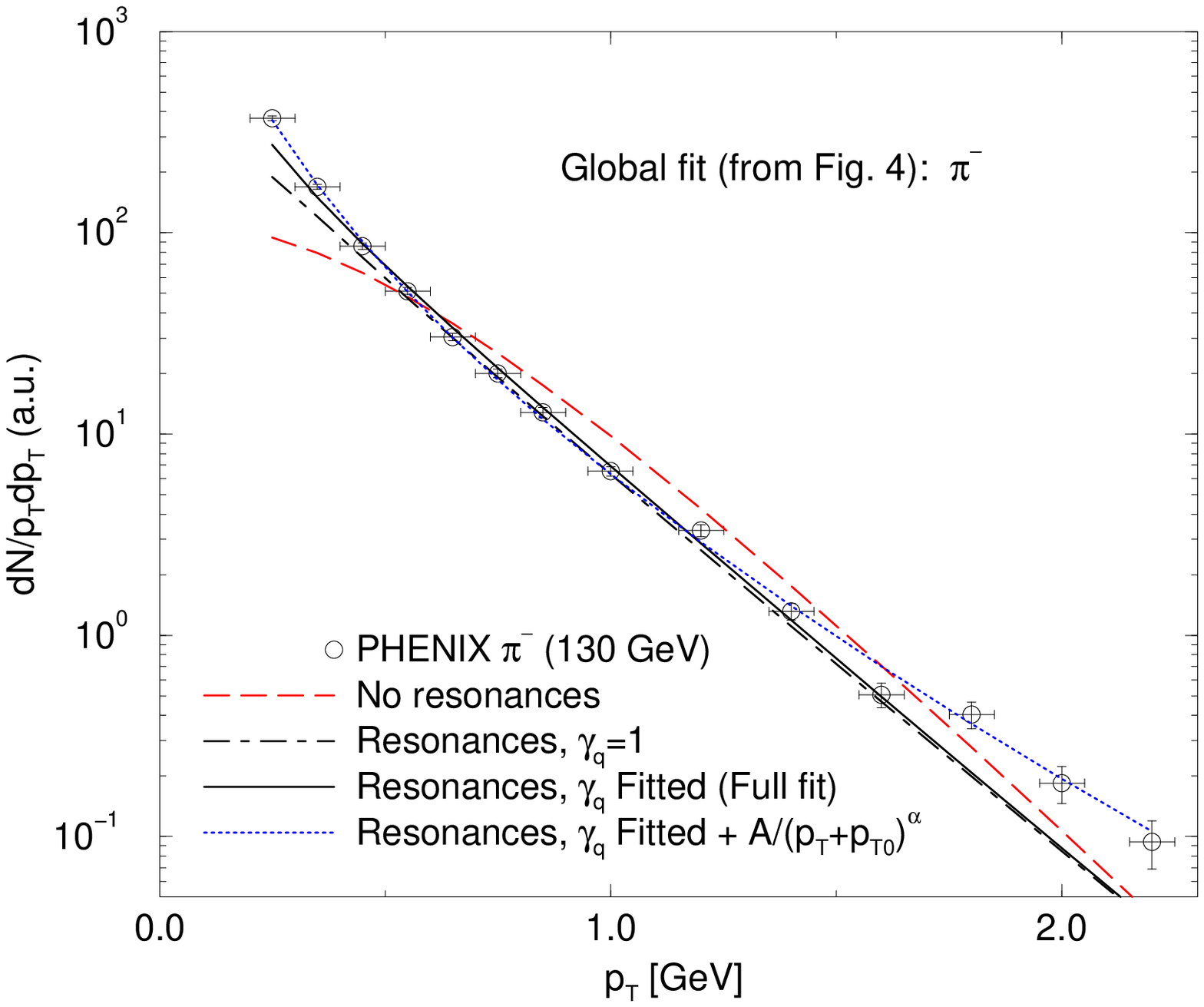}
\end{center}
\caption{(color online)Left:  $\Xi$ and $\Omega$ $m_T$ distributions, within a global fit including resonances and chemical potentials 
(Fig.~\ref{fitdata}, solid black) as well as no resonances or chemistry ( Fig.~\ref{fitdata},dashed red).  The model with chemical potentials was used to predict the
$\Sigma^*$ $m_T$ distribution using the Monte-Carlo.
Right: PHENIX $\pi^-$ $p_T$ distribution, within the global fit from Fig.~\ref{fitdata}.
As can be seen, both resonances and $\gamma_q$ help, but are not  sufficient to explain the pion distribution fully.
\label{twoplots} }
\end{figure}

Fig.~\ref{twoplots} shows the hyperon and pion spectra from the global fit 
of Fig.~\ref{fitdata}.   A comparison of the fits on the left panel confirms that a model
with no chemistry is about as good at fitting particle spectra as a model
with resonances and chemical potentials.    However, the second fit also has
predictive power:  We have used the fitted parameters to predict the $m_T$ spectrum for the $\Sigma^*$.
Unsurprisingly, we found that the $\Sigma^*$ should have roughly the same slope as the $\Xi$s, but its total multiplicity should be about three times as big.
We therefore suggest that a greater sample of spectra, in particular more spectra of heavy resonances taken within the same centrality bin as light particles (to make sure
both flow and emission volume are the same for each particle)  would help in
establishing whether chemical potentials are a good way to normalize
hadron spectra or not.

The only spectrum which presents a significant systematic deviation for most
models is the $\pi^-$.    As Fig.~\ref{twoplots} (right panel) shows, most models fail to catch the upward dip of the low momentum pions, and indeed simple transverse expansion predicts the reverse trend \cite{burward-hoy,vanleeuwen,NA57spectra,castillo}.   Including resonances, and allowing
for $\gamma_q > 1$ helps (the latter is equivalent to postulating a pion
``chemical potential''  \cite{heinzpions})
However, to fully account for the lowest momentum pions, even addition of resonances and $\gamma_q > 1$ are not enough.
One has to add a power-law component to the pion spectrum
\begin{equation}
\label{powerlow}
E\frac{dN}{d^3 p}=\left(E\frac{dN}{d^3 p} \right)_{cf}+\frac{A}{(p_T+p_{T0})^{\alpha}}
\end{equation}
This contribution (roughly $6 \%$ of the total pion yield in the best fit) also accounts for the highest $p_T$ pions.
Such a parametrization has been justified \cite{powerlaw} and successfully used \cite{peitzmann} before.

\begin{figure}[h]
\begin{center}
\epsfig{width=6.5cm,clip=,figure=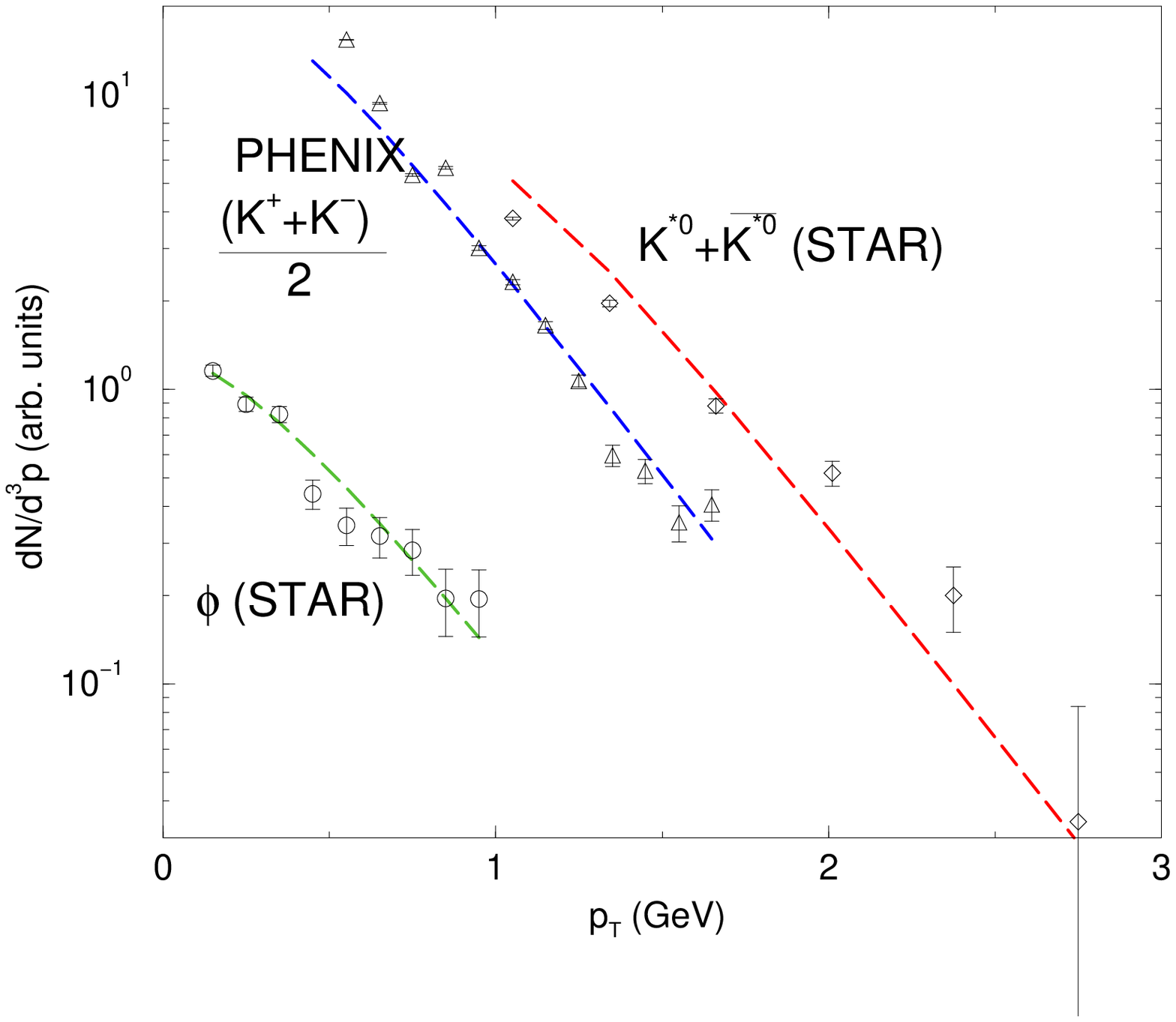}
\epsfig{width=6.5cm,clip=,figure=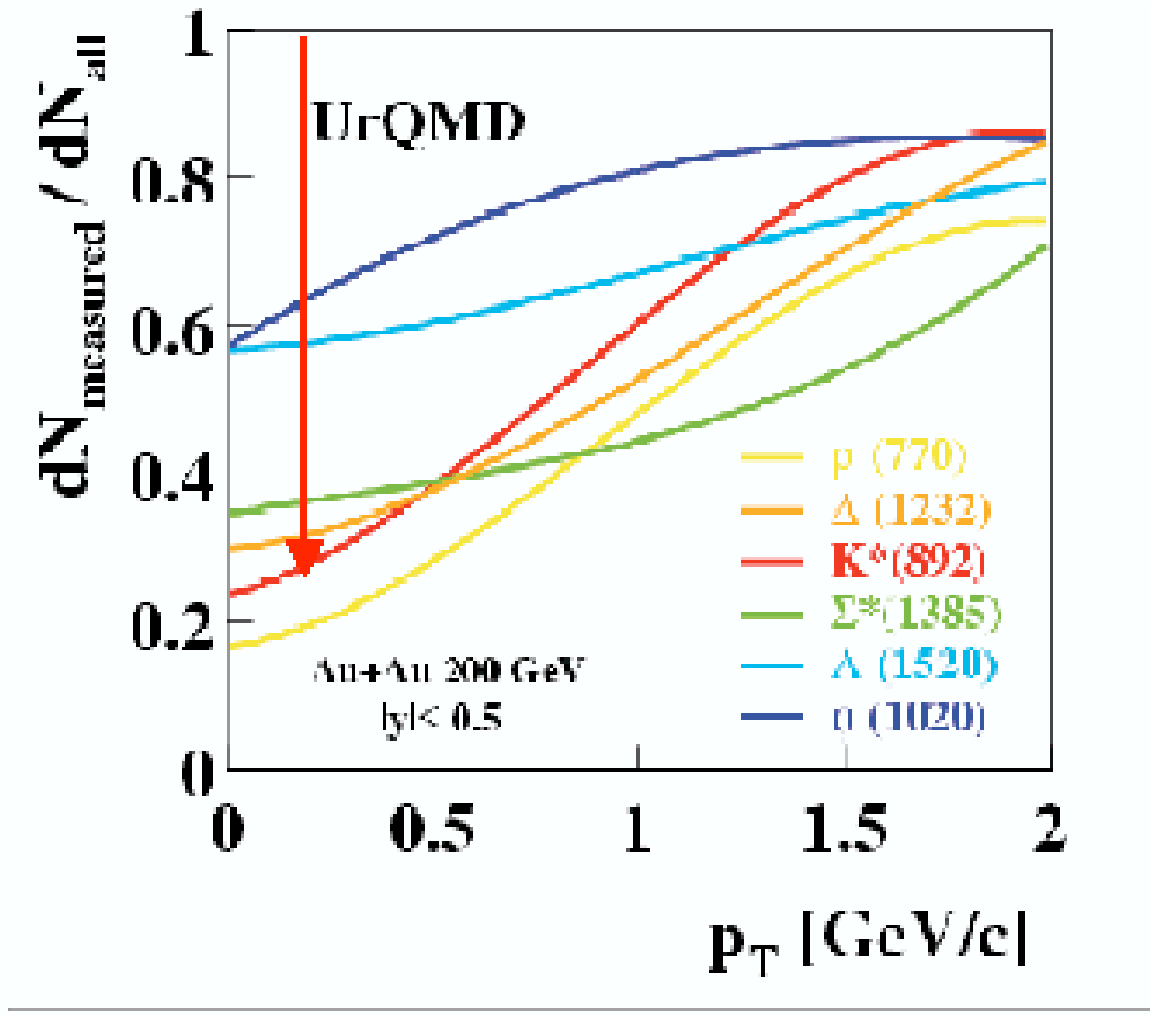}
\end{center}
\caption{(color online)(left) $K^*$, $K^-$ and $\phi$ normalized 
spectra within the global fit using non-equilibrium sudden freeze-out  ( Fig.~\ref{fitdata}, black).
(right) momentum dependence of $K^*$ suppression through rescattering of decay products, calculated in the microscopic hadronic
model uRQMD 
\label{kstar}.}
\end{figure}
Finally, Fig.~\ref{kstar} (left) shows the Blast wave fit to the $K^*$ momentum distribution.
The fact $K^*$ can be fitted to the same temperature and flow as the stable hadrons is, in itself, remarkable, since
it is a short lived resonance detected by invariant mass reconstruction, which potentially decays before freeze-out and 
whose decay products can be rescattered
(and made undetectable) by other particles in the medium.
We shall not dwell on this for too long, since the next chapter is mainly devoted to a discussion on directly
detectable short-lived resonances.   However, it should be mentioned that, as  Fig.~\ref{kstar} (right) shows, 
microscopic models predict the rescattering of decay products to be non-trivial and strongly momentum dependant \cite{bleicher}.
Given this, the relative goodness of the fit to $K^*$ data is a strong boost for the sudden freeze-out picture.

The acceptable $K^*$ description within a fit which includes $K^+$ and $K^-$ is encouraging for another reason:
Resonances and absolute normalization in fits with many particles remove the correlation between temperature and flow:  It is not possible
to adjust the slopes by shifting the temperature and transverse flow along the same contour, since in this
case normalization is also affected.    However, temperature is still strongly correlated with chemical potentials
(expecially $\gamma_s$ and $\gamma_q$).    
However, if the fit includes $K^*$s and Ks, this correlation is removed, since both spectra share a common
chemical potential.    Shifting the temperature from the minimum will result in a shift of one of these spectra
away from the data, no matter what chemical potential is used.
For this reason, a common tempeature minimum in a fit which includes $K^*$ and Ks is an important test
the sudden freeze-out model successfully passed.    We hope further tests, in the form of more resonance
spectra, are forthcoming.

In conclusion, we have given an overview of different hadronization models, and described how they arise 
from different freeze-out scenarios of a system forming
from a thermalized quark gluon plasma.
We have also shown, in the boost invariant limit, how the various hadronization ansatzes give rise
to quantitative differences in observed particle spectra.
We have then used MonteCarlo simulated data to study the sensitivity to model choice of presently
available experimental data, and have evaluated different models ability to
fit presently existing RHIC data.
While data slightly favors a chemical non-equilibrium explosive freeze-out, there is not enough evidence to make a definitive conclusion
about this issue.   We hope the forthcoming 200 ${\rm GeV}$ results will clarify this further.

\setcounter{figure}{0}
\setcounter{equation}{0}
\setcounter{table}{0}
\chapter{Direct detection of short-lived resonances}
\label{cha:resonances1}
\subsection{Introduction and motivation}

\begin{figure}[tb]
\centerline{\includegraphics[width=012cm]{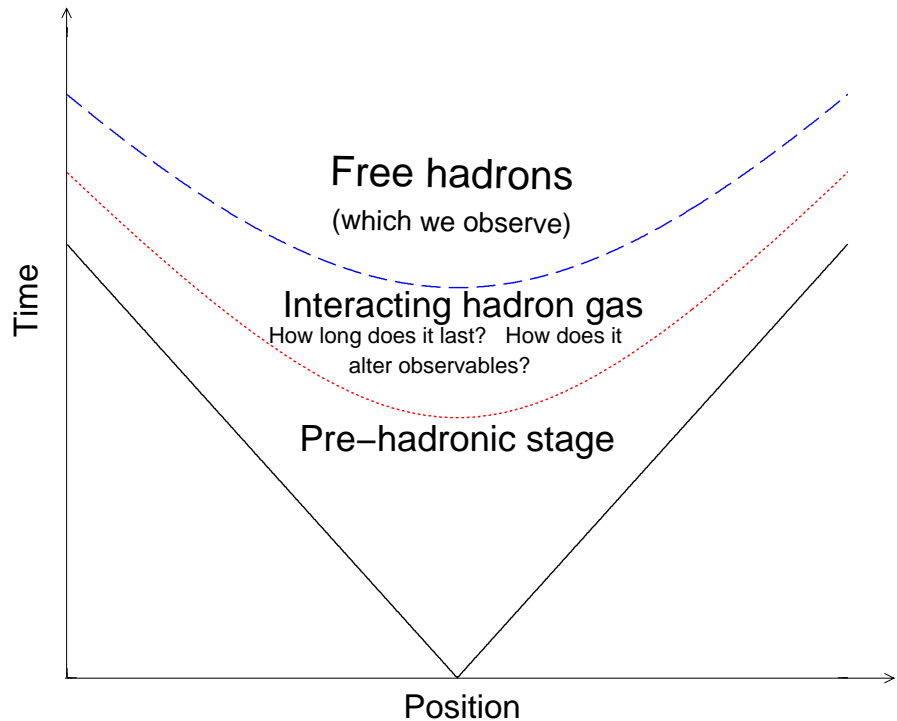}}
\caption{ Stages of the space-time evolution of a
heavy ion collision.  At a certain moment in proper time (known as chemical freeze-out), hadrons emerge.
The system then evolves as an interacting hadron gas, until thermal
freeze-out, the point at which all elastic interactions cease. \label{problem}}
\end{figure}

The experimental measurement of short-lived hadron resonances can potentially
be very useful in clarifying some of the least understood aspects of heavy
ion collisions.
In general, evolution of a hot hadronic system proceeds according
to Fig.~\ref{problem}.
When mesons and baryons emerge from a pre-hadronic state, presumably
quark gluon plasma, their abundances are
expected to be fixed by hadronization temperature and chemical fugacities.
This stage of fireball evolution is commonly
known as chemical freeze-out.  
After initial hadronization, the system may evolve as an interacting
hadron gas.
At a certain point (which can vary according to particle
species), thermal freeze-out, where hadrons stop interacting, is reached.

A quantitative understanding of the above picture is crucial
for any meaningful analysis of the final state particles.
Many probes of deconfinement are most sensitive when 
the dense hadron matter fireball breakup is sudden and
re-interaction time short or non-existent.
Final state particles could, however also emerge remembering relatively
little about their primordial source, having been subject
to re-scattering in purely hadronic gas phase.
In fact, theoretical arguments have been advanced in support of
both a sudden reaction picture  \cite{sudden1,sudden2} and
a long re-interaction timescale \cite{bdmsps,cern_evidence}.
Both pictures have been applied to phenomenological fits of hyperon
abundances and distributions \cite{PBM01,Grassi,torrieri_sps1}.\\
\vskip -0.5cm
It is apparent that hyperon resonances can be crucial in resolving this 
ambiguity:       
their initial abundance, compared to stable particles
with the same quark composition, depends primarily to the temperature
at hadron formation.
However, the observed  
abundance will be potentially quite different,
and will strongly depend on the lifetime of the interacting hadron gas
phase.
Resonances can only be observed via invariant mass reconstruction, and
their short lifetime means that they can decay within the interacting hadron
gas (Fig.~\ref{rsc_fig}).
In this case, the decay products
can undergo re-scattering, and emerge from the fireball with no memory about 
the parent resonance.
Thus, the observable 
resonance abundance is sensitive to precisely those parameters
needed to distinguish between the sudden and gradual freeze-out models.

In this chapter,
we start with a review of presently available experimental data,
and the open questions which arise.
We then proceed to describe how to calculate the initial resonance
abundance and the effect of re-scattering in terms of the hadronization
temperature and the lifetime of the interacting hadron gas.    
We show how these two parameters can be extracted
from the experimental observations. Finally, we discuss
possible answers to experimental challenges raised in the first section, and
suggest ways by which these questions could be resolved by further
measurements.
Most of this write-up is based on recently
published experimental measurements \cite{fri01,fat01,fat02,mar01,xu01,zh04,salur_sigma}
and theoretical papers \cite{torrieri_reso1,torrieri_reso2,torrieri_reso3}
\newpage
\begin{figure}[h]
\begin{center}
\psfig{width=6.5cm,clip=1,figure=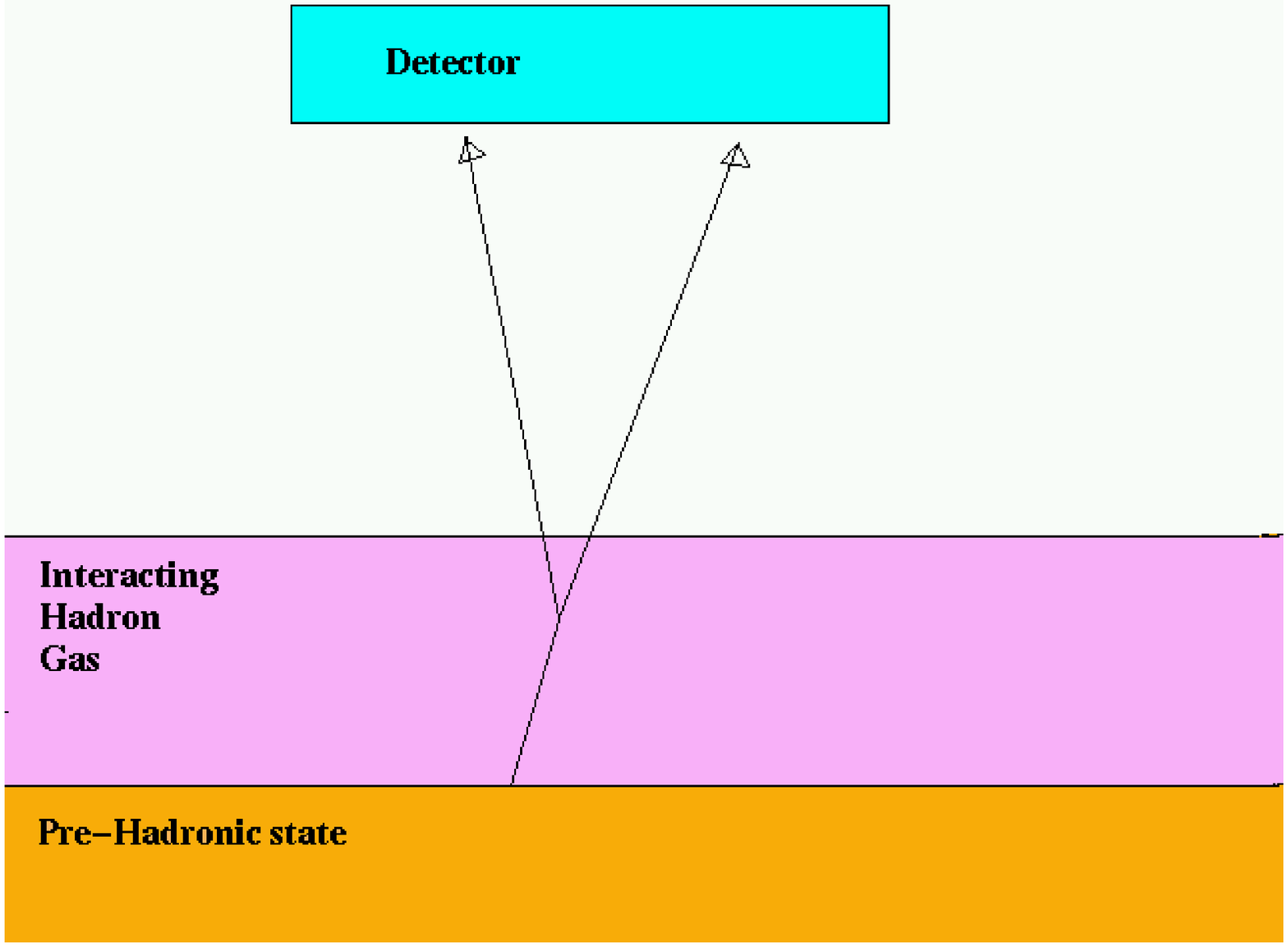}
\psfig{width=6.5cm,clip=1,figure=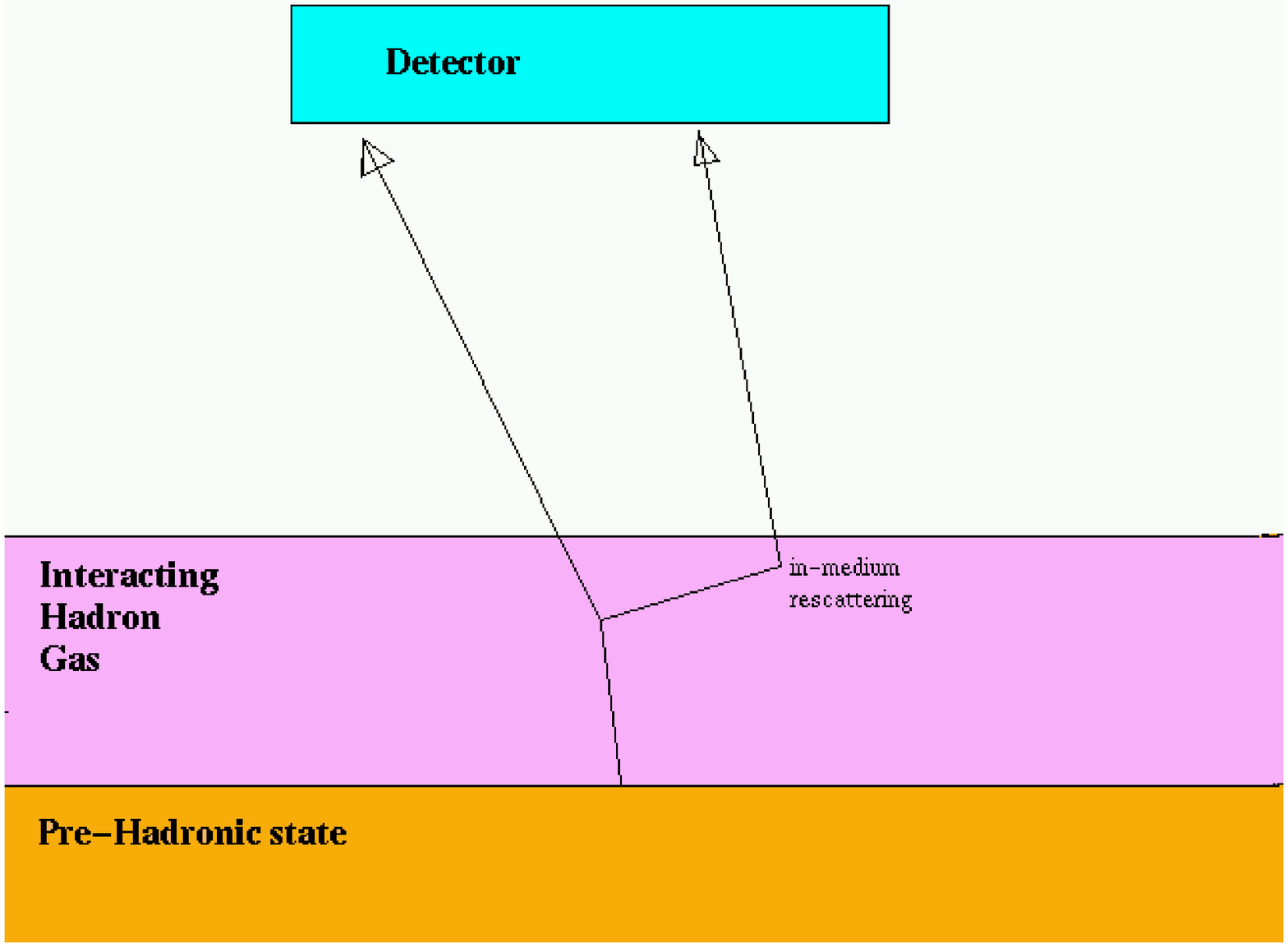}
\end{center}
\end{figure}
\begin{center}
 $\Downarrow$ .....................................................$\Downarrow$
\end{center}
\begin{figure}[h]
\begin{center}
\psfig{width=6.5cm,clip=1,figure=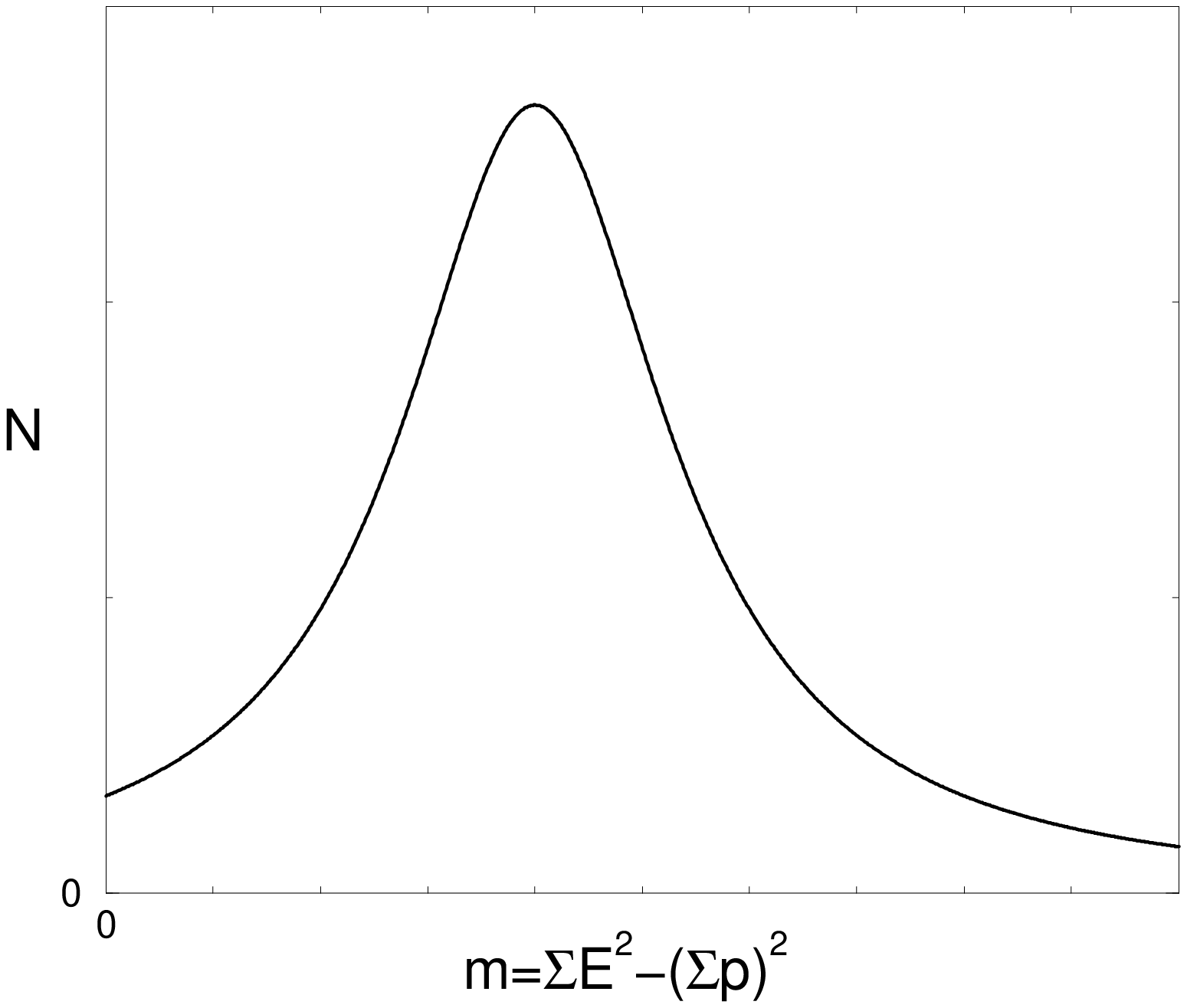}
\psfig{width=6.5cm,clip=1,figure=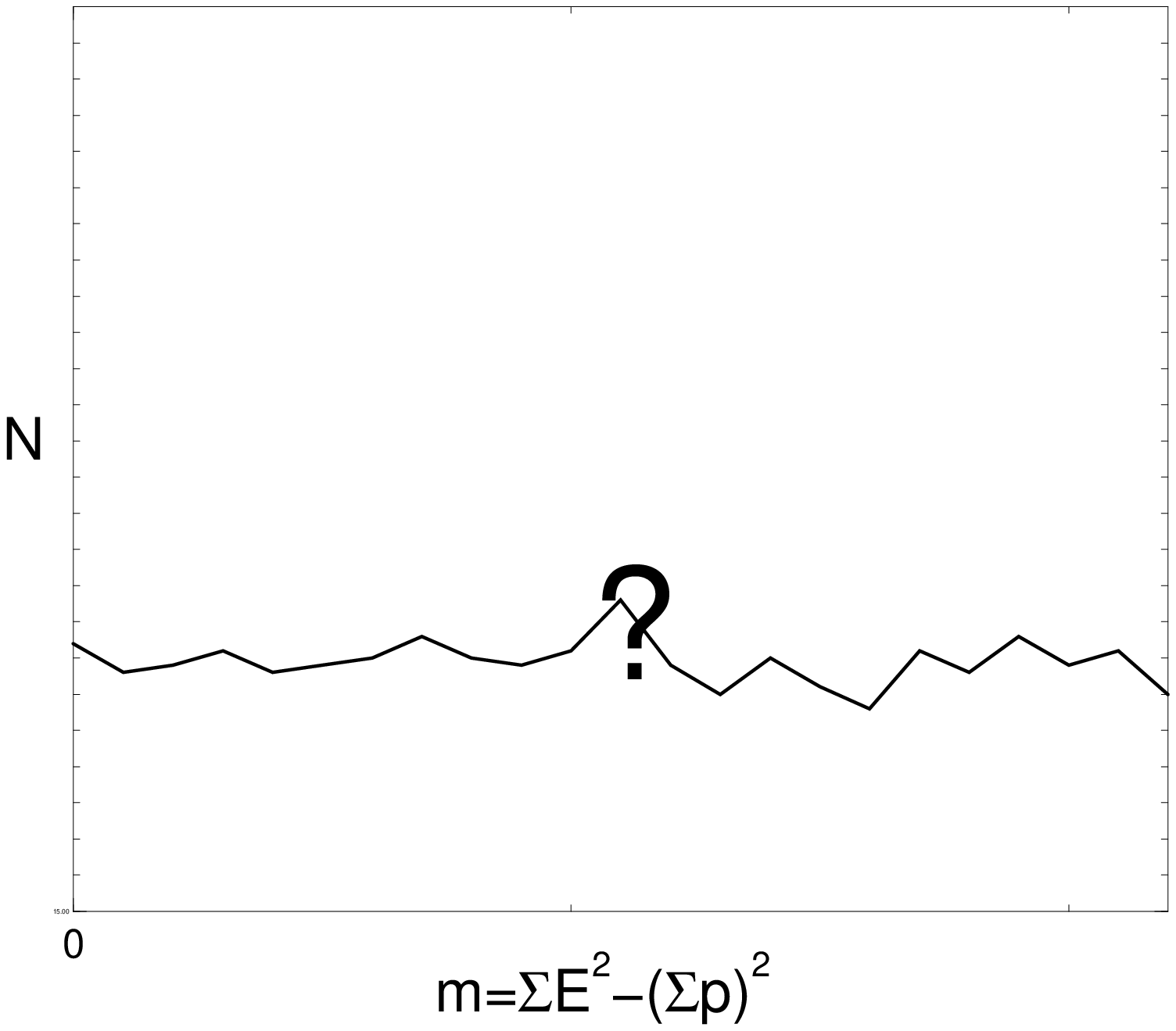}
\caption{How re-scattering can inhibit resonance reconstruction.
If the resonance decay products reach the detector without further interactions, their invariant mass distribution should yield a clear peak at the
resonance mass.   However, if these decay products undergo re-scattering
before reaching the detector, the signal may be indistinguishable from
the background caused by unrelated particle pairs.
 \label{rsc_fig}}
\end{center}
\end{figure}

\section{Can short-lived resonances be explained by a statistical model?}

\begin{figure}[t]
\vskip 0.3cm
\centerline{
\psfig{width=6.5cm,figure=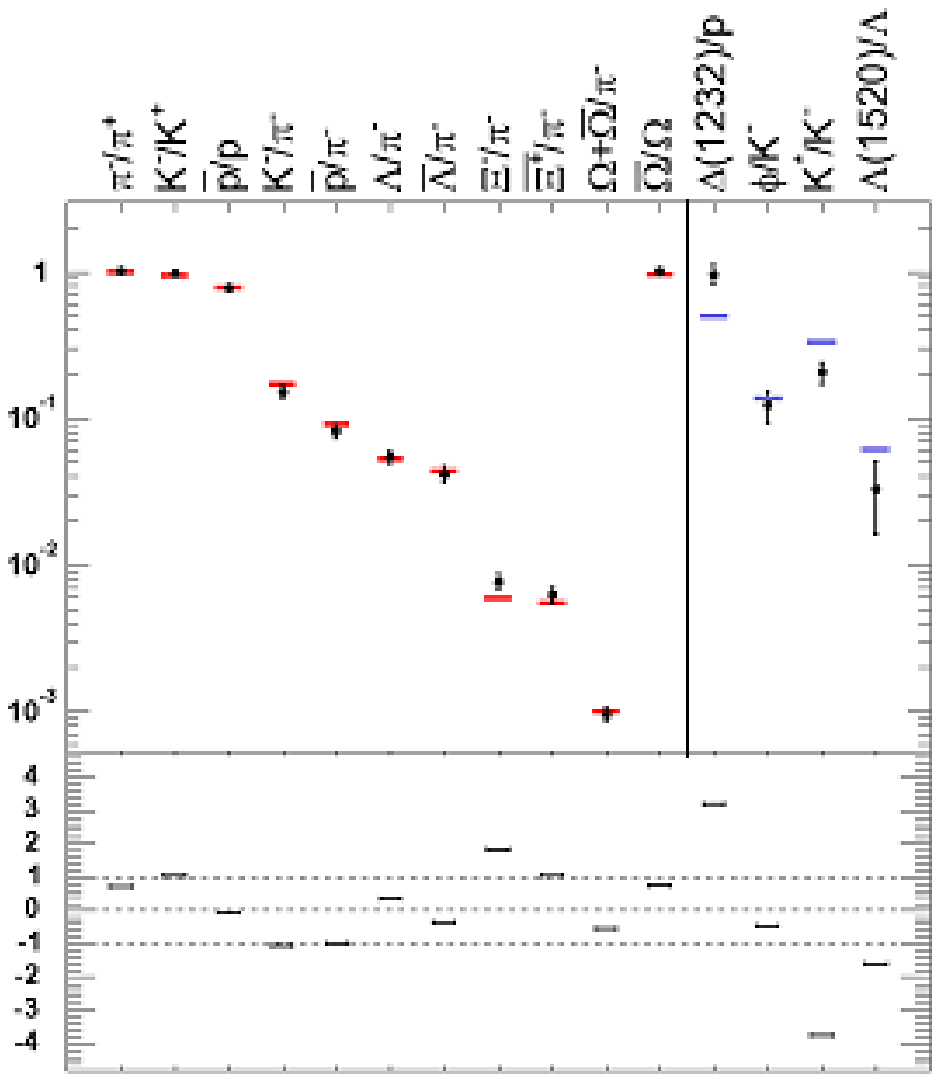}
\psfig{width=6.5cm,figure=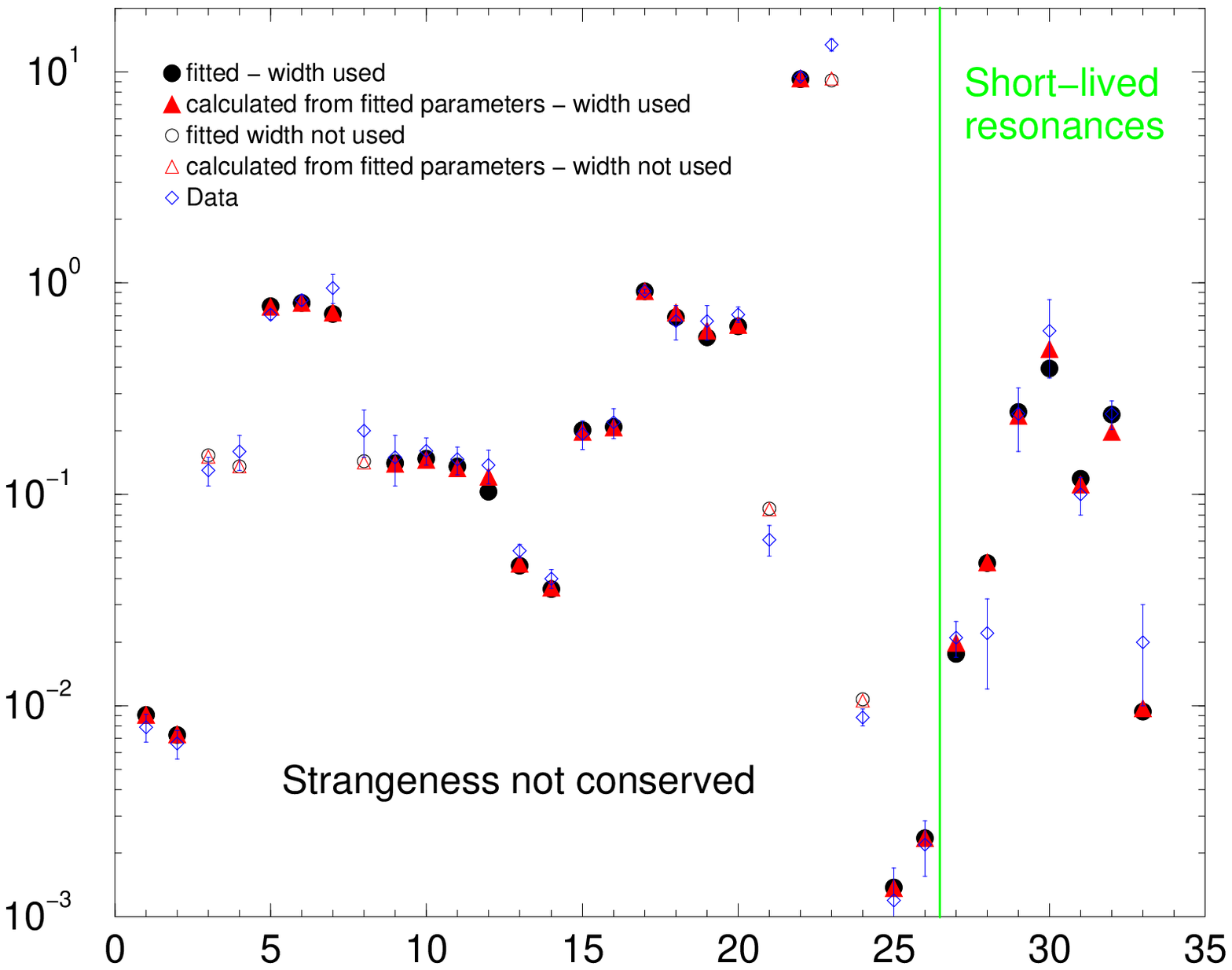}}
\caption{\label{resofit} (Color on-line)  Fits to particle
ratios containing several short-lived resonances, using an equilibrium (left) and a non-equilibrium (right) formalism.
The fit in (right), was obtained using the SHARE program (See Appendix D).}
\end{figure}
Since its beginning, the RHIC heavy ion program, and in particular the STAR collaboration, has managed to
measure an impressive amount of resonance data.   In addition to the $K^*$ and the $\Lambda(1520)$, the $\rho$, $f_0 (980)$, $\Delta$, $\Xi^*$
and $\Sigma^*$ have been measured in p-p, and sometimes Au-Au \cite{allres,zh04}.  Further results in Au-Au collisions are expected soon.
Including these particles in a global fit, of the type described in section 2.2.1 places a tight constraint on the temperature, since
many of the resonances have the same chemical composition but a different mass from observed ``light'' particles.
Fig.~\ref{resofit} shows what happens when such a fit is done.
A chemical equilibrium model \cite{bdmrhic} fails to describe short-lived resonances.
Under chemical non-equilibrium, however (Fig.~\ref{resofit} right), both short-lived and long-lived resonances
are described with equal acceptability.   In a sense, the proponents of both a long freeze-out stage (which would
pull short-lived resonances out of equilibrium ) and sudden freeze-out can be satisfied by this result:  Proponents of hadronization
at T=170 ${\rm MeV}$ can rely on the hypothesis that in-medium interactions will make short-lived resonances unobservable, while
the short-lived data points are well described by the non-equilibrium model, at the same temperature and $\gamma_{q,s}$ as found in earlier
fits \cite{Zak03}.
We can try to see which of the pictures is right by concentrating on a few particularly simple resonances, and attempting to quantify the
effect of a post-hadronization hadron gas phase.   This topic will be addressed in the next section.

\section{Modeling resonances in heavy ion collisions}

\subsection{\label{prodrat} Direct production at hadronization}

We assume that at hadronization, a volume element
will be at thermal and chemical equilibrium in its local rest frame.
As shown in section 2.2.1, this means that details of flow and freeze-out hypersurface cancel out
when the global (4 $\pi$) ratio, or a mid-rapidity ratio in a Bjorken-type fireball, are
considered.   

In the Boltzmann approximation (applicable here since $\lambda_{particle} \sim O(1)$ and $m>>T$), the chemical potential will then cancel
out when two particles of the same quark composition are compared.
The ratio of the heavier resonance $N^*\rightarrow N \pi$ to the lighter resonance $N$ will then be given by Eq.~(\ref{Nk}))
\begin{eqnarray}
\label{direct}
\frac{N^*}{N} =\frac{g^* m^{*2} K_2 \left( \frac{m^*}{T}  \right)}{g^* m^{*2} K_2 \left( \frac{m^*}{T}  \right)+g  m^2  K_2 \left( \frac{m}{T}  \right)}
\end{eqnarray}
(Here $g$ is the statistical degeneracy).
Ratios where the Boltzmann approximation is applicable and feed-down from particles
of different quark constituents is small include
$\Sigma^{*}/\Lambda$ and  $\Lambda(1520)/\Lambda$.
$K^{*0}/K^{+}$ and $\overline{K^{*0}}/K^{-}$ has a larger correction due to $\phi \rightarrow KK$\footnote{Of course, the $K^{*0}/K^-$ ratio
can be corrected for $\phi \rightarrow K^+ K^-$ feed-down, since the NA49 and STAR experiments have published yields for both.   We recommend that such
a correction be done when presenting future $K^*$ results, as their interpretation becomes considerably more model-independent.},
while in other, recently measured ratios ($\Delta/p,\rho/\pi$ \cite{zh04}) the feed-down from particles
with a different chemical potential is simply too large to be neglected.  

If less than 4 $\pi$ acceptance is observed, we can obtain the equivalent of Eq.~(\ref{direct}) by integrating Eq.~(\ref{f_sph2}) and
Eq.~(\ref{after_int}) across a finite $y-p_T$ range.
As will be shown, the finite acceptance effects will cancel out to a very good approximation if particles of a comparable mass are considered.
Hence, Eq.~(\ref{direct}) can still be used for ratios with a finite acceptance detector.

Table~\ref{processes} summarizes the decay processes 
considered in our analysis and their parameters (Clebsh-Gordon 
coefficients have been used to estimate decays such as
($N^{*0} \rightarrow N^+ \pi^-)/(N^{*0} \rightarrow N^0 \pi^0)$).

\begin{table}[h!]
\begin{center}
\caption{Resonances contributing to $\Lambda$ and K production, with their
degeneracies, rest-frame momentum ($p^*$) and possibility for experimental
reconstruction.}
\begin{tabular}{ccccc} 
\hline
\tablehead{1}{g} & \tablehead{1}{Reaction} & \tablehead{1}{$p^{*}$}  & 
\tablehead{1}{branching} & \tablehead{1}{visible?} \\ 
\hline
$\approx 4$ & $\Sigma^{* 0}(1385) \rightarrow \Sigma^{0} \pi^{ 0}$ 
& 127 & $ \approx 4 \%$   & No
\\  \hline  
8 & $\Sigma^{* \pm}(1385) \rightarrow \Lambda \pi^{\pm}$ &
208 & $88 \%$  & Yes  
\\  \hline  
4 & $\Sigma^{* 0}(1385) \rightarrow \Lambda \pi^{0}$ &
208 & $88 \%$   & No
\\  \hline  
2 & $\Sigma^{0} \rightarrow \Lambda \gamma$ & 74  & $100 \%$ & No
\\  \hline  
4 & $\Lambda(1520) \rightarrow N \overline{K}$ & 244 & $45 \%$ & Yes
\\ \hline \hline
3 & $K^{*\,0}(892) \rightarrow K^+ \pi^-$ & 291  & $67 \%$ & Yes \\ \hline
\end{tabular}
\label{processes}
\end{center}
\end{table}

\begin{figure}[tb]
\vspace*{-2cm}
\hspace*{.1cm}
\centerline{
\psfig{width=10cm,clip=1,figure=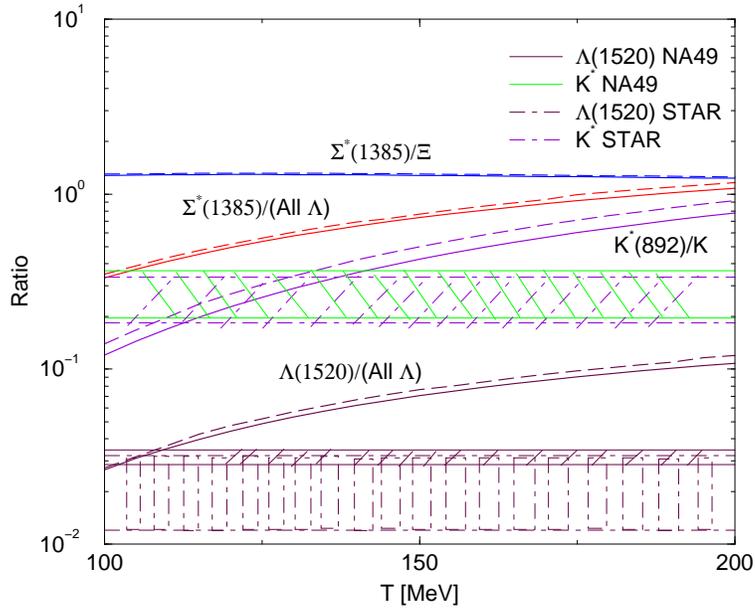   }}
\vspace*{-0.21cm}
\caption{ Temperature dependence of ratios of $\Sigma^{*}$, $K^{*\,0}$
and $\Lambda(1520)$ to the total number of observed $K$
$\Lambda$ s and $\Xi$ s. Branching ratios are included.
Dashed lines show the result for a measurement
 at central rapidity $\Delta y=\pm0.5$.
The experimental measurements included in the diagram are
discussed in the first part of this paper.
It should be noted that the $K^*/K$ ratio is actually $\overline{K^*}/K^-$ for
the NA49 measurement, and an average $(K^*+\overline{K^*})/(K^{+}+K^{-})$
for STAR. \label{prodratios} 
 }
\end{figure}

In Fig.~\ref{prodratios} we show the relative thermal  production ratios
at chemical freeze-out  over the entire spectrum
of rapidity and $m_{T}$  (solid lines) and central rapidity range 
(dashed lines). 
The sensitivity of resonance yields to hadronization
temperature is apparent for all resonances under consideration.
In particular, the $\Sigma^*$ emerges as a very promising candidate
for further study.   
For example,  at the lowest current estimates ($T\simeq 100$ ${\rm MeV}$)
of the final break up temperature in 158$A$ ${\rm GeV}$ SPS collisions 
$33 \%$ of $\Lambda$ s are actually primary $\Sigma^{*}$ s, the percentage
rises to slightly more than $50 \%$ if chemical freeze-out
occurs at $T= 190$ ${\rm MeV}$. 
For all of these cases,  the $\Sigma^*$ primary yield should be considerably
bigger than the $\Xi$ (if the chemical potentials in \cite{Zak03} are used
for the prediction), and the angle between the decay products should be about the same
as in the $\Omega \rightarrow \Lambda K$ decay.
The NA49 detector in the  SPS  and the STAR detector at RHIC are both capable
to measure hyperon and resonance yields well within the precision
required to make a good measurement of the hadronization temperature.

We also note that while the experimentally observed $K^*$ yield is
compatible with the produced ratio in the case of non-equilibrium hadronization \cite{Zak03},
the $\Lambda(1520)$ seems very suppressed.
To discuss this further, however, an estimation of the effect of re-scattering
on resonance abundances is necessary.

\vskip -0.5cm

\section{Rescattering}

As explained in the introduction, direct observation of resonances relies
on invariant mass reconstruction.
Therefore, to calculate the observed resonance abundances, 
rescattering after hadronization will have to be taken into account.
This can be looked at within a microscopic model
of hadronic matter such as uRQMD.
Here, we present such a study using a ``back of the envelope'' model which 
nevertheless seems to provide an acceptable quantitative description of how
the propagation of resonances and their decay products through opaque matter deviates
from an equilibrium prediction.

We first note in Fig.~\ref{prodratios}, that the 
relative $\Sigma^{*}/\Xi$ signal is remarkably independent
(within  5 \%) of Temperature.    This is because 
the $\Xi^*(1530)$ contribution cancels nearly exactly the thermal
suppression of the $\Xi$ originating in the $\Xi-\Sigma^{*}$ mass difference.
This effect  could be used for a direct estimate of the
$\Sigma^{*}$ lost through rescattering,
 even without knowledge of the freeze-out 
temperature, should  the chemical
parameters ($\lambda_{\mathrm{i}}$ and $\gamma_{\mathrm{i}}$) be
independently known (They can be obtained from a fit of the stable particle ratios). A simple test of sudden hadronization model 
consists in measurement of the ratio $\Sigma^{*}/\Xi$.
If it is significantly smaller than unity,
we should expect a re-equilibration mechanism to be present. Otherwise 
sudden hadronization probably applies, since $\Sigma^*$ emerge from
chemical freeze-out without undergoing interactions. 

We can, however, go further and use  the suppression
of the considered resonances  as a tool capable
of estimating conditions at particle freeze-out. 
We consider the decay of a generic resonance $N^{*}$
\begin{equation}
\label{decay}
N^{*} \rightarrow Y \pi \,,
\end{equation}
in a gas of pions and nucleons.
We shall assume that one interaction of either Y or $\pi$ 
is sufficient for that resonance to be undetectable, and that
the decay products travel through the medium with speed $\mathrm{v}_{\mathrm{i}}$
(where $i$ can mean either $Y$ or $\pi$).
The interaction probability is proportional to $v_{\mathrm{i}}$, the interaction
cross-section of the decay product
with each particle in the hadronic medium 
($\sigma_{\mathrm{i j}}(v_{\mathrm{i}})$, where $j$ can
refer to either pions, Kaons, nucleons or antinucleons. 
Note that the cross-section itself depends, in a generally 
complicated way, on the incident momentum, and
hence on $\mathrm{v}_{\mathrm{i}}$),
and the particle density in the fireball $\rho_j$.
$\rho_j$ is increased 
by a factor $\gamma_{\mathrm{i}}=1/\sqrt{1-\mathrm{v}_{\mathrm{i}}^2}$ due to Lorentz-contraction, and decreases
as time passes because of the fireball's collective expansion (parametrized by the
flow velocity $\mathrm{v_{flow}}$, assumed to be of the order of the
relativistic sound speed $c/\sqrt{3}$.)
The time dependence of the densities will thereflore be
\begin{equation}
\label{expansion}
\rho_{j}(t)=\gamma_{\mathrm{i}} \rho_{0j} \left( \frac{R}{R+\mathrm{v_{flow}} t}\right)^3,
\end{equation}
and $\rho_{0j}$, the density of j at hadronization, can be calculated
from the chemical freezeout temperatures and chemical potentials.
Putting everything together, the rescattering reaction rate  is
\begin{equation}
\label{scatter}
P_{\mathrm{i}} = \sum_{\mathrm{v_i}}
\left[ \sigma_{i \pi}(\mathrm{v_i}) \rho_{0 \pi}+\sigma_{i K}(\mathrm{v_i}) \rho_{0 K}+\sigma_{i N} (\mathrm{v_i}) \rho_{0 N} +\sigma_{i \overline{N}}(\mathrm{v_i})  \rho_{0 \overline{N}}) \right] (\gamma \mathrm{v})_{\mathrm{i}}
\left( \frac{R}{R+\mathrm{v_{flow}} t} \right)^3,
\end{equation}
If we use the average,
\begin{eqnarray}
\label{approx}
\sum_{\mathrm{v_i}} \sigma (\mathrm{v_i}) \mathrm{v_{i}} \gamma_{\mathrm{i}}  \simeq <\sigma><\gamma_{\mathrm{i}} \mathrm{v_i}>=<\sigma> \frac{p_{\mathrm{i}}}{m_{\mathrm{{i}}}},
\end{eqnarray}
(where p and m are the resonance's momentum and mass)
eq. (\ref{scatter}) becomes,
\begin{equation}
\label{simplescatter}
P_{\mathrm{i}} = \left[
\left< \sigma_{i \pi} \right>  \rho_{0 \pi}+\left<\sigma_{i K}\right> \rho_{0 K}+\left< \sigma_{i N} \right> \rho_{0N} + \left< \sigma_{i \overline{N}} \right>\rho_{0 \overline{N}}) \right]  \frac{p_{\mathrm{i}}}{m_{\mathrm{{i}}}}
\left( \frac{R}{R+\mathrm{v_{flow}} t} \right)^3,
\end{equation}
Neglecting in-medium resonance regeneration and particle escape from the 
fireball, the population equation describing the scattering loss abundance
($N_{\mathrm{i}}$)  is:
\begin{eqnarray}
 \frac{d N_{\mathrm{i}}}{d t}&=& \frac{1}{\tau} N_{N^{*}} -N_{\mathrm{i}} P_{\mathrm{i}} \,, \quad i=1,2 \\ 
 \frac{d N_{N^*}}{d t} &=& -\frac{1}{\tau} N_{N^{*}}   \,    
 \label{model}
\end{eqnarray}
The required nucleon and antinucleon density $\rho_{N}$ is obtained through Eq.~(\ref{nanal})
\begin{equation}\label{relboltz}
\rho_{0N}=\frac{ g}{(2 \pi \hbar c)^3} 4 \pi m^2 (\lambda_q \gamma_q)^3 T K_2 (\frac{m}{T}) \,,
\end{equation}
We consider the nucleons to have a mass of $\simeq$1 ${\rm GeV}$, and a degeneracy of
6, to take the p,n and the thermally suppressed but higher degeneracy
$\Delta$ contributions into account.
the pion density is computed in the massless particle limit, leading to
\begin{equation}\label{pidens}
\rho_{0 \pi}= \sum \frac{3  g}{(2 \pi \hbar c)^3} 4 \pi m_{\pi}^2 \gamma_q)^{2(n+1)} \frac{T}{n} K_2 (\frac{n m}{T})
\end{equation}
The model presented here is remarkably insensitive to the individual 
cross-sectional areas.
The values we used in the calculation are given in table
\ref{parameters},but order-of-magnitude variations of the more uncertain
cross-sections did not produce variations of more than $30 \%$.
Similarly, the value of the initial fireball radius $R_0$ (which is constrained
by the entropy per baryon) does not
significantly affect the final ratios.
This reassures us that had we used a more exact approach than
the approximations in Eq.~(\ref{approx}), the qualitative features of our
model would not have changed.
\begin{figure}[h]
\begin{center}
\epsfig{width=6.5cm,clip=,figure=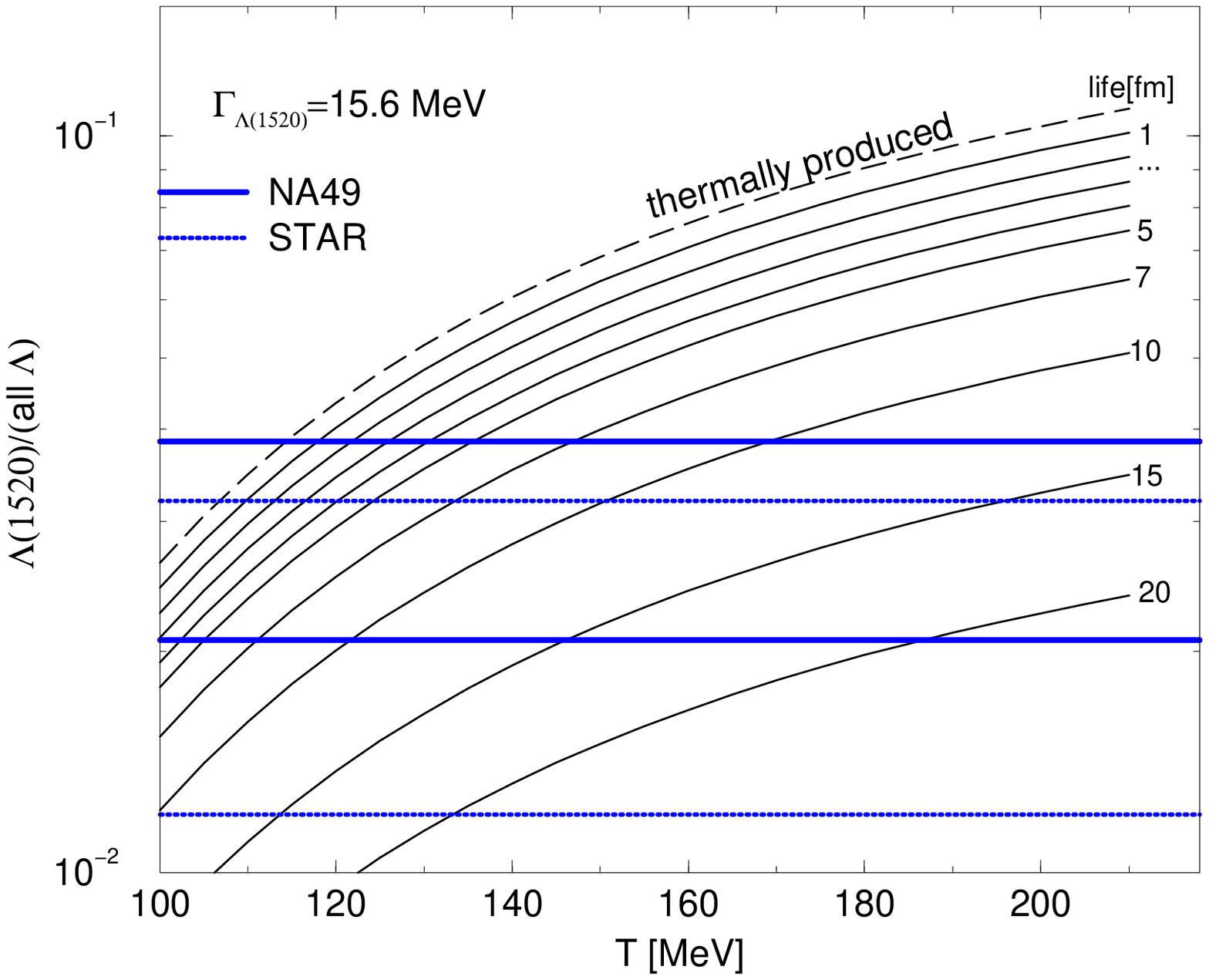}
\epsfig{width=6.5cm,clip=,figure=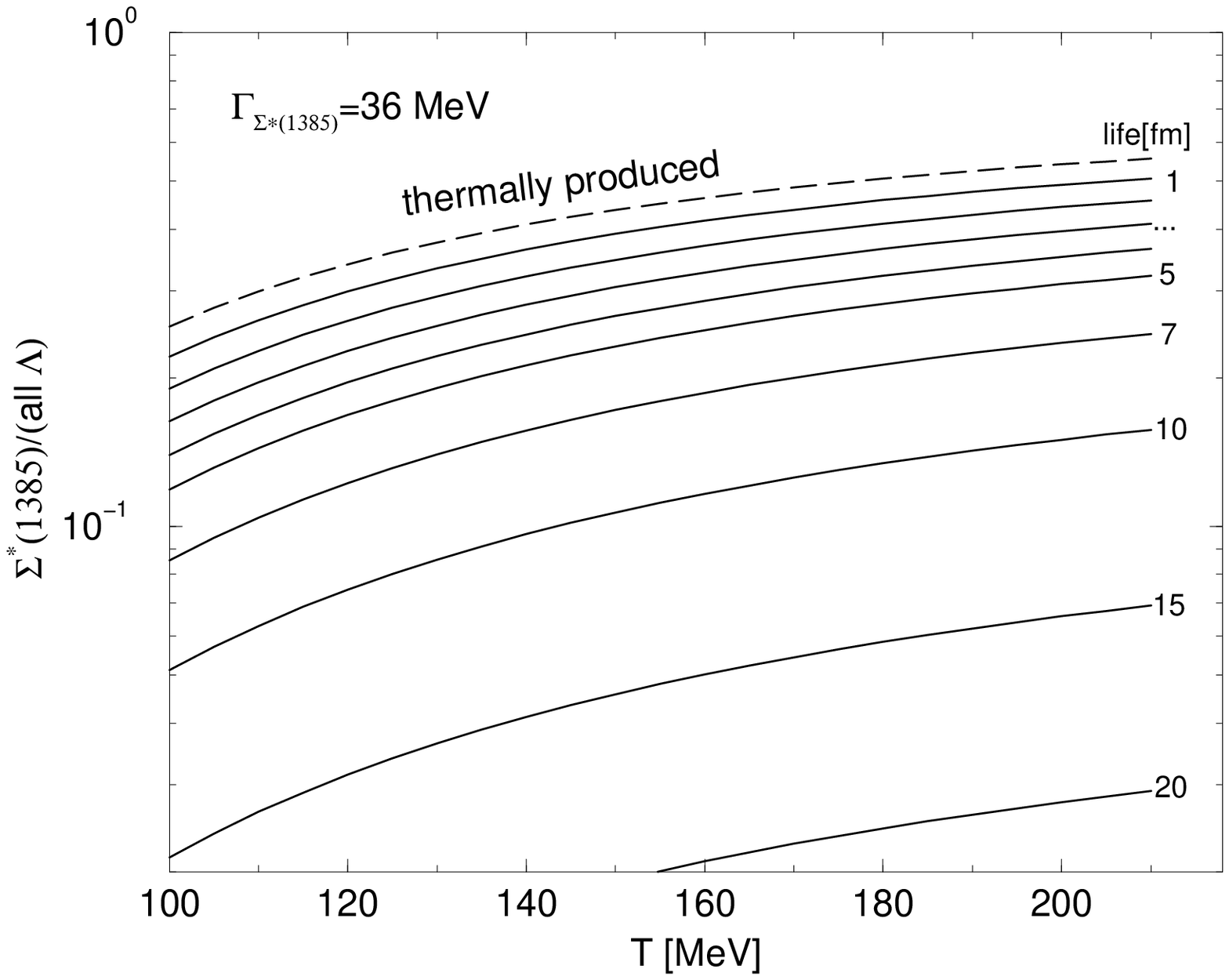}\\
\epsfig{width=6.5cm,clip=,figure=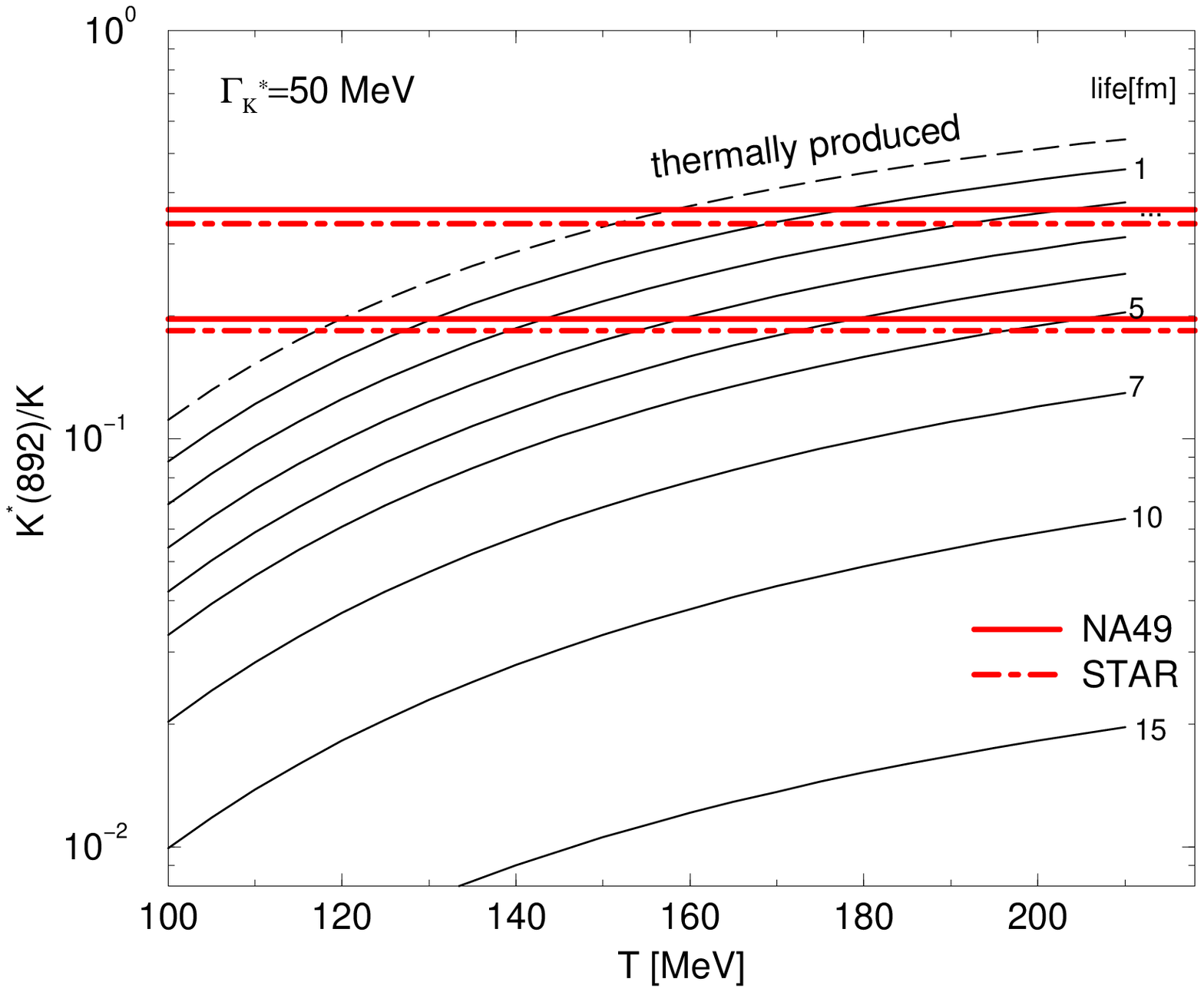}
\epsfig{width=6.5cm,clip=,figure=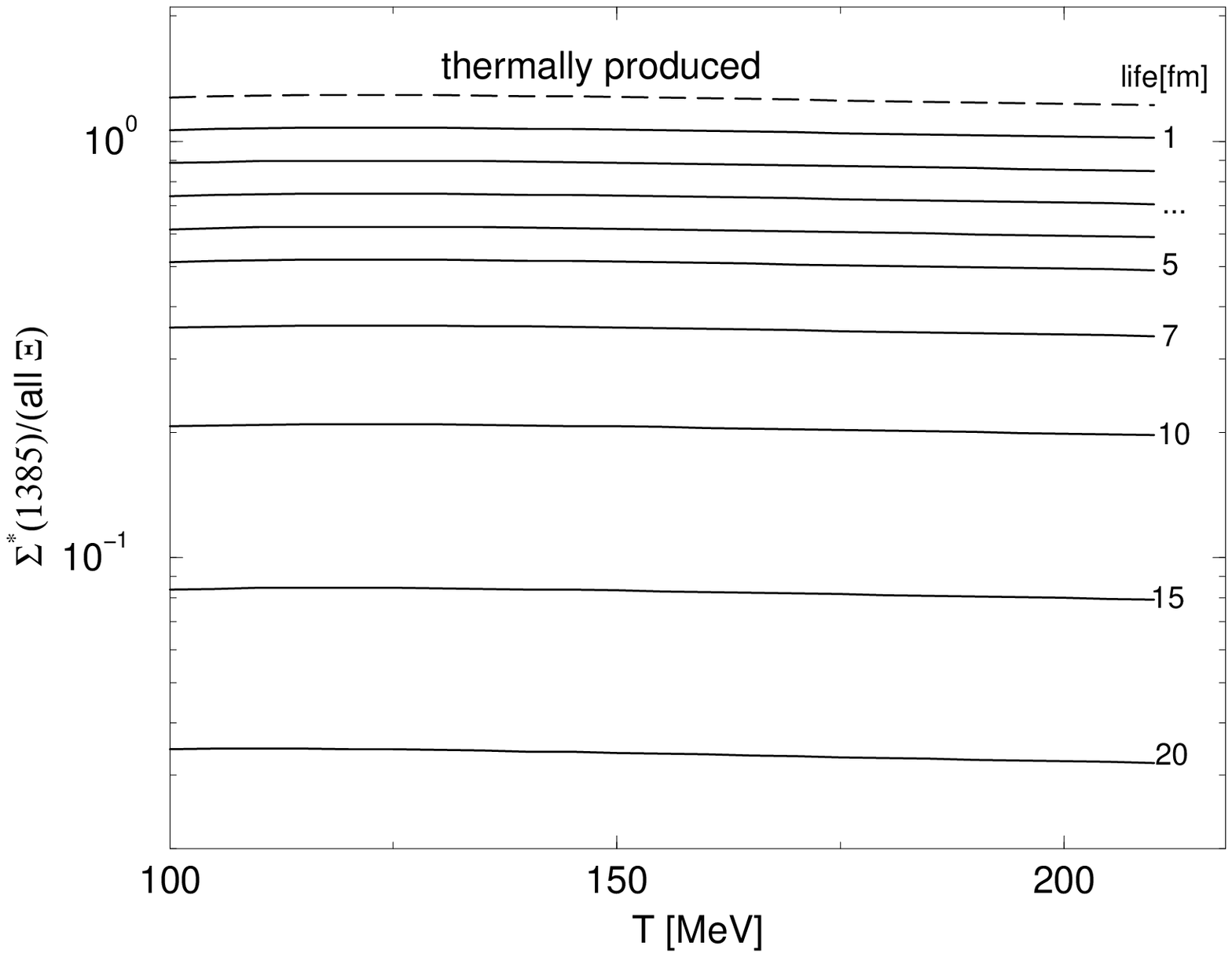}
 \caption{Produced (dashed line) and observable (solid lines) ratios
$\Lambda(1520)$/(total $\Lambda$),$\Sigma^{*}$/(total $\Lambda$),
$K^{*0}/K$ and $\Xi$/(total $\Lambda$).
The solid lines correspond to evolution after chemical freeze-out of
1,2,3,4,5,7,10,15,20 fm/c, respectively.  The values at time zero (chemical
freezeout) were taken from
Fig.~\ref{prodratios}.
See Fig.~\ref{prodratios} caption for the meaning of $K,K^*$.}
\label{obsratios}
\end{center}
\end{figure}
The results, however, exhibit a very strong dependence on both the Temperature 
(which fixes the initial resonance yield as well as the hadron
density of the fireball) and fireball lifetime (in a short-lived fireball
not many resonances decay, so their products do not get a chance to rescatter).
Fig.~\ref{obsratios} shows the dependence of the $\Lambda(1520)/\Lambda$,
$\Sigma^{*}/\Lambda$ and $K^{*\,0}(892)/K$
on the temperature and lifetime of the  interacting phase.
It is clear that, given a determination of the respective signals
to a reasonable precision, a qualitative distinction between the high temperature chemical freeze-out scenario followed by a rescattering phase and the low
temperature sudden hadronization scenario can be made. We also note that
despite the shorter lifetime of the $\Sigma^{*}$ and
higher pion interaction cross section, 
more $\Sigma^{*}$ decay products should be reconstructible
than in the $\Lambda(1520)$ case, at all but the highest temperatures
under consideration.   
This reinforces our proposal that the
$\Sigma^*$ is a very good candidate for further measurement.

\begin{figure}[tb]
\begin{center}
\psfig{width=6.5cm,clip=1,figure=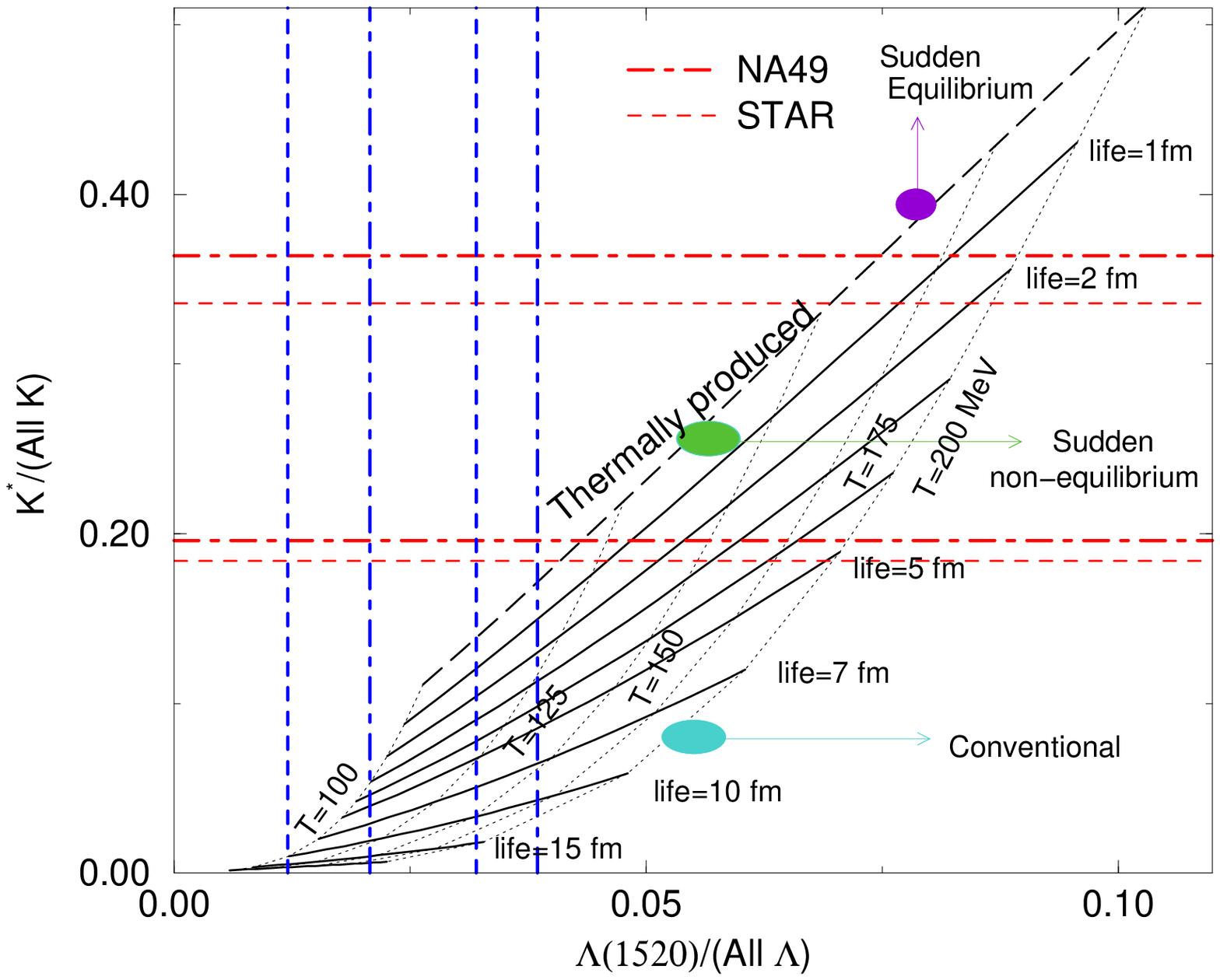}
\psfig{width=6.5cm,clip=1,figure=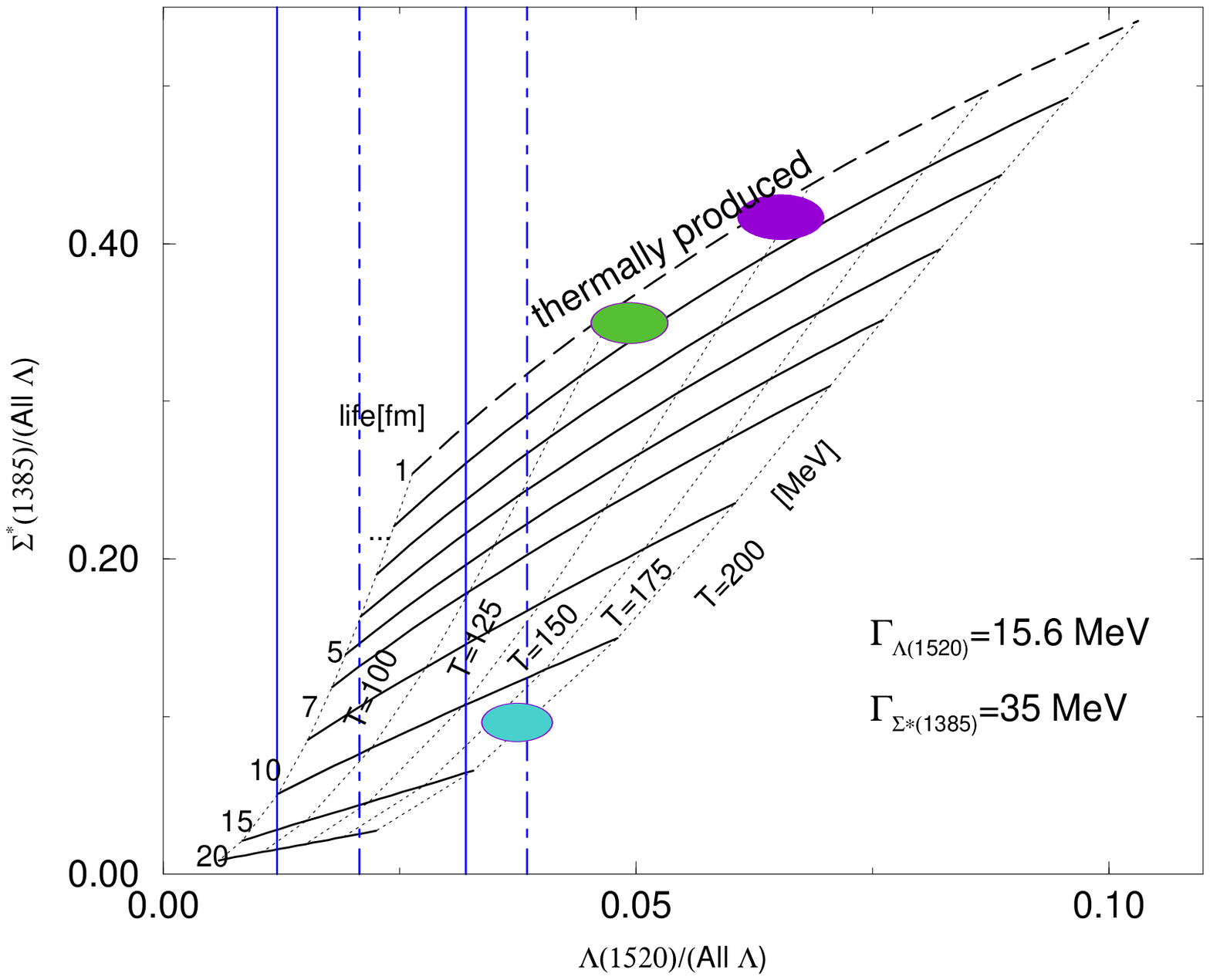 }\\
\psfig{width=6.5cm,clip=1,figure=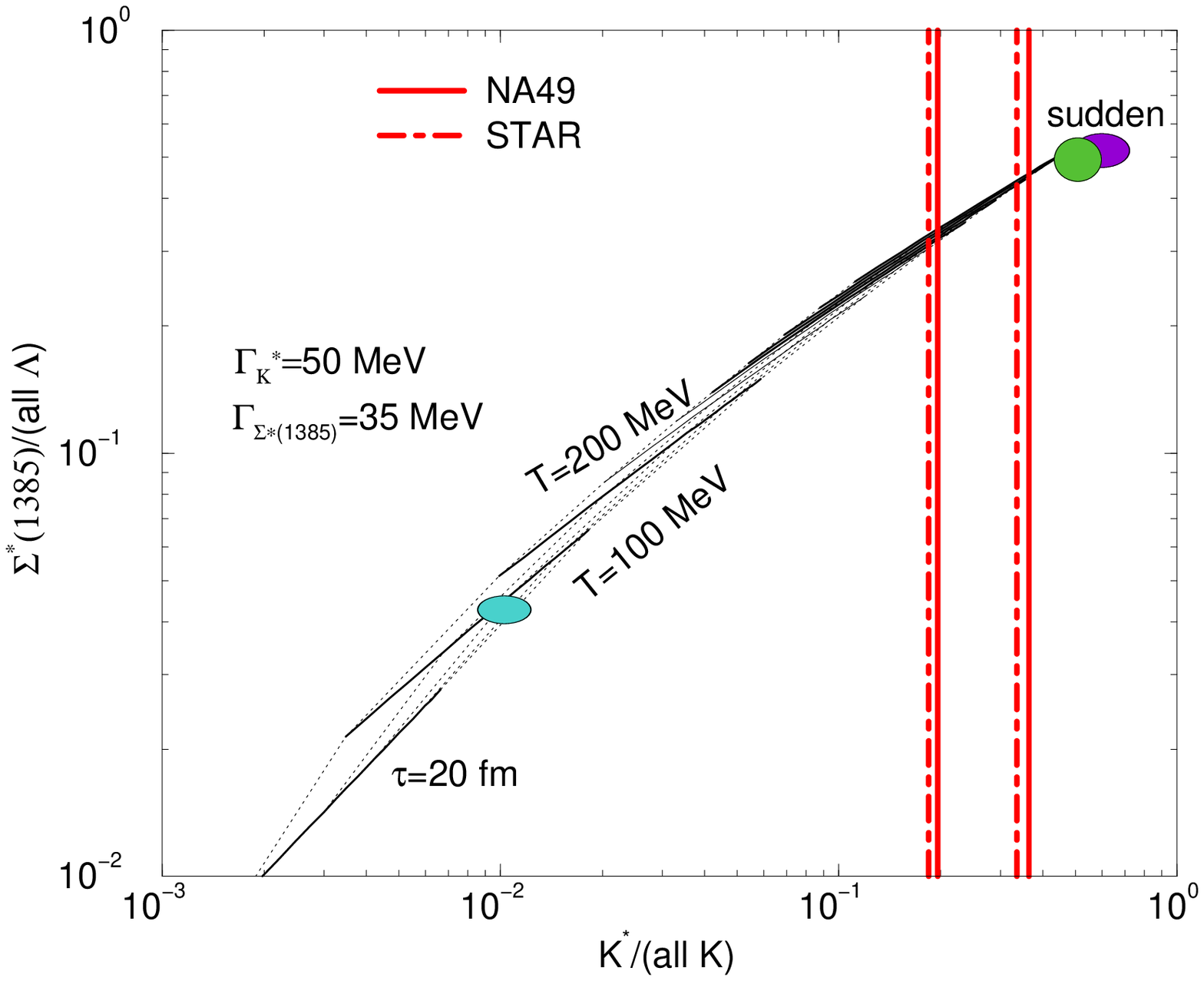 }
\psfig{width=6.5cm,clip=1,figure=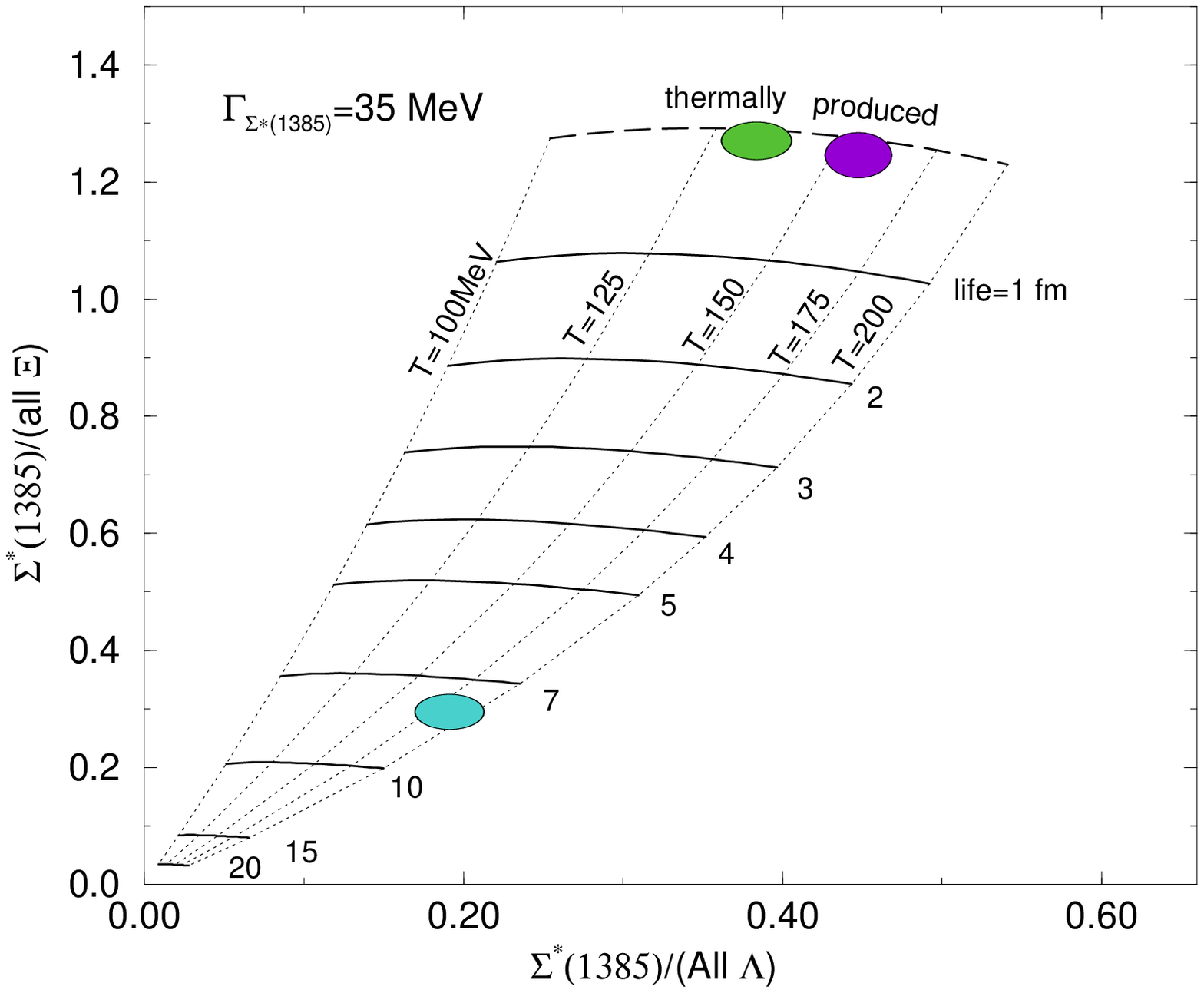 }
\caption{How temperature and fireball lifetime
decouple when two resonances of different masses
and widths are compared.   A point on any of the above diagrams is potentially
sufficient to fix  both of these quantities.
The experimental measurements are discussed in the first section.
See Fig.~\ref{prodratios} caption for the meaning of $K,K^*$.}
\label{projdiag}
\end{center}
\end{figure}

Diagrams such as those in Fig.~\ref{obsratios} still contain an ambiguity
between temperature and lifetime of the interacting hadron gas phase.
A low observed ratio can either mean a low freeze-out temperature or
a lot of rescattering in a long re-interaction phase.
However, this ambiguity can be resolved by looking at
a selection of resonances, with different masses and lifetimes.
Fig.~\ref{projdiag} shows how the initial temperature
and the lifetime of the re-interaction phase decouple when two resonance
ratios are measured simultaneusly.
A data point on diagrams such as those in Fig.~\ref{projdiag} is enough to
measure both the hadronization temperature and to
distinguish between the sudden freeze-out scenario and a long
re-interaction phase.

The plots in Fig.~\ref{projdiag}
can also be used as consistency checks for the model:
For example, the near independence of $\Sigma^{*}/\Xi$ on temperature
means that the equal temperature lines in the $\Sigma^{*}/\Xi$ vs
 $\Sigma^{*}/\Lambda$ diagram are 
nearly insensitive to the details of the rescattering model.
Moreover the mass differences and lifetimes of the $\Sigma^*$, $K^*$ 
combine in such a way as to make the
$\Sigma^{*}$/(all $\Lambda$) vs $K^{*0}(892)/$(all K-) diagram fold
into a very narrow band.
Any serious shortcoming within our rescattering model would be revealed
if the observed particle ratios stray from this band.
\begin{table}
\begin{center}
\centering
\caption{Scattering model parameters}
\begin{tabular}{cccccc} 
\hline
$\sigma_{\pi N}$ (mb) & $\sigma_{K N}$ & $\sigma_{\pi \pi}$ &
$\sigma_{\pi K}$ &  $\sigma_{N N}$ & $\sigma_{\overline{N} N}$  \\ \hline
24 & 20 & $40$ & $20$ & 24 & 50 \\ \hline
\multicolumn{2}{c} {$\Gamma_{\Sigma^{*}}$ } & 
\multicolumn{2}{c} {$\Gamma_{\Lambda(1520)}$}    &
\multicolumn{2}{c} {$\Gamma_{K^{*0}(892)}$}       \\ \hline
\multicolumn{2}{c} {35 ${\rm MeV}$}& 
\multicolumn{2}{c} {15.6 ${\rm MeV}$} & 
\multicolumn{2}{c} {50 ${\rm MeV}$} \\ \hline
\multicolumn{3}{c} {escape rate (fm$^{-3}$)} &  
\multicolumn{3}{c} {negligible}   \\ \hline
\multicolumn{3}{c} {v}          & 
\multicolumn{3}{c} {0.5}  \\ \hline
\multicolumn{3}{c} {R(fm)}      & 
\multicolumn{3}{c} {8\,145/$T$ [${\rm MeV}$]}   \\ \hline
\multicolumn{3}{c} {$\mu_b$} & 
\multicolumn{3}{c} {$220$ ${\rm MeV}$}   \\ \hline
\end{tabular}
\label{parameters}
\end{center}
\end{table}

To understand the low $\Lambda(1520)$ multiplicity measured by NA49 and STAR
it should be realized that the $\Lambda(1520)$ is a very unusual particle
\cite{L1520disc}.
 Unlike most other hadronic resonances 
($\Sigma^*,K^*,\Delta,\rho$ ecc.), its high spin is due not to valence quark spin configuration
but to the fact that the  $\Lambda(1520)$ s valence quarks are (using the
constituent quark picture) in an $L=1^{-}$ state.
The dominant decay (N K), however, has to go through a relative momentum L=2
(d-wave) process, through a channel which is very close to threshold production.
All this conspires to reduce the $\Lambda(1520)$'s width: 
Isospin conservation reduces the number
of channels it can decay (most notably $\Lambda(1520) \rightarrow \Lambda \pi$ is not allowed), 
while the high relative angular momentum (and negative space parity) and threshold suppress the decay phase space for both the dominant ($N K$) and additional ($\Sigma \pi,\Lambda \pi \pi$) $\Lambda(1520)$ decay modes.

\begin{figure}[tb]
\begin{center}
\psfig{width=10cm,clip=1,figure=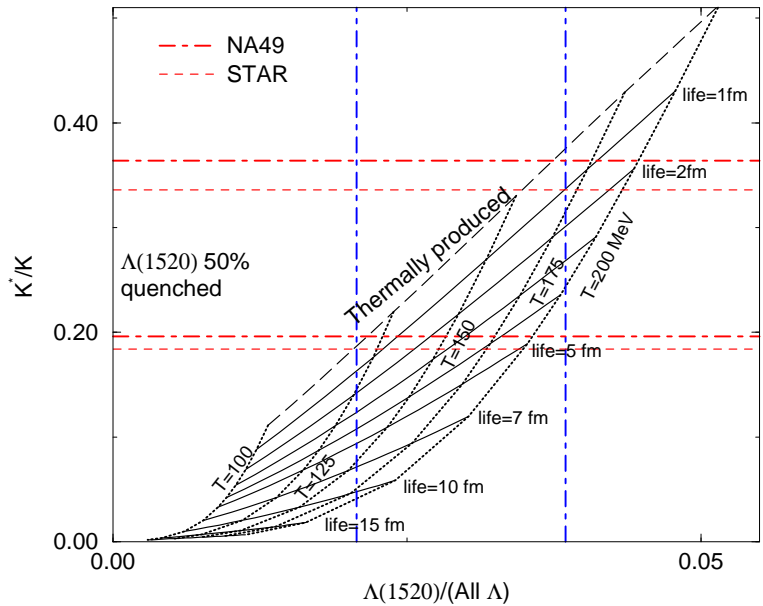 }
\caption{$K^*/K$ vs $\Lambda(1520)/\Lambda$ assuming half of the $\Lambda(1520)$ s are suppressed in-medium.   See Fig.~\ref{prodratios} caption for the meaning of $K,K^*$.  }
\label{quenched}
\end{center}
\end{figure}

Therefore, measurement of the $\Lambda(1520)$ presents both experimental challenges (The high partial wave
component introduces an angular/spin dependence of the decay products which is difficoult to measure in an unpolarized
high multiplicity experiment) and theoretical uncertainities (perhaps new in-medium physics is at work in $L \ne 0$
resonances \cite{l15201,l15202}). 
It should be noticed that, as Fig.~\ref{quenched} shows, a $50 \%$ suppression of the $\Lambda(1520)$
signal at hadronization would mean the data is perfectly compatible
with the sudden freeze-out model described in \cite{sudden2}.  This strongly suggests that other resonances should be used to constrain
the fireball freeze-out properties.

The non-suppression of the $K^*$ makes it
likely that other hadron resonances can also be detectable.
The diagrams shown in the previous section make it clear 
that the $\Sigma^*$ should be abundantly produced, and it has many
characteristics which would make it a logical next step in the study
of resonances produced in heavy ion collisions.
$\Xi^*/\Xi$ will also be a candidate for this kind of study.

Focusing on these resonances will also make regeneration less likely.
In particular, regeneration is less likely to affet 
the $\Sigma^*$ detectability:
$95 \%$ of $\Sigma^*$ decay through the $p$-wave $\Sigma^* \rightarrow \Lambda \pi$ channel.    However, regenerating $\Sigma^*$ s in a gas of $\Lambda$ s and
$\pi$ s is considerably more difficoult, since $\Lambda \pi$ scattering
will be dominated by the s-wave
$\Lambda \pi \rightarrow \Sigma^{\pm}$.   
For these reasons, a measurement of the $\Sigma^*/\Lambda$ and $\Xi^*/\Xi$ (which is also unlikely
to regenerate due to its small width) should be able to distinguish a sudden hadronization model
from a slow sequential freeze-out with certainity.

\section{Production of pentaquarks}

The discovery   by the NA49 collaboration~\cite{Alt03}of a new $\Xi^{--}(1862)$ $I=3/2$ narrow $\Gamma<5$ ${\rm MeV}$ resonance 
in their $pp$ background, ($\sqrt{s_{\rm NN}}=17.2$ ${\rm GeV}$) 
demonstrates that heavy ion detectors are capable of discovering states missed at
previous energy scans.     Statistical hadronization and strangeness enhancement mean that these states should be
produced abundantly in heavy ion collisions.   The enhanced production rate, together with experimental
capabilities, makes the Heavy ion experiments promising probes for the discovery of new hadronic states.

This newly discovered  
hadron resonance has, given the mass and charge, an  exceedingly narrow width.
This  feature is common with  $\Theta^+(1540)$, another recently reported 
resonance~\cite{Nak03,Bar03,Step03}, which
decays into the channel with quark content $uudd\bar s$ and  $I=0$. This is
believed to be  the predicted~\cite{Dia97}, lowest mass, pentaquark state.
The  $\Xi^*(1862)$ can be interpreted as its  most massive  isospin quartet member
$ssdd\bar u,\, ssud\bar u,\, ssud\bar d,\, ssuu\bar d$ 
with electrical  charge varying, respectively, from $-2$ to +1, in units of $|e|$.

Appearance of these new resonances can have many consequences in the field
of heavy ion collisions. 
We at first explore  how the introduction into the family of hadronic particles
of these two  new resonances,   $\Theta^+(1540)$ and $\Xi^*(1862)$,
influence the results of statistical hadronization fit to relativistic  heavy ion
hadron production experimental results. 
We use the same data set as has been employed in\ Ref~\cite{Zak03}
and obtain predictions of how the relative abundances of these
new resonant states vary as function of the heavy ion collision energy. 

Importantly, only the two already identified states with $I=0$, and $I= 3/2$ 
of the anti decuplet, which also includes the  $I=1/2$, and $I= 1$ states 
are of relevance in the study
of the statistical hadronization fits. Thus, in our analysis, we do not depend 
on the unknown masses of  $I=1/2$, and $I=1$ states. However, 
 the interpretation of the newly discovered narrow  states as pentaquarks
enters our considerations decisively.
In our approach~\cite{Zak03},
 as in other recent work~\cite{becattini},  the chemical equilibrium 
and non-equilibrium is considered.
Accordingly, we allow  quark pair phase space 
occupancies, for  light quarks $\gamma_q\ne 1$, and/or 
strange quarks $\gamma_s\ne 1$.

The pentaquark (if this is indeed a pentaquark) valance quark content 
enters  the assigned chemical fugacities and 
phase space occupancies in a different way than either mesons or baryons.
Hence, predictions for the pentaquark yield are likely to be very different
whether equilibrium is assumed or not.    In particular, if the phase
space is saturated above equilibrium, this will yield to a large ($\sim \gamma^3$) enhancement
of pentaquarks with respect to hadrons of a similar mass.
And nature has been generous enough to provide us with detectable hadrons of a comparable
mass and width: The $\Lambda(1520)$ and $\Xi(1530)$, with mass and width nearly identical
to $\Theta^+$.

Since we 
study at SPS the  total particle multiplicities, and at RHIC 
the  central yields which can be considered  produced by 
rapidity-localized fireballs of matter, we  require in
our fits  balance in the
 strange and antistrange quark content.
 
There are two independent fit parameters when we assume complete chemical
equilibrium, the 
chemical freeze-out temperature $T$ and the light quark fugacity 
$\lambda_q=\sqrt{\lambda_u\lambda_d}=e^{\mu_b/(3T)}$. The baryochemical
 potential $\mu_b$ is the physical parameter controlling baryon density. 
Strangeness conservation fixes the strange quark fugacity $\lambda_s$ 
(equivalently, strangeness chemical potential, for more details 
 see, e.g., \cite{Zak03}).
Adding the possibility that the number of strange quark pairs is not
in chemical equilibrium, $\gamma_s\ne 1$, we  have  3 parameters, and
allowing also that light quark pair number is not in chemical 
equilibrium, we have 4 parameters. These three alternatives 
will be coded as open
triangles, open squares and filled squares, respectively,  in
all  results we present graphically.

 We find that  the  new resonance  $\Theta^+(1540)$  influences significantly the 
statistical hadronization fit to particle production at the lowest SPS 
energies. In a  baryon rich environment 
the introduction into the fit of  $\Theta^+(1540)$,  a 
$b=1$ baryon with `wrong' strangeness influences the strangeness 
balance condition, and thus indirectly the individual yields of all
strange hadrons.   
This leads to a   reduction in the statistical fit error 
for our hadronization study of the 40$A$ ${\rm GeV}$ Pb--Pb  reactions 
where we see a significant change in the relative yield of
kaons and $\Lambda$. We also find  changes in the details of the 
statistical fit parameters. In comparison to~\cite{Raf03ma}, 
aside  of the introduction 
of the new resonances, we also have harmonized our hadron
decay table with those used by the Krak\'ow group~\cite{florkowski}.
The improvement of the particle yield fit is both,  a   
theoretical confirmation of the validity of the statistical hadronization
model of particle production, and its applicability at low SPS energies.
\begin{figure}[t]
\vskip 0.3cm
\hspace*{-.2cm}\centerline{\psfig{width=8.5cm,figure=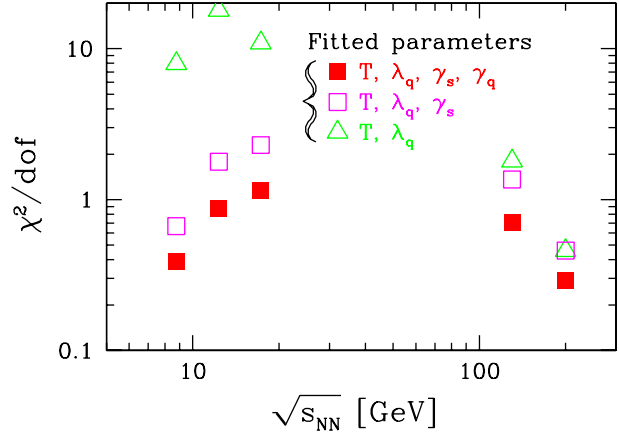}}
\caption{\label{PLError} (Color online)
$\chi^2$/dof for statistical hadronization fits at SPS and RHIC:
results are shown for 40, 80, 158$A$  ${\rm GeV}$ Pb on stationary Pb
target  collisions  and at
RHIC for 65+65 and 100+100$A$ ${\rm GeV}$ Au--Au head on interactions.
}
\end{figure}
We show how the fit error evolves in Fig.~\ref{PLError},
which is also presented in the bottom lines of 
tables ~\ref{table_raf1sps} and~\ref{table_raf1} along with the number of 
data points and resulting degrees of freedom. Considering  the small
number of degrees of freedom at SPS, we need
 $\chi^2/$dof $<1$ to have good significance of the fit\footnote{See appendix B}.
The  errors seen in Fig.~\ref{PLError} are, for the chemical nonequilibrium case (filled squares), 
sufficiently  small to allow us to conclude that  
the introduction of  $\Theta^+(1540)$  assures that the  statistical
hadronization works well down to the lowest SPS energies. 
To compare with  earlier results on $\chi^2/$dof, obtained prior 
to the discovery of these new resonances,  see Ref.~\cite{Raf03ma}, figure 16.

 An interesting point, seen in Fig.~\ref{PLError}, is that  
the chemical equilibrium fit $\gamma_s=1,\,\gamma_q= 1$ is rendered 
unacceptable  at all SPS energies in presence of 
the new resonances. The semi-equilibrium fit, which 
allows a varying strangeness saturation, but assumes light quark 
equilibrium is generally resulting in twice as large  $\chi^2$
compared to the full non-equilibrium approach.  In a study of 
$\chi^2$ profile as function of $\gamma_q$ we find a clear 
and strong minimum for $\gamma_q\to \gamma_q^{\rm max}\equiv e^{m_\pi/(2T)}$.
Acquisition by the fit of this limiting value implies that there is 
no fitting error in the $\gamma_q$ presented below.

\begin{figure}[!bt]
\vskip -0.5cm
\centerline{
\hspace*{-.2cm}\psfig{width=7.cm,figure=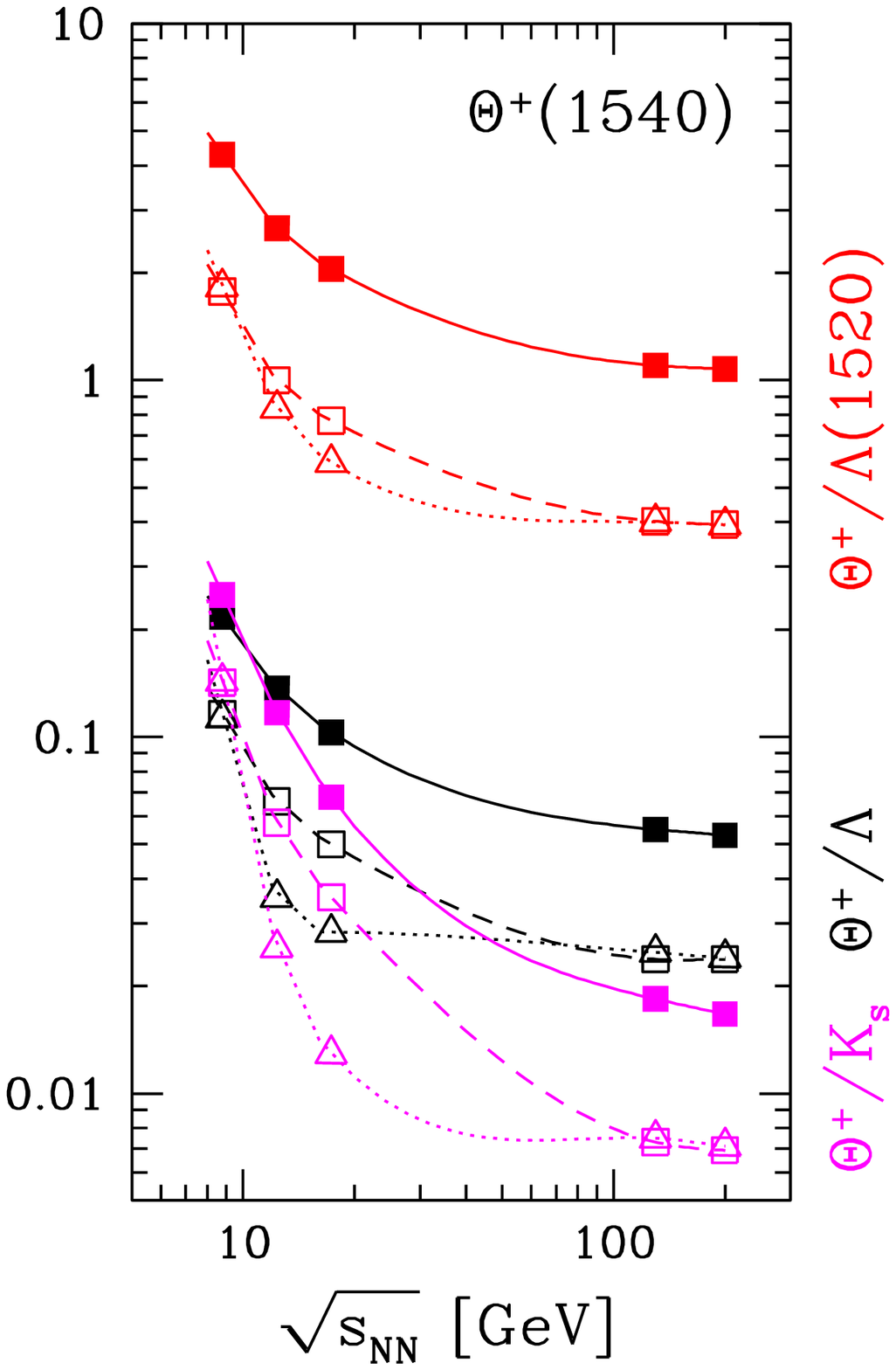}
\hspace*{-.2cm}\psfig{width=7.cm,figure=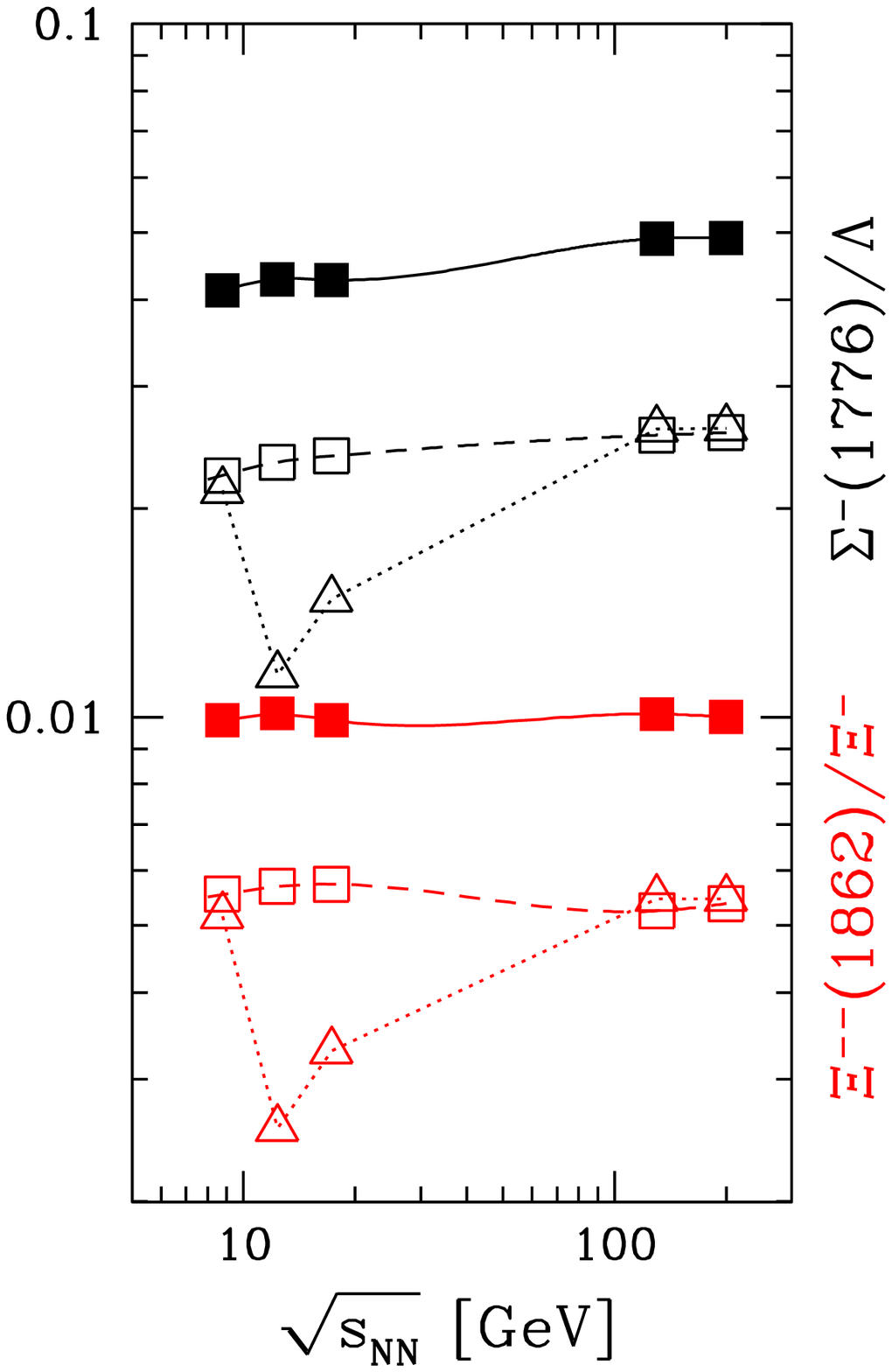}}

\vspace*{-.2cm}
\caption{\label{PLQ1RATIO} (Color online)
Yield of $\Theta^+(1540)$ in relativistic heavy ion collisions, based on
statistical hadronization fit to hadronization parameters 
at SPS and RHIC  40, 80, 158$A$ ${\rm GeV}$ Pb on stationary Pb
target  collisions  and at 
RHIC for 65+65 and 100+100$A$ ${\rm GeV}$ Au--Au head on interactions. 
Relative yields with $K_s,\,\Lambda,$ and $\Lambda(1520)$ are shown
from bottom to top. 
}
\vskip -0.3cm
\end{figure}

The chemical freeze-out parameters of the fits considered play a
very important role in predicting the (relative) yield of hadronic particles,
and this dependence is even stronger for many pentaquark states, due to their
unusual quantum numbers.
These fit parameters for RHIC are shown in table~\ref{table_raf1}, and for 
SPS in table~\ref{table_raf1sps} along with 
the freeze-out temperature. We note that for the full chemical
non-equilibrium, the freeze-out temperature is found to be
smaller than for semi-equilibrium case. This reduction is over-compensated
in pentaquark  yields by the significantly increased  value of $\gamma_q$.

We now consider the relative
yields of the new resonances in figures~\ref{PLQ1RATIO} (left) and~\ref{PLQ1RATIO} (right). 
These yields vary strongly with collision energy 
for the case of  $\Theta^+(1540)$  in Fig.~\ref{PLQ1RATIO} (left), but 
 are rather constant in Fig.~\ref{PLQ1RATIO} (right). Certainly our result 
differs greatly from expectations arising from an earlier study of the 
statistical model production of the $\Theta^+(1540)$ resonance~\cite{Ran03fq}
where the  decisive variation of the particle yield with chemical potentials
was not explored. Moreover, the hadron yields, presented in \cite{Ran03fq},  did not
include the contributions from decay of short lived hadron resonances.
We checked that the relative particle yields shown  in~\cite{Ran03fq}
for zero chemical potentials and varying temperature 
 are mathematically correct, also as a cross check of our program.

In Fig.~\ref{PLQ1RATIO} (left), we show  (from top to bottom) the relative yields
 $\Theta^+(1540)/\Lambda(1520),\,\Theta^+(1540)/\Lambda,\,\Theta^+(1540)/K_s $
for chemical nonequilibrium (solid lines), semi-equilibrium
($\gamma_q=1$, dashed lines) and chemical equilibrium (dotted lines). 
The yields of $\Lambda$ used here include 50\% weak interaction cascade from $\Xi$.

The reason that the chemical nonequilibrium is  leading to 
greater than equilibrium yields is that the lower
hadronization temperature is overcompensated by the chemical
factors, e.g., 
$\Theta^+(1540)/\Lambda(1520) \simeq 1/2\, \gamma_q^2(\lambda_q/\lambda_s)^2$
ignoring the small mass difference. The factor $1/2$  is due to the difference in 
spin degeneracy. The actually observed  yield ratio $\Theta^+(1540)/\Lambda(1520)$
is probably going to be still greater:
As explained in the previous section,  $\Lambda(1520)$ is seen at 50\% of the expected
statistical hadronization yield in heavy ion collisions.
As the $\Theta^+(1540)$ is an L=0 resonance, we predict that its yield should
more closely match the statistical model calculation.

In Fig.~\ref{PLQ1RATIO} (left), we also recognize that
the reason that there is such a significant impact
at low SPS energies of   $\Theta^+(1540)$ is that it
 is produced at the level
of +10--20\% of $\Lambda$ in  fits at 40$A$ ${\rm GeV}$. 
This is due to the large prevailing baryochemical density. Clearly, this 
is the environment in which one would want to study the properties 
of this new resonance in more detail. However, at all energies considered, we
find that at chemical non-equilibrium the $\Theta^+(1540)$ is more abundant compared to $\Lambda(1520)$
and thus this new resonance could become an important probe of the
 hadronization
dynamics, provided detection issues, such as different branching ratios
 and experimental acceptances for the analyzed decays,
($\Lambda(1520) \rightarrow p K^+$ and $\Theta^+ \rightarrow p K_S$) 
are resolved.

The observation of the  pattern of relative yield of 
$\Theta^+(1540)$,  seen in Fig.~\ref{PLQ1RATIO} (left),
would firmly confirm the 4-quark, one anti quark content of this state. Namely,
were for example the $\Theta^+(1540)$ another tri-quark baryon state, the
yield ratio with (strange) baryons would be quite flat as function
of collision energy. We further note that  the absolute
magnitude of the relative yield, seen in  Fig.~\ref{PLQ1RATIO} (left),
will be of help in establishing the degree of chemical equilibration.

In Fig.~\ref{PLQ1RATIO} (right), we show at the bottom the expected relative 
yield of the $\Xi^{--}(1862)[ssdd\bar u]$ relative to $\Xi^-[ssd]$.
The 
$\Xi^*(1862)$ adds at the percentile level to the yield of observed $\Xi$
 and thus it  is less
influential in the statistical hadronization approach. 
The absence of variation of the relative yield with collision energy
is due to cancellation of chemical factors.
This relatively small relative yield 
at all collision energies here considered shows that indeed the $pp$ 
environment, where it has been identified by the NA49 collaboration, is 
most suitable. The dotted lines, in Fig.~\ref{PLQ1RATIO} (right),
are visibly breaking the trend in some of the results, indicating that
the large $\chi^2$ chemical equilibrium 
fit generates  unreliable statistical model parameters.

We also show, in Fig.~\ref{PLQ1RATIO} (right) on the top, 
the yield of the pentaquark state $\Sigma(1776)[sddu\bar u]$
which for purpose of this study is assumed at the mass indicated. 
Again due to cancellation of key chemical factors in  ratios shown 
in Fig.~\ref{PLQ1RATIO} (right), 
both being proportional to $\gamma_q^2 \lambda_d/\lambda_u$, the
ratio is flat (except for the failed fit chemical equilibrium results).
Considering that  $\lambda_d\simeq \lambda_u$ and 
$\gamma_q\simeq 1.6$, the magnitude of relative yields 
seen in Fig.~\ref{PLQ1RATIO} (right)  is primarily due to the hadron
mass, and degeneracy. 

We have shown that inclusion of the  pentaquark states in the study of
particle production in heavy ion collisions improves the quality of our
fits to experimental data. We find that  
$\Theta^+(1540)$ state influences the low energy SPS particle yield fit results.
It can be expected that it will  be detectable, in particular 
 at  low heavy 
ion collision energies, and thus should become a new probe of 
hadronization dynamics. The other pentaquark states will be hard to 
observe in heavy ion collisions. 

\section{Momentum dependence of the resonance-particle ratios as a freeze-out probe}
\label{res}
In section 5.4, we have shown that the measurement of resonances
can probe both the hadronization temperature, and
the lifetime of the interacting hadron gas phase
\cite{torrieri_reso1,torrieri_reso2}.
Ratios of a generic resonance (henceforward called $N^*$) to the 
light particle (which we will refer to as $N$) with an identical
number of valence quarks are particularly sensitive  probes of freezeout temperature
because chemical dependence cancels out within the ratio. 
If we examine this ratio within a given  $m_T>m_{N^*}$ range,
 we expect to disentangle flow and freeze-out conditions,
since the ratio $N^*/N$ should not depend on $m_T$ for a purely static and 
 thermal source (if there is no flow, $N^{*} \propto e^{m_T \cosh(y)}$ so
$(N^*/N)_{m_T>m^*}$ is constant.

We therefore take the most general Boost-invariant freeze-out hypersurface
in the Boltzmann limit (see Table~\ref{hypsurf}, Eq.~(\ref{after_int}))
\begin{equation}
\label{blastgen}
\frac{dN}{dm_{T}^2} \propto S(m_T,p_T)=\int_{\Sigma} r dr S (m_T,p_T,r),
\end{equation}
where
\begin{equation}
\label{sform}
S(m_T,p_T,r) = 
m_T K_1 (\beta m_T) I_{0} (\alpha p_T)
- \frac{\partial t_f}{\partial r} p_T K_0 (\beta m_T) I_1
 (\alpha p_T),
\end{equation}
with
\begin{eqnarray}
\label{alphabeta}
\beta  = \frac{\cosh[y_T(r)]}{T},\quad 
\alpha = \frac{\sinh[y_T(r)]}{T}
\end{eqnarray}
and use it to calculate the ratio between two particles with the same
chemical composition.
The chemical factors cancel out, and we are left with
\begin{equation}
\label{blastratio}
\frac{N^*}{N} = \left(\frac{g^*}{g}\right) 
\frac{S(m_T,p_T^*)}{S(m_T,p_T)},
\end{equation}
where $g^*$ and $g$ refer to each particle's degeneracy and
the function $S(m_T,p_T)$ is given by Eq.~(\ref{sform}).
(Note that $m_T$ is the same for $N^*$ and $N$, but $p_T$ varies).

\begin{figure}
\centerline{\resizebox*{!}{0.32\textheight}{
\includegraphics{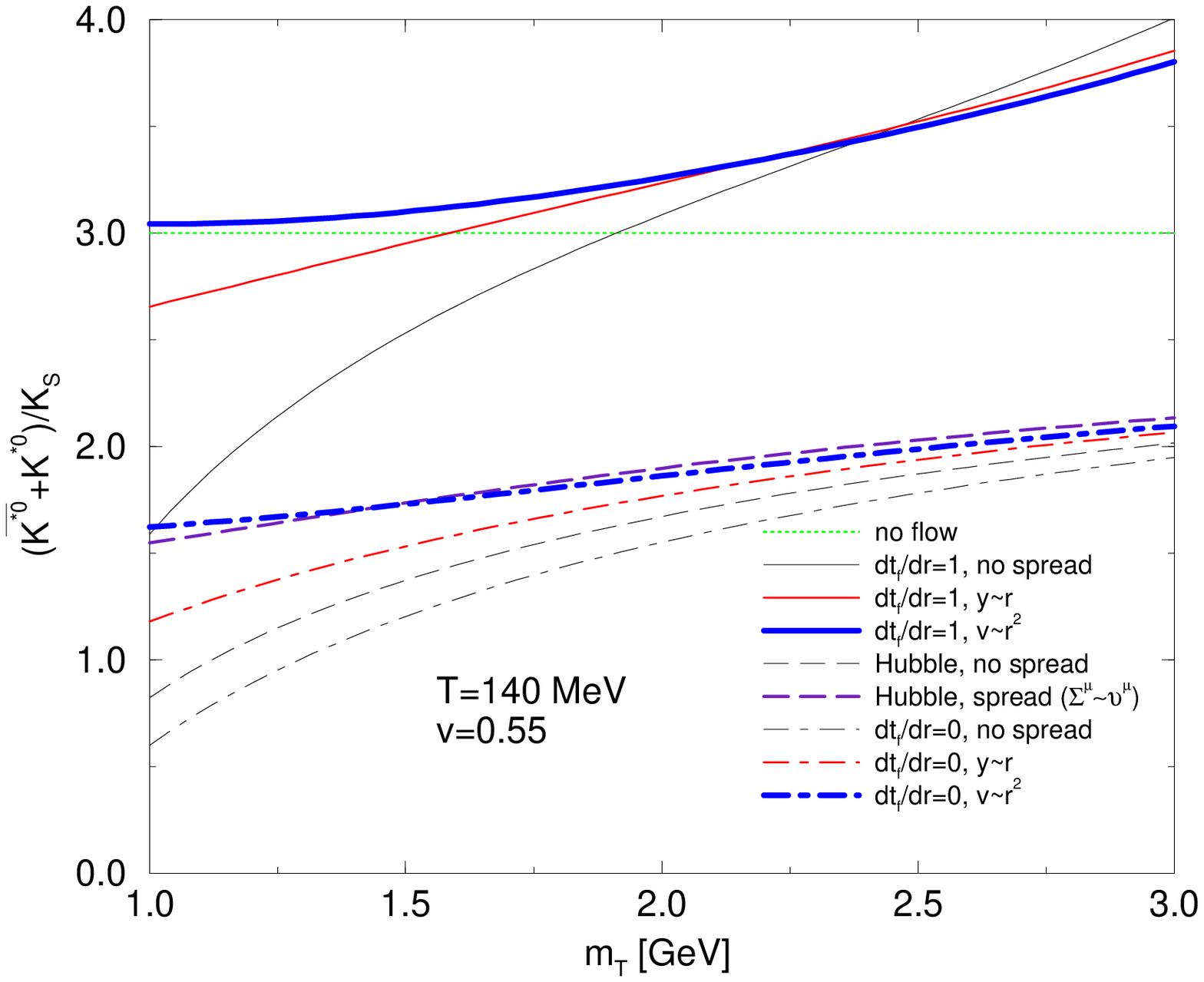}
}
\resizebox*{!}{0.32\textheight}{
\includegraphics{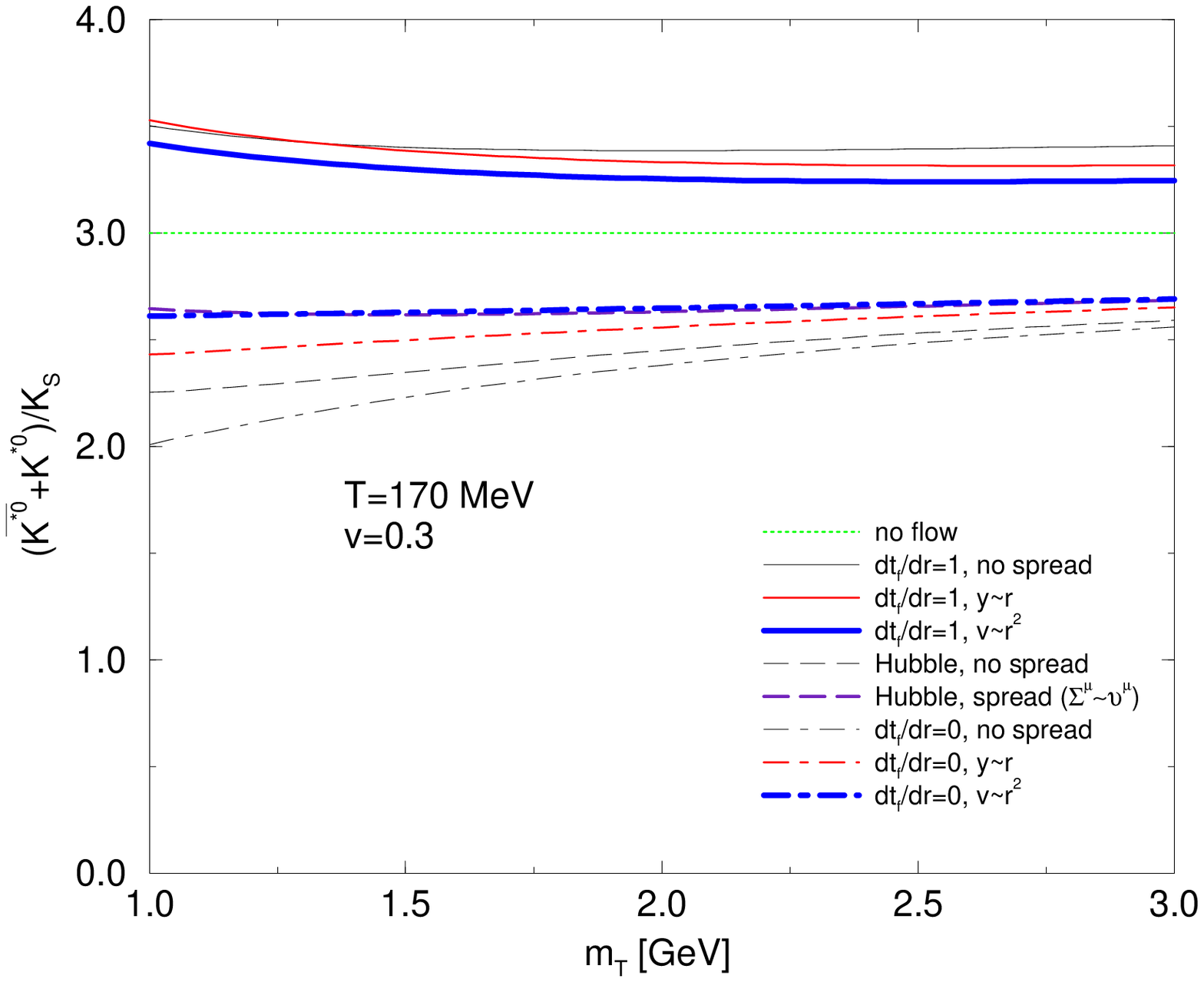}
}}
\centerline{\resizebox*{!}{0.32\textheight}{
\includegraphics{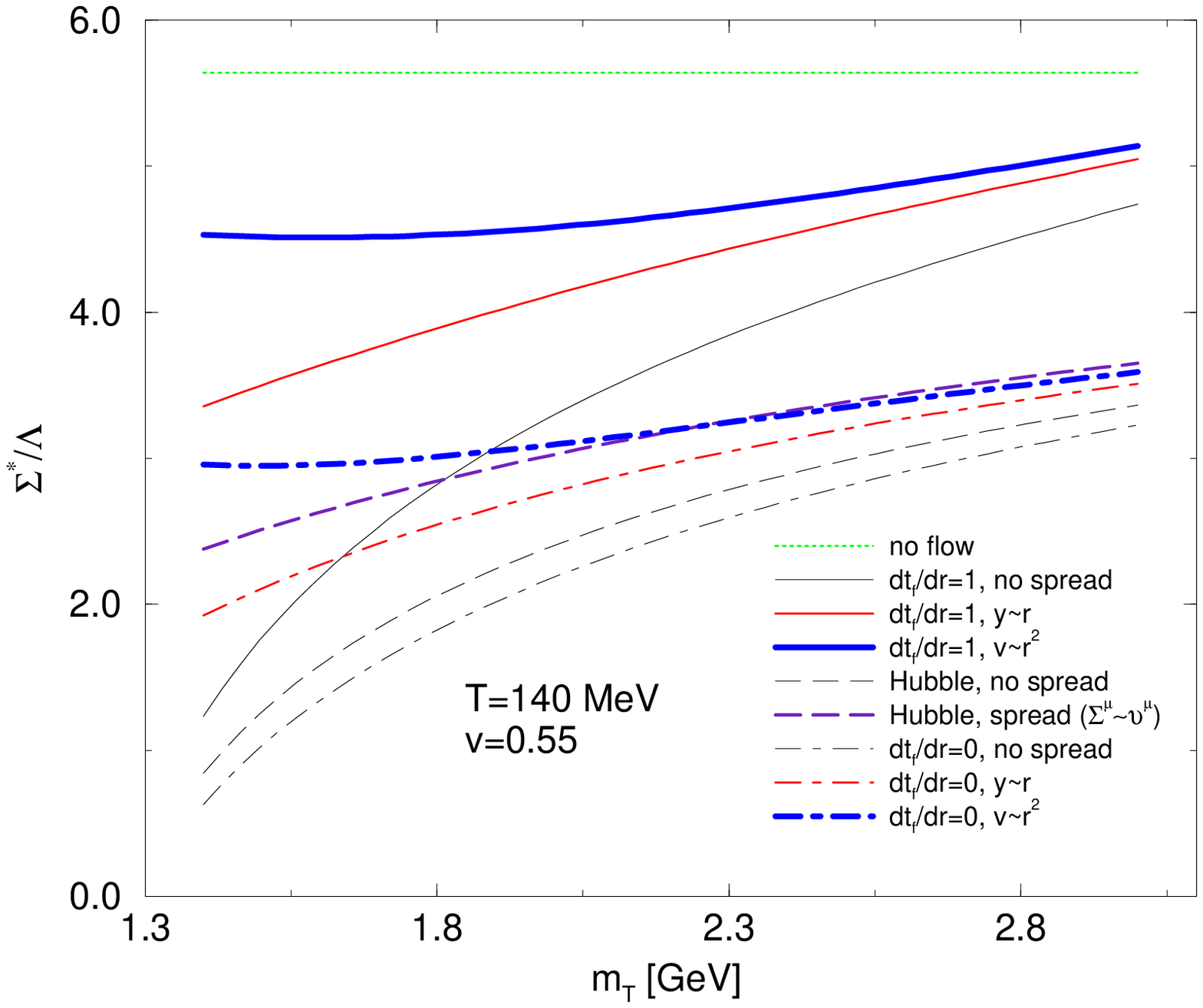}
}
\resizebox*{!}{0.32\textheight}{
\includegraphics{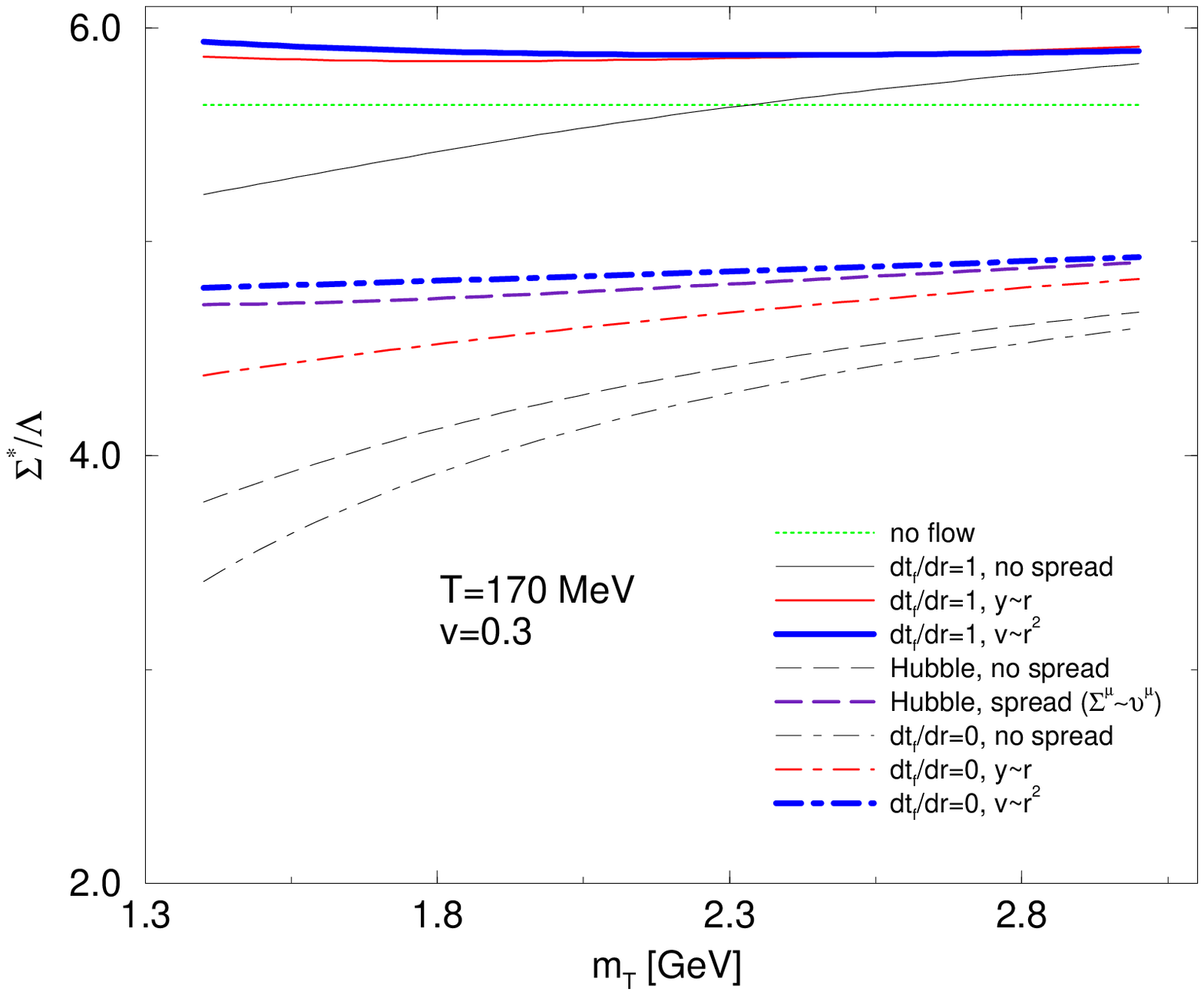}
}}
\centerline{\resizebox*{!}{0.32\textheight}{
\includegraphics{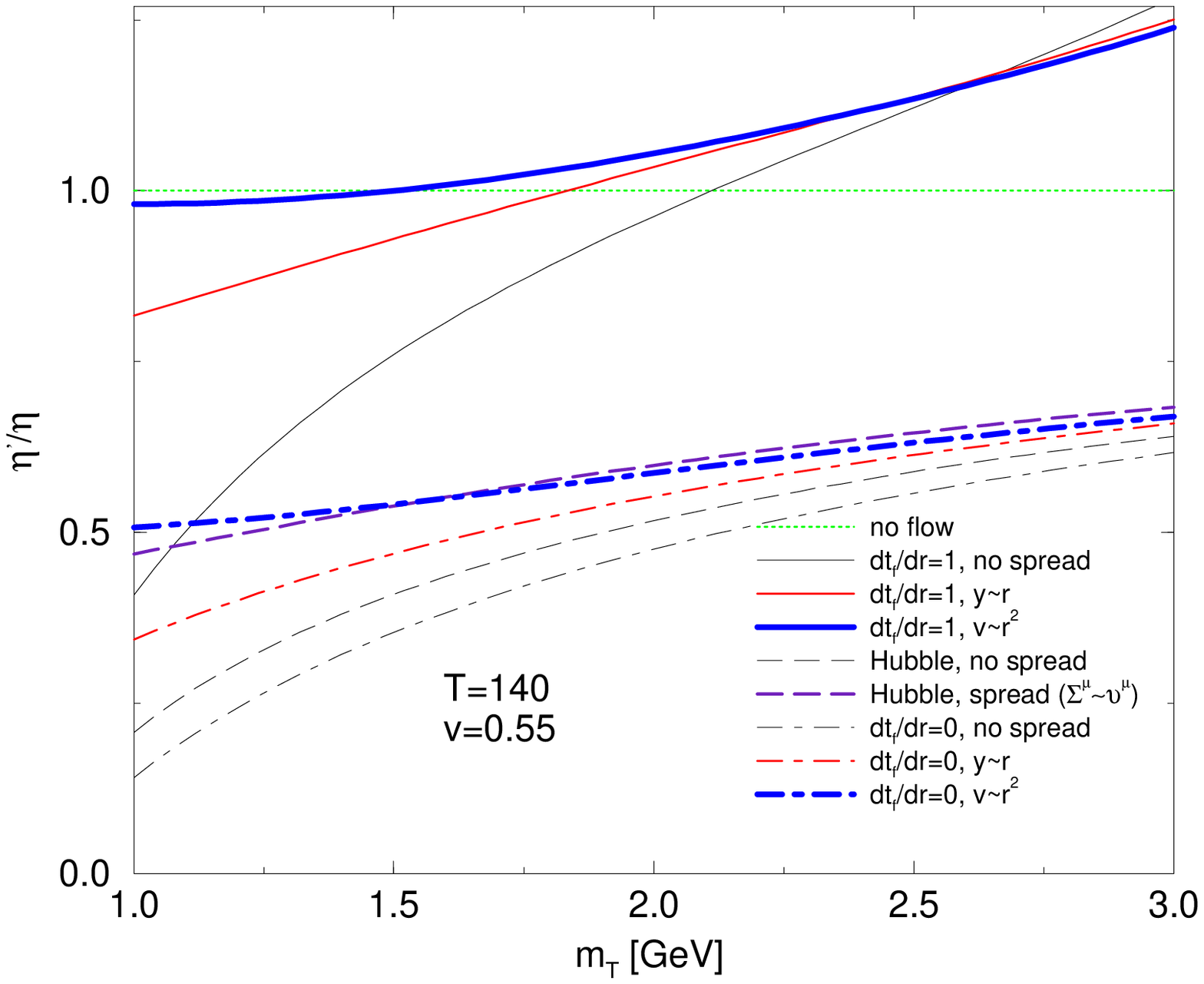}
}
\resizebox*{!}{0.32\textheight}{
\includegraphics{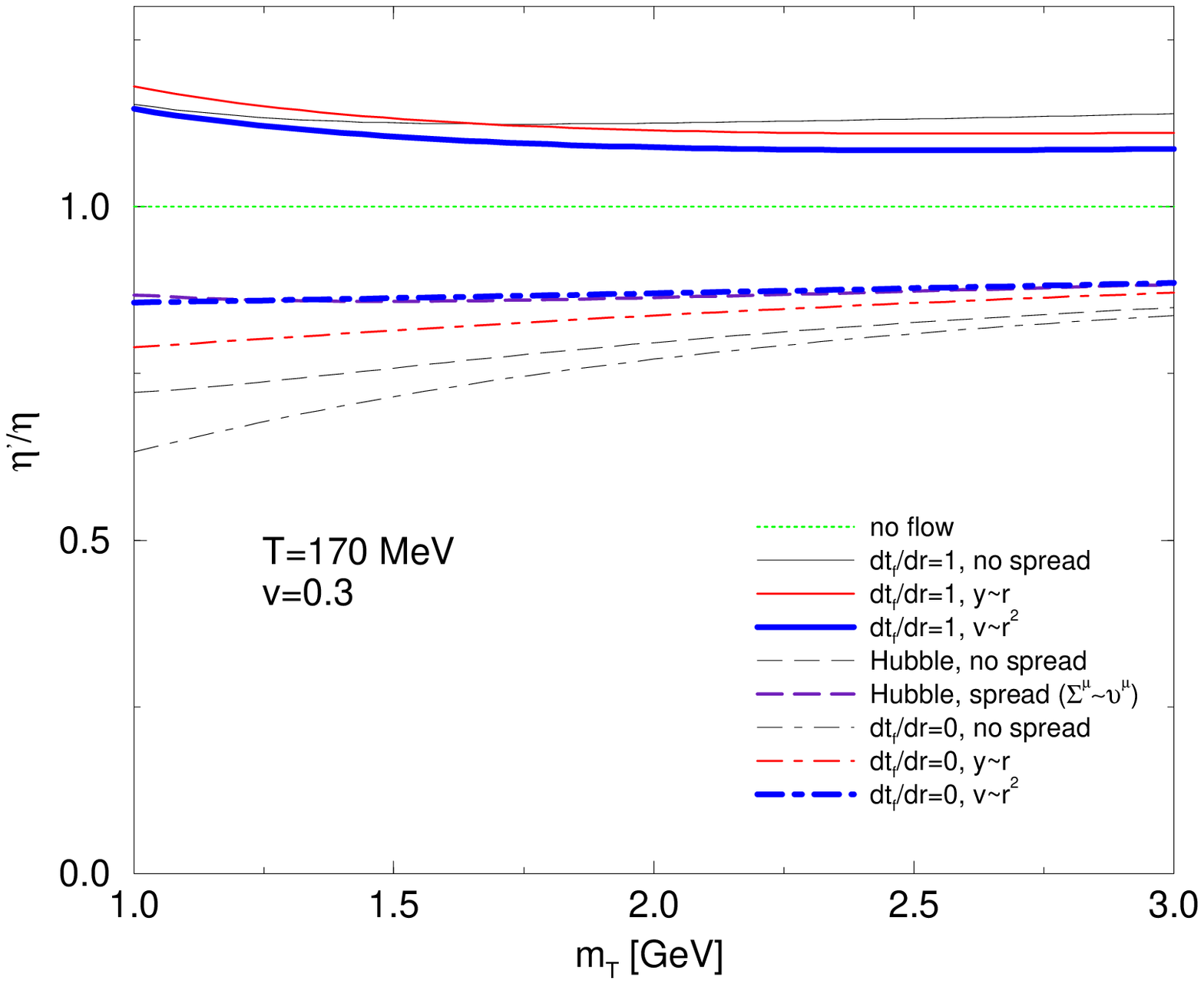}
}}
\caption{(Color online) Dependence of the  $K^*/K$, $\Sigma^*/\Lambda$ and $\eta'/eta$ on
the Freeze-out model  \label{diagres}.}
\end{figure}

Fig.~\ref{diagres}  shows the application of this procedure
to the cases  $(K^*+\overline{K^*})/(K_S)$ (top), $\Sigma^*(1385)/\Lambda$ (middle), and
$\eta'/\eta$ (bottom) at two freeze-out temperatures and flows:
$T=140 \mbox{{\rm MeV}}, v_{max}/c=0.55$ on left and
$T=170 \mbox{{\rm MeV}}, v_{max}/c=0.3$ on the right.  Significant deviations 
from  simple constant values are observed, showing the sensitivity of the
ratio to freeze-out geometry and dynamics. 
The analytically simple result  in Eq.~(\ref{blastratio}) is
 valid only if the light particle $Y$ has
been corrected for feed down from resonances, including $N^*$.
In other words, Eq.~(\ref{blastratio}) as well
as Fig.~\ref{diagres} require that decay products from reconstructed
$N^*$ do not appear on the bottom of the ratio.
Experiments usually do not
do such feed down corrections~\cite{fat01,mar01,fri01},
 since this would
increase both statistical and systematic error on the ratio, and 
it is not always possible
to do such corrections at all (undetected decays) 
 or in the full range of experimental sensitivity.

Introducing the feed down corrections into Eq.~(\ref{blastratio}),
we obtain
\begin{equation}
\label{blastratiofeed}
\frac{N^*_{observed}}{N_{observed}} =\frac{g^* S(m_T,p_T^*)}{g S(m_T,p_T) + \sum_{i} g^{*}_{i} b_{N^*_i \rightarrow N} R(m_T,p_{Ti})  }.
\end{equation}
Here, $S(m_T,p_T)$ describes the directly
produced particles and has the form given by Eq.~(\ref{blastgen})
and each term $R(m_T,p_{Ti}^{*})$ describes a feed down contribution, in the form given in 
section 2.3.1 (In particular Eq.~(\ref{decayphase}) and following)

\begin{figure}
\begin{center}
\centerline{\resizebox*{!}{0.32\textheight}{
\includegraphics{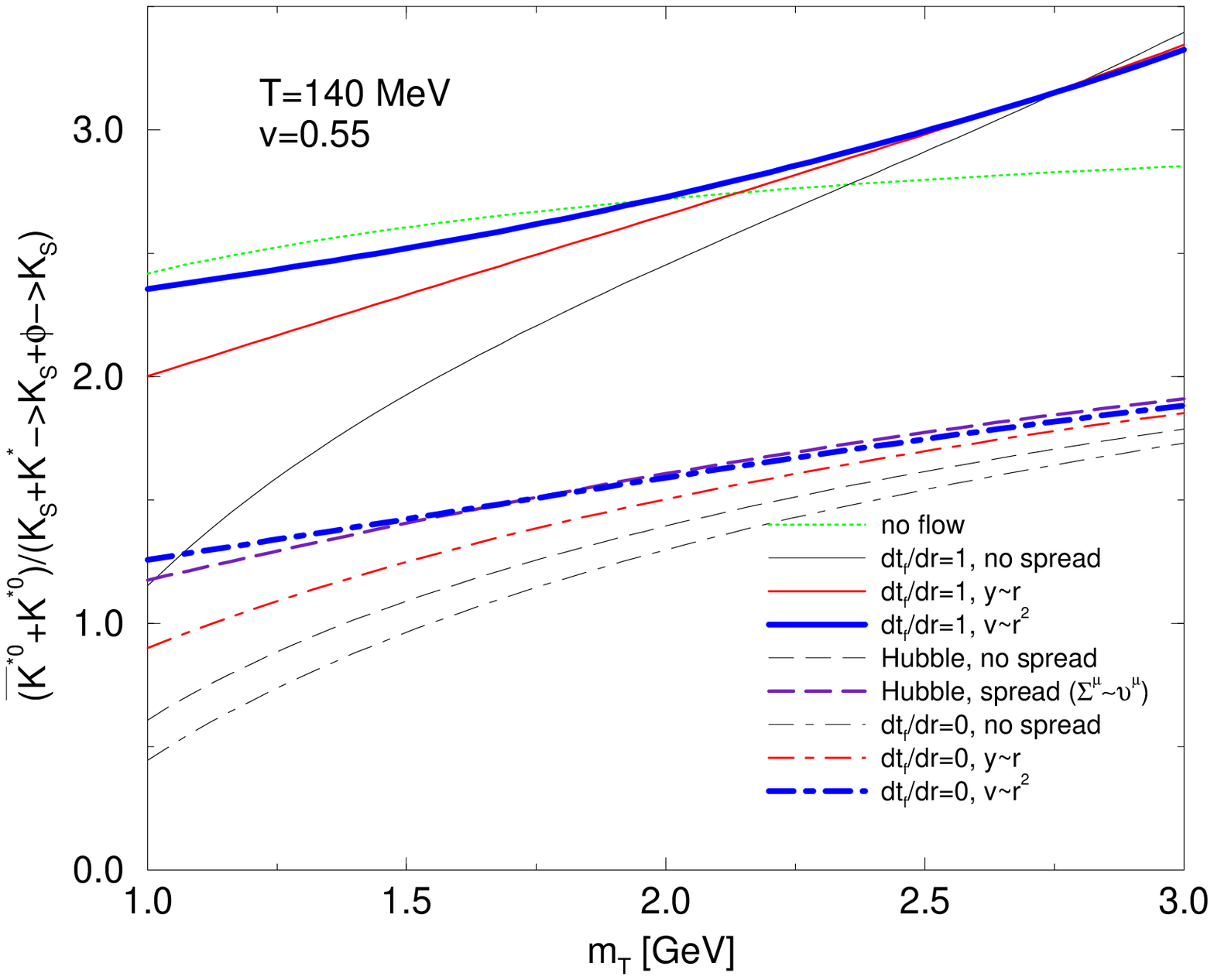}
}
\resizebox*{!}{0.32\textheight}{
\includegraphics{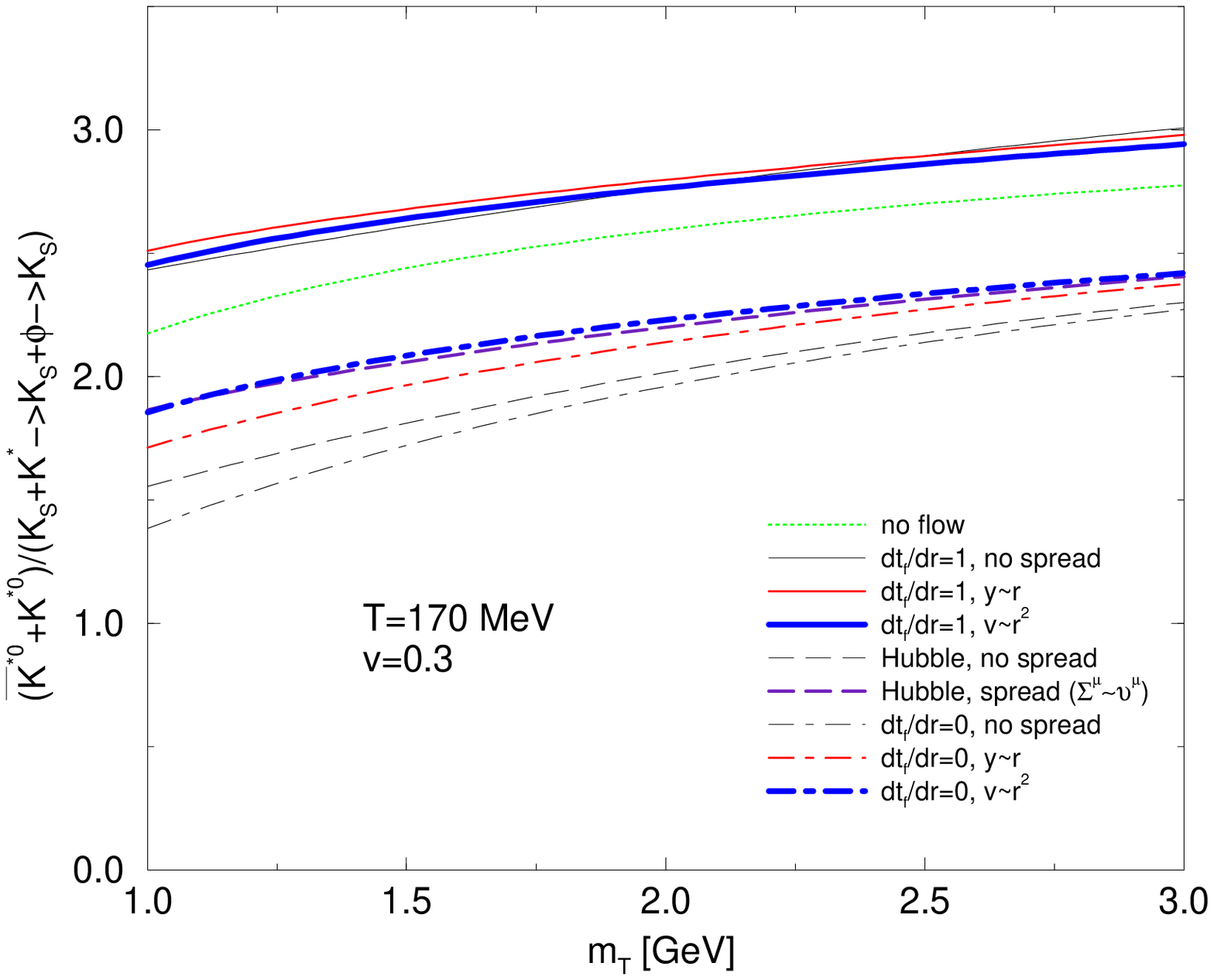}
}}
\centerline{\resizebox*{!}{0.32\textheight}{
\includegraphics{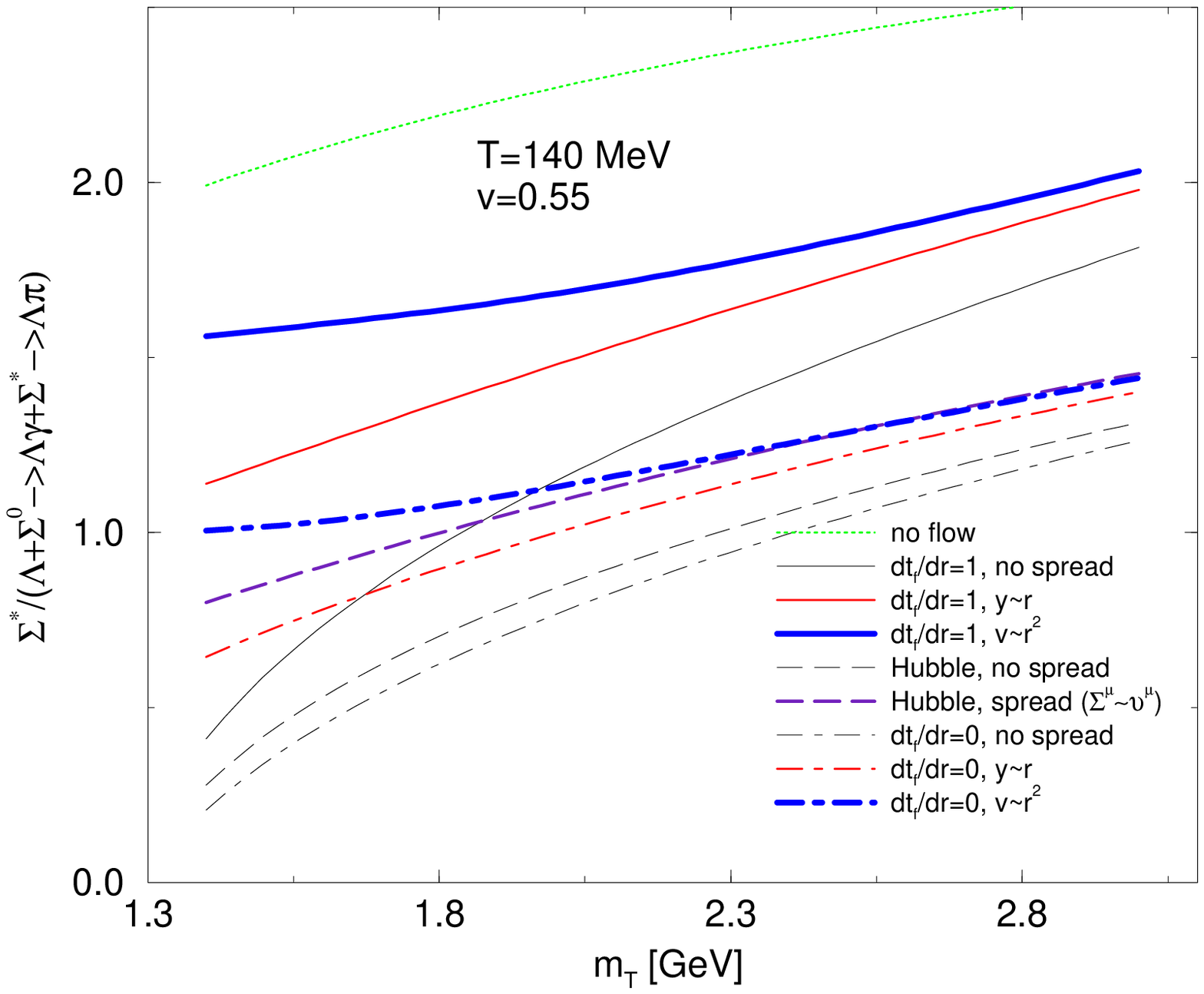}
}
\resizebox*{!}{0.32\textheight}{
\includegraphics{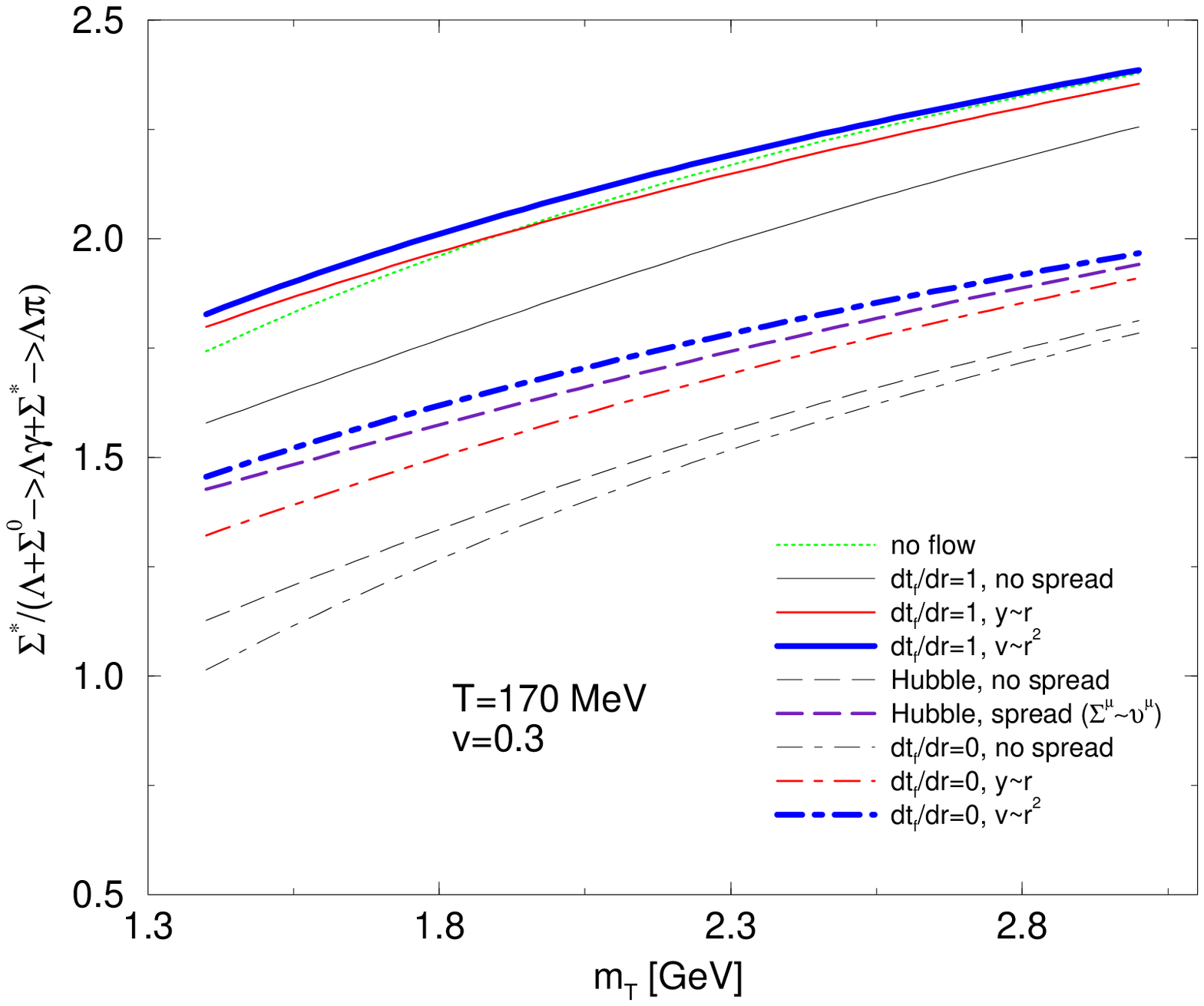}
}}
\centerline{\resizebox*{!}{0.32\textheight}{
\includegraphics{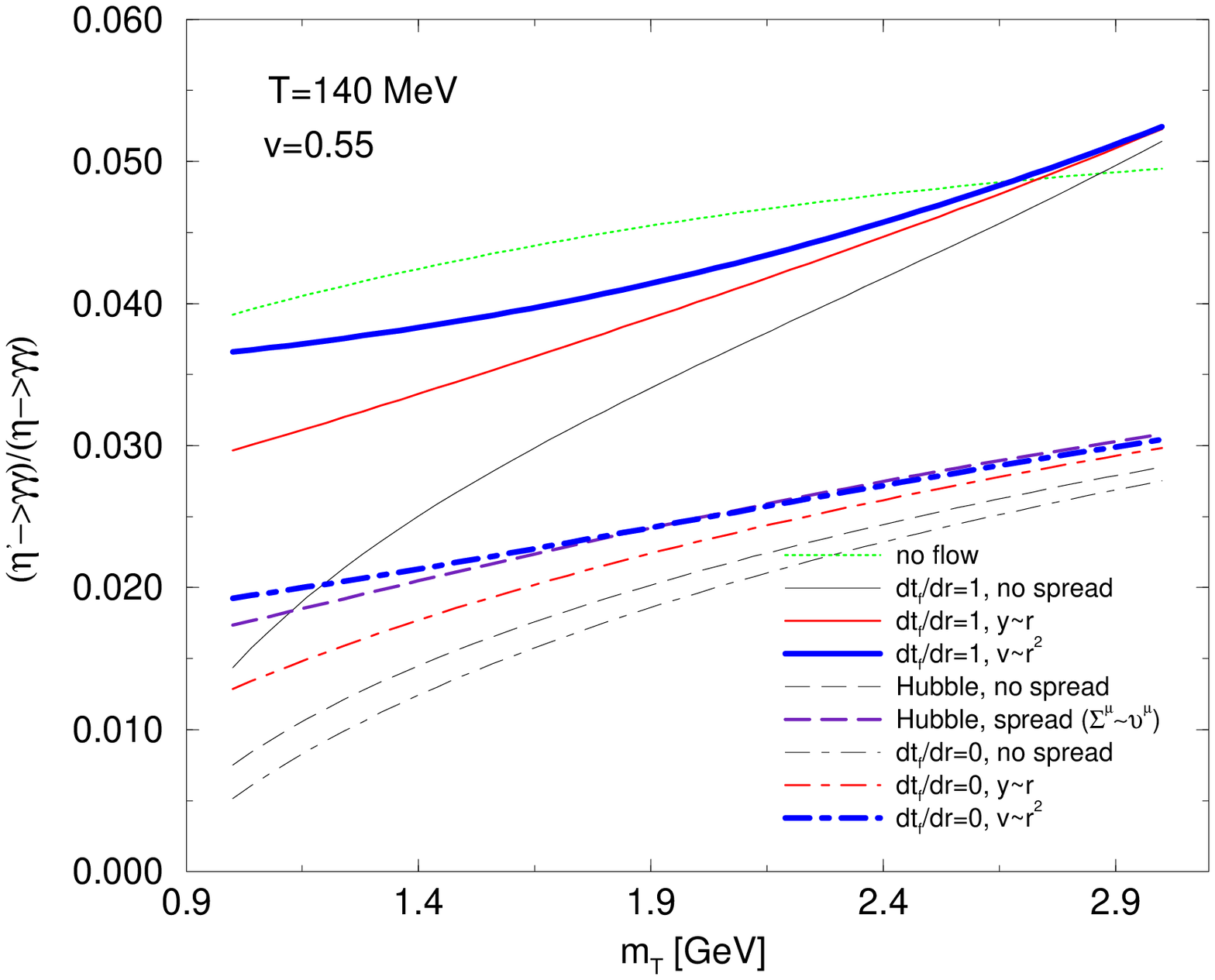}
}
\resizebox*{!}{0.32\textheight}{
\includegraphics{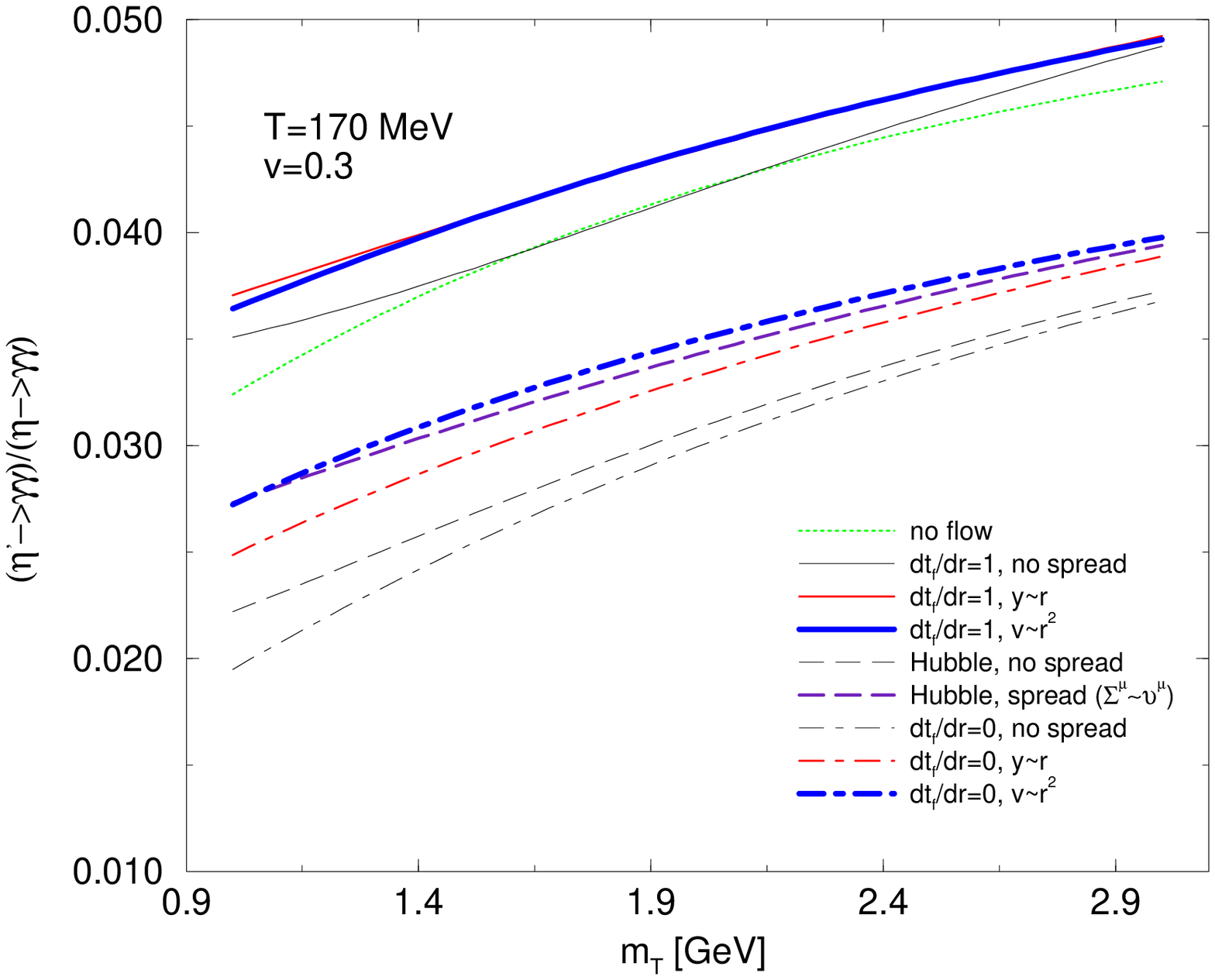}
}}
\end{center}
\caption{(Color online) $(K^*+\overline{K}^*)/($all $K_S)$, $\Sigma^*(1385)/($all $\Lambda)$ 
and $\eta'/($all $\eta)$ ratios, including feed down from resonances.
\label{diagfeed} }
\end{figure}

Fig.~\ref{diagfeed} shows the ratios, including feed down
of resonances, for the same particles and statistical hadronization 
conditions  as were studied in Fig.~\ref{diagres}.
In the $\Sigma^*/($all $\Lambda)$ case we omitted the feed down from $\Xi$ to 
$\Lambda$ which is usually corrected for (if this is not done
the ratio  $\Sigma^*/($all $\Lambda)$ would depend strongly on the chemical potentials).
We did allow for the $\phi \rightarrow K_S K_L$ feed down, since it is a 
strong decay that cannot so easily be corrected for\footnote{See footnote in subsection~\ref{prodrat}}. We note that 
the feed down from particles with a different chemical composition
cannot always be corrected for, and thus some resonances ratios will also acquire a 
(mild) dependence on the chemical potentials.
This is even  true for ratios such as $\eta'/($all $\eta)$, given 
different $s \overline{s}$ content.
In this chapter, these type chemical corrections were set equal to unity.

\begin{figure}[tb]
\begin{center}
\centerline{\resizebox*{!}{0.3\textheight}{
\includegraphics{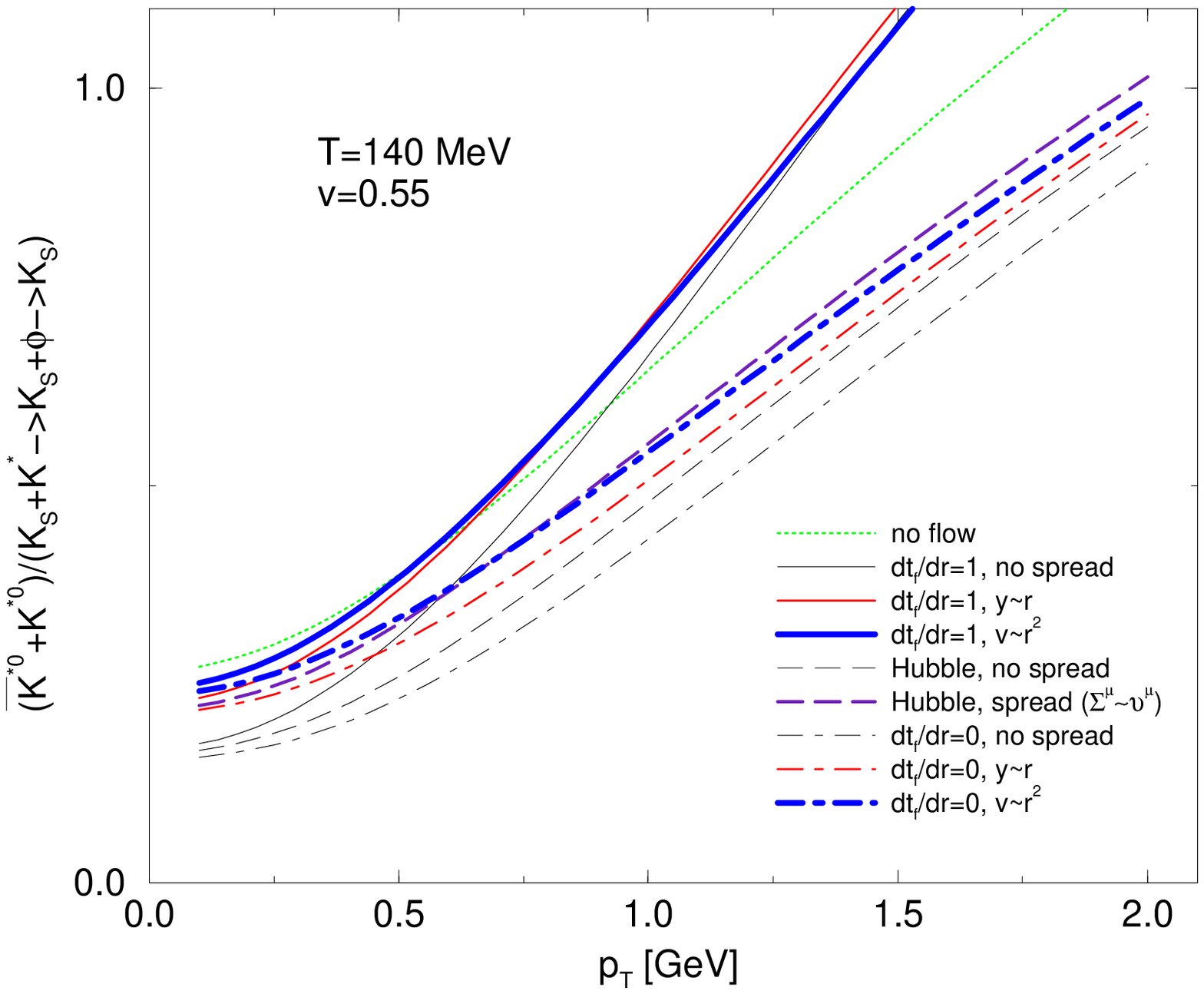}
}
\resizebox*{!}{0.3\textheight}{
\includegraphics{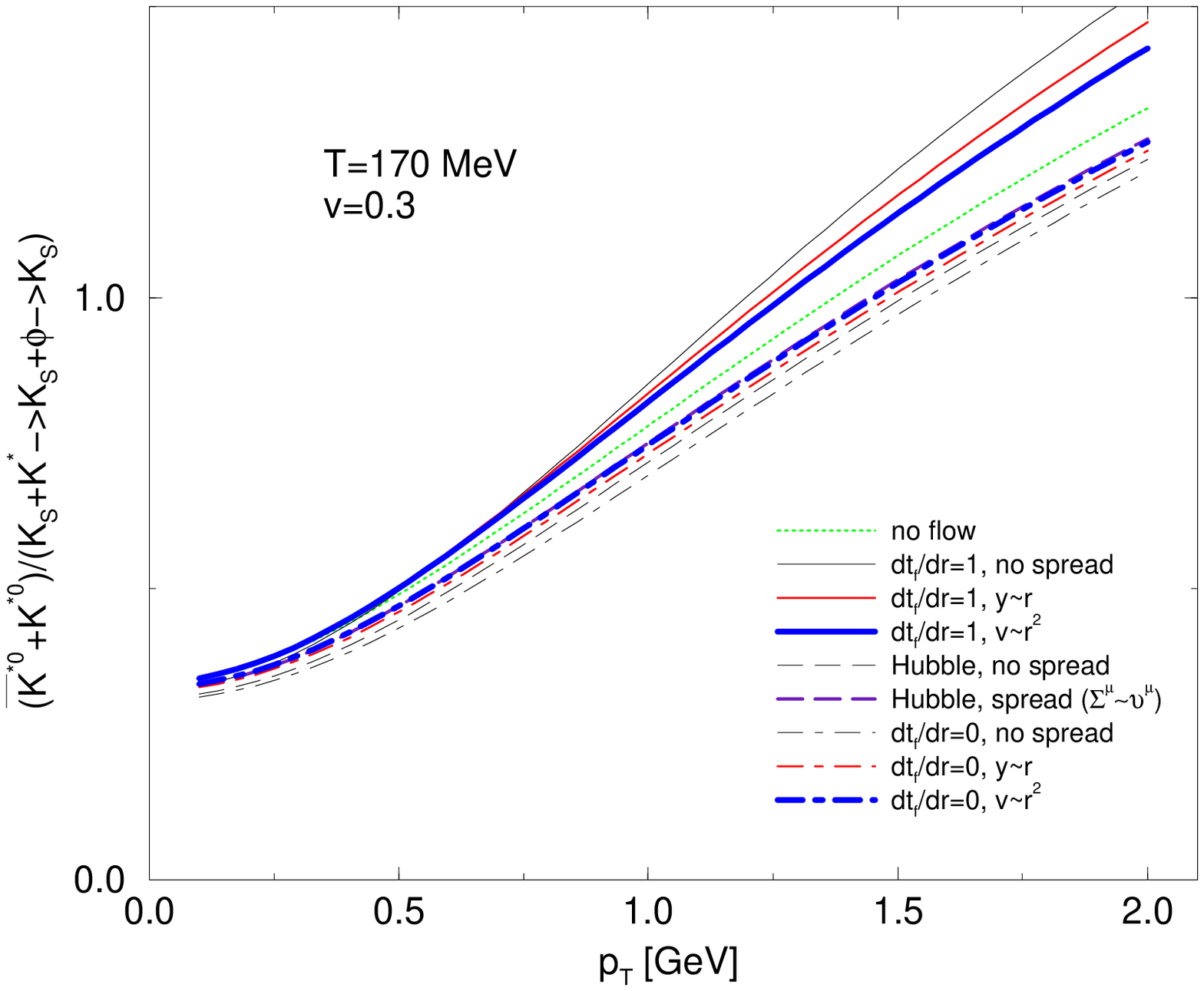}
}}
\centerline{\resizebox*{!}{0.3\textheight}{
\includegraphics{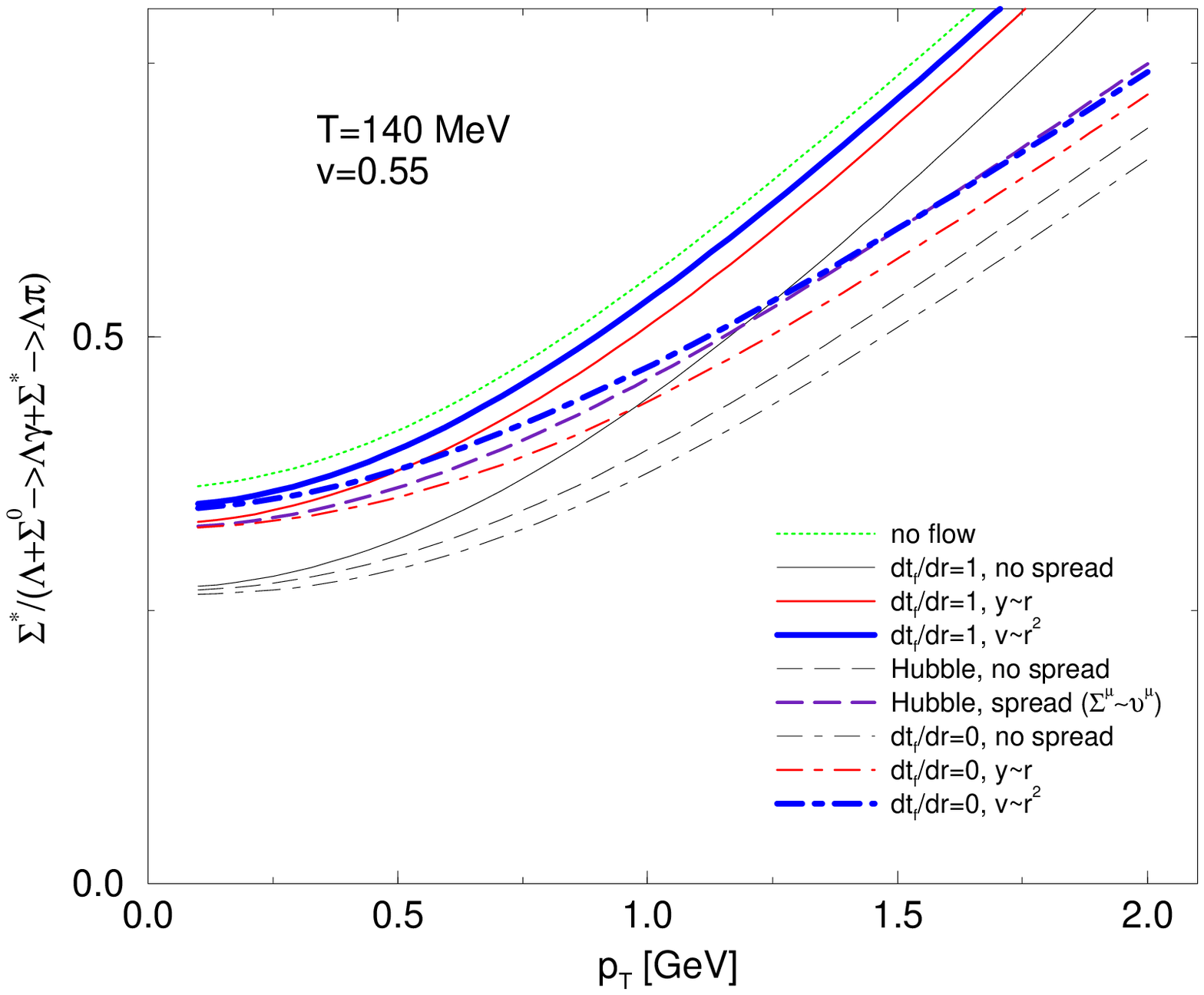}
}
\resizebox*{!}{0.3\textheight}{
\includegraphics{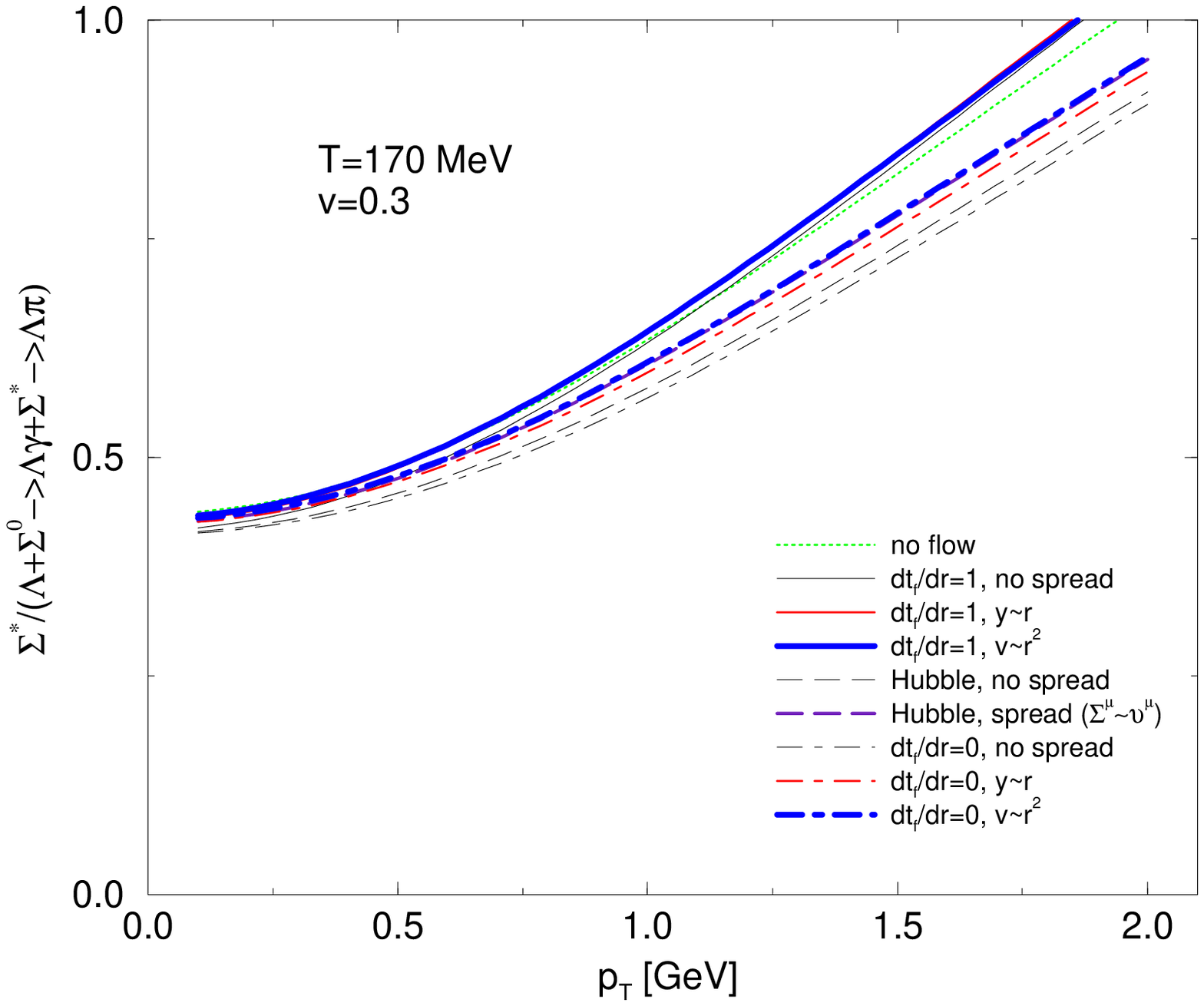}
}}
\centerline{\resizebox*{!}{0.28\textheight}{
\includegraphics{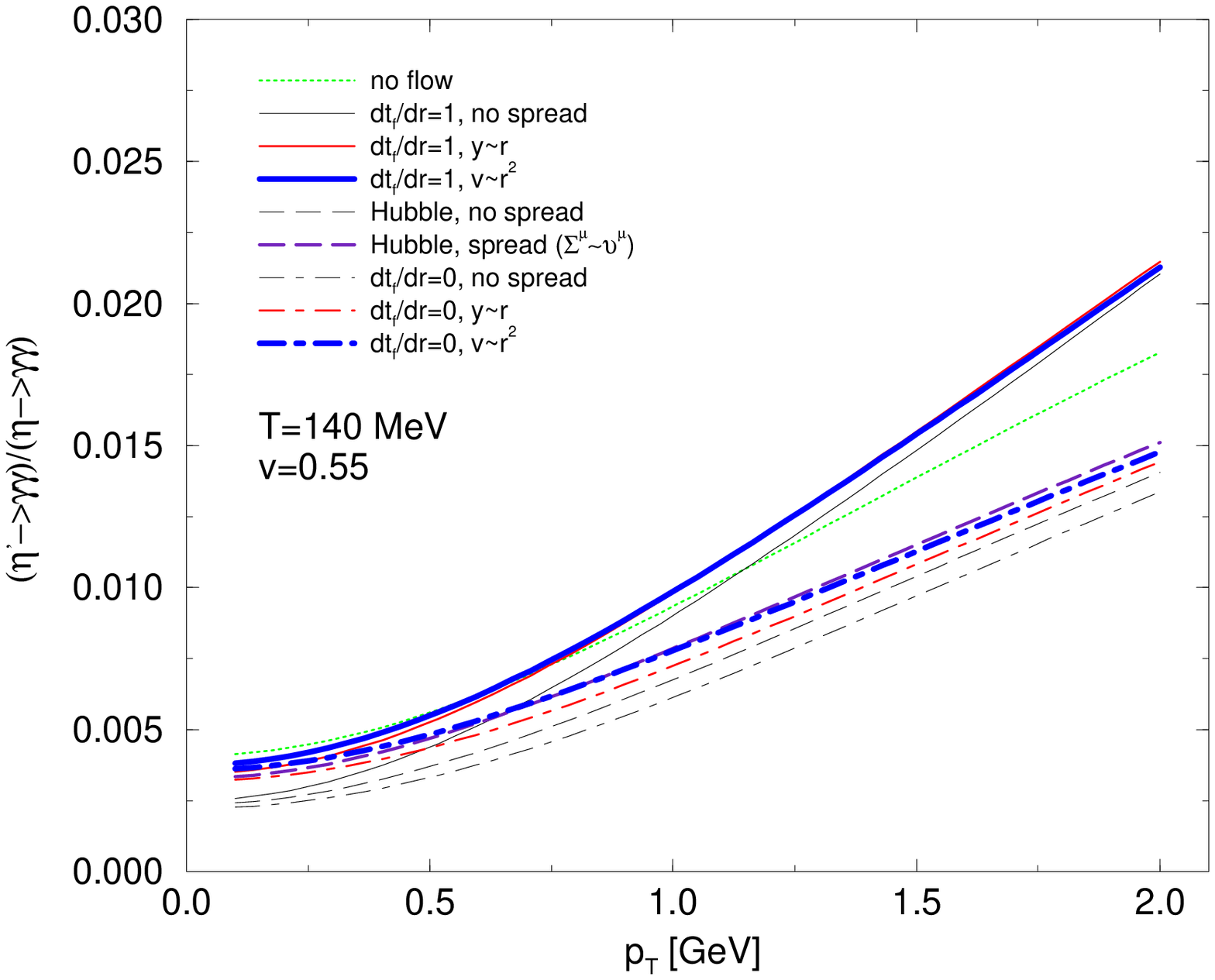}
}
\resizebox*{!}{0.28\textheight}{
\includegraphics{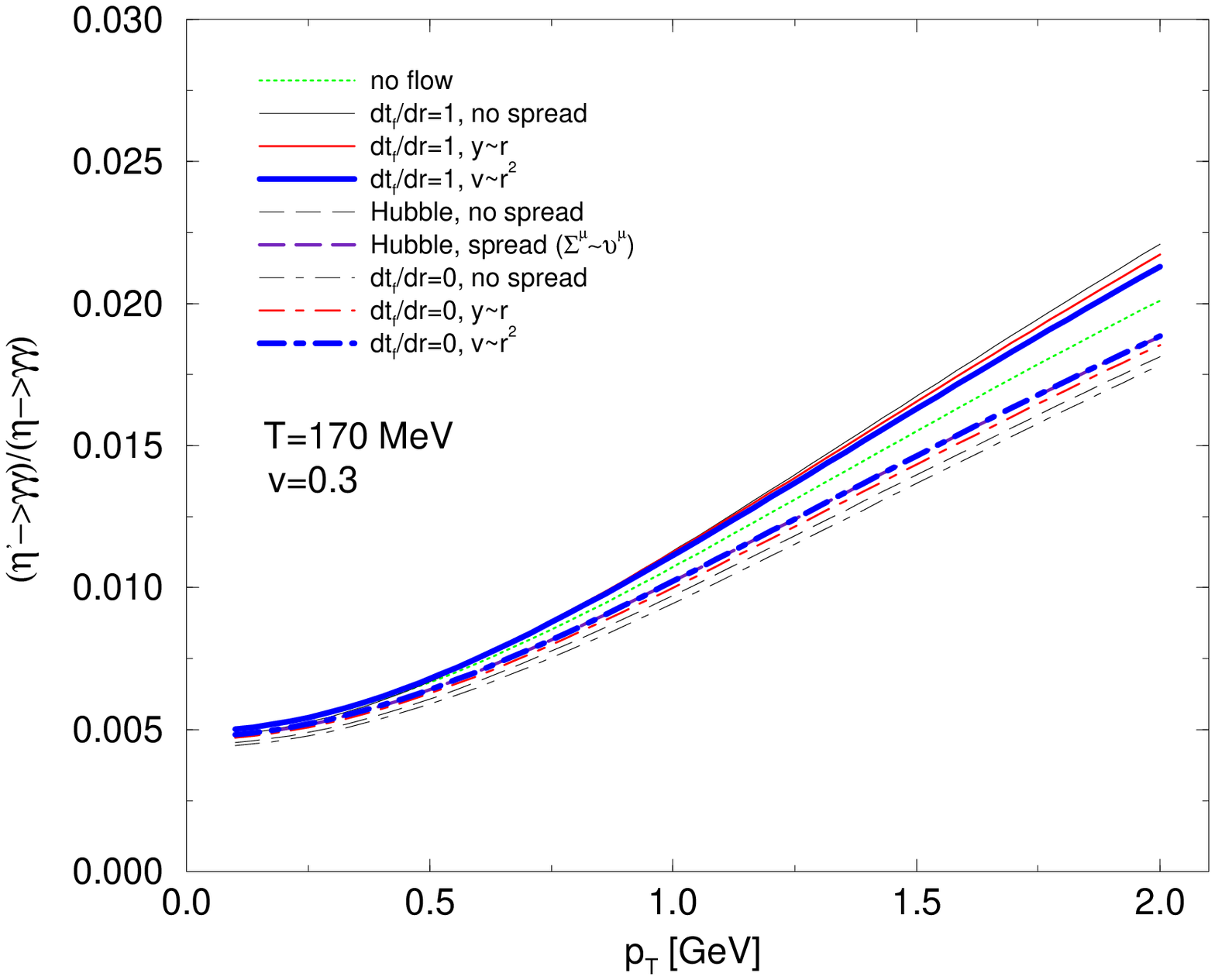}
}}
\end{center}
\caption{(Color online) $p_T$ dependence of $(K^*+\overline{K}^*)/($all $K_S)$, $\Sigma^*(1385)/($all $\Lambda)$ 
and $\eta'/($all $\eta)$ ratios, including feed down from resonances.
\label{diagfeedpt} }
\end{figure}

To further study the sensitivity of resonance-particle $m_T$-ratio 
to freeze-out dynamics, we also present the (feed down corrected)  case 
as a function of $p_T$ rather than $m_T$ in Fig.~\ref{diagfeedpt}.
Unsurprisingly, we see grossly different behaviors, with many of the results
coalescing. This of course  is an expression of the fact that 
$N^*$ and $N$  have dramatically different $p_T$ at the same $m_T$ and vice versa.
We believe  that the $m_T$ ratio will in general be more sensitive to freeze-out 
dynamics, since its dependence on $m_T$ is dominantly  due to freeze-out geometry 
and dynamics. 
However, the $p_T$ dependence seen in Fig.~\ref{diagfeedpt} provides
 an important self-consistency check for our previous results. We have found that
the $m_T$ ratios are often greater than unity even though  
there must be more ground state
 particles than resonances. Now it can be  seen in the  
$p_T$ ratio,  that this requirement is satisfied. 

\section{Discussion}
In general the the $m_T$ and $p_T$ dependence of the ratios in Fig 3 and, respectively, Fig
4 depends on freeze-out geometry, temperature and flow velocity.
The introduction of a steeper flow profile will further  raise 
all of the considered ratios,
since a considerable fraction of particles will be produced in regions that do
not flow as much.
The influence of freeze-out dynamics will generally go in the same direction
as freeze-out approaches the explosive limit ($dt_f/dr\rightarrow1$).
However, both the magnitude and the qualitative features of the two 
effects (flow and freeze-out velocity) 
will be considerably different. Especially, when more than one ratio is 
measured, it would appear that we will be able to 
determine the freeze-out condition. 
This is in contrast to the $m_T$ distributions in Fig.~\ref{mcplots}, where 
the effects discussed in this paper result in linear corrections, which tend to
compete, making the task of extracting the freeze-out dynamics much more
ambiguous. Thus, there is considerable 
potential of resonance-particle $m_T$-ratios as a freeze-out 
probe.
 
The presence of a long living hadronic gas rescattering phase
can distort our freeze-out probe.
In particular, the apparent  $N^*/N$ ratio will be altered due to the depletion of
the detectable resonances through the rescattering of their decay products.
Its dependence on $m_T$  will  be affected in a non-trivial
way,  since faster (higher
$p_T$) resonances will have a greater chance to escape the fireball
without decaying, thus avoiding the rescattering phase altogether.
Regeneration of resonances in hadron scattering may add another  
$m_T$ dependence which is different for the $\Sigma^*/\Lambda$ and
the $K^*/K$ ratios~\cite{bleicher}.
While, as discussed in the last section, evidence that
a rescattering phase  plays a great role in particle distributions is lacking,
it would seem that the ``safest''  probes for freeze-out are
 the particles  and resonances most unlikely to rescatter.

For this reason we have included the $\eta'/\eta$ ratio in our considerations. 
$\eta\rightarrow \gamma \gamma$ and $\eta'\rightarrow \gamma \gamma$ have very 
different branching ratios,
but  have the same degeneracies and similar but rather  small partial  widths.
The electromagnetic decay mode is practically  insensitive to posthadronization
dynamics. Regeneration effects are suppressed since the hadronic two body decay channel
is suppressed. 
All these features make these particles interesting  probes, allowing 
for the analysis considered here.
$\eta,\eta'$ mesons have been measured  at SPS energies 
in the $\gamma\gamma$ decay channel~\cite{eta1,eta2}, and detectors such as PHENIX
are capable of reconstructing the same decays at RHIC.  

In summary, we have presented an overview of the different statistical 
freeze-out models used to fit heavy ion data. We have shown how the 
freeze-out geometry and freeze-out dynamics influences the hadron spectra.
Our primary result is the finding that the 
 $m_T$ dependence of the resonance-particle  ratios is a probe 
of freeze-out. We have presented these ratios for three particle species and 
two freeze-out conditions and have considered how our results could 
be altered by posthadronization phenomena. 

\setcounter{figure}{0}
\setcounter{equation}{0}
\setcounter{table}{0}
\chapter{Azimuthal anisotropy}
\label{cha:v_2}

\section{Introduction}
Matter flow azimuthal  anisotropy has long been considered a 
promising soft hadron  observable in the study of heavy ion collisions.
This anisotropy is considered an important evidence for 
collective matter flow \cite{ollitraut1,voloshin1}, as it indicates thermalization
early in the system's evolution \cite{sorge1}.

Generally one simplifies the complex transport problem and 
 considers hydrodynamic model, i.e. 
dynamics of  first moment of the momentum distribution  
of particles. Assuming 
in the local reference frame an asymmetric thermal momentum 
distribution we evaluate below the actual final state  particle
 distributions. Azimuthal anisotropy in hydrodynamics 
arises from anisotropic  gradient of pressure (force field) 
when the collision between nuclei is not exactly head-on central. In this 
way hydrodynamic model leads to  the azimuthal momentum dependent  asymmetry of 
final state particle distributions. 

Anisotropy in the final state 
can be quantified in terms of a Fourier decomposition of the 
momentum distribution \cite{poskanzer1},
\begin{equation}
E \frac{dN}{d^3 p} =  \frac{1}{2 \pi}\frac{dN}{p_T dp_T dy} \left[ 1+\sum 2 v_n \cos(n \phi) \right],
\label{vn_definition}
\end{equation}
where the angle $\phi$ is defined with respect to the reaction plane
\cite{ollitraut2} in an event-by-event analysis (Fig.~\ref{config}).

The second coefficient $v_2$, usually called  
elliptic flow, has been subject to a considerable amount of 
experimental investigation at AGS, SPS and RHIC energies \cite{exp1,exp2,exp3,exp4}.
 The fourth coefficient $v_4$ has recently been measured at RHIC \cite{expv4}.
(odd components, such as the directed flow $v_1$, disappear in the Boost-Invariant limit) 

Generally, $v_n$ is a function of momentum. 
It has been found that $v_2(p_T)$ at low $p_T$ and central rapidity agrees  with hydrodynamical
predictions \cite{exphydro,hirano1}.  However, RHIC experiments have shown that at
 $p_T \sim 1.5$ ${\rm GeV}$ $v_2$ rise  saturates to a particle-dependent 
limit \cite{exp3,exp4}, something
so far  hydrodynamic approach could not explain 
 Since this $p_T$ value is also
seen as being too soft for $v_2(p_T)$ to be determined by 
  ``hard'' perturbative processes, this
behavior has not been fully understood. 

However, coalescence of partons obeying perturbative dynamics 
has been shown to lead to saturation of  $v_2(p_T)$ \cite{molnar1}. 
While this approach does not as yet cover consistently the low $p_T$ range \cite{coalescence} and hence
is yet to undergo detailed quantitative testing,
its success suggests that, more generally, the process of quark-gluon hadronization 
can influence significantly the resulting $v_2(p_T)$. Here we address this
possibility  systematically, but qualitatively in that we do not introduce resonance
decays and flow profile of matter. Our 
objective is to identify physical  mechanisms 
rather than to explain the 
experimental data. 

Our study will show that there are $v_2(p_T)$ 
saturation mechanisms based solely on rapid 
hadronization of a hydro-dynamically evolving 
opaque system, such as quark-gluon plasma.
We shall use the Cooper-Frye approach of Eq.~(\ref{cf}) \cite{cf}.
\begin{equation}
 E \frac{dN}{d^3 p} = \int d^3 \Sigma^{\mu} p_{\mu} f(u^{\mu} p_{\mu},T,\mu).
\end{equation}
As we have argued in the previous chapters,  this formula is particularly relevant if
most observed particles are emitted when the system undergoes a phase
transition to a gas of particles having a much larger mean free path (QGP $\rightarrow$ expanding  
Hadron Gas).   In this case, 
the Cooper-Frye formula, in particular the hadronization hypersurface
$d^3 \Sigma_{\mu}$, ceases to be just a computational
prescription to generate particles from a continuum, but acquires physical significance as the 
representation of the hadronizing QGP in position space.  Thus
a more thorough exploration of how it can modulate the shape  
of  $v_2(p_T)$ is required .   The form of $d^3 \Sigma_{\mu}$
is determined by the dynamics of the QGP $\rightarrow$ HG phase transition.

Several theoretical studies and also general behavior of 
experimental results suggest that  hadronization happens 
rapidly, perhaps through viscous 
fingering of the vacuum \cite{sudden1},
driven by a mechanical instability at the point when
the pressures (including that of flow) of the two 
vacua balance ,  but the velocity still point outwards \cite{sudden2}. 
In this situation it is natural to expect
a ``burning log'' type emission \cite{torrieri_freeze}, with a fast ($\sim c$) moving 
emission surface rapidly consuming the fireball from outside in. In contrast
to this dynamical hadronization picture,  a ``blast-wave'' freeze-out model in which 
the freeze-out time does not depend on the transverse  freeze-out radius is 
presently popular among experimentalists, perhaps due to the simplicity 
of its use in a fitting procedure \cite{burward-hoy,castillo,van-leuween,NA57spectra}.

As we will show  these two scenarios (burning log and non-dynamic 
blast wave) give considerably
 different $v_2(p_T) $ for the same initial matter  flow anisotropy.
In particular, we find that a burning-log type freeze-out dynamics 
leads in qualitative terms to the observed $v_2$ saturation.

\section{$v_n$ in the Cooper-Frye approach}
\begin{figure*}
\centerline{\resizebox*{!}{0.3\textheight}
{
\includegraphics{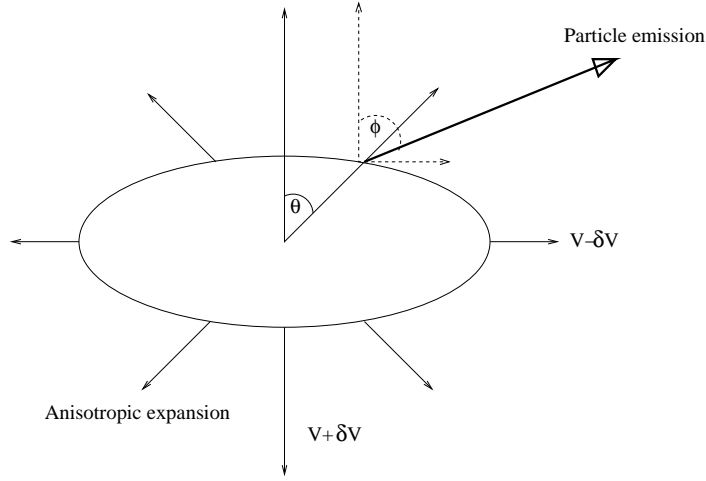}}}
\caption{A general freeze-out surface in configuration space \label{config}.}
\end{figure*}

We shall use the approach taken in chapter 4, using boost-invariance
to construct our flow and $\Sigma_{\mu}$.
Hence, in the most general boost-invariant case
\begin{equation}
\Sigma^{\mu}=\left( \begin{array}{c} t_f \cosh(y_L) \\ r_f \cos(\theta) \\
r_f \sin(\theta)\\ t_f \sinh(y_L) \end{array} \right),  \;
u^{\mu} = \gamma_{T} \left( \begin{array}{c} \cosh(y_L) \\ v_T \cos(\theta) \\
v_T \sin(\theta) \\ \sinh(y_L) \end{array} \right).
\end{equation}
Where $\theta$ is the angle of the emission point in configuration space
( Fig.~\ref{config}, not to be confused with the emission angle $\phi$).
However, we forgo cylindrical symmetry, and Fourier expand both the flow
and hadronization hypersurface, keeping only even terms given the 
symmetry,
\begin{eqnarray}
t_f (r,\theta)= \sum_{n=0}^{\infty} \Delta_{2n} t_f (r) \cos(2 n \theta),\\
r_f (r,\theta)= \sum_{n=0}^{\infty} \Delta_{2n} r_f (r) \cos(2 n \theta),\\
v_T (r,\theta)= \sum_{n=0}^{\infty} \Delta_{2n} v (r) \cos(2 n \theta).
\end{eqnarray}
We assume a purely elliptic fireball ($n=0,1$ only), and moreover a
``fast'' freeze-out so as to neglect $t_f$ $\theta$-dependence. 
If the fireball is approximately circular in transverse space, the elliptical
freeze-out hypersurface becomes an ellipse (with a correction of $O(\delta r^2)$),
\begin{equation}
\Sigma^{\mu}=\left( \begin{array}{l} t_f(r) \cosh(y_L) \\ 
(r+\delta r) \cos(\theta) \\
r \sin(\theta)\\ 
t_f(r) \sinh(y_L) 
\end{array} \right), 
\end{equation}
\begin{equation}
u^{\mu} = \gamma_{T}(r,\theta) \left( \begin{array}{l} \cosh(y_L) \\ 
(v+\delta v \cos(\theta)) \cos(\theta) \\
(v+\delta v \cos(\theta)) \sin(\theta) \\
 \sinh(y_L) \end{array} \right),
\end{equation}
\begin{equation}
\label{gammaT}
\gamma_T=\frac{1}{\sqrt{1-(v+\delta v \cos(\theta))^2}}.
\end{equation}
We can now parametrize the freeze-out 
hypersurface element in terms of $r,\theta,y_L=$ (radius, angle and longitudinal
rapidity $y_L$)
\begin{eqnarray}
 d^3 \Sigma^{\mu}& =& \epsilon^{\mu \nu \alpha \beta} 
\frac{\partial \Sigma_{\nu}}{\partial r } 
\frac{\partial \Sigma_{\alpha}}{\partial \theta} 
\frac{\partial \Sigma_{\beta} }{\partial y_L} \nonumber \\[0.2cm]
&=&\left( 
\begin{array}{l}
\cosh (y_L) (r+\delta r \sin^2 (\theta))\\[0.1cm]
\frac{\partial t_f}{\partial r} r \cos( \theta )\\[0.2cm]
\frac{\partial t_f}{\partial r} (r+\delta  r) \sin( \theta )\\[0.2cm]
 \sinh (y_L) (r+\delta r \sin^2 (\theta))
\end{array} \right) d y_L dr d\theta.\
\end{eqnarray}
If we combine the obtained $u^{\mu}$ and $d^3 \Sigma_{\mu}$ with the usual parametrization for the emitted particle's 4-momentum
\begin{equation}
p^{\mu}= \left( \begin{array}{c} m_T \cosh(y) \\ p_T \sin(\phi) \\ p_T \cos(\phi)\\ m_T \sinh(y) \end{array} \right)
\end{equation} 
we obtain an expression of the boost-invariant momentum distribution in terms
of $\phi$,$p_T$,$m_T$ and $y$:
\begin{equation}
\begin{array}{l}
 E \frac{dN}{d^3 p} = \int_0^{R_{max}} r dr \int_{-\infty}^{\infty} dy_L \int_{0}^{2 \pi} d\theta
 \left[ m_T \cosh(y-y_L)\times  \right. \\[0.3cm] \left.\times 
\left( 1+ \frac{\delta r}{r} \sin^2 (\theta) \right)
-p_T \frac{\partial t_f}{\partial r} ( \cos(\theta-\phi)+ \frac{\delta r}{r} 
\sin(\theta)) \right]  \\[0.3cm]
  f (\frac{\gamma_{T}}{T} [ m_T \cosh(y-y_L) - 
p_T (v+\delta v \cos(2 \theta)) \cos(\theta-\phi)])
\end{array}
\end{equation}
Where $f(x)=e^{-x}$ in the Boltzmann approximation (good for all particles except pions) and for pions:
$$f(x) = \sum_{n=0}^{\infty} (-1)^{n} \lambda_{\pi}^{n+1} e^{-nx}.$$ 
 
Each $v_n$ can now be calculated using
\begin{equation}
v_n =\frac{\int_0^{2 \pi} \cos( n \phi) E \frac{dN}{d^3 p} }{2 \int_0^{2 \pi}  E \frac{dN}{d^3 p}}
\end{equation}
the integrals over $\phi$ and $y_L$ can be done analytically in the boost-invariant limit, using the modified Bessel Functions
\begin{eqnarray}
\label{bessel}
 I_n (x) &=&  \int_0^{2 \pi} \frac{d \phi}{2 \pi} \cos(n \phi) e^{ x \cos( \phi)}\\
 K_n (x) &=& \int_{-\infty}^{\infty} d y_L \cosh(n y_L) e^{ - x \cosh( y_L)}
\end{eqnarray}
and the following result 
\begin{eqnarray}
\int_0^{2 \pi} \frac{d \phi}{2 \pi} \cos(n \phi) e^{a \cos(\phi)+b \sin(\phi)} 
=T_n \left( \frac{a}{\sqrt{a^2+b^2}} \right) I_n \left( \sqrt{a^2+b^2} \right)\\
\int_0^{2 \pi} \frac{d \phi}{2 \pi} \sin(n \phi) e^{a \cos(\phi)+b \sin(\phi)} 
=S_n \left( \frac{a}{\sqrt{a^2+b^2}} \right) I_n \left( \sqrt{a^2+b^2} \right)
\end{eqnarray}
where $T_n(x)$ and $S_n(x)$ are defined in terms of Chebyshev polynomials
\begin{eqnarray}
 T_n (x) &=& \cos(n \cos^{-1} (x))   \nonumber \\
S_n (x) &=& \sin(n \cos^{-1} (x))=\sqrt{1- T_n^2 (x)}   \nonumber
\end{eqnarray}
Putting everything together, we get 
\begin{equation}
v_n(p_T)=\frac{\int r dr \int d\theta J_n
}{2 \int r dr \int d \theta J_0},
\label{cfvn}
\end{equation} 
where ($\alpha= p_{T} (v+\delta v \cos(\theta))$)
\begin{equation}
\label{vn_explicit}
\begin{array}{l}
 J_{n>0}=\left[ K_{1} (\frac{\gamma_{T} m_T}{T}) (1+\frac{\delta r}{r} \sin(\theta)^2) - p_T \frac{\partial t_f}{\partial r} K_{0} (\frac{\gamma_{T} m_T}{T}) \right] T_n (\cos[\theta]) I_n (\alpha)  - p_T \frac{\partial t_f}{\partial r}  K_{0} (\frac{\gamma_{T} m_T}{T})  \\
 \left[
 \frac{1}{2} \cos[\theta] \left(  T_{n+1} (\cos[\theta]) I_{n+1} (\alpha) +  T_{n-1} (\cos[\theta]) I_{n-1} (\alpha) \right)  + \right. \\
  \frac{1}{2} \sin(\theta) \left(  S_{n+1} (\cos[\theta]) I_{n+1} (\alpha) -  S_{n-1} (\cos[\theta]) I_{n-1} (\alpha) \right) \left. \right] 
\end{array}
\end{equation}
and 
\begin{equation}
\label{v0_explicit}
\begin{array}{l}
 J_0=\left[ K_{1} (\frac{\gamma_{T} m_T}{T}) (1+\frac{\delta r}{r} \sin(\theta)^2) - 
 p_T \frac{\partial t_f}{\partial r} K_{0} (\frac{\gamma_{T} m_T}{T}) \right] I_0 (\alpha)  - p_T \frac{\partial t_f}{\partial r}  K_{0} (\frac{\gamma_{T} m_T}{T}) I_0 (\alpha)
\end{array}
\end{equation}
For a quantitative fit to be physically meaningful, resonance contributions ($\sim 50 \%$ of the total $K_S$,$\Lambda$, more for $\pi$ \cite{torrieri_mc}) need to be taken into account.  As we have seen in the previous chapters, calculating these contributions is numerically intensive \cite{florkowski}, and
 work to perform such a quantitative fit, together
with analysis of particle spectra \cite{share}, is currently in progress.

\section{Results and discussion}
\begin{figure}
\centerline{
\resizebox*{!}{0.26\textheight}
{
\includegraphics{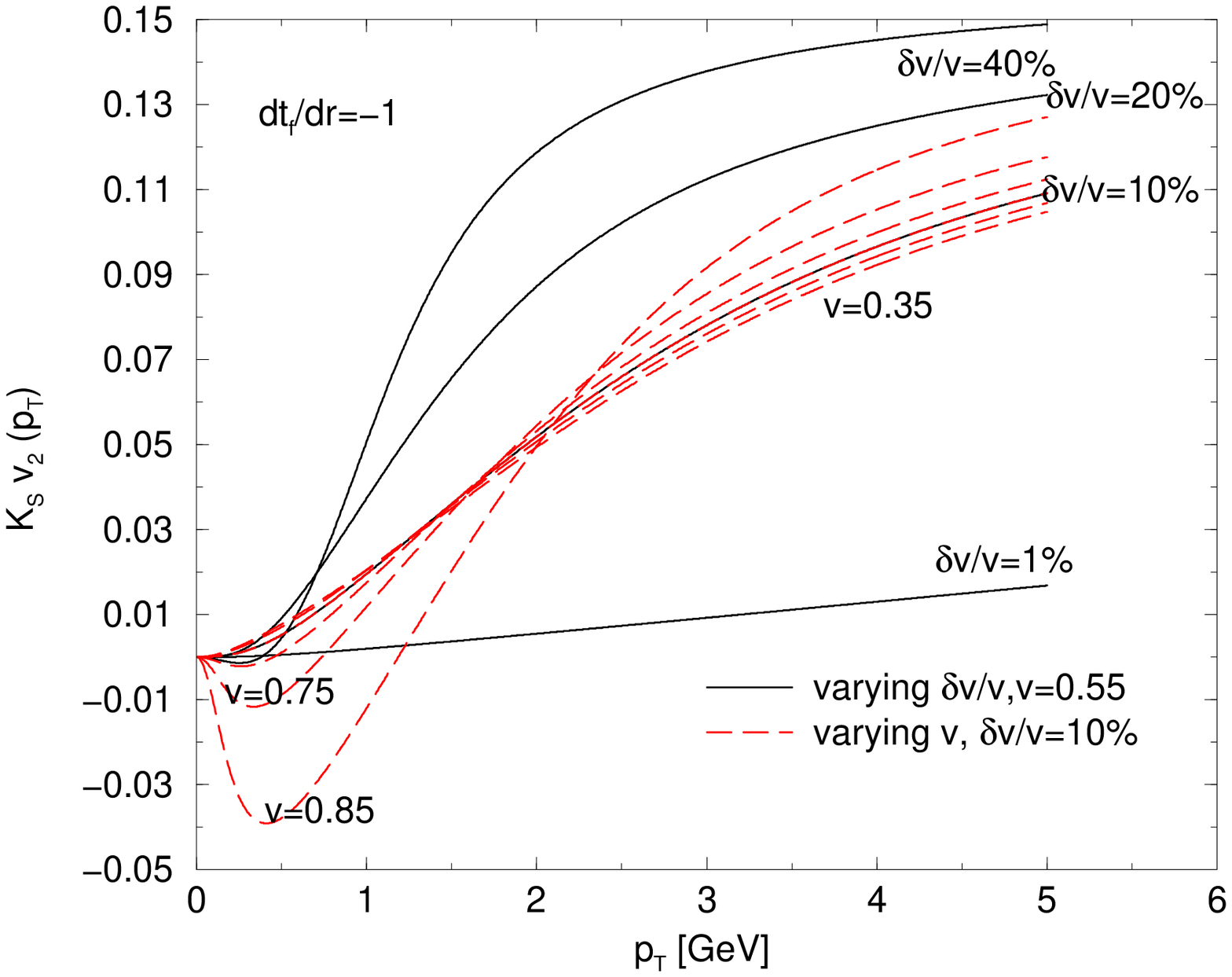}}
\resizebox*{!}{0.26\textheight}
{
\includegraphics{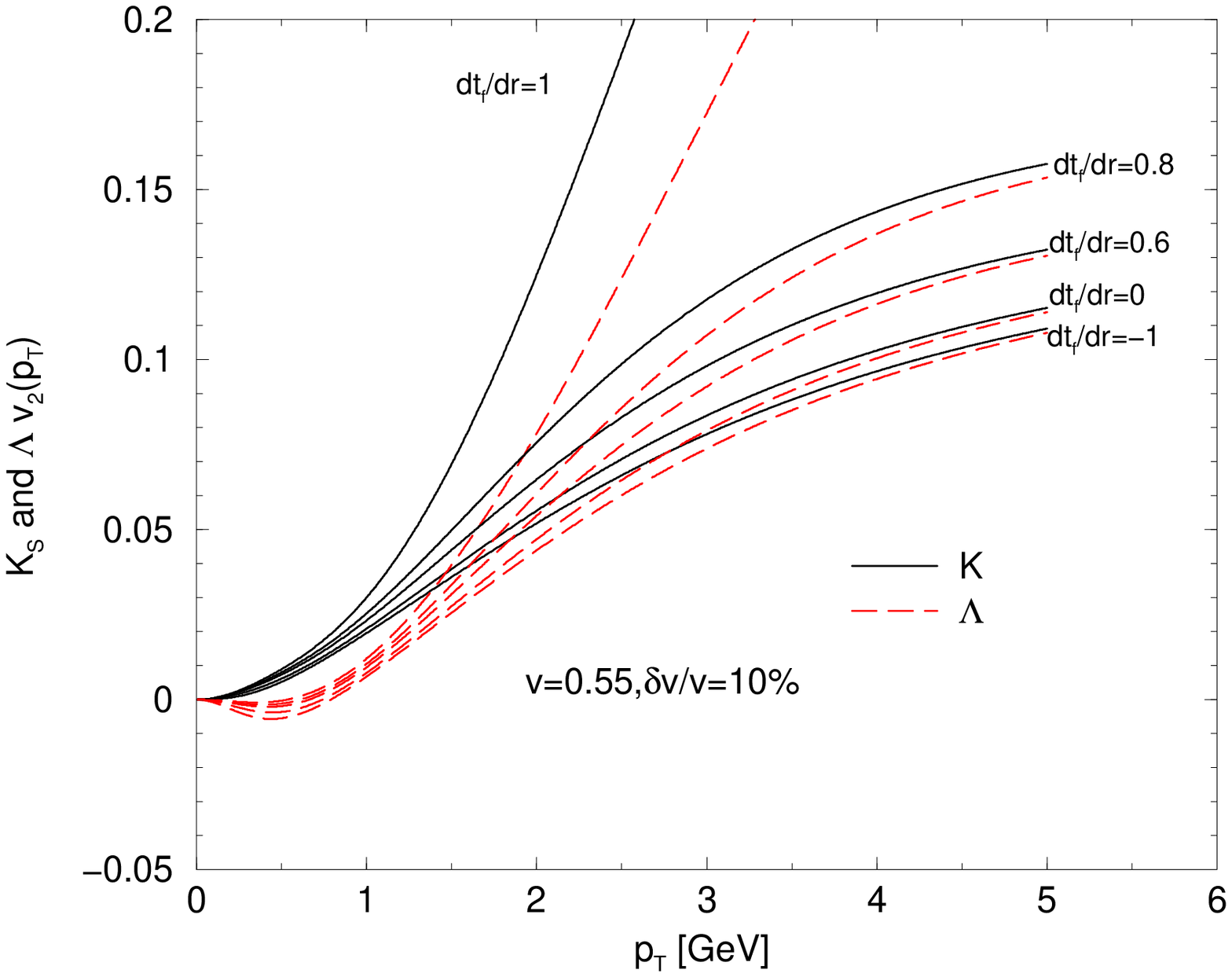}}}
\caption{ Left:  Varying flow anisotropy.  Right: Varying freeze-out dynamics \label{dt_dv}.}
\end{figure}

We have used Eq.~(\ref{cfvn})
 to qualitatively explore the dependence 
of $v_2 (p_T)$ for $K_S, \Lambda$ and $\pi$ on the parameters of statistical freeze-out.
It is apparent that the freeze-out dynamics parameter $\partial t_f /\partial r$ has a
non-negligible  effect on $v_2 (p_T)$ (Fig. \ref{dt_dv} right).
In particular, by changing the freeze-out hypersurface towards the ``burning log'' ($\partial t_f/\partial r=-1$) limit, it is possible
to raise and lower the saturation scale and $p_T$ at which the saturation occurs.
Varying $\partial t_f/\partial r$ also controls the mass dependence of
the observed $v_2(p_T)$, with an increased $\partial t_f/\partial r$ 
leading to a different saturation scare for the $K^0$ and $\Lambda$, as observed
by experiment.
Comparing the right panels of Fig. \ref{dt_dv} left and right, it becomes
clear that a simultaneous fit for the $v_2 (p_T)$ of several different
particles would constrain both the flow anisotropy and the freeze-out
dynamics.

\begin{figure}
\centerline{\resizebox*{!}{0.26\textheight}
{
\includegraphics{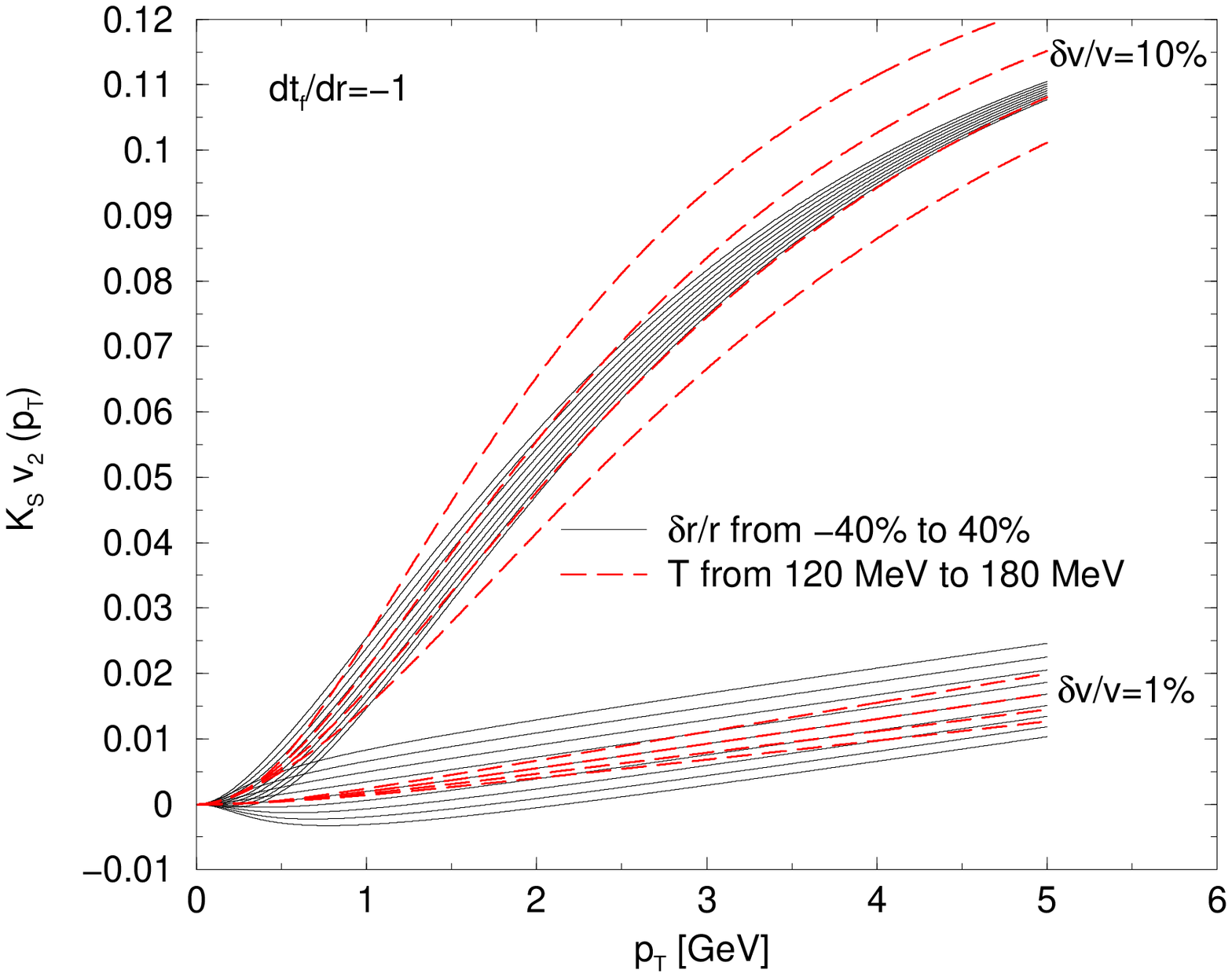}}
\resizebox*{!}{0.26\textheight}
{
\includegraphics{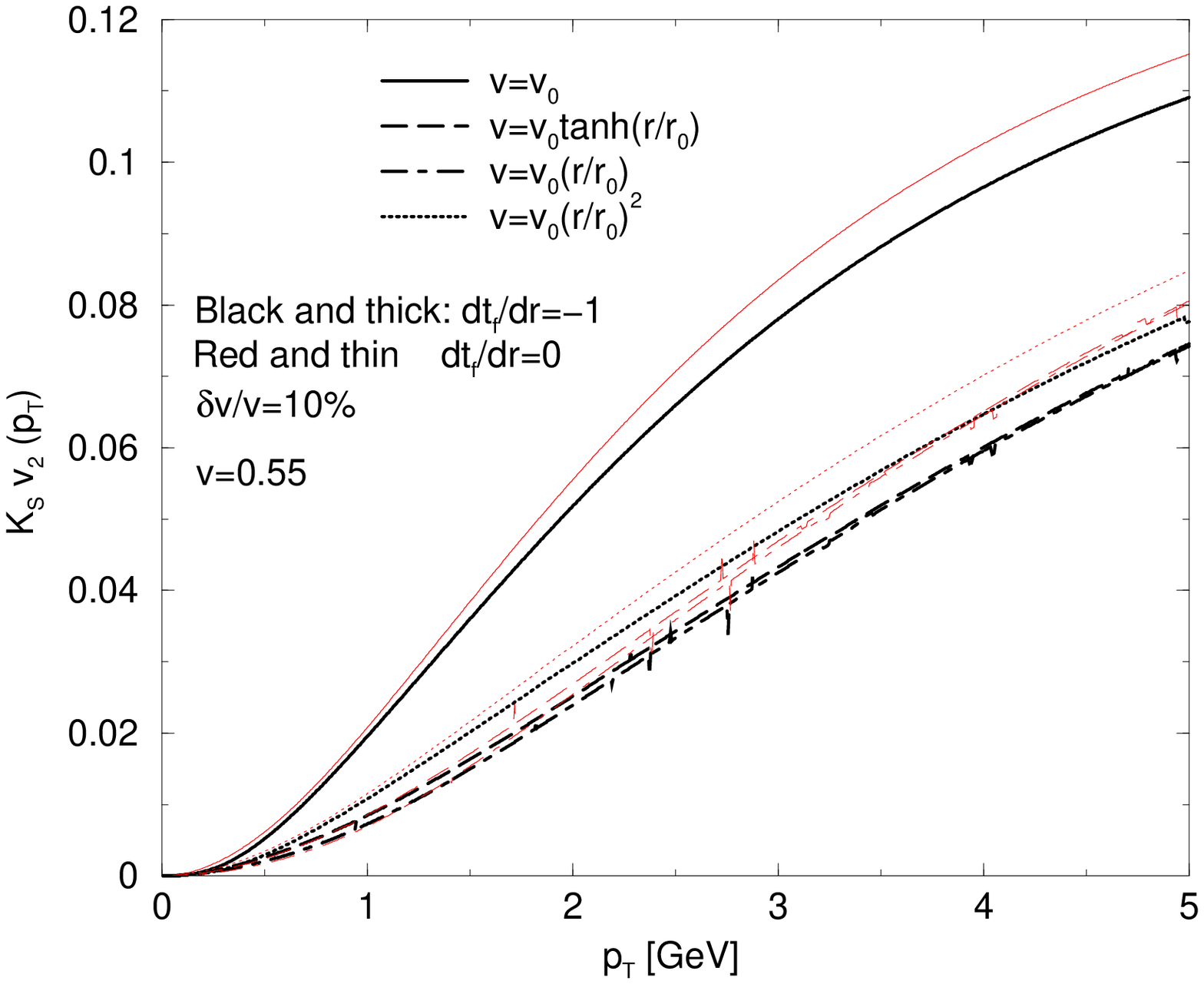}}}
\caption{ Left:  Varying geometrical anisotropy.  Right: Varying flow spread  \label{dr_Tv}.}
\end{figure}

Of course, a physical fireball is not characterized by a single flow, but
by a spatially varying flow profile.
Fig. \ref{dr_Tv} (right) explores the effect of a range of physically reasonable profiles on the observed $v_2$.
Unsurprisingly, adding a flow profile diminishes the observed $v_2$, while maintaining the saturation scale determined by $dv/v$ and $\partial t_f/\partial r$.

\begin{figure}
\centerline{\resizebox*{!}{0.26\textheight}
{
\includegraphics{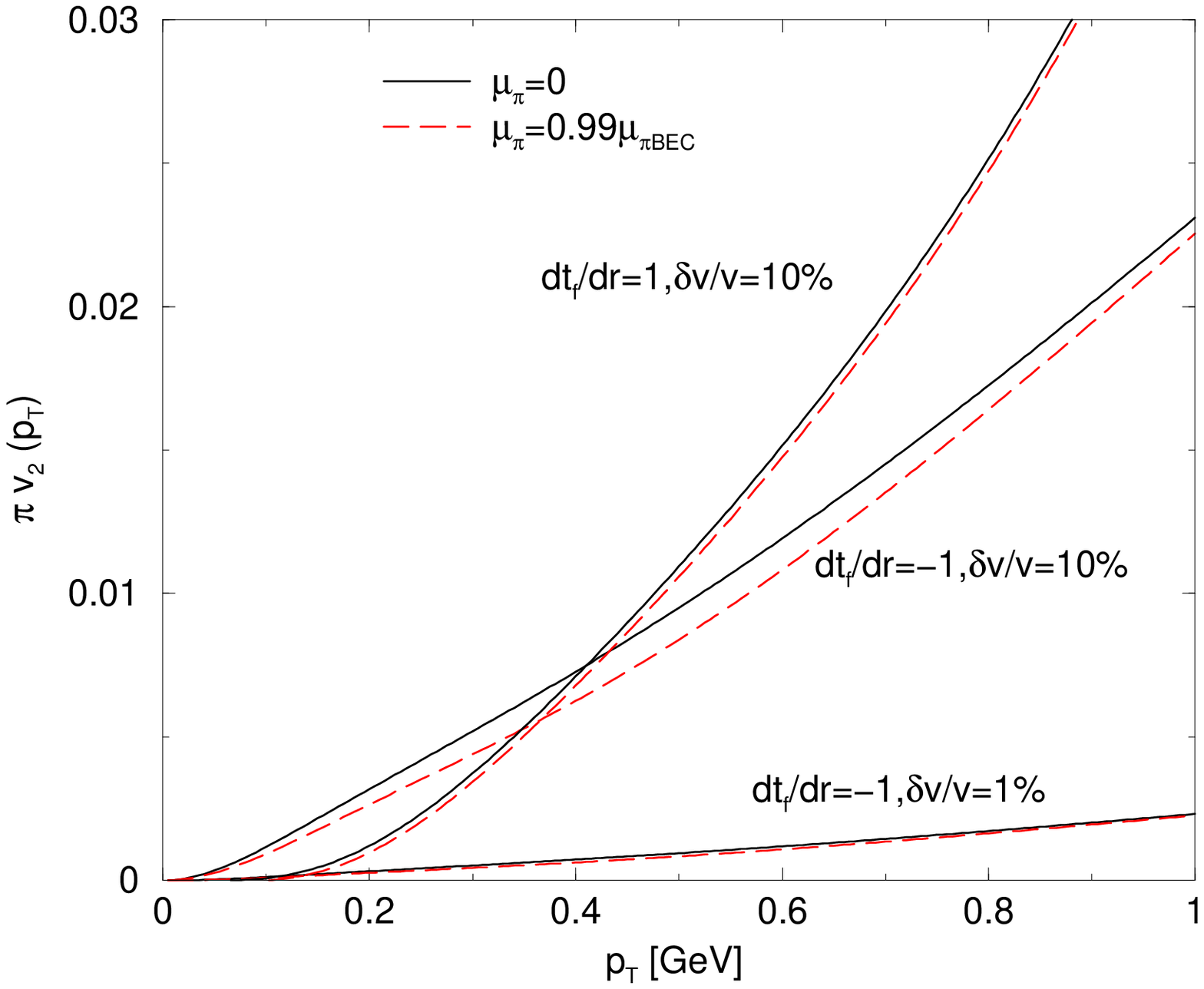}}
\resizebox*{!}{0.26\textheight}
{
\includegraphics{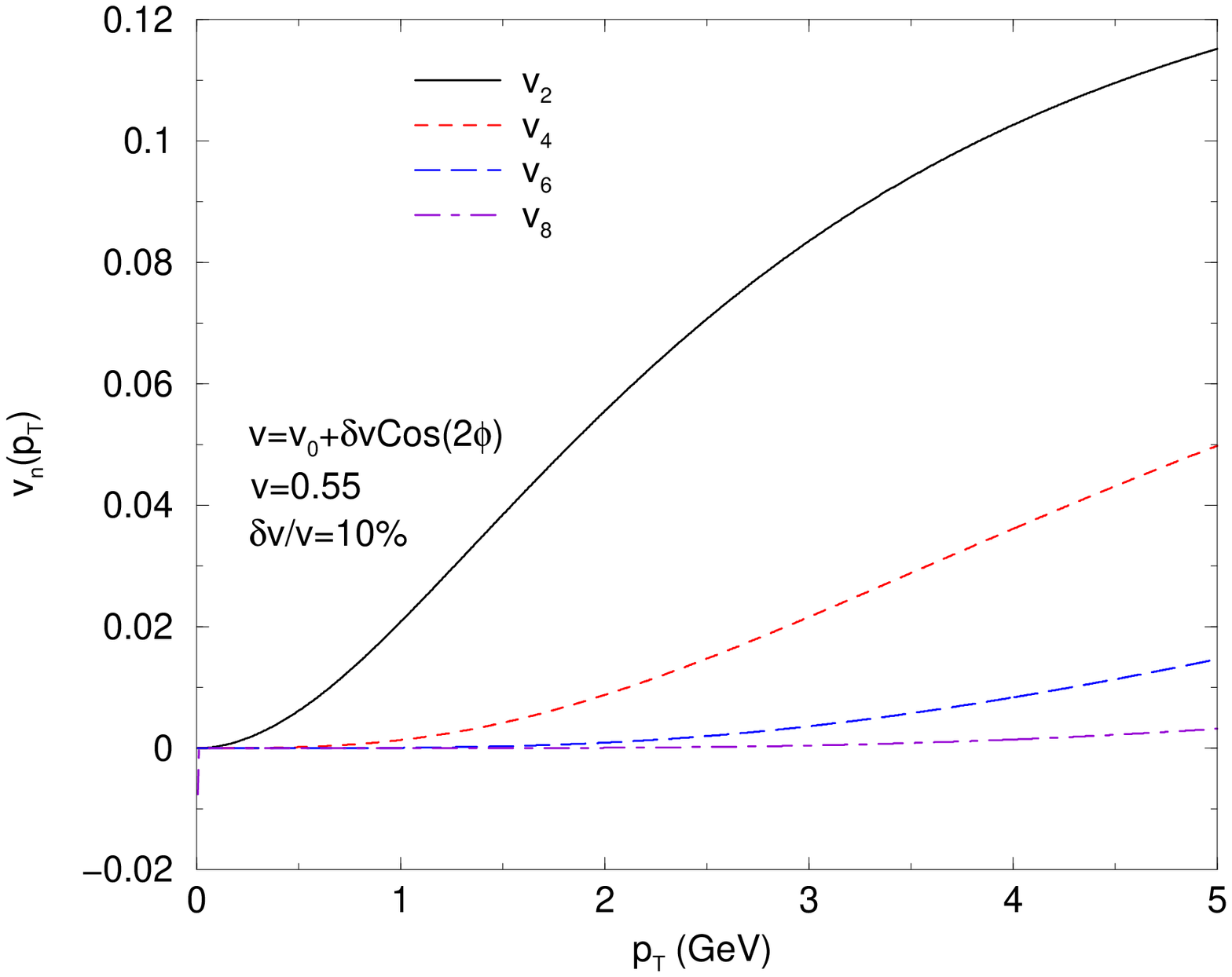}}}
\caption{ Left:  Varying flow profile.  Right: Higher $v_n$ s from an elliptic fireball \label{prof_vn}.}
\end{figure}

Finally, we note that corrections due to azimuthal anisotropy
of the freeze-out hypersurface and chemical non-equilibrium are also very small, even as the chemical potential approaches the Bose-Einstein condensation limit.(Fig.~\ref{dr_Tv} left, Fig.~\ref{prof_vn} left). 
Temperature corrections are more noticeable (Fig.~\ref{dr_Tv} left).

Eq.~(\ref{vn_explicit}) makes it apparent that the interference between space and flow anisotropy means that even a purely elliptic fireball will
acquire $v_{n>2} \ne 0$ for all even numbers
Fig.~\ref{prof_vn} (right) shows the contributions of $v_n$ arising from the freeze-out of a purely elliptical fireball.
It can be seen that, for $n=4$, this contribution is not negligible, and is in fact comparable to that arising from Hydro \cite{kolbv4}. 

This result, together with the recent experimental observation of directed flow \cite{expv4}, suggests that $v_3$ and $v_5$ away from central rapidity could also be promising observables:
Our model finds  $v_{3,5}=0$ since we assume boost-invariance.    However, a full 3-fluid calculation, together with Eq.~(\ref{vn_explicit}), would
yield these quantities easily.  
Given the significant observed directed flow \cite{expv4}, as well as the $v_4$ calculated in Fig.~\ref{prof_vn}, odd $v_{n>1}$ s could be within experimental
observation.
These coefficients make for particularly interesting probes, since they  arise solely from freeze-out effects and deviations from boost-invariance.

In conclusion, we have shown that observed $v_2 (p_T)$ depends strongly on the freeze-out hypersurface, and that a burning-log type freeze-out can explain
the observed $v_2 (p_T)$ saturation.
We have further shown that $v_n (p_T)$ is an extremely sensitive freeze-out probe, capable of constraining statistical freeze-out models very tightly.
As experimental data in this area becomes richer and more precise, more realistic numerical studies of both spectra and $v_n (p_T)$ might establish
themselves as crucial analysis tools.   

\setcounter{figure}{0}
\setcounter{equation}{0}
\setcounter{table}{0}
\chapter{Summary and outlook}
\label{cha:conclusion}
\section{What chapters 3-6 have shown}
The statistical hadronization model, together with hydrodynamic flow,
are capable of describing a large amount and variety of experimental data
with comparatively few fit parameters.
However, the variety of statistical models on ``the market'', which
manage to describe the data acceptably (to at least part of the heavy ion community) somewhat 
invalidates statistical hadronization as a physical picture.

A ``slow'' staged equilibrium freeze-out starting at T=170 $\mathrm{MeV}$ \cite{cern_evidence}, A sudden
equilibrium freeze-out at T=165 $\mathrm{MeV}$ \cite{florkowski}, and a fast 
explosive freeze-out starting at T=140 $\mathrm{MeV}$ \cite{Zak03} are not ``the same model''.
They are profoundly different physical pictures, which lead to different
conclusions about what the QGP-HG phase transition should look like.

That all three have been successfully employed to describe experimental
data simply means that two (or possibly all three!) of these pictures
are the heavy ion equivalent of epicycles:   Models with no connection
with reality, which are lucky enough to fit some of the experimental data.

In chapter 2, we have presented several candidates for a 
probe capable of distinguishing the epicycles from the real physics 
Chapters 3-6 examined each of the candidates in detail, in light of the experimental evidence.
To what extent have we succeeded at isolating the true physical picture?
\subsection{Does non-equilibrium sudden hadronization work?}

Sudden hadronization scenarios are considerably easier to falsify than a complicated
staged freeze-out phase.
In both the equilibrium and non-equilibrium scenarios, particle
yields place strong constraints on temperature and chemical potential.
Particle spectra place equally strong constraints on temperature and flow.
If the temperature which fixes the yields (``chemical freeze-out'') differs
from the temperature which fixes the spectra (``thermal freeze-out''), a fit to
normalized yields should conclusively fail.

That such a fit does not fail is, perhaps, the most firm conclusion to be drawn from this thesis.
In chapters 3 and 4, we have shown that fits based on the same temperature for yields and spectra
succeed , or at least are acceptable, both at SPS and RHIC energies, if the non-equilibrium
sudden hadronization ansatz is used.    

If, however, hadronization happens in equilibrium,
we find that the $\chi^2/\mathrm{DoF}$ increases beyond meaningfulness, a conclusion enforced by the lack
of definite minima in equilibrium $\chi^2$ profiles.        Fig.~\ref{resofit}
shows that the equilibrium scenario fails to describe short-lived resonance yields (which it does not
aim to describe anyways):   The scarcity of these resonances drives the fit to a temperature
at which its impossible to describe ratios such as $\Lambda/\Xi$ without introducing $\gamma_q$ and $\gamma_s$.

The checks performed in this thesis are not, however, the only ones possible.
In particular, in the following two years experiments should publish normalized spectra of unstable resonances.
These spectra pose much tighter constraints because such resonances have the same quark composition as
light particles.     Hence, their normalization, with respect to these light particles, is controlled
by temperature alone.

Resonances and absolute normalization remove the correlation between temperature and flow:  It is not possible
to adjust the slopes by shifting the temperature and transverse flow along the same contour, since in this
case normalization suffers.    However, temperature is still strongly correlated with chemical potentials
(especially $\gamma_s$ and $\gamma_q$).    Using spectra of short-lived resonances will remove this
correlation, and show once and for all if there is a meaningful temperature minimum.

It should be noted that spectra of $\phi$ and $K^*$ were already modeled at RHIC, and are well described
by the single freeze-out ansatz.   Much more, however, is yet to come.
\subsection{Have we ruled out staged hadronization?}
Ruling out Staged hadronization is considerably more difficult, since this model is not very quantitative beyond
the ``Two temperatures, one for chemical and one for thermal freeze-out'' requirement.

One can argue that a good description of both abundancies and spectra of particles with very different
interaction channels within the same temperature and flow is, by itself, a piece of evidence
against staged hadronization.
If staged hadronization occurs, why should the $\Omega$, with no s-channel interaction with the $\pi$, have
the same freeze-out temperature and flow as the $\pi$, which interacts via the s-channel with pretty much every
other particle in the hadron gas.

Why should the $\Lambda$ and $\overline{\Lambda}$, or  $\Xi$ and $\overline{\Xi}$ be described
by the same temperature and flow at the SPS, given that in a baryon-rich medium particles
and antiparticles have very different reaction channels ($\Lambda p \rightarrow \Lambda p$.  $\overline{\Lambda} p
\rightarrow  \pi K, ...$).
This is even more true in the case of spectra of short lived resonances:  As shown in Fig.~\ref{kstar}, microscopic models
predict a very $p_T$ dependent re-scattering of decay products, which should bring the spectra of short-lived resonances away from
hydro-dynamically predicted values.

However, this evidence is not really enough to rule out the model,  considering that, as proven in chapter
4, a fit for quite a large range of particles can be made with either a ``sudden freeze-out+resonances'' ansatz, or a ``no resonances
and arbitrary normalization'' description.  The  $\chi^2$ and the fitted temperature comes out remarkably similar in these two approaches.   

Direct detection has made clear that the presence of resonances in spectra is an experimental fact rather than a model assumption.
The analysis of resonance yields performed in chapter 5 points to a short interacting hadron gas phase, as well as a low chemical
freeze-out temperature.    Of course, proponents of staged hadronization would argue that the effect of out-of-equilibrium
regeneration would invalidate this conclusion \cite{bleicher}.

\subsection{Have we constrained freeze-out dynamics?}
Here, the answer must be ``not yet'':  Our fits to spectra have shown that freeze-out dynamics, parametrized for central and
near-central collisions by $\frac{\partial t_f}{\partial r}$, is an important factor in determining inverse slopes, and can be extracted
if enough particles are fitted.
We have presented attempts at such extraction at both SPS and RHIC energies.
A minimum in the $\chi^2$ profile was found, but it is not fully convincing.   One way of constraining this minimum further is through fitting more
low $p_T$ particles, since including the low $p_T$ region makes the fit considerably more sensitive to the variation of $ \frac{\partial t_f}{\partial r}$ 
(The phase space volume of the emitted hadron $\sim E - p_T \frac{\partial t_f}{\partial r}$).
PHOBOS recently lowered the observed $p_T$ range by two orders of magnitude \cite{PHOBOSlowpt}, and more particles (certainly $K$, hopefully
protons) will be coming.  

However, a deeper problem remains:   As we have shown, the choice of flow profile correlates with the choice of freeze-out dynamics.
This is not surprising, since both quantities are determined by the condition chosen for freeze-out, as can be seen by comparing
different hydrodynamic models \cite{hydro_shuryak,hydroheinz}.
However, it means that extracting conclusively $\Sigma^{\mu}$ from fits becomes considerably more computationally intensive, which makes
exploring other probes, more capable of extricating freeze-out dynamics separately from flow profile, necessary.

Some probes were proposed in chapters 5 and 6.   Resonances ratios as a function of $m_T$ are a promising analysis tool, but experimental results
have yet to be presented.   $v_2$, described in chapter 7, is also a promising probe, but once again resonance contributions are needed to
describe existing experimental data \cite{v2_reso}.

Finally, there is the traditional probe for exploring freeze-out in configuration space: Hanbury-Brown-Twiss interferometry (HBT) \cite{HBTreview}.   Since this thesis did not include HBT calculations, we have not discussed the current state of affairs in any detail beyond the problems faced by statistical models in fitting HBT
\cite{HBTpuzzle}.   The solution to these problems might be tied to sudden freeze-out \cite{sudden3}, or to the fact that assumptions going into
HBT models might be, for many reasons, overly simplistic \cite{HBTrqmd,HBTreso,HBTint}.
\section{Outlook on the heavy ion experimental program}
This thesis was written well before all of the RHIC data has been analyzed.
Indeed, the most interesting and surprising experimental results might well still be in the future.
As noted elsewhere, while the resonance program has collected a convincing signal for a huge array
of particles (with more to come), detailed quantitative
observations, such as yields, ratios and spectra for all detected resonances and exotica, are yet to be published.

Charm and bottom production is another topic which awaits most of its experimental data.
Heavy quarks will probably never be chemically equilibrated at any energy, and the great majority of their yields
will always be produced during the initial high $Q^2$ collisions \cite{charm3,thews2}.
However, a quark gluon plasma will equilibrate them thermally, and hadronization will distribute them in a
radically different way than microscopic hadronic processes.
For sudden freeze-out to be consistent, it should describe quarks independently of their mass.
This means that charmed particles ($\psi,\psi',D$ etc), beauty states and mixtures ($B_c$) should be produced according to their statistical weights, and their
momentum distribution should be modeled by the same flow as that of light quarks.
We shall see if this is really the case, as the predictions of statistical models can differ drastically
($\sim$ order of magnitude) from the purely hadronic ones \cite{thews2}.

Beyond the current RHIC program, we can look forward to the TeV energy range accessible to the LHC.
While hadronization conditions will probably not be that different from RHIC (the chemical potentials
at mid-rapidity should both be close to zero), the initial conditions and equilibrated evolution are likely to be
markedly different.    In particular, initial collisions are likely to generate a number of heavy quarks
comparable to the strange quarks created at SPS and RHIC.
If the system thermalizes as rapidly as the RHIC system seems to be, its temperature during the initial
stage of the hydrodynamic evolution is likely to be well in the perturbative QGP range.   Perhaps, if
this phase lasts a long time, we will also see interesting electroweak effects (another phase transition
accessible to experimental study?).    At present it is not clear what the effect of this initial evolution will be, and
whether any hadronic signals will survive subsequent cooling.

The most interesting physics, however, might not come so much from pushing the energy boundary higher and higher, as from
comparing as many different collision energies, system sizes and chemical conditions as possible.
The SIS, AGS and SPS have explored conditions in a wide range of nuclei at energies ranging from 1 to 20 $\mathrm{GeV}$.
RHIC has pushed this boundary to the hundreds of $\mathrm{GeV}$, the LHC will push it into the TeV range, and
the future SIS facility will look for low energy but high-density systems.
Taking all these experiments together results in a large region in temperature, chemical potential and system
size which needs to be systematically analyzed.
At what point do qualitatively new features (eg equilibration, or $\gamma_{q,s} \ne 1$ ) come in?
What is the critical system size or collision energy needed for ``most of the system'' to become thermalized?
Where exactly is the phase transition boundary, and are there any critical points?    

We do not yet know the answer, but can say convincingly that soft physics will play an important role in obtaining it.
As remarked in chapter 1, hard probes (such as jet quenching) are useful to tell us some bulk properties
of the matter produced in highly energetic heavy ion collisions.
If, however, one wants to explore how these bulk properties change with energy, or even relate
the measured values to thermodynamics, hard processes, on their own become much less useful.
It is only soft physics which will be able to determine at what point a qualitative change in microscopic
degrees of freedom has occurred.
To accomplish this, however, requires more than the new experimental data.
Statistical hadronization itself needs to be understood better, both at the phenomenological
and fundamental level.
\section{Outlook on statistical hadronization}
As the previous sections have shown, there is a lot more data which need to be analyzed in the statistical
model framework to be able to tell convincingly which statistical model reflects physics.
Fortunately, as the previous chapters have mentioned, the soft physics data taking at both SPS and RHIC is
far from over.    The range of particles analyzed in this theses, and many more, need to be systematically
examined at the large range of collision energies and system size accessible to modern day and future experiments.

For this reason, the immediate next step as far as the work described in this thesis is concerned is the development
of a statistical hadronization code to be made available open-source to experimental collaborations.
This will provide experimentalists with an open and universal ``standard statistical
model'' to which to compare all their data, as well as a guarantee against errors (which, as we found, are very easy
to make, given the thousands of particles and decay modes which need to be calculated).
We have been developing such a code (named SHARE) together with the Krakow group \cite{florkowski}.
The first part of SHARE, which calculates $4 \pi$ yields, has recently
been released \cite{share}.   Further modules (covering spectra, $v_2$ and HBT) are currently in development.

Another possible direction for statistical hadronization research is investigating its mathematical
consistency, and improving its formalism.
At present, there is no indication that either sudden freeze-out or staged freeze-out are self-consistent.
Would inelastic interactions of a hadron gas really freeze-out at T=170 $\mathrm{MeV}$, as staged freeze-out assumes?
Would elastic interactions of an explosively hadronizing plasma at T=140 $\mathrm{MeV}$ really be negligible?
If the answer to both these questions is no, to what extent does departure from equilibration
impact observables?  What is the quantitative systematic error such effects give to fitted thermal
and flow parameters?

One way to answer these questions is to feed statistical model output into microscopic models, such as uRQMD.
At present, uRQMD is incapable of fully describing any of the RHIC observables (strange particle yields,
collective flow, and resonance production).   However, this does not preclude statistical production
followed by hydrodynamic evolution.   The Monte-Carlo module described in chapter 3 is being developed and
incorporated into SHARE, as a ``plug'' for uRQMD.   Hopefully, this will lead to some of the above questions
being answered.

On a more fundamental level, it is clear that the statistical hadronization formalism needs developing.
Reducing all particles to Lorentz-scalar objects emitted from a 3D locus in space, as the Cooper-Frye
formula does, is a rough approximation which might miss important physical features of the hadronizing system.
In particular, interaction of the two phases of matter at hadronization might have non-negligible consequences
\cite{bugaev}.    Enforcing conservation laws and entropy non-decrease locally on each hadronizing volume element
might lead to non-trivial changes of the statistical parameters across the phase transition.
$\gamma_q>1$, discussed extensively in this dissertation, is
a simple example.  It might not be the only parameter which changes significantly as the system
undergoes a rapid phase transition.  (other workers have proposed a change in T \cite{sudden4}).

More ambitiously, the above questions might be answered by tying the statistical model 
to a more fundamental picture of hadronization, possibly involving a microscopic effective theory. 
Perhaps insights in this direction might be obtained by examining at what energy range ($p_T$ and
rapidity) the statistical model ceases to provide a good physical description of the data, and what takes
its place.
Recently, a lot of interest has been aroused by the fact that several intermediate energy observables
($p_T\sim 1 \mathrm{GeV}$, below the scale described by pQCD) can be described in terms of parton coalescence
\cite{coalescence,molnar1}.  While this model's ability to describe the data is compelling, its failings
on a fundamental level (energy-momentum conservation and entropy non-decrease) mean that is a limiting case of
a larger process.   In particular, it should be possible to combine coalescence and statistical
hadronization as two limits of a single framework (this is especially true for non-equilibrium
statistical hadronization:  After all, it can be argued that statistical hadronization
with $\gamma_{q,s} \ne 1$ is another way of viewing parton coalescence into massive states, with probabilities given
by thermal weights).

It should be reiterated, however, that these advances in theoretical understanding are
very unlikely unless the right statistical hadronization model is experimentally selected.
Max Planck found that black body radiation fits a $\frac{1}{e^{\frac{\omega}{T}}+1}$ distribution years before
he found a fundamental description of this phenomenon.
Similarly, one hopes our rough phenomenological fits will help elucidate one of the greatest unsolved mysteries in modern physics:  The structure of that strange, complex and unpredictable medium which we have naively been calling empty space.

\appendix


\setcounter{figure}{0}
\setcounter{equation}{0}
\setcounter{table}{0}
\chapter{Coordinates and units}
\label{cha:appendix_A}
Throughout this dissertation we use the so called ``natural'' system of
units, in which
\begin{equation} \hbar=c=k_B=1 \end{equation}
Given a unit of energy, the conversion $(\hbar c)$ will make it an inverse
of the distance.   Thus,
\begin{equation}  \hbar c = 197  \mathrm{MeV} \mathrm{fm}, \;  \;  1 GeV = 5.4 \mathrm{fm}^{-1}\end{equation}
When dealing with a collision at ultra-relativistic energies, it
is convenient to split the momentum into components parallel and perpendicular to the collision and express the parallel part as an energy-momentum
rapidity
\begin{equation}  y=\frac{1}{2}\ln \left(\frac{E+p_L}{E-p_L}\right)  \end{equation}
it is easy to show that y transforms linearly under a Lorentz boost
\begin{equation}  y \rightarrow y+y_0 \end{equation}
it is also easy to show that 
\begin{equation} E = m_T \cosh(y)  \;  \; \;  p_L = m_T \sinh(y) \end{equation}
where we defined the ``transverse mass'' $m_T=\sqrt{p_T^2+m^2}$. 
Hence, a particle's 4-momentum can be parametrized as
\begin{equation}
p^{\mu} = \left( \begin{array}{c} m_T \cosh(y) \\ p_T \cos(\phi) \\ p_T \sin(\phi) \\ m_T \sinh(y)  \end{array} \right)\end{equation}
Experimentally measuring rapidity can be problematic, as it requires particle identification.
Hence, many experiments measure pseudo-rapidity
\begin{equation} \eta = \frac{1}{2}\ln \left(\frac{p+p_L}{p-p_L}\right) = 
-\ln \left[ 
\tan \left( \frac{\theta}{2} \right)\right]  \end{equation}
where $\theta = \tan(p_T/p_L)$.
in the ultra-relativistic limit $m<<p \rightarrow E$
\begin{equation}
y \approx \eta-\frac{m^2 }{2 p} \frac{2 p_L}{p^2 -p_L^2}+... 
\end{equation}
so the approximation works much better in the central rapidity ($p_L << p$) region.

\setcounter{figure}{0}
\setcounter{equation}{0}
\setcounter{table}{0}
\chapter{Fitting}
\label{cha:fitting}
Suppose we have a theory which relates variables
y and x through a parameter $\alpha$
\begin{equation}
\label{function}
  y = f(x,\alpha)  
\end{equation}
 and a set of uncorrelated experimental data points  $(y_i,x_i)$ with an experimental error $\sigma_i$.
The probability that the ``true'' value y gives the experimental result $y_i$ is given by some probability density function
$P(y,y_i,\sigma_i)$.
Hence, the probability that N data points $y$  were measured as $y_i$ is
given by the likelihood function
\begin{equation}
P_{tot} = \prod_{i=1}^{N} P\left( f(x_i,\alpha),y_i,\sigma_i \right)
\end{equation}
by maximizing $P_{tot}$, or (for mathematical simplicity)
\begin{equation}
\log(P_{tot})=  \sum_{i=1}^{N} \log \left( P\left( f(x_i,\alpha),y_i,\sigma_i \right)  \right)
\end{equation}
we can find an estimate of the parameter $\alpha$.

In particular, if the error function is a Gaussian
\begin{equation}
P(x,x_i,\sigma_i) = \frac{1}{\sqrt{2 \pi \sigma_i}} e^{-\frac{(x-x_i)^2}{2 \sigma_i^2}}
\end{equation}
the Maximum likelihood corresponds to the minimum $\chi^2$ where
\begin{equation}
\chi^2 = \sum_{i=1}^{N} \left( \frac{ f(x_i,\alpha)-y_i}{\sigma_i} \right)^2 
\end{equation}
By Taylor-expanding $\chi^2$ in the region of the minimum
\begin{equation}
\chi^2 \approx \chi^2_{min} + \frac{d^2 \chi^2}{d^2 \alpha} (\alpha - \alpha_{min})^2 = 
\end{equation}
and remembering that $P_{tot} \sim e^{\chi^2}$ the  error on the parameter $\alpha$
can be found by 
\begin{equation}
\Delta \alpha= \frac{2}{\sqrt{\frac{d^2 \chi^2}{d^2 \alpha}}}
\end{equation}
This procedure can be generalized from one parameter $\alpha$ to a
set of parameters $\alpha_i$ (In this thesis, minimizations with 13 parameters are performed).
The error then becomes the determinant of a matrix of correlation coefficients
\begin{equation}
\sigma_{\alpha_{i,j}}= \frac{2}{\sqrt{\left[ \frac{\partial^2 \chi^2}{\partial \alpha_{i} \partial \alpha_j}\right]^{-1}}}
\end{equation}
However, in this case the structure of the minimum in parameter space becomes non-trivial, with contours, plateaus,
saddle points etc. making finding of the ``true minimum'' potentially very hard.     
In this thesis, we rely on the MINUIT numerical minimization package \cite{minuit} to minimize
$\chi^2$, but we test the dependence on the parameters by explicitly calculating
$\chi^2$ profiles ($\chi^2$ as a function of one parameter) and contours ($\chi^2$ as a function of two parameters).

\begin{figure}
\centering
\label{conf_lev}
\epsfig{figure=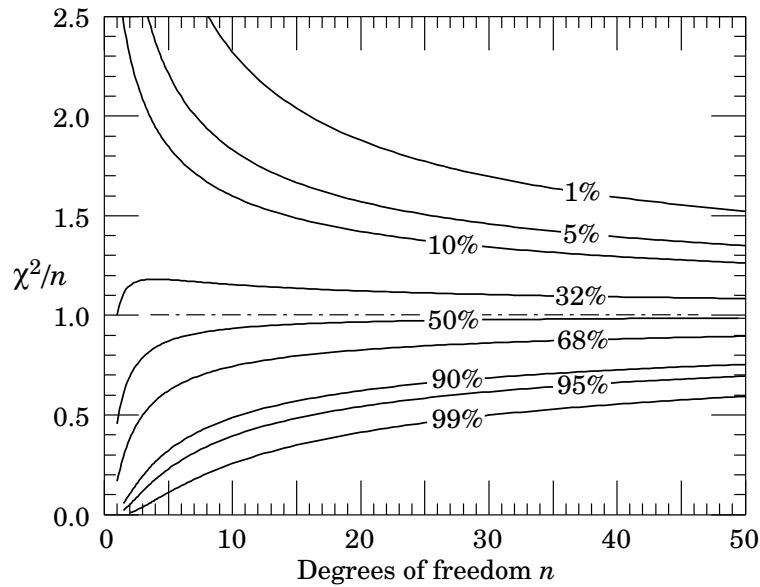}
\caption{Confidence levels as a function of degrees of freedom.}
\end{figure}

The $\chi^2$ fit is also a test of how well does the theory  $y = f(x,\alpha)$ describe the
experimental data.  After $\alpha$ is found, the minimized $\chi^2$ can be used to obtain the
``confidence level'' of the fit, ie the probability, given $f(x,y,\alpha_{min})$, that the total 
$\chi^2$ obtained in an experiment is of that value or higher.
\begin{equation}
CL(\chi^2) = \int_{\chi^2}^{\infty} \prod_{i}^{n} P(x,x_i,\sigma_i) \approx \int_{\chi^2}^{\infty} e^{- z^2/2} dz
\end{equation}
the quantitative dependence of $CL(\chi^2)$ as a function of $\chi^2$ and the degrees of freedom is
shown in Fig.~\ref{conf_lev} \cite{pdg}.
As can be seen, the confidence levels tend to an asymptotic limit as the degrees of freedom increase, corresponding
to an expectation of $\chi^2 \simeq \mathrm{DoF}$.  This limit is valid in the fits to particle spectra performed in chapters
3 and 4.
However, if the degrees of freedom in the fit are limited (as in the fits to particle yields discussed in chapter 2), the confidence
level depends strongly on the number of degrees of freedom, and a $\chi^2/\mathrm{DoF}$ considerably less than 1 is required for the fit to be statistically significant. 

It should be remembered that the confidence level plot is only significant when
all errors are Gaussian.  This is generally not true if large systematic errors are present.
In this case, when these errors can be estimated precisely they should be taken into account
by summing systematic and statistical errors
\begin{equation}
\label{chisys}
\chi^2 = \sum_{i=1}^{N} \left( \frac{ f(x_i,\alpha)-y_i}{\sigma_i^{statistical}+\sigma_i^{systematic}} \right)^2. 
\end{equation}
If, however, this can not be done statistical stops being a quantitative measure of the fit's validity.
In chapter 3 we encounter precisely this problem with $K_s$.

\setcounter{figure}{0}
\setcounter{equation}{0}
\setcounter{table}{0}
\chapter{Relativistic phase space}

The relativistic phase space for a physical system is the volume in 4-momentum space corresponding to the system's invariant mass (which we call m.  For many-particle systems, it corresponds to $\sqrt{s}$)
\begin{equation}
\label{phasespace}
\Omega= \int d^4 p \delta(p_{\mu} p^{\mu} - m^2) = \int d^3 p \int_0^{\infty} d E \delta \left({\sqrt{E^2-p^2-m^2}}\right)
\end{equation}
This integral can be simplified keeping in mind that
\[\  \delta\left(f(x)\right) = \frac{1}{\left[\frac{df}{dx}\right]}_{x=x_0} \delta(x-x_0)  \]
differentiating the delta-function in Eq.(~\ref{phasespace}) by E it can be seen, immediately, that
\begin{equation}
\label{phasespace}
\Omega= \int d^3 p \int_0^{\infty} \frac{d E}{2 E} \delta\left(E-\sqrt{p^2-m^2}\right) = \int \frac{d^3 p}{2 E} \delta\left( E-\sqrt{p^2-m^2}\right)
\end{equation}
It can be seen that this expression is relativistically invariant by Lorentz-transforming a volume of phase space
explicitly  (without loss of generality, in the z direction)
\[\  dp'_x dp'_y dp'_z  = dp_x dp_y \gamma (d p_z - v dE) = dp_x dp_y dp_z  \gamma \left( 1-v \frac{dE}{d p_z} \right)  \]
Since 
\[\ \frac{dE}{d p_z} =\frac{d}{d p_z} \sqrt{m^2 + p_z^2} = \frac{p_z}{E}   \]
we get
\[\ dp_x dp_y dp_z \left( 1-v \frac{dE}{d p_z} \right) =   dp_x dp_y dp_z \gamma \left( 1-v \frac{p_z}{E} \right) = d p_x dp_y dp_z \frac{E'}{E} \]
Hence, we have proven that
\[\ \frac{dp'_x dp'_y dp'_z}{E'} = \frac{dp_x dp_y dp_z}{E}   \]
as required.

We have therefore derived the relativistic phase space for a particle of momentum p, energy E and mass m.
For a system of N particles with momentum p and energy E, 
such as those discussed in Chapter 2, it is easy to generalize our result to
\[\ \frac{d^3 p}{E} \rightarrow \int \prod_{i=1}^{N}   \frac{d^3 p_i}{E_i} \delta \left( \sum_{i=1}^{N} p_i -p \right) \delta \left(  \sum_{i=1}^{N} E_i -E \right)  \]

\setcounter{figure}{0}
\setcounter{equation}{0}
\setcounter{table}{0}
\chapter{Publications}
Several parts of this thesis have been published in peer-Reviewed Journals.
We shall give a listing of publications contained in this thesis, together with the chapter
which follows the publication most closely.

\begin{itemize}
\item \cite{torrieri_mc,torrieri_sps1,torrieri_sps2} \textbf{Chapter 3}\\
\textit{Paper \cite{torrieri_sps1} was chosen for the IoP select article collection. See http://www.iop.org/Select/}
\item \cite{torrieri_reso1,torrieri_reso2,torrieri_reso3,torrieri_penta} \textbf{Chapter 4}
\item \cite{torrieri_v2} \textbf{Chapter 5}
\item Part of chapter 2 (the inclusion of width in particle yields) \cite{share}has recently been submitted to Communications in Physical Computing.
\item Chapter 6 \cite{torrieri_v2} is part of a continued research effort which
will eventually be published.
\end{itemize}

\setcounter{figure}{0}
\setcounter{equation}{0}
\setcounter{table}{0}
\chapter{Further acknowledgments}

This section can only start with Dr Johann Rafelski, my advisor.
Through four hard years, he did his best to make me into a research scientist.
I am humbly grateful to him, and sincerely hope to make the best of his teachings.

This dissertation, in its current form, would also have been impossible without
the generous help and support given by my committee.   Therefore, I would like to thank
Drs Keith Dienes, Mike Shupe, Robert Thews and Ubirajara van Kolck for
the time taken to read the dissertation, as well as their assistance and advice.
I also thank Jean Letessier for his extensive assistance all throughout these four years, from
computational problems to his help with this dissertation.

Part of the research summarized in this dissertation has been performed during
my stay in Krakow in the summer of 2003.  I therefore thank Dr Wojciech Broniowski and Wojciech Florkowski of the Niewodniczanski Nuclear Physics
Institute for their hospitality and collaboration.    

As well as the grant-giving bodies mentioned earlier, I would also like
to thank the organizers of Strangeness 2000, 2001 and 2003, as well as
Quark Matter 2001, 2002, 2004, for their travel support which enabled
me to attend these conferences.    Support from the Pan American Advanced
Studies Institute (PASI), as well as the Nuclear Physics Summer School (NPSS) is also greatly appreciated.

The interaction with experimentalists throughout my work is also greatly appreciated.
In particular, I thank Sevil Salur for taking an interest in my work, keeping me continuously on my toes with her questions, and, in general, for her friendship.
I also greatly benefited by the interaction I had with Christina Markert and Patricia Fachini.

This department has been the most socially cohesive and friendly physics department I have ever come across.  
I would like to thank everyone who contributes in maintaining this, and urge future students to keep it up.
Of course, the physicists and honorary physicists 
of the LSS crowd deserve a mention here.  Mike, Mikey, Anne, Tommy, Gabe, Torsten, James, Jessica, Hermann, David, Delphi, John and many others have made the ride here infinitely more
bearable. 

Despite what many physicists tell you, life does not end at the exit of the physics department.  I wish to credit the wonderful people who assisted me in life
outside of physics for the last four and a half years.    I sincerely hope I learned something from them.
Thank you Celeste, Carrie, Sarmad, Baha, Nesreen, Paul, Rachel, Hanna, Will, Czarina, Jessica, Allen, Pietro, Susan and many many others.

I should also acknowledge that I am lucky enough to be able to devote 28 years
of my life to full-time education, and make a living
by satisfying my intellectual curiosity in subjects unrelated to 
everyday survival.
This is a privilege which only a small subset of humanity, in space and time,
can enjoy.    I hope some of my activities, at the University of Arizona
and elsewhere, have contributed to increase that subset.

Finally, the most important acknowledgment of all:  This dissertation brought me to the other side of the world from my wonderful parents, who raised me, taught me 
curiosity and determination, and had to endure the painful separation so I could achieve my dreams.  

\noindent Thank you all.\\
I could not have done it without you.


\end{document}